\documentclass[amsmath,amssymb,aps,showpacs,onecolumn,superscriptaddress,letterpaper,longbibliography]{revtex4-2}

\usepackage{graphicx}
\usepackage{dcolumn}
\usepackage{bm}
\usepackage{hyperref}
\usepackage{lineno}
\linenumbers\relax 
\usepackage{natbib} 
\usepackage{amsmath, amssymb} 
\usepackage{color}
\usepackage[dvipsnames]{xcolor}
\usepackage{comment} 
\usepackage{hyperref} 
\usepackage{float} 
\usepackage{makecell} 
\usepackage{cleveref}
\usepackage{tikz}

\hypersetup{breaklinks = true, colorlinks = true, citecolor = blue, linkcolor = blue, urlcolor = blue}

\begin{document}

\nolinenumbers
\title{Measurement of nuclear effects in neutrino-argon interactions using generalized kinematic imbalance variables with the MicroBooNE detector}

\newcommand{\ANL}{Argonne National Laboratory (ANL), Lemont, IL, 60439, USA}
\newcommand{\Bern}{Universit{\"a}t Bern, Bern CH-3012, Switzerland}
\newcommand{\BNL}{Brookhaven National Laboratory (BNL), Upton, NY, 11973, USA}
\newcommand{\UCSB}{University of California, Santa Barbara, CA, 93106, USA}
\newcommand{\Cambridge}{University of Cambridge, Cambridge CB3 0HE, United Kingdom}
\newcommand{\CIEMAT}{Centro de Investigaciones Energ\'{e}ticas, Medioambientales y Tecnol\'{o}gicas (CIEMAT), Madrid E-28040, Spain}
\newcommand{\Chicago}{University of Chicago, Chicago, IL, 60637, USA}
\newcommand{\Cincinnati}{University of Cincinnati, Cincinnati, OH, 45221, USA}
\newcommand{\CSU}{Colorado State University, Fort Collins, CO, 80523, USA}
\newcommand{\Columbia}{Columbia University, New York, NY, 10027, USA}
\newcommand{\Edinburgh}{University of Edinburgh, Edinburgh EH9 3FD, United Kingdom}
\newcommand{\FNAL}{Fermi National Accelerator Laboratory (FNAL), Batavia, IL 60510, USA}
\newcommand{\Granada}{Universidad de Granada, Granada E-18071, Spain}
\newcommand{\Harvard}{Harvard University, Cambridge, MA 02138, USA}
\newcommand{\IIT}{Illinois Institute of Technology (IIT), Chicago, IL 60616, USA}
\newcommand{\Indiana}{Indiana University, Bloomington, IN 47405, USA}
\newcommand{\KSU}{Kansas State University (KSU), Manhattan, KS, 66506, USA}
\newcommand{\Lancaster}{Lancaster University, Lancaster LA1 4YW, United Kingdom}
\newcommand{\LANL}{Los Alamos National Laboratory (LANL), Los Alamos, NM, 87545, USA}
\newcommand{\Louisiana}{Louisiana State University, Baton Rouge, LA, 70803, USA}
\newcommand{\Manchester}{The University of Manchester, Manchester M13 9PL, United Kingdom}
\newcommand{\MIT}{Massachusetts Institute of Technology (MIT), Cambridge, MA, 02139, USA}
\newcommand{\Michigan}{University of Michigan, Ann Arbor, MI, 48109, USA}
\newcommand{\MSU}{Michigan State University, East Lansing, MI 48824, USA}
\newcommand{\Minnesota}{University of Minnesota, Minneapolis, MN, 55455, USA}
\newcommand{\Nankai}{Nankai University, Nankai District, Tianjin 300071, China}
\newcommand{\NMSU}{New Mexico State University (NMSU), Las Cruces, NM, 88003, USA}
\newcommand{\Oxford}{University of Oxford, Oxford OX1 3RH, United Kingdom}
\newcommand{\Pitt}{University of Pittsburgh, Pittsburgh, PA, 15260, USA}
\newcommand{\Rutgers}{Rutgers University, Piscataway, NJ, 08854, USA}
\newcommand{\SLAC}{SLAC National Accelerator Laboratory, Menlo Park, CA, 94025, USA}
\newcommand{\SDSMT}{South Dakota School of Mines and Technology (SDSMT), Rapid City, SD, 57701, USA}
\newcommand{\Maine}{University of Southern Maine, Portland, ME, 04104, USA}
\newcommand{\Syracuse}{Syracuse University, Syracuse, NY, 13244, USA}
\newcommand{\TelAviv}{Tel Aviv University, Tel Aviv, Israel, 69978}
\newcommand{\Tennessee}{University of Tennessee, Knoxville, TN, 37996, USA}
\newcommand{\UTA}{University of Texas, Arlington, TX, 76019, USA}
\newcommand{\Tufts}{Tufts University, Medford, MA, 02155, USA}
\newcommand{\UCL}{University College London, London WC1E 6BT, United Kingdom}
\newcommand{\VTech}{Center for Neutrino Physics, Virginia Tech, Blacksburg, VA, 24061, USA}
\newcommand{\Warwick}{University of Warwick, Coventry CV4 7AL, United Kingdom}
\newcommand{\Yale}{Wright Laboratory, Department of Physics, Yale University, New Haven, CT, 06520, USA}

\affiliation{\ANL}
\affiliation{\Bern}
\affiliation{\BNL}
\affiliation{\UCSB}
\affiliation{\Cambridge}
\affiliation{\CIEMAT}
\affiliation{\Chicago}
\affiliation{\Cincinnati}
\affiliation{\CSU}
\affiliation{\Columbia}
\affiliation{\Edinburgh}
\affiliation{\FNAL}
\affiliation{\Granada}
\affiliation{\Harvard}
\affiliation{\IIT}
\affiliation{\Indiana}
\affiliation{\KSU}
\affiliation{\Lancaster}
\affiliation{\LANL}
\affiliation{\Louisiana}
\affiliation{\Manchester}
\affiliation{\MIT}
\affiliation{\Michigan}
\affiliation{\MSU}
\affiliation{\Minnesota}
\affiliation{\Nankai}
\affiliation{\NMSU}
\affiliation{\Oxford}
\affiliation{\Pitt}
\affiliation{\Rutgers}
\affiliation{\SLAC}
\affiliation{\SDSMT}
\affiliation{\Maine}
\affiliation{\Syracuse}
\affiliation{\TelAviv}
\affiliation{\Tennessee}
\affiliation{\UTA}
\affiliation{\Tufts}
\affiliation{\UCL}
\affiliation{\VTech}
\affiliation{\Warwick}
\affiliation{\Yale}

\author{P.~Abratenko} \affiliation{\Tufts}
\author{O.~Alterkait} \affiliation{\Tufts}
\author{D.~Andrade~Aldana} \affiliation{\IIT}
\author{L.~Arellano} \affiliation{\Manchester}
\author{J.~Asaadi} \affiliation{\UTA}
\author{A.~Ashkenazi}\affiliation{\TelAviv}
\author{S.~Balasubramanian}\affiliation{\FNAL}
\author{B.~Baller} \affiliation{\FNAL}
\author{G.~Barr} \affiliation{\Oxford}
\author{D.~Barrow} \affiliation{\Oxford}
% ...

\collaboration{The MicroBooNE Collaboration}
\thanks{microboone\_info@fnal.gov}\noaffiliation


\begin{abstract}
\noindent

We present a set of new generalized kinematic imbalance variables that can be measured in neutrino scattering.
These variables extend previous measurements of kinematic imbalance on the transverse plane, and are more sensitive to modeling of nuclear effects.
We demonstrate the enhanced power of these variables using simulation, and then use the MicroBooNE detector to measure them for the first time.
We report flux-integrated single- and  double-differential measurements of charged-current muon neutrino scattering on argon using a topolgy with one muon and one proton in the final state as a function of these novel kinematic imbalance variables.
These measurements allow us to demonstrate that the treatment of charged current quasielastic interactions in GENIE version 2 is inadequate to describe data.
Further, they reveal tensions with more modern generator predictions particularly in regions of phase space where final state interactions are important.
\end{abstract}

\maketitle


\section{Introduction}

All current and upcoming accelerator neutrino oscillation experiments rely on the precise modeling of neutrino-nucleus interactions to perform high-accuracy measurements~\cite{pdg2018,T2KNature20,DUNE1:2016oaz,DUNE2:2016oaz,DUNE3:2016oaz,HK}.
The experimental sensitivity of these measurements can be limited by interaction modeling uncertainties related to nuclear effects~\cite{NAGU2020114888,dieminger2023uncertainties}. 
Significant progress has been made in understanding these effects and improving their modeling in neutrino event generators.
Yet, outstanding tensions between measurement and theory still remain unresolved~\cite{Tensions2019,ALVAREZRUSO20181,BETANCOURT20181}.
A major challenge in the study of neutrino-nucleus interactions originates from the wide-band accelerator neutrino beams, since nuclear effects cannot be easily disentangled when averaged over a broad energy spectrum. 
Examples of these are Fermi motion, final state interactions (FSI), and nucleon-nucleon correlations.
However, it has been shown that kinematic imbalance variables in the plane transverse to the neutrino direction of travel are powerful tools that can be used to separate these nuclear effects, while minimizing the correlation to the neutrino energy~\cite{PhysRevD.103.112009}.

In this work, we extend these measurements to generalized kinematic imbalance (GKI) variables by considering the longitudinal components along the beam direction.
We illustrate with generator-level studies that these generalized variables achieve improved sensitivity to nuclear effects.
Furthermore, we present the first flux-integrated single- and double-differential cross-section measurements for muon-neutrino charged-current (CC) reactions on argon ($\nu_{\mu}$-Ar) as a function of these generalized variables.
Here we focus on reactions where a single muon-proton pair is reconstructed with no additional detected particles, similar to previous measurements ~\cite{RefPRL,RefPRD}
We refer to these events as CC1p0$\pi$, and they are dominated by quasielastic (QE) interactions as it is required that there are no pions above the detection threshold.
The results reported here use the MicroBooNE detector~\cite{Acciarri:2016smi} with an exposure of $6.79 \times 10^{20}$ protons on target from the Booster Neutrino Beam (BNB)~\cite{AguilarArevalo:2008yp} at Fermi National Accelerator Laboratory.

In Sec.~\ref{sec:formalism} we define the GKI variables. 
Sec.~\ref{sec:generators} shows the enhanced sensitivity of GKI variables to nuclear effects by presenting comparisons to their transverse plane equivalents.
In Sec.~\ref{sec:MicroBooNE} we present the first flux-integrated single- and double-differential cross section measurements in these variables using $\nu_{\mu}$-Ar CC1p0$\pi$ interactions recorded by the MicroBooNE detector.
Finally, conclusions are presented in Sec.~\ref{sec:discussion}.


\section{Observables} \label{sec:formalism}

Variables based on the transverse kinematic imbalance are powerful discriminators between interaction models as reported by multiple experiments using a number of final states~\cite{PhysRevC.94.015503,PhysRevLett.121.022504,PhysRevD.101.092001,Bathe-Peters:2022kkj,Abe:2018pwo,PhysRevD.103.112009,PhysRevD.102.072007,RefPRL,RefPRD}.
The simplest case is for charged current quasielastic (CCQE) interactions, where the final state can be characterized by a muon with transverse momentum $\vec{p}_T^{\,\mu}$ and a proton with transverse momentum $\vec{p}_T^{\,p}$.
To extend to other hadronic final states, the proton's momentum is replaced with the momentum of the combined hadronic system.
In this work, however, we assume a final state containing only a muon and a proton.
The kinematics of the two particle final state on the transverse plane can be fully characterized by a magnitude $\delta p_{T}$ = $|\delta \vec{p}_T| = |\vec{p}_T^{\,\mu} + \vec{p}_T^{\,p}|$, as well as two angles,

\begin{align}
\delta \phi_T &= \cos^{-1} \left(\frac{-\vec{p}_T^{\,\mu} \cdot \vec{p}_T^{\,p}}{|\vec{p}_T^{\,\mu}| |\vec{p}_T^{\,p}|}\right), \label{TKIDeltaPhiT} \\
\delta \alpha_T &= \cos^{-1} \left(\frac{-\vec{p}_T^{\,\mu} \cdot \delta \vec{p}_T}{|\vec{p}_T^{\,\mu}| |\delta\vec{p}_{T}|}\right). \label{TKIDeltaAlphaT}
\end{align}
The muon transverse momentum ($\vec{p}^{\,\mu}_{T}$) is equal and opposite to the transverse component of the momentum transfer to the nucleus ($\vec{q}_{T}$).
The vector $\delta\vec{p}_T$ is the struck nucleon transverse momentum, and should be zero in the absence of initial state nucleon transverse momentum and FSI.
Any non-zero value of $\delta\vec{p}_T$ can therefore be described as the missing transverse momentum (or rather, the negative thereof).
Fermi motion inside the nucleus produces non-zero values of $\delta\vec{p}_T$ for QE interactions.
However, more complex interactions, namely meson exchange currents (MEC), resonance interactions (RES) and deep inelastic scattering events (DIS), can still yield events with only a muon and a proton.
This can be the case due to FSI, such as pion absorption, and such non-QE events populate the region above the Fermi momentum.

\begin{figure}[htb!]
\centering
\begin{tikzpicture} \draw (0, 0) node[inner sep=0] {
\includegraphics[width=0.6\textwidth]{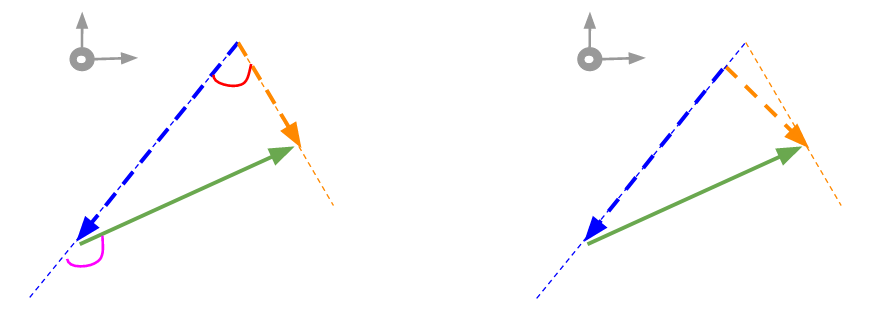}
};
\draw (-3, -2.5) node {(a)};
\draw (-4.1, 1.8) node {\textcolor{gray}{$y$}};
\draw (-4.6, 1.) node {\textcolor{gray}{$z$}};
\draw (-3.65, 1.) node {\textcolor{gray}{$x$}};
\draw (-1.5, 0.8) node {\textcolor{orange}{\Large{$p^{p}_{T}$}}};
\draw (-2.5, 0.5) node {\textcolor{red}{\Large{$\delta\phi_{T}$}}};
\draw (-4., 0.0) node {\textcolor{blue}{\Large{$q_{T}$}}};
\draw (-2.8, -0.9) node {\textcolor{ForestGreen}{\Large{$\delta p_{T}$}}};
\draw (-3.8, -1.7) node {\textcolor{magenta}{\Large{$\delta\alpha_{T}$}}};
\draw (3.2, -2.5) node {(b)};
\draw (2.1, 1.8) node {\textcolor{gray}{$y$}};
\draw (1.5, 1.) node {\textcolor{gray}{$z$}};
\draw (2.6, 1.) node {\textcolor{gray}{$x$}};
\draw (4.8, 0.8) node {\textcolor{orange}{\Large{$\delta p_{Tx}$}}};
\draw (1.9, 0.0) node {\textcolor{blue}{\Large{$\delta p_{Ty}$}}};
\draw (3.5, -0.9) node {\textcolor{ForestGreen}{\Large{$\delta p_{T}$}}};
\end{tikzpicture}
\caption{
(a) Representation of the kinematic imbalance variables on the transverse plane and (b) alternative representation using the projections parallel and perpendicular to the transverse momentum transfer vector.
The finely dashed lines correspond to the direction of the momentum transfer (blue) and the proton (orange).
The $z$ axis corresponds to the neutrino direction of travel.
}
\label{fig:transversePlaneDiagram}
\end{figure}
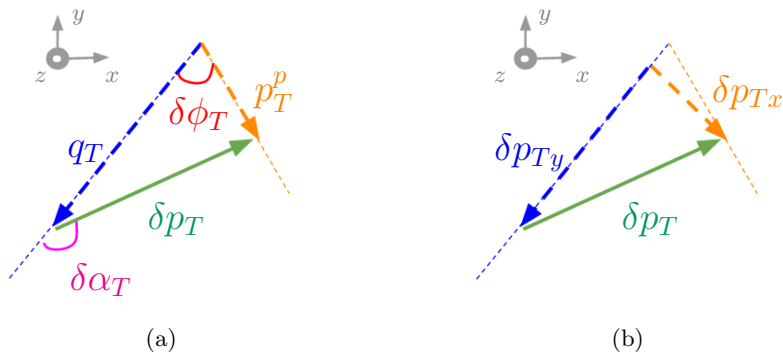

The variable $\delta \phi_T$ corresponds to the angle between the transverse momentum transfer vector and the final state transverse proton momentum ($\vec{p}_T^{\,p}$).
These vectors would be aligned in the case of a free stationary nucleon target and this angle would be zero.
Small values of $\delta \phi_T$ are produced by initial state motion, while larger values are indicative to FSI.
The angle $\delta \alpha_T$ is defined between the transverse momentum transfer vector (--\,$\vec{p}_T^{\,\mu}$) and the transverse missing momentum vector ($\delta \vec{p}_T$).
The angle is sensitive to FSI but much less sensitive to initial state nucleon motion.
In the absence of FSI, $\delta \alpha_T$ does not have a preferred orientation and yields a fairly flat distribution.
A representation of the kinematic imbalance variables on the transverse plane is shown in Fig.~\ref{fig:transversePlaneDiagram}(a).

A more recent investigation identifies an alternative representation of the transverse missing momentum vector using the projection components perpendicular and parallel to the momentum transfer~\cite{PhysRevD.101.092001} [see Fig.~\ref{fig:transversePlaneDiagram}(b)], given by

\begin{align}
\label{ptxLabel}
\delta p_{T,x} &= (\hat{z} \times \hat{p}_{T}^{\mu}) \cdot \delta \vec{p}_{T},\\
\label{ptyLabel}
\delta p_{T,y} &= - \hat{p}_{T}^{\mu} \cdot \delta\vec{p}_{T}.
\end{align}
These are the components of $\delta\vec{p}_{T}$ perpendicular and parallel, respectively, to the transverse momentum transfer vector.

The interpretation of these variables on the transverse plane can be generalized to their three-dimensional equivalents.
To do this, the longitudinal components of the missing momentum and momentum transfer vectors are required, and therefore
an assumption of the incoming neutrino energy has to be made.
An initial attempt to perform this generalization is reported in Ref.~\cite{Furmanski2016} and first measured in Ref.~\cite{PhysRevLett.121.022504}, which assumes that a neutrino interacts via a CCQE interaction with a bound stationary neutron at rest inside a nucleus.
This formalism was expanded to other final states in~\cite{PhysRevC.99.055504,WCLu}.
This is then used to obtain an estimate for the neutrino energy and the three components of the missing momentum vector, labeled $\vec{p}_n$.
Related studies that consider the longitudinal components are included in~\cite{Abe:2018pwo,PhysRevD.74.052002}.

We present a slightly different formalism which relies on conservation of energy and momentum.
For the notation presented below, the speed of light $c$ is assumed to be unity.
For a massless neutrino, if the entire final state is visible, then the total visible energy and the longitudinal momentum will be the same.
The ``calorimetric visible energy" is constructed as the primary neutrino energy estimator

\begin{align}
E_{\text{cal}} = E_{\mu} + K_{p} + B,
\end{align}
where $E_{\mu}$ is the total muon energy, $K_{p}$ is the proton kinetic energy, and $B$ the argon removal energy set to 30.9 MeV~\cite{BodekCai2019}.

The second estimator of the neutrino energy is the total momentum along the neutrino direction of travel.
This should be equal to the calorimetric energy if all final state particles have been observed and the target nucleon is at rest.
Therefore the difference between the total longitudinal momentum and the calorimetric energy provides the longitudinal component of the missing momentum~\cite{PWIA},

\begin{align}
p_L = p_{L}^{\mu} + p_{L}^{p} - E_{\text{cal}},
\end{align}
where $p_{L}^{\mu (p)}$ is the longitudinal component of the muon (proton) momentum vector.
This definition of $p_L$ is numerically very close to that in Ref.~\cite{Furmanski2016}, but it enables a trivial generalization to other final states without having to make an assumption about the underlying interaction.

\begin{figure}[htb!]
\centering
\begin{tikzpicture} \draw (0, 0) node[inner sep=0] {
\includegraphics[width=0.6\textwidth]{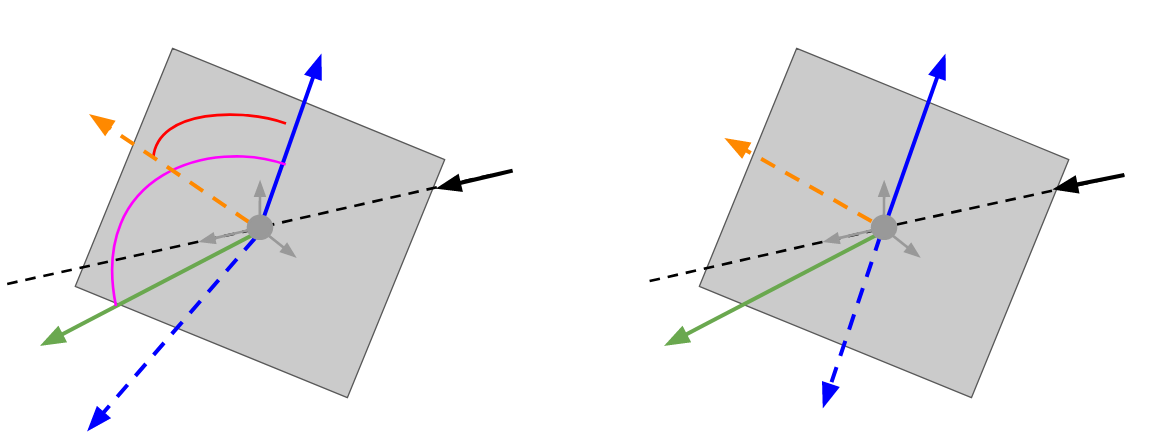}
};
\draw (-3.5, -2.8) node {(a)};
\draw (-3.1, 0.4) node {\textcolor{gray}{$y$}};
\draw (-3.6, 0.) node {\textcolor{gray}{$z$}};
\draw (-2.5, -0.3) node {\textcolor{gray}{$x$}};
\draw (-0.4, 0.5) node {$p_{\nu}$};
\draw (-4.7, 0.6) node {\textcolor{orange}{\Large{$p^{p}$}}};
\draw (-3.5, 1.3) node {\textcolor{red}{\Large{$\phi_{3D}$}}};
\draw (-2.1, 1.5) node {\textcolor{blue}{\Large{$q$}}};
\draw (-4.8, -1.4) node {\textcolor{ForestGreen}{\Large{$p_{n}$}}};
\draw (-4.2, -2) node {\textcolor{blue}{\Large{$p^{\mu}$}}};
\draw (-5., -0.1) node {\textcolor{magenta}{\Large{$\alpha_{3D}$}}};
\draw (2.7, -2.8) node {(b)};
\draw (2.7, 0.4) node {\textcolor{gray}{$y$}};
\draw (2.2, 0.) node {\textcolor{gray}{$z$}};
\draw (3.4, -0.3) node {\textcolor{gray}{$x$}};
\draw (5.3, 0.4) node {$p_{\nu}$};
\draw (1.2, 1.) node {\textcolor{orange}{\Large{$p_{n\perp}$}}};
\draw (3.6, 1.5) node {\textcolor{blue}{\Large{$q$}}};
\draw (0.8, -1.4) node {\textcolor{ForestGreen}{\Large{$p_{n}$}}};
\draw (2.5, -2.) node {\textcolor{blue}{\Large{$p_{n\parallel}$}}};
\end{tikzpicture}
\caption{
(a) Representation of the generalized kinematic imbalance variables and (b) alternative representation using the projections parallel and perpendicular to the missing momentum vector.
The $z$ axis corresponds to the neutrino direction of travel.
}
\label{fig:projDiagram}
\end{figure}
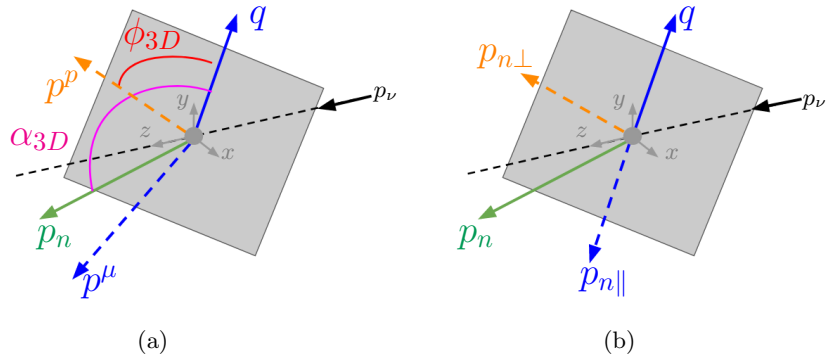

The missing momentum vector $\vec{p}_n$ is obtained as the vector sum of the transverse missing momentum $\delta\vec{p}_T$ and the longitudinal component $p_{L}$.
Under the assumption that FSI are weak or absent, the definitions given here produce a vector that aligns with the initial struck neutron momentum, consistent with the definitions used in previous work~\cite{Furmanski2016}.
Assuming a neutrino traveling in the $z$ direction of a detector coordinate system, the momentum transfer $\vec{q}$ of the interaction is derived as the difference between the inferred neutrino and the muon momentum vectors

\begin{align}
\label{qLabel}
\vec{q} &= E_{\text{cal}} \hat{z} - \vec{p}_{\mu}.
\end{align}
With these definitions, we generalize the transverse missing momentum variables to three dimensions [see Fig~\ref{fig:projDiagram}(a)] to be

\begin{align}
\label{GKIpn}
p_n &= |\vec{p}_n| = \sqrt{p_{L}^{2} + \delta p_{T}^{2}},\\
\label{GKIphi3D}
\phi_{3D} &= \cos^{-1} \left(\frac{\vec{q} \cdot \vec{p}_{\,p}}{|\vec{q}\,| |\vec{p}_{\,p}|}\right), \\
\label{GKIalpha3D}
\alpha_{3D} &= \cos^{-1} \left(\frac{\vec{q} \cdot \vec{p}_{n}}{|\vec{q}\,| |\vec{p}_{n}|}\right).
\end{align}
The vector $\vec{p}_n$ accounts for the missing momentum and is an estimate of the initial struck nucleon momentum when all the particles are reconstructed;
$\phi_{3D}$ is the opening angle between the total momentum transfer vector $\vec{q}$ and the proton momentum vector $\vec{p}^{\,p}$;
and $\alpha_{3D}$ is the angle between the momentum transfer vector $\vec{q}$ and the missing momentum vector $\vec{p}_n$.
These variables are the three-dimensional analogues to the ones on the transverse plane $\vec{p}_T$, $\delta \alpha_T$, and $\delta \phi_T$.

Likewise, the alternative representation on the transverse plane can also be extended to three dimensions [Fig.~\ref{fig:projDiagram}(b)], in the form

\begin{align}
\label{pnperpxLabel}
p_{n \perp,x} &= (\hat{q}_T \times \hat{z}) \cdot \vec{p}_n,\\
\label{pnperpyLabel}
p_{n \perp,y} &= (\hat{q} \times (\hat{q}_T \times \hat{z})) \cdot \vec{p}_n,\\
\label{pnperpLabel}
p_{n \perp} &= \sqrt{(p_{n \perp,x})^2 + (p_{n \perp,y})^2} = |p_n| \sin(\alpha_{3D}),\\
\label{pnparLabel}
p_{n \parallel} &= \hat{q} \cdot \vec{p}_{n} = |p_n| \cos(\alpha_{3D}).
\end{align}
Here, $p_{n\parallel}$ ($p_{n\perp}$) is the component of the missing momentum vector parallel (perpendicular) to the momentum transfer vector, $p_{n \perp,x}$ ($p_{n \perp,y}$) is the component of $p_{n\perp}$ in the neutrino-muon scattering plane (perpendicular to the neutrino-muon scattering plane), $\hat{q}$ is the unit vector aligned with $\vec{q}$, $\hat{q}_T$ is the unit vector aligned with the transverse component of $\vec{q}$, and $\hat{z}$ is the unit vector aligned with the neutrino direction of travel (the z-axis).

These variables have been defined for a CCQE interaction with only a muon and a proton in the final state, however this is easily extended to other final states by summing the hadron momenta, and adjusting the assumed binding energy depending on the nucleon multiplicity.
It is worth noting that Eqn.~\ref{GKIalpha3D}-\ref{pnparLabel} are introduced for the first time in the literature.


\section{Generator Comparisons} \label{sec:generators}

To demonstrate the power of these novel variables, we use the CC1p0$\pi$ signal definition included in Refs.~\cite{RefPRL,RefPRD} with several commonly-used neutrino interaction generators and model configurations, convoluted with the MicroBooNE $\nu_{\mu}$ flux prediction~\cite{AguilarArevalo:2008yp}.
The CC1p0$\pi$ signal definition used in this analysis includes all $\nu_{\mu}$-Ar scattering events with a final-state muon with momentum 0.1 $< p_{\mu}<$ 1.2\,GeV/$c$, and exactly one proton with 0.3 $< p_{p} <$ 1\,GeV/$c$.
Events with final-state neutral pions at any momentum are excluded.
Signal events may contain any number of protons below 300 MeV/$c$ or above 1\,GeV/$c$, neutrons at any momentum, and charged pions with momentum lower than 70 MeV/$c$.
A number of simulation predictions used for comparison correspond to $\texttt{GENIE}$ model configurations, an event generator commonly used by Fermilab-based experiments.
Our comparisons include an older $\texttt{GENIE}$ version which was extensively used in the past, as well as more modern $\texttt{GENIE}$ tunes currently used by MicroBooNE~\cite{RefPRD,RefPRL,GENIE_tune}.
Additionally, comparisons to alternative event generators used by other neutrino experiments, such as T2K, or in the theory community, are included.
Overflow (underflow) values are included in the last (first) bin.
Table~\ref{names} lists a summary of the abbreviations of the generators and configurations used in this analysis, which are presented in more detail below.
The relevant samples have been processed via the Nuisance framework~\cite{Stowell_2017}.

\begin{center}
\begin{table}[H]
\centering
\caption{Generators and configurations abbreviations used.}
\begin{tabular}{ | c | c | }
  \hline
  \makecell{Name} & \makecell{Generator / Configuration} \\ 
  \hline
  \hline
$\texttt{Gv2}$  & $\texttt{GENIE v2.12.10}$~\cite{Andreopoulos:2009rq,Andreopoulos:2015wxa}\\
$\texttt{G18}$  & $\texttt{GENIE v3.0.6 G18\_10a\_02\_11a}$~\cite{geniev3highlights}\\
$\texttt{G18T}$  & $\texttt{G18}$ with tune~\cite{GENIE_tune}\\
$\texttt{G21}$  & $\texttt{GENIE v3.2.0 G21\_11b\_00\_000}$~\cite{geniev3highlights}\\
$\texttt{GiBUU}$  & $\texttt{GiBUU 2021}$~\cite{Mosel:2019vhx}\\
$\texttt{NuWro}$  & $\texttt{NuWro v19.02.1}$~\cite{GolanNuWro:2008yp}\\
$\texttt{NEUT}$  & $\texttt{NEUT v5.4.0}$~\cite{Hayato:2008yp}\\
\hline
 \end{tabular}
 \label{names}
\end{table} 
\end{center}

The $\texttt{GENIE}$ configurations used are:
\begin{itemize}

\item $\texttt{GENIE v2.12.10 (Gv2)}$~\cite{Andreopoulos:2009rq,Andreopoulos:2015wxa}: This version corresponds to a historical reference extensively used in the earliest MicroBooNE cross section analyses~\cite{PhysRevLett.125.201803,PhysRevD.102.112013} but superseded afterwards~\cite{GENIE_tune}. 
$\texttt{Gv2}$ includes the Bodek-Ritchie Fermi Gas model, the Llewellyn Smith CCQE scattering prescription~\cite{LlewellynSmith:1971uhs}, the empirical MEC model~\cite{Katori:2013eoa}, a Rein-Sehgal RES and coherent (COH) scattering model~\cite{Rein:1980wg}, the Bodek-Yang DIS model~\cite{PhysRevLett.82.2467} coupled $\texttt{PYTHIA}$~\cite{Sjostrand:2006za} for the hadronization part, and a data driven FSI model denoted as ``hA''~\cite{Mashnik:2005ay}. 
More modern $\texttt{GENIE}$ versions include improvements over $\texttt{Gv2}$ related to FSI issues, ground state modeling, and lepton-hadron correlations.

\item $\texttt{GENIE v3.0.6 G18\_10a\_02\_11a (G18)}$~\cite{geniev3highlights}: This more modern model configuration uses the local Fermi gas (LFG) model~\cite{Carrasco:1989vq}, the Nieves CCQE scattering prescription~\cite{Nieves:2012yz} which includes Coulomb corrections for the outgoing muon~\cite{Engel:1997fy}, and random phase approximation (RPA) corrections~\cite{RPA}.
Additionally, it uses the Nieves MEC model~\cite{Schwehr:2016pvn}, the Kuzmin-Lyubushkin-Naumov Berger-Sehgal RES~\cite{Berger:2007rq,PhysRevD.104.072009,Kuzmin:2003ji}, Berger-Sehgal COH~\cite{Berger:2008xs}, and Bodek-Yang DIS~\cite{PhysRevLett.82.2467} scattering models with the $\texttt{PYTHIA}$~\cite{Sjostrand:2006za} hadronization part, and the hA2018 FSI model~\cite{Ashery:1981tq}.

\item $\texttt{G18T}$~\cite{GENIE_tune}: Corresponds to the same $\texttt{G18}$ configuration with additional MicroBooNE-specific tuning.

\item $\texttt{GENIE v3.2.0 G21\_11b\_00\_000}$ ($\texttt{G21}$)~\cite{geniev3highlights}. 
This configuration includes the SuSAv2 prediction for the QE and MEC scattering modes~\cite{PhysRevD.101.033003} and the hN2018 FSI model~\cite{hN2018}.	
The modeling options for RES, DIS, and COH interactions are the same as for $\texttt{G18}$.
\end{itemize}

The alternative event generator predictions are:
\begin{itemize}

\item $\texttt{GiBUU 2021 (GiBUU)}$~\cite{Mosel:2019vhx}: Uses similar models to $\texttt{GENIE}$, but they are implemented in a coherent way by solving the Boltzmann-Uehling-Uhlenbeck transport equation~\cite{Mosel:2019vhx}. 
The modeling includes the LFG model~\cite{Carrasco:1989vq}, a standard CCQE expression~\cite{Leitner:2006ww}, an empirical MEC model, and a dedicated spin dependent resonance amplitude calculation following the $\texttt{MAID}$ analysis~\cite{Mosel:2019vhx}. 
The DIS model is from $\texttt{PYTHIA}$~\cite{Sjostrand:2006za}.
$\texttt{GiBUU}$'s FSI treatment propagates the hadrons through the residual nucleus in a nuclear potential consistent with the initial state.

\item $\texttt{NuWro v19.02.1 (NuWro)}$~\cite{GolanNuWro:2008yp}: Includes the LFG model~\cite{Carrasco:1989vq}, the Llewellyn Smith  model for QE events~\cite{LlewellynSmith:1971uhs}, the Nieves model for MEC events~\cite{ValenciaModel}, the Adler-Rarita-Schwinger formalism to calculate the $\Delta$ resonance explicitly~\cite{Graczyk:2007bc}, the BS COH~\cite{Berger:2008xs} scattering model, an intranuclear cascade model for FSI~\cite{ValenciaModel}, and a coupling to $\texttt{PYTHIA}$~\cite{Sjostrand:2006za} for hadronization.

\item $\texttt{NEUT v5.4.0 (NEUT)}$~\cite{neut}: Corresponds to the combination of the LFG model~\cite{Carrasco:1989vq,PWIA}, the Nieves CCQE scattering prescription~\cite{Nieves:2012yz}, the Nieves MEC model using a lookup table~\cite{Schwehr:2016pvn}, the Berger Sehgal RES~\cite{Berger:2007rq,Graczyk:2007bc,Auchincloss} and BS COH~\cite{Berger:2008xs} scattering models, FSI with medium corrections for pions~\cite{Andreopoulos:2009rq,Andreopoulos:2015wxa}, and $\texttt{PYTHIA}$~\cite{Sjostrand:2006za} purposes.

\end{itemize}

\begin{figure}[htb!]
\centering
\begin{tikzpicture} \draw (0, 0) node[inner sep=0] {
\includegraphics[width=0.45\textwidth]{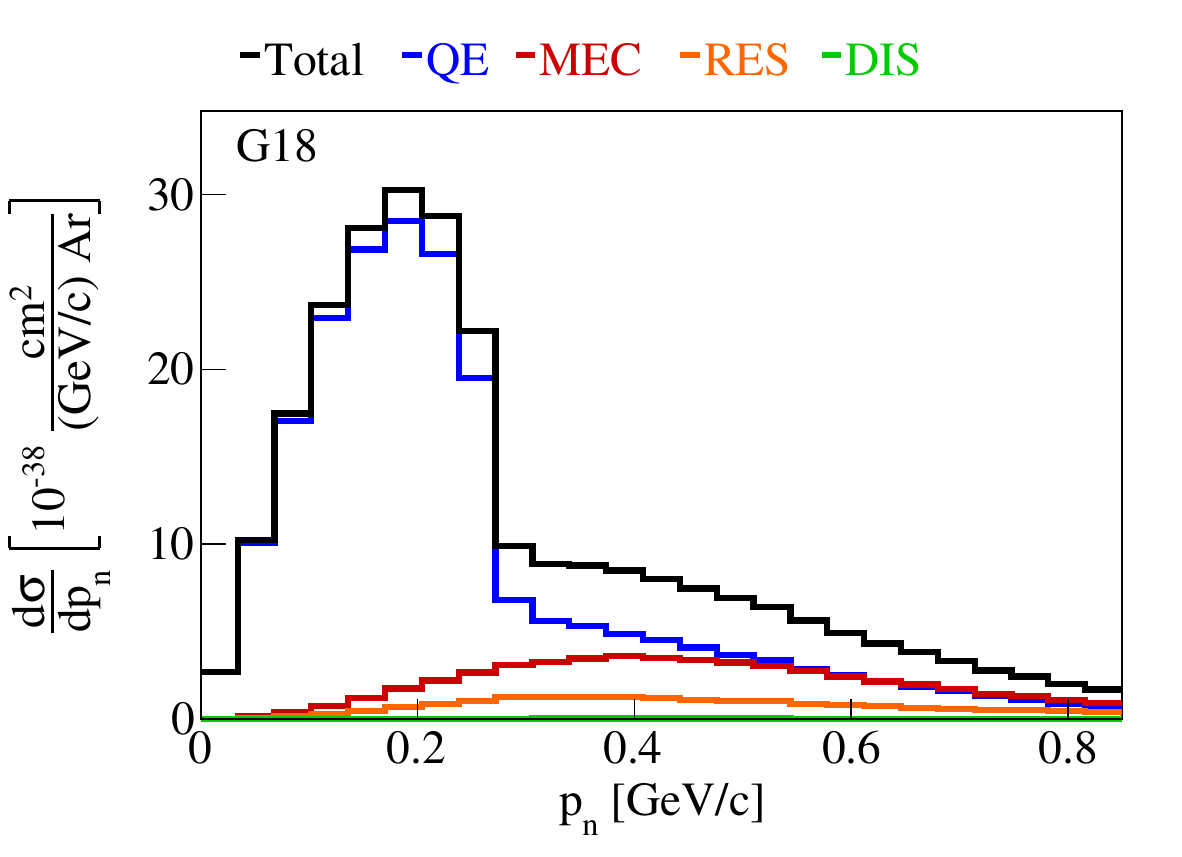}	
};
\draw (0.4, -3.2) node {(a)};	
\end{tikzpicture}
\hspace{0.05 \textwidth}
\begin{tikzpicture} \draw (0, 0) node[inner sep=0] {
\includegraphics[width=0.45\textwidth]{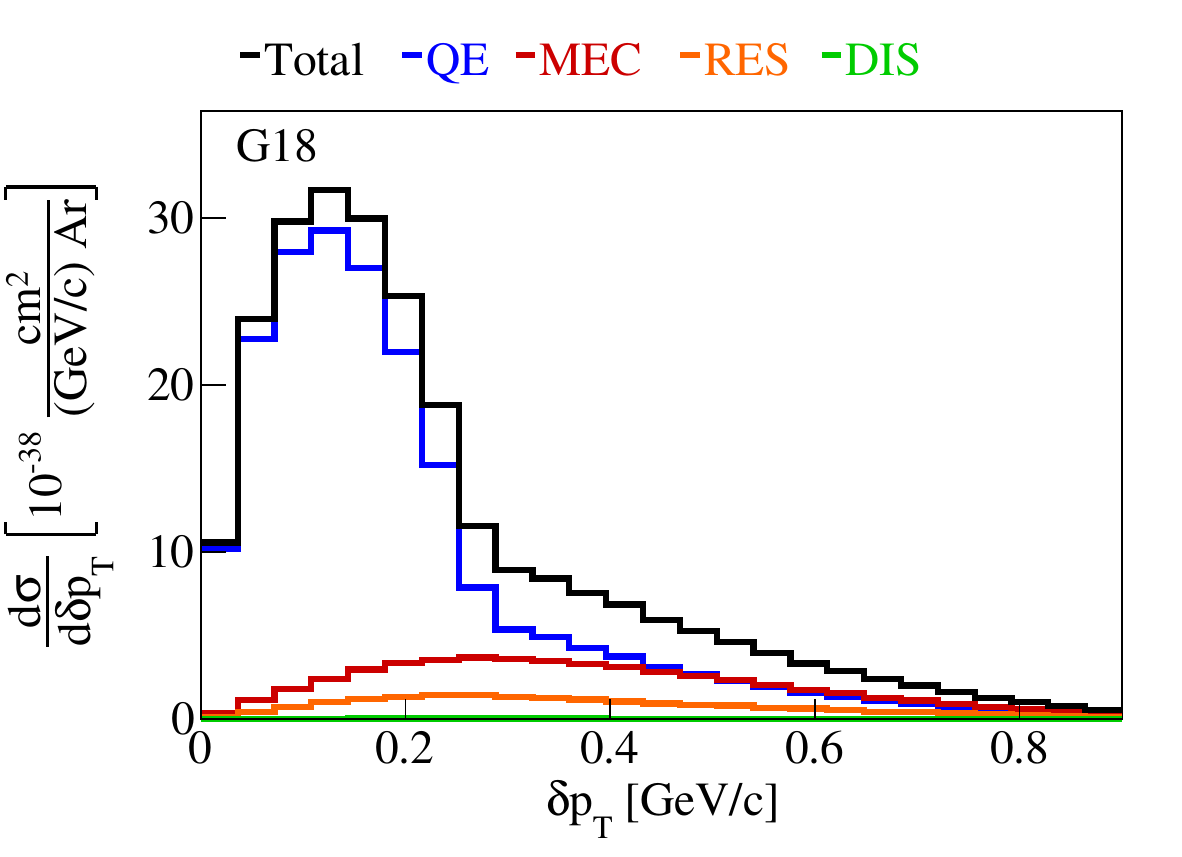}
};
\draw (0.4, -3.2) node {(b)};	
\end{tikzpicture}
\caption{The flux-integrated single-differential cross section interaction breakdown as a function of (a) $p_{n}$ and (b) $\delta p_{T}$ for the selected CC1p0$\pi$ events.
Colored lines show the results of theoretical cross section calculations using the $\texttt{G18}$ prediction for QE (blue), MEC (red), RES (orange), and DIS (green) interactions.}
\label{fig:ptvspn}
\end{figure}

If the interacting neutrino scatters off a stationary neutron and all final state particles are observed, $p_n$ and $\delta p_T$ become zero.
Within the dense nuclear medium of a heavy nucleus like argon, these variables follow a broad distribution due to the struck nucleon motion before the interaction, with a high missing momentum tail from non-CCQE events.
Figure~\ref{fig:ptvspn} shows the $p_{n}$ distribution for interaction types using $\texttt{G18}$ and shows the corresponding breakdown for $\delta p_{T}$ for the selected CC1p0$\pi$ events.
Both results illustrate a CCQE dominance that is driven by the CC1p0$\pi$ signal definition.
The non-CCQE events in the $p_n$ distribution are concentrated at higher momentum values than in the $\delta p_{T}$ distribution.
Therefore, the missing momentum $p_n$ illustrates enhanced discrimination capabilities between CCQE and non-CCQE events when compared to $\delta p_{T}$, as already demonstrated in Ref.~\cite{Furmanski2016}.

\begin{figure}[htb!]
\centering
\begin{tikzpicture} \draw (0, 0) node[inner sep=0] {
\includegraphics[width=0.45\textwidth]{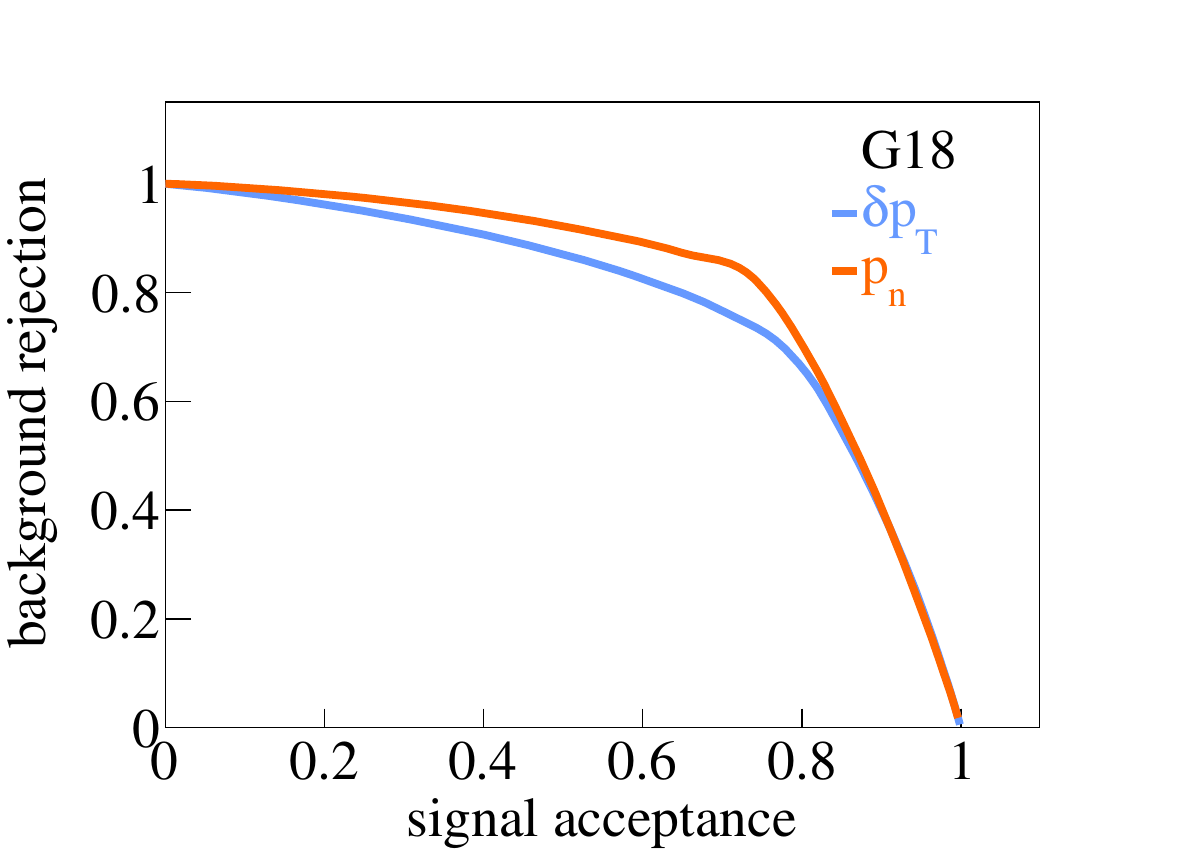}	
};
\draw (0.4, -3.2) node {(a)};	
\end{tikzpicture}
\hspace{0.05 \textwidth}
\begin{tikzpicture} \draw (0, 0) node[inner sep=0] {
\includegraphics[width=0.45\textwidth]{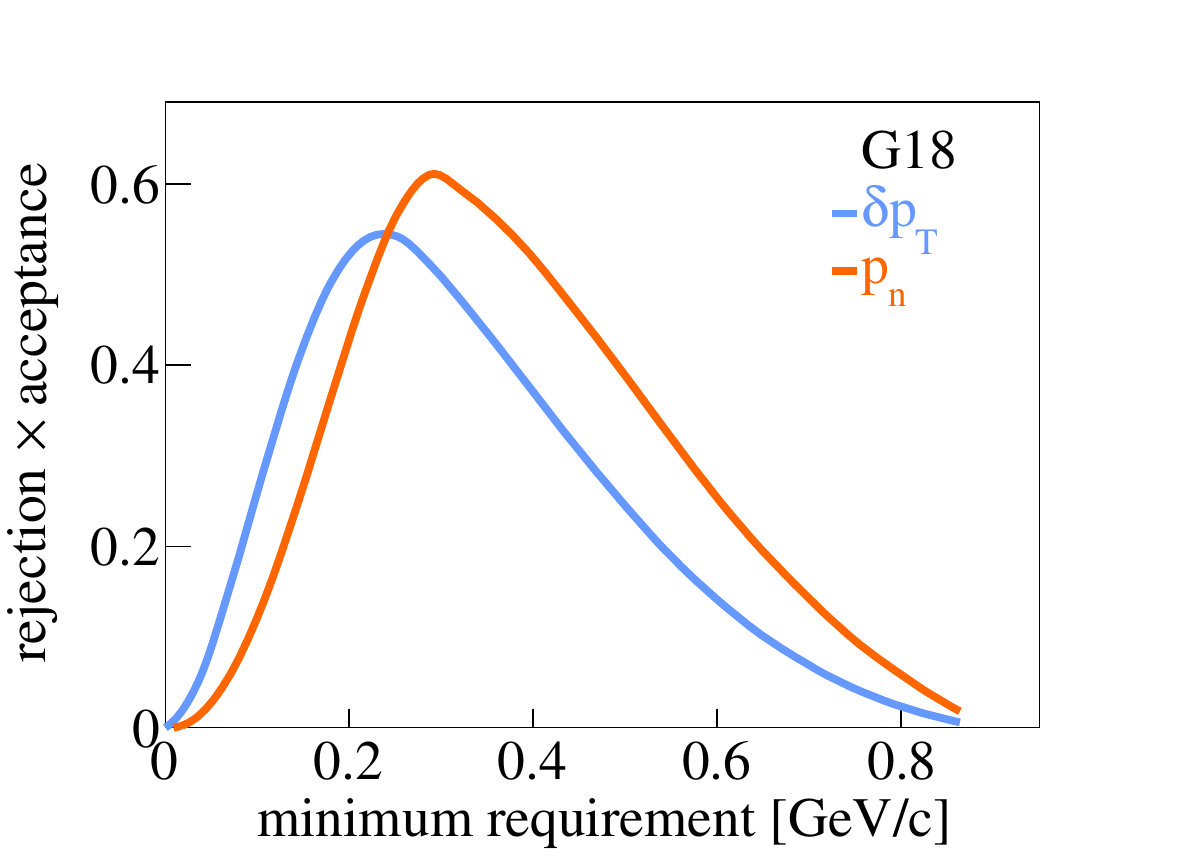}
};
\draw (0.4, -3.2) node {(b)};	
\end{tikzpicture}
\caption{
(a) Signal acceptance fraction vs background rejection fraction as a function of the minimum requirement on either the missing momentum $p_{n}$ (orange) or the transverse missing momentum $\delta p_{T}$ (blue) for CC1p0$\pi$ events using the G18 prediction.
(b) Evolution of the product between the signal acceptance and the background rejection denoted as ``rejection $\times$ acceptance'' as a function of the minimum requirement.
}
\label{fig:rejection}
\end{figure}

To quantify this improved separation ability, we form signal acceptance-background rejection curves for both kinematic imbalance variables using $\texttt{G18}$, as shown in Fig.~\ref{fig:rejection}(a).
Assuming events with a value of $p_{n}$ less than a given ``cut value'' are retained, the fraction of true CCQE events accepted and the fraction of true non-CCQE events rejected can be calculated as a function of this cut value.
An improved ability to isolate true CCQE events is observed for $p_n$ compared to $\delta p_{T}$, while rejecting a larger fraction of non-CCQE background events.
The evolution of the product between the signal acceptance and the background rejection denoted by ``rejection $\times$ acceptance'' as a function of the cut value yields an optimal requirement at $p_{n}$ $\approx$ 0.3\,GeV/c and at $\delta p_{T}$ $\approx$ 0.2\,GeV/c [Fig.~\ref{fig:rejection}(b)].
The application of the corresponding selection criteria results in a CCQE loss of 25.3\% for $p_{n}$ and 29.7\% for $\delta p_{T}$.
The looser requirement on $p_{n}$ results in a slightly more pure CCQE sample (95\% CCQE purity) with higher statistics compared to the equivalent case with $\delta p_{T}$ (92\% CCQE purity).
The same behavior is observed across all event generators and is shown in the Supplemental Material.

\begin{figure}
\centering
\begin{tikzpicture} \draw (0, 0) node[inner sep=0] {
\includegraphics[width=0.45\textwidth]{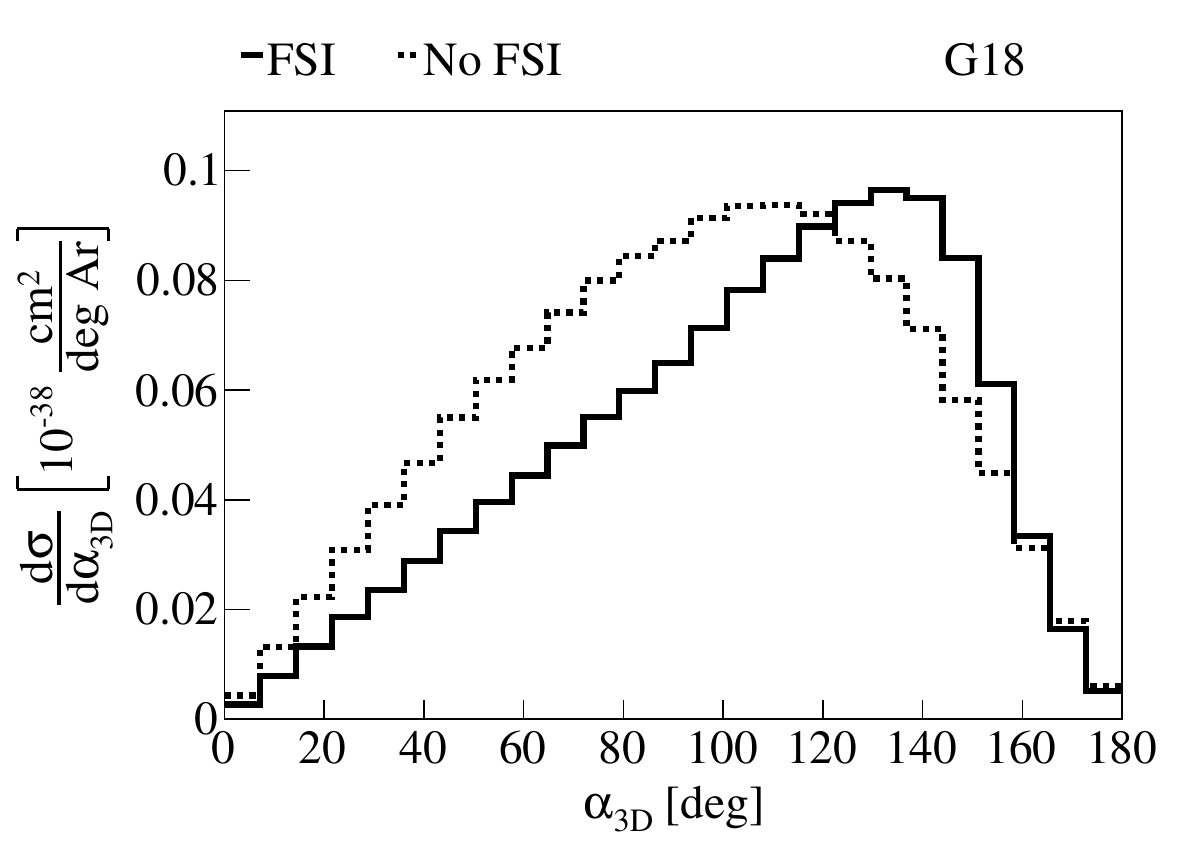}	
};
\draw (0.4, -3.2) node {(a)};	
\end{tikzpicture}
\hspace{0.05 \textwidth}
\begin{tikzpicture} \draw (0, 0) node[inner sep=0] {
\includegraphics[width=0.45\textwidth]{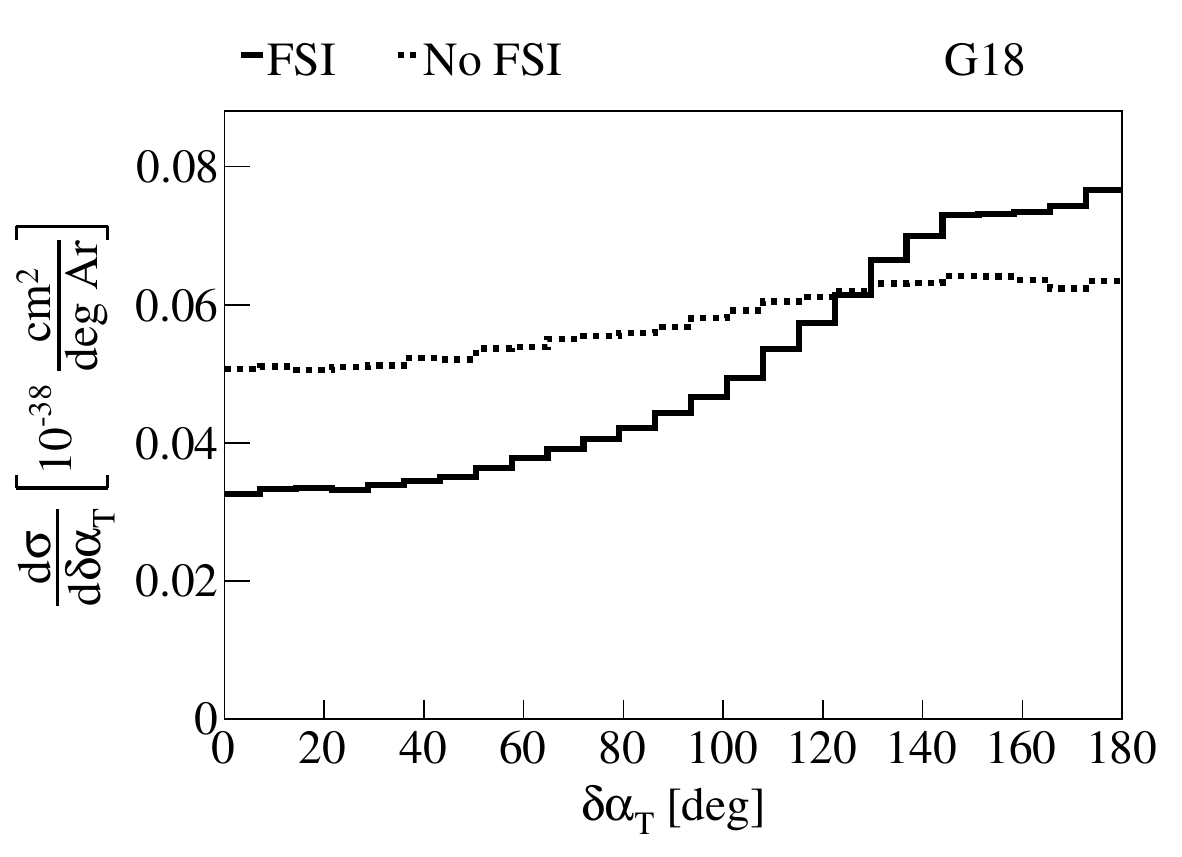}
};
\draw (0.4, -3.2) node {(b)};	
\end{tikzpicture}
\caption{
Comparison of the flux-integrated single-differential cross section as a function of (a) $\alpha_{3D}$ and (b) $\delta \alpha_T$ with (solid) and without (dashed) FSI effects using the $\texttt{G18}$ prediction for the selected CC1p0$\pi$ events.
}
\label{fig:alphavs3D}
\end{figure}

\begin{figure}[htb!]
\centering
\begin{tikzpicture} \draw (0, 0) node[inner sep=0] {
\includegraphics[width=0.45\textwidth]{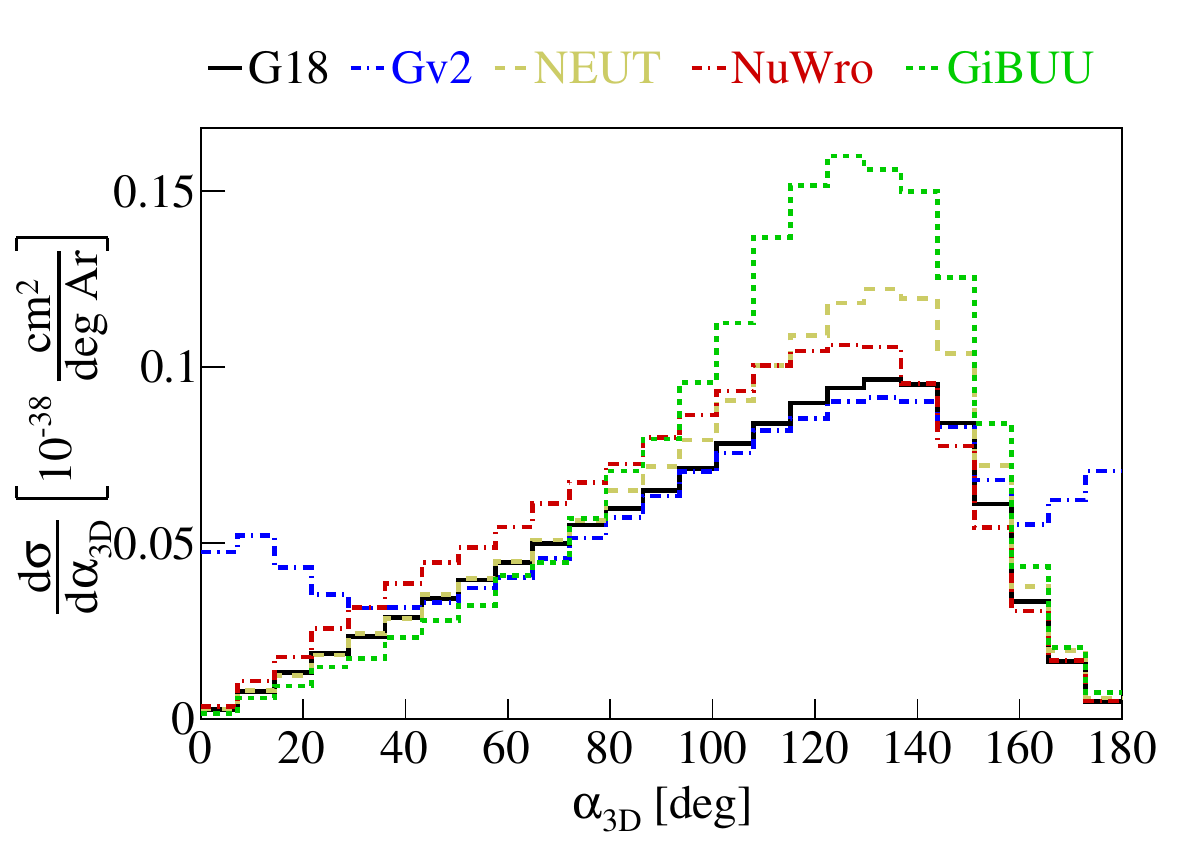}	
};
\draw (0.4, -3.2) node {(a)};	
\end{tikzpicture}
\hspace{0.05 \textwidth}
\begin{tikzpicture} \draw (0, 0) node[inner sep=0] {
\includegraphics[width=0.45\textwidth]{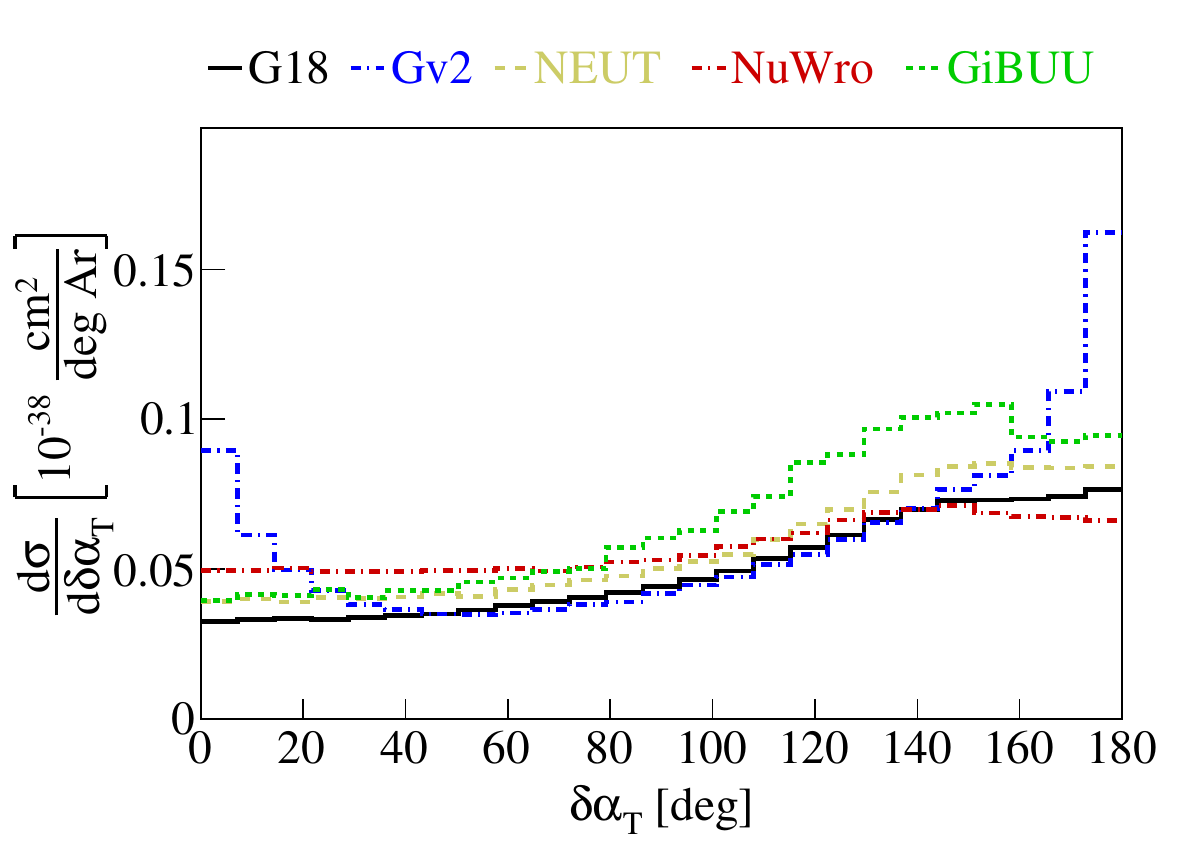}
};
\draw (0.4, -3.2) node {(b)};	
\end{tikzpicture}
\caption{
The flux-integrated single-differential cross section as a function of (a) $\alpha_{3D}$ and (b) $\delta\alpha_{T}$ for the selected CC1p0$\pi$ events.
Colored lines show the results of cross section calculations using the $\texttt{G18}$ (solid black), $\texttt{Gv2}$ (blue), $\texttt{NEUT}$ (dashed pink), $\texttt{NuWro}$ (red) and $\texttt{GiBUU}$ (green) predictions.
}
\label{fig:DeltaAlpha3DqDeltaAlphaTGene}
\end{figure}

If a neutrino scatters off a moving, but unbound, neutron, $\alpha_{3D}$ becomes the angle of the struck nucleon direction before the interaction relative to the momentum transfer vector.
There is no directional preference since the nucleus is at rest, so $\alpha_{3D}$ follows an approximately sinusoidal curve due to the phase space for a randomly distributed three-dimensional direction.
However, FSI~\cite{hN2018} in the nucleus introduce missing momentum, which is transferred from the hadronic system to the residual nucleus, reducing the magnitude of $\vec{p}^{\,p}$ and enhancing the magnitude of the components transverse to $\vec{q}$.  
This effect enhances the contribution of events with higher values of $\alpha_{3D}$.
Like $\alpha_{3D}$, the angular orientation $\delta \alpha_T$ has been shown to be sensitive to FSI effects~\cite{PhysRevD.103.112009}.
The distribution of $\delta \alpha_T$ illustrates a transition from a uniform angular orientation in the absence of FSI to one that peaks close to 180$^{\circ}$ due to the reduction of $\vec{p}^{\,p}$ in the presence of FSI. 
Figure~\ref{fig:alphavs3D} shows the distribution of $\alpha_{3D}$ and $\delta \alpha_T$ for the G18 predictions with and without FSI, illustrating the impact of FSI on the shapes.
Figure~\ref{fig:DeltaAlpha3DqDeltaAlphaTGene} shows the simulation predictions using several event generators for $\alpha_{3D}$ and $\delta\alpha_{T}$.
The $\texttt{Gv2}$ model predicts a significantly different shape that peaks at the edges of the distribution.
The $\texttt{GiBUU}$ distribution is more sharply peaked in $\alpha_{3D}$ than the other generators and shows a larger discrepancy.
All other event generators yield consistent predictions.

\begin{figure}[htb!]
\centering
\begin{tikzpicture} \draw (0, 0) node[inner sep=0] {
\includegraphics[width=0.45\textwidth]{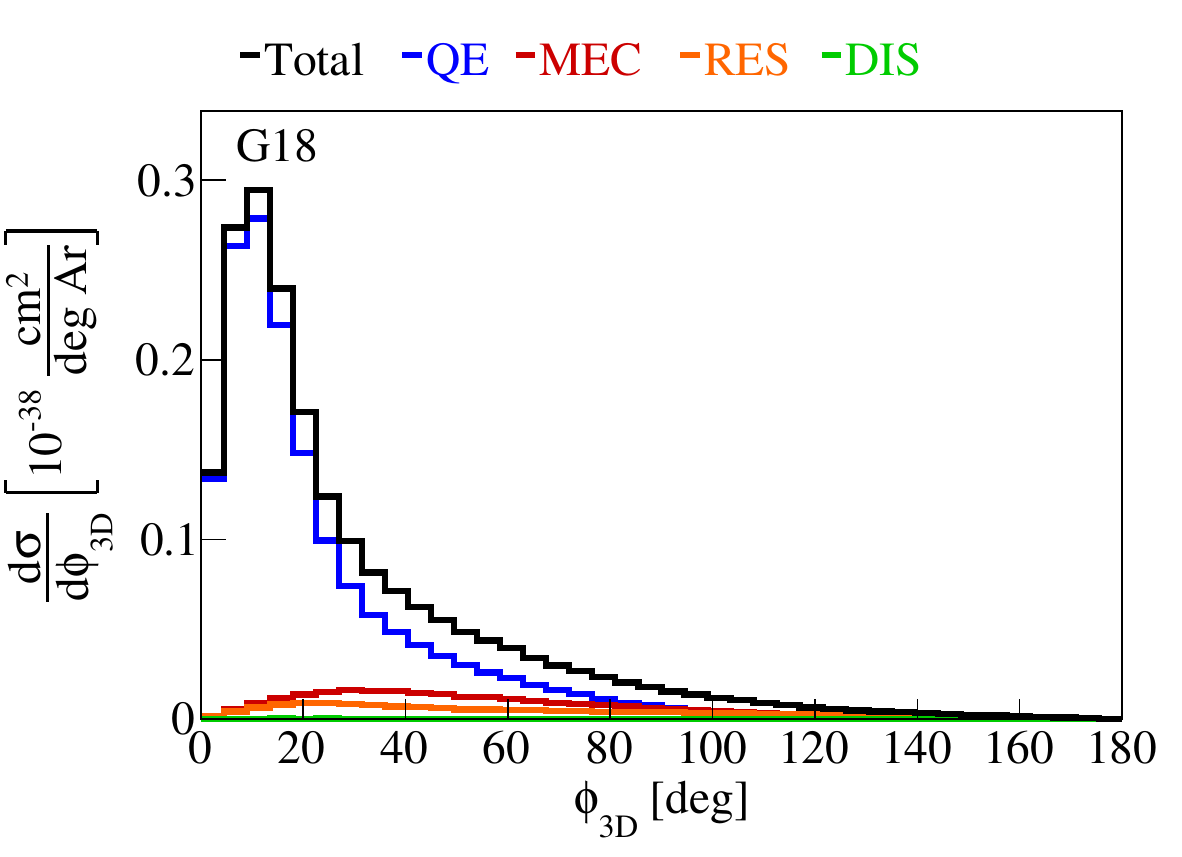}	
};
\draw (0.4, -3.2) node {(a)};	
\end{tikzpicture}
\hspace{0.05 \textwidth}
\begin{tikzpicture} \draw (0, 0) node[inner sep=0] {
\includegraphics[width=0.45\textwidth]{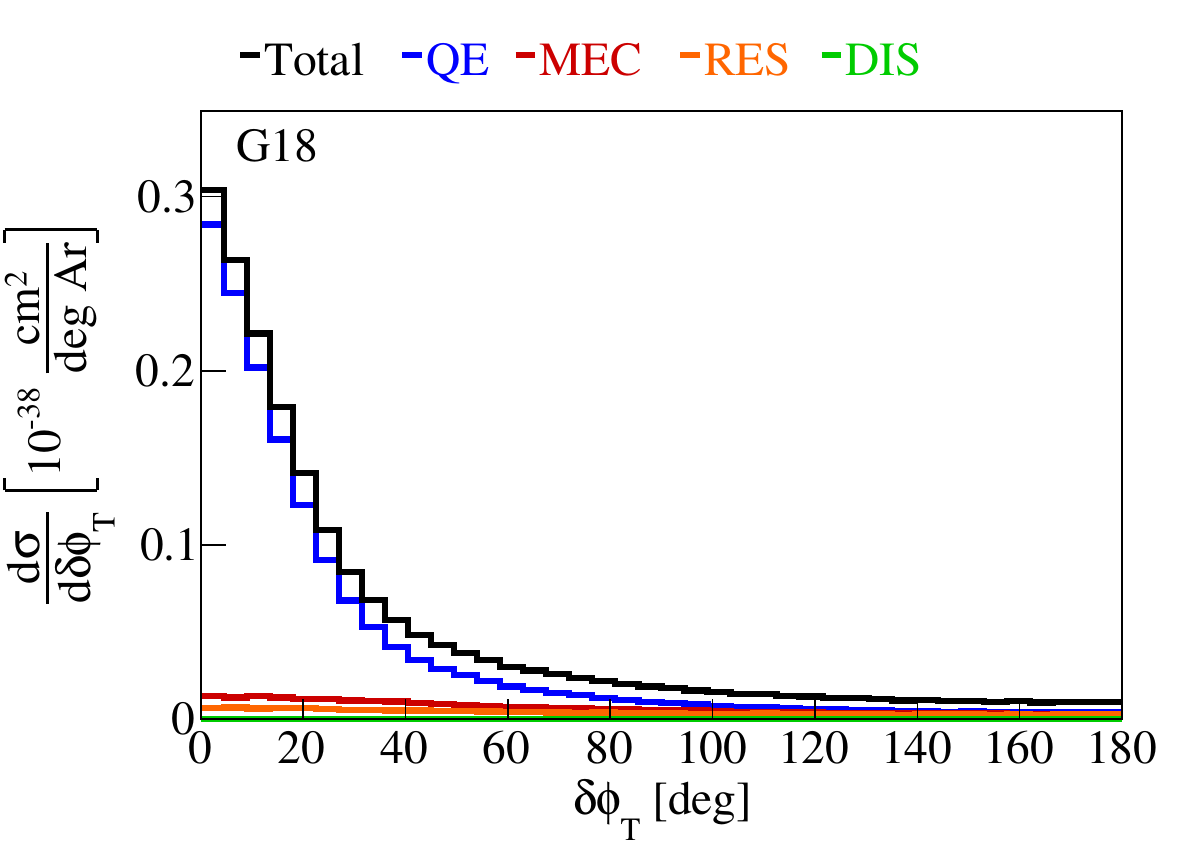}
};
\draw (0.4, -3.2) node {(b)};	
\end{tikzpicture}
\caption{
The flux-integrated single-differential cross section interactions as a function of (a) $\phi_{3D}$ and (b) $\phi_T$ for the selected CC1p0$\pi$ events.
Colored lines show the results of cross section calculations using the $\texttt{G18}$ prediction for QE (blue), MEC (red), RES (orange), and DIS (green) predictions.
}
\label{fig:phivs3D}
\end{figure}

The third transverse kinematic imbalance variable, $\delta \phi_{T}$, has the benefit of not depending on the magnitude of the particles' momenta, but only their direction.
Therefore, it is often more precisely measured.
The other two variables in the transverse plane rely on an accurate momentum reconstruction for the muon and the hadronic system.  
Unlike $\delta \phi_{T}$, the definition of $\phi_{3D}$ requires an estimation of the momentum transfer.
Figure~\ref{fig:phivs3D} shows the interactions using the $\texttt{G18}$ prediction for $\phi_{3D}$ with a turnover at low values.
The corresponding $\delta\phi_{T}$ distribution monotonically decreases at higher values.

\begin{figure}[htb!]
\centering
\begin{tikzpicture} \draw (0, 0) node[inner sep=0] {
\includegraphics[width=0.45\textwidth]{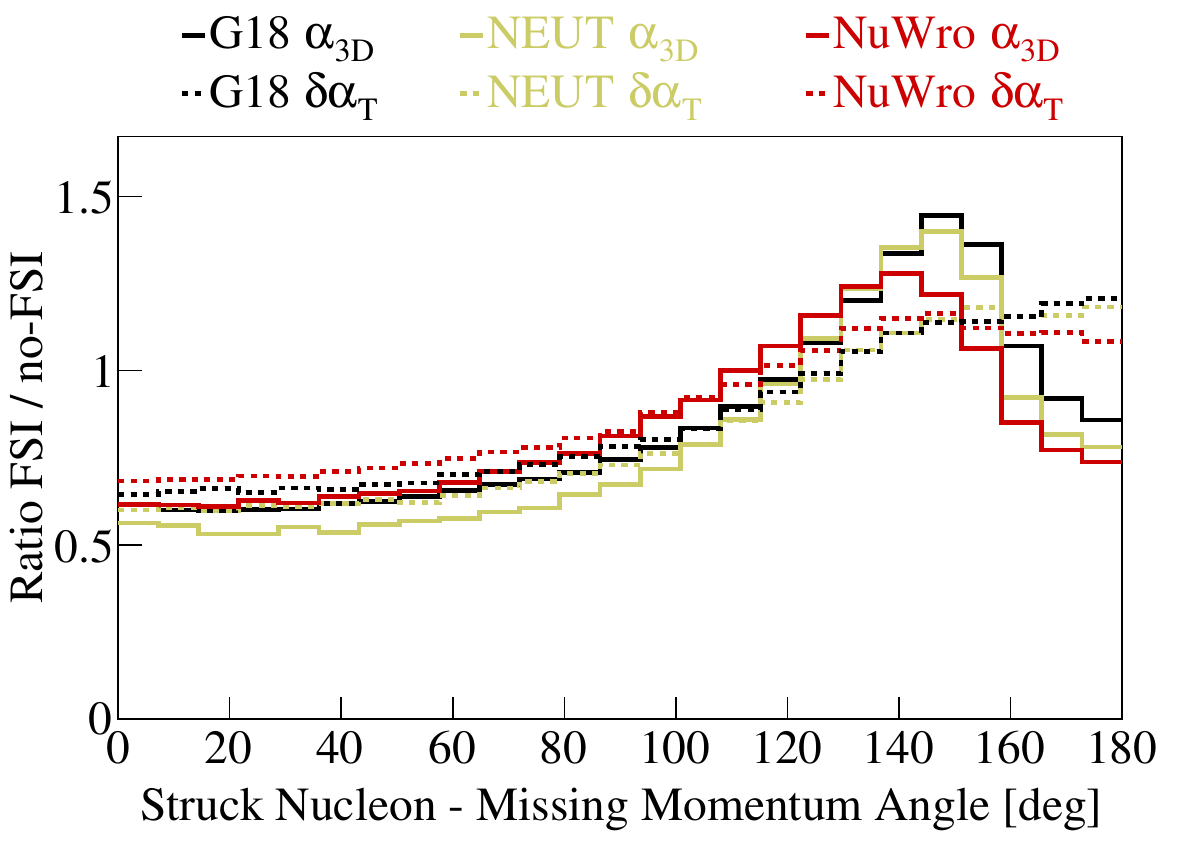}	
};
\draw (0.4, -3.2) node {(a)};	
\end{tikzpicture}
\hspace{0.05 \textwidth}
\begin{tikzpicture} \draw (0, 0) node[inner sep=0] {
\includegraphics[width=0.45\textwidth]{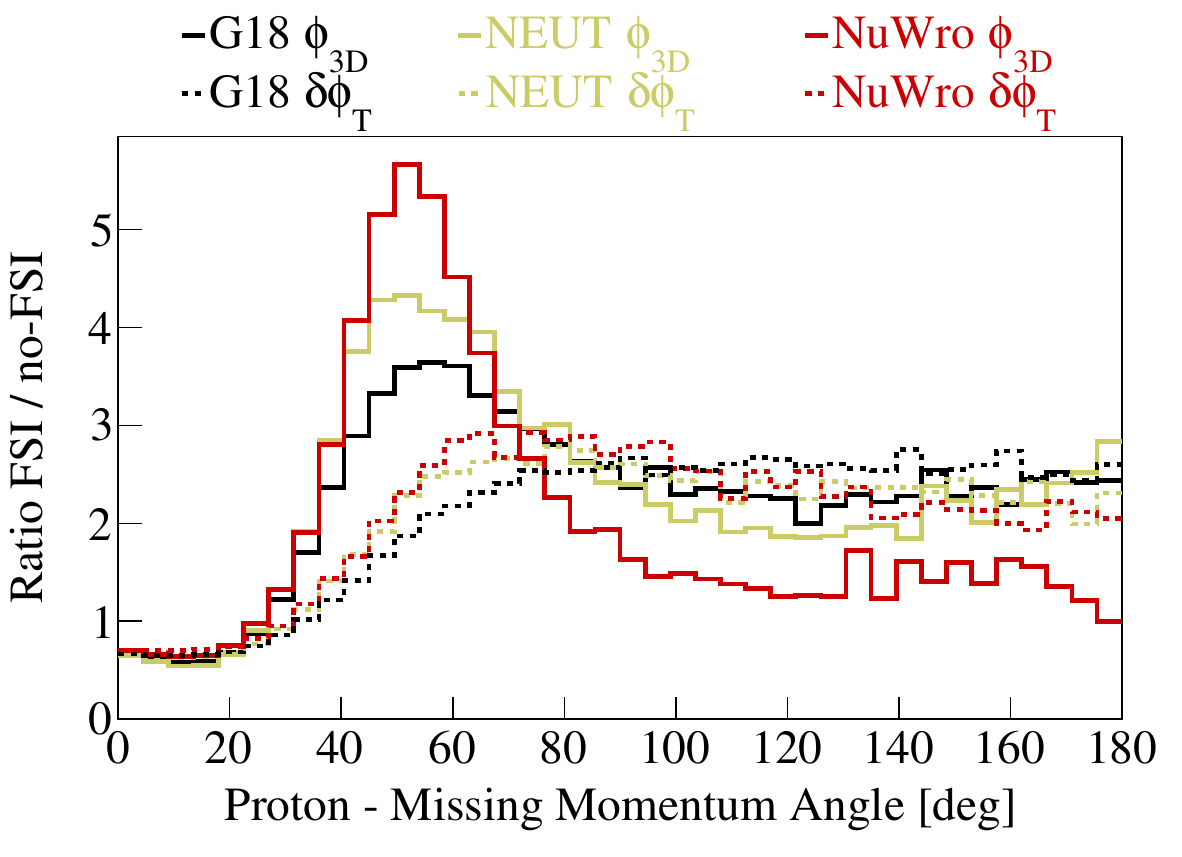}
};
\draw (0.4, -3.2) node {(b)};	
\end{tikzpicture}
\caption{
Ratios with and without FSI for $\texttt{G18}$, $\texttt{NEUT}$, and $\texttt{NuWro}$  predictions as a function of (a) the struck nucleon-missing momentum angle ($\delta \alpha_T$ or $\alpha_{3D}$) and (b) the proton-missing momentum angle ($\delta \phi_T$ or $\phi_{3D}$) for the selected CC1p0$\pi$ events.
}
\label{fig:ratio}
\end{figure}

The ratios with and without FSI are shown in Fig.~\ref{fig:ratio} for the stuck nucleon-missing momentum opening angles ($\delta \alpha_T$ or $\alpha_{3D}$) and the proton-missing momentum opening angles ($\delta \phi_T$ or $\phi_{3D}$) using several event generator predictions.
Shape differences become more pronounced and yield a larger range of ratios when the GKI variables are used when compared to the equivalent TKI results, indicating greater sensitivity to the details of FSI modeling.

\begin{figure}[htb!]
\centering
\begin{tikzpicture} \draw (0, 0) node[inner sep=0] {
\includegraphics[width=0.45\textwidth]{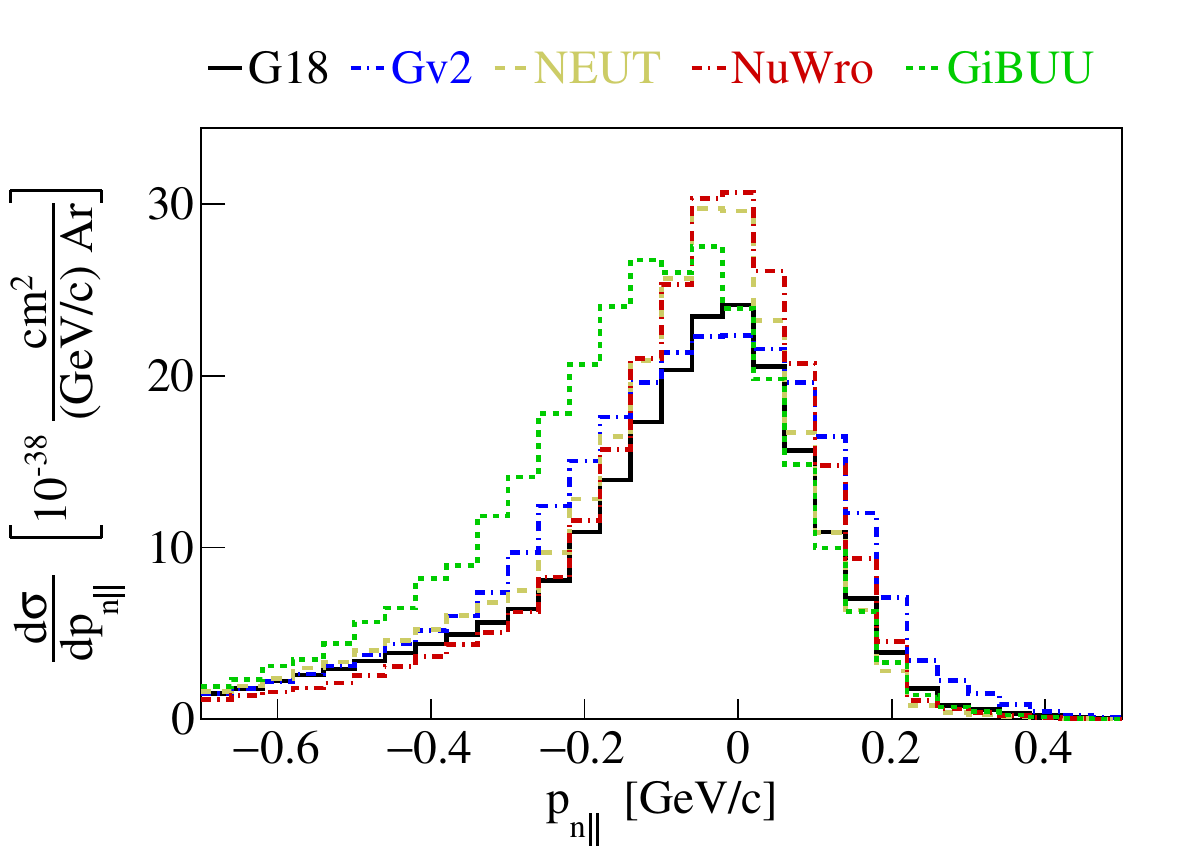}	
};
\draw (0.4, -3.2) node {(a)};	
\end{tikzpicture}
\hspace{0.05 \textwidth}
\begin{tikzpicture} \draw (0, 0) node[inner sep=0] {
\includegraphics[width=0.45\textwidth]{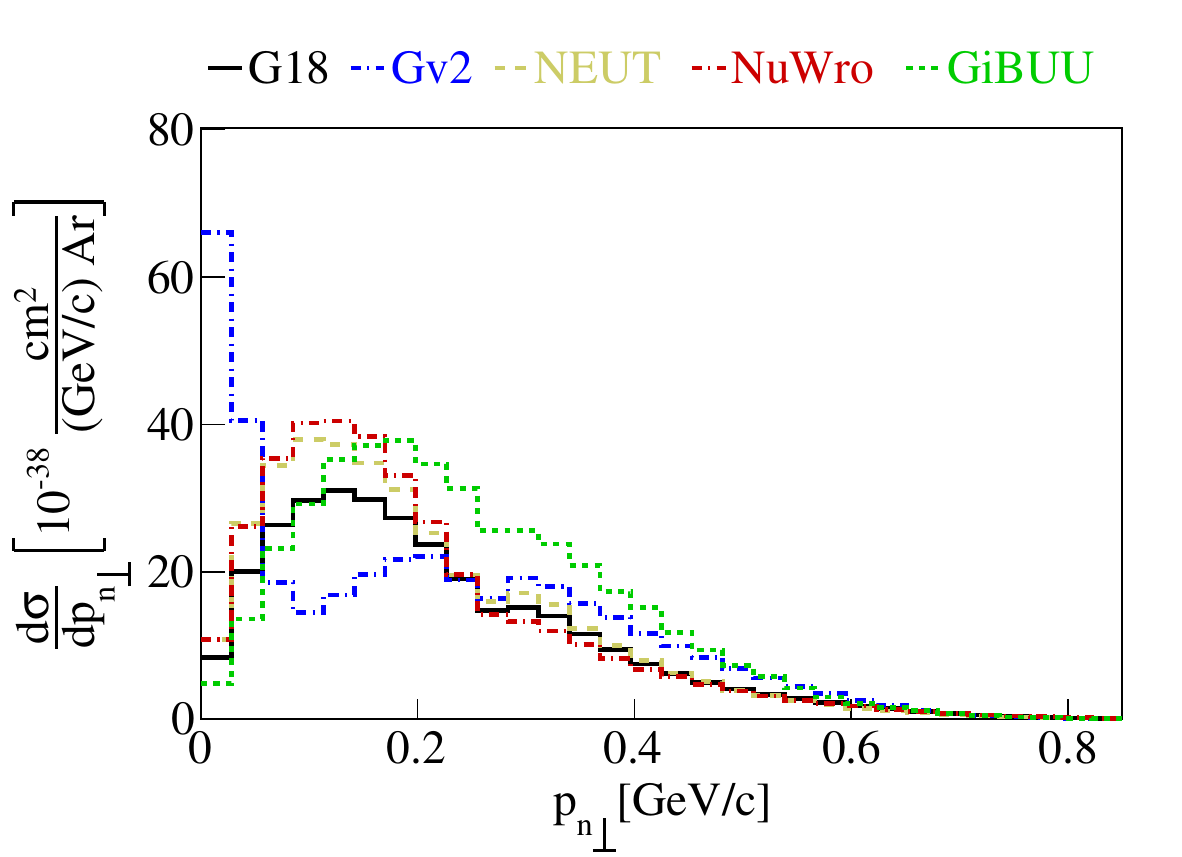}
};
\draw (0.4, -3.2) node {(b)};	
\end{tikzpicture}
\caption{
The flux-integrated single-differential cross section as a function of (a) $p_{n\,\parallel}$ and (b) $p_{n\,\perp}$ for the selected CC1p0$\pi$ events.
Colored lines show the results of cross section calculations using the $\texttt{G18}$ (solid black), $\texttt{Gv2}$ (blue), $\texttt{NEUT}$ (dashed pink), $\texttt{NuWro}$ (red) and $\texttt{GiBUU}$ (green) generators.
}
\label{fig:projfig}
\end{figure}

\begin{figure}[htb!]
\centering
\begin{tikzpicture} \draw (0, 0) node[inner sep=0] {
\includegraphics[width=0.45\textwidth]{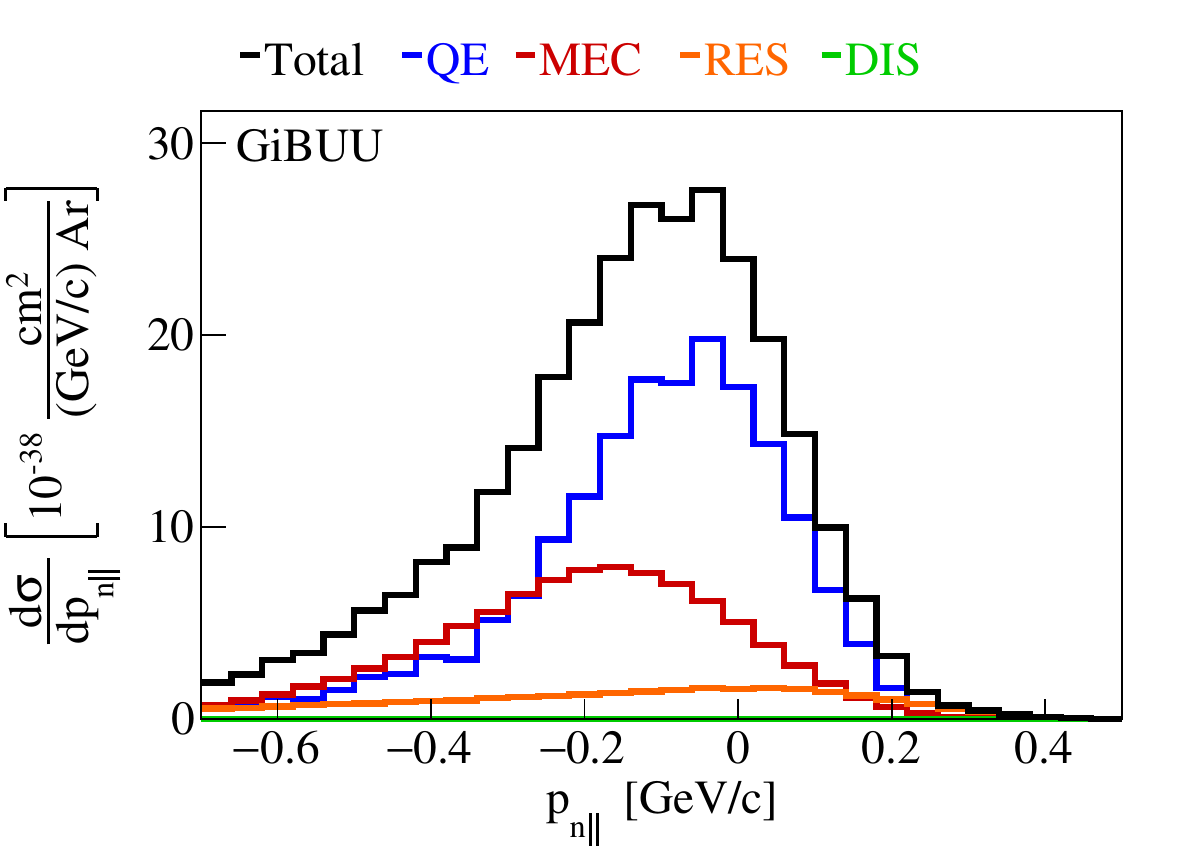}	
};
\draw (0.4, -3.2) node {(a)};	
\end{tikzpicture}
\hspace{0.05 \textwidth}
\begin{tikzpicture} \draw (0, 0) node[inner sep=0] {
\includegraphics[width=0.45\textwidth]{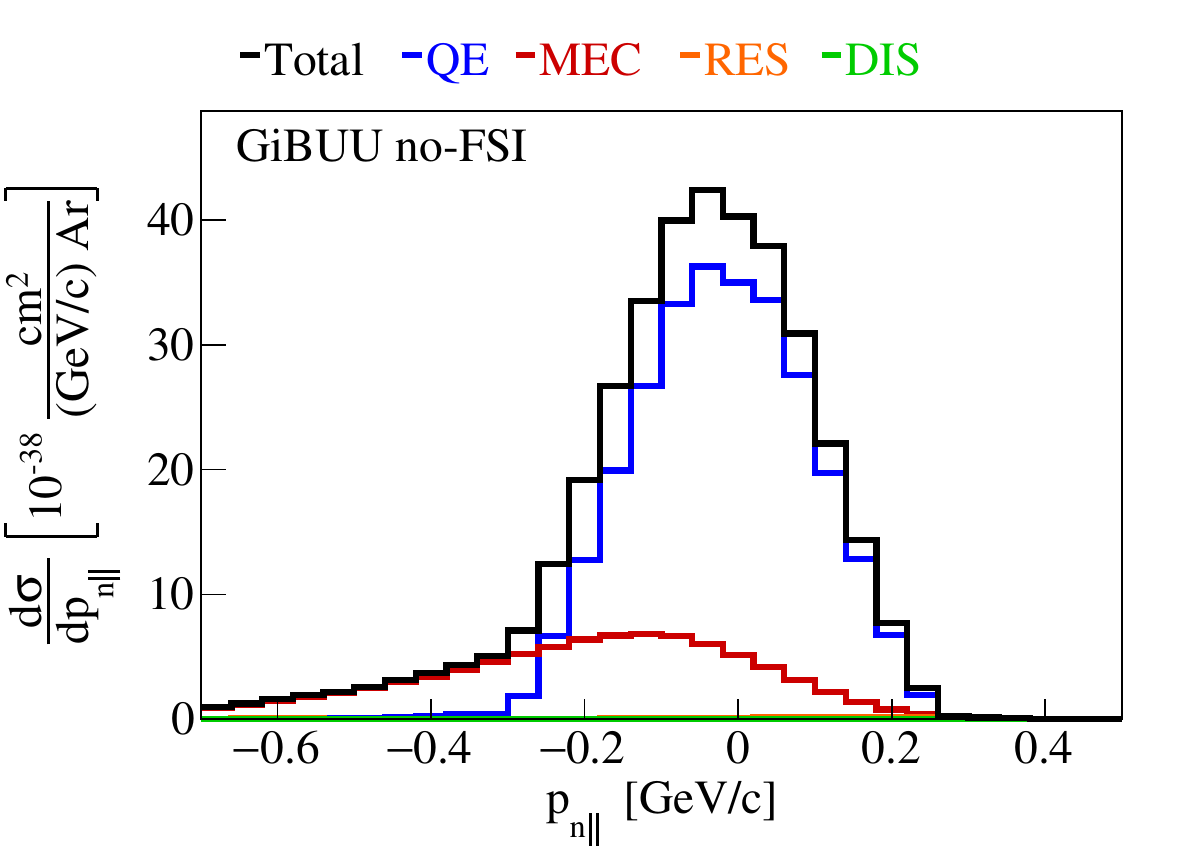}
};
\draw (0.4, -3.2) node {(b)};	
\end{tikzpicture}
\caption{
The flux-integrated single-differential cross section interaction breakdown as a function of $p_{n\parallel}$ (a) with FSI and (b) without FSI for the selected CC1p0$\pi$ events.
Colored lines show the results of cross section calculations using the $\texttt{GiBUU}$ prediction for QE (blue), MEC (red), RES (orange), and DIS (green) generators.
}
\label{fig:GiBUUPnPara}
\end{figure}

The projection variables $p_{n \parallel}$ and $p_{n \perp}$ vary when studied using several generators.
Like its transverse equivalent, $\delta p_{T_y}$~\cite{RefPRD,PhysRevD.101.092001}, $p_{n\parallel}$ shows an asymmetric behavior due to the tendency for FSI to decelerate reinteracting hadrons [Fig.~\ref{fig:projfig}(a)].
In addition, $\texttt{GiBUU}$ exhibits an offset by 0.15\,GeV/c to smaller values, unlike any of the other event generators where the peak is centered around 0\,GeV/c.
The $\texttt{GiBUU}$ interaction breakdown in Fig.~\ref{fig:GiBUUPnPara}(a) shows that the shift is driven both by the QE and MEC contributions.
In the absence of FSI, the QE offset is no longer present and the asymmetric behavior is driven by the MEC contribution.

$\texttt{GiBUU}$ shows a shift to higher values compared to the other generators for the $p_{n\perp}$ (Fig.~\ref{fig:projfig}).
The $\texttt{Gv2}$ distributions are significantly different compared to the other more modern generators, especially at lower values.
In this older version, the hadronic and leptonic kinematics are generated independently which violates required correlations.
This is modified for QE interactions in more recent versions (such as $\texttt{G18}$), resulting in smooth distributions that are similar to the other generators.
The corresponding $p_{n\perp,x}$ and $p_{n\perp,y}$ distributions predict a much sharper peak around 0\,GeV/c for $\texttt{Gv2}$ compared to the other generators (Fig.~\ref{fig:projcomp}).
The results without FSI are presented in the Supplemental Material.

The two perpendicular projection variables, $p_{n\perp,x}$ and $p_{n\perp,y}$, show some shape differences which naively might not be expected.
Due to the way these variables are defined, $p_{n\perp,y}$ is always perpendicular to the neutrino-muon scattering plane, which contains the vector $\vec{q}$. 
As there is no directional preference for nuclear motion, this is symmetric.
Similarly, $p_{n\perp,x}$ is also perpendicular to $\vec{q}$, so naively it should also be symmetric.
However, the variable is defined in the scattering plane.
In this case missing energy can lead to small differences between the estimated direction of $\vec{q}$ and the true momentum transfer vector.
This leads to a slightly wider distribution with a tail at negative $p_{n\perp,x}$ values.

\begin{figure}[htb!]
\centering
\begin{tikzpicture} \draw (0, 0) node[inner sep=0] {
\includegraphics[width=0.45\textwidth]{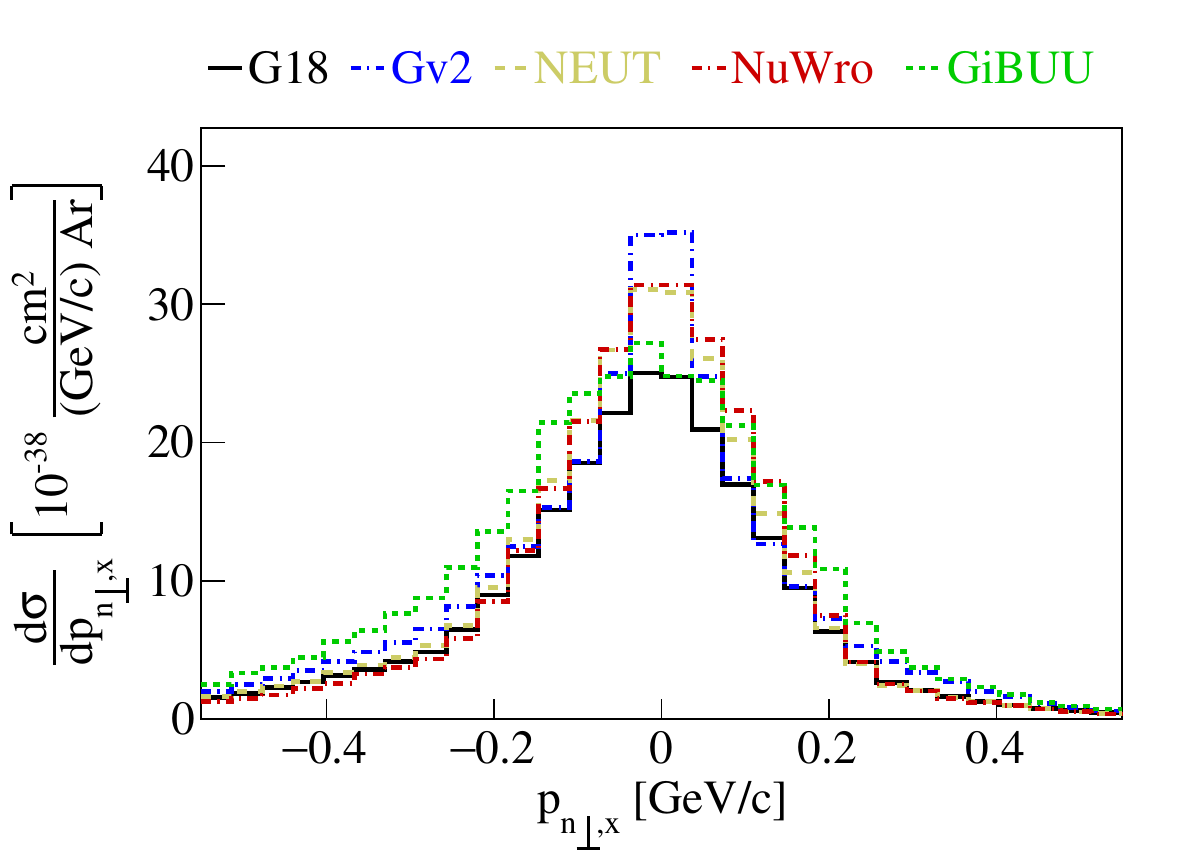}	
};
\draw (0.4, -3.2) node {(a)};	
\end{tikzpicture}
\hspace{0.05 \textwidth}
\begin{tikzpicture} \draw (0, 0) node[inner sep=0] {
\includegraphics[width=0.45\textwidth]{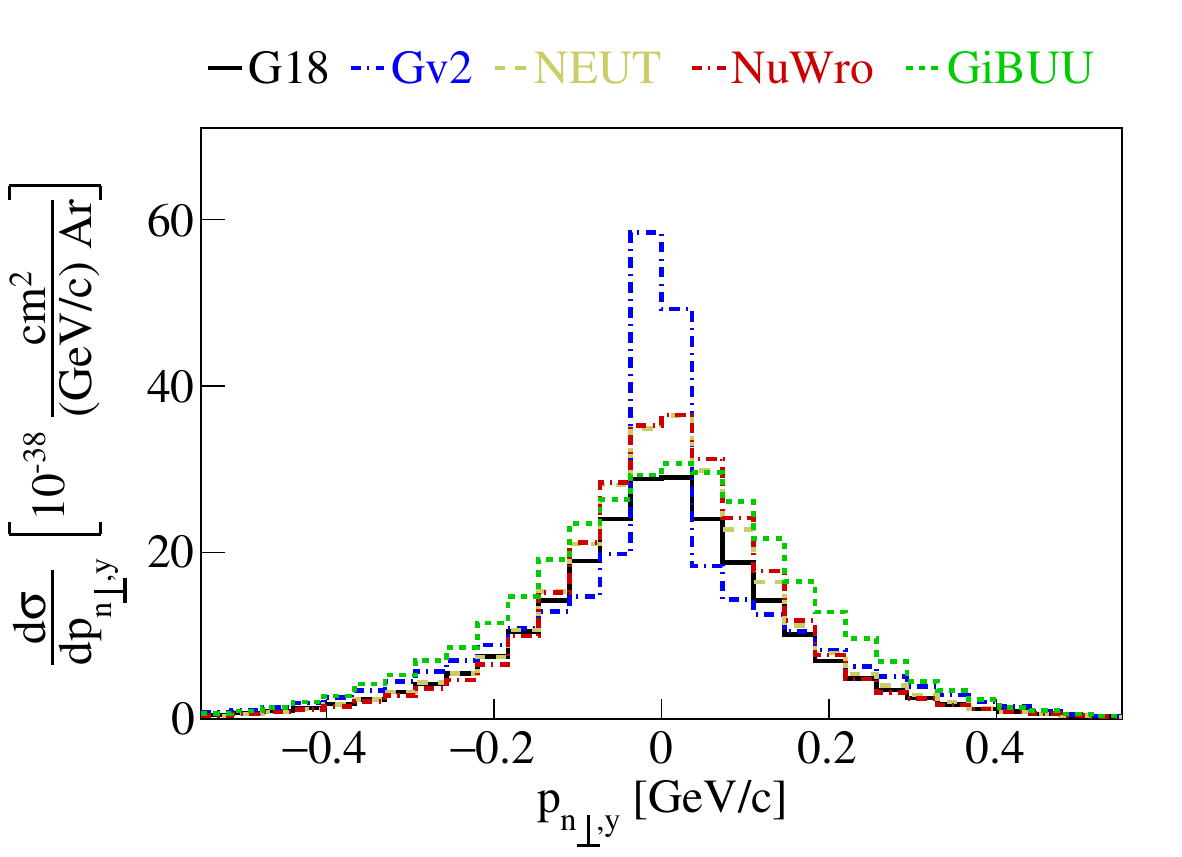}
};
\draw (0.4, -3.2) node {(b)};	
\end{tikzpicture}
\caption{
The flux-integrated single-differential cross section as a function of (a) $p_{n\,\perp,x}$ and (b) $p_{n\,\perp,y}$ for the selected CC1p0$\pi$ events.
Colored lines show the results of cross section calculations using the $\texttt{G18}$ (solid black), $\texttt{Gv2}$ (blue), $\texttt{NEUT}$ (dashed pink), $\texttt{NuWro}$ (red) and $\texttt{GiBUU}$ (green) generators.
}
\label{fig:projcomp}
\end{figure}

\begin{figure}[htb!]
\centering
\begin{tikzpicture} \draw (0, 0) node[inner sep=0] {
\includegraphics[width=0.45\textwidth]{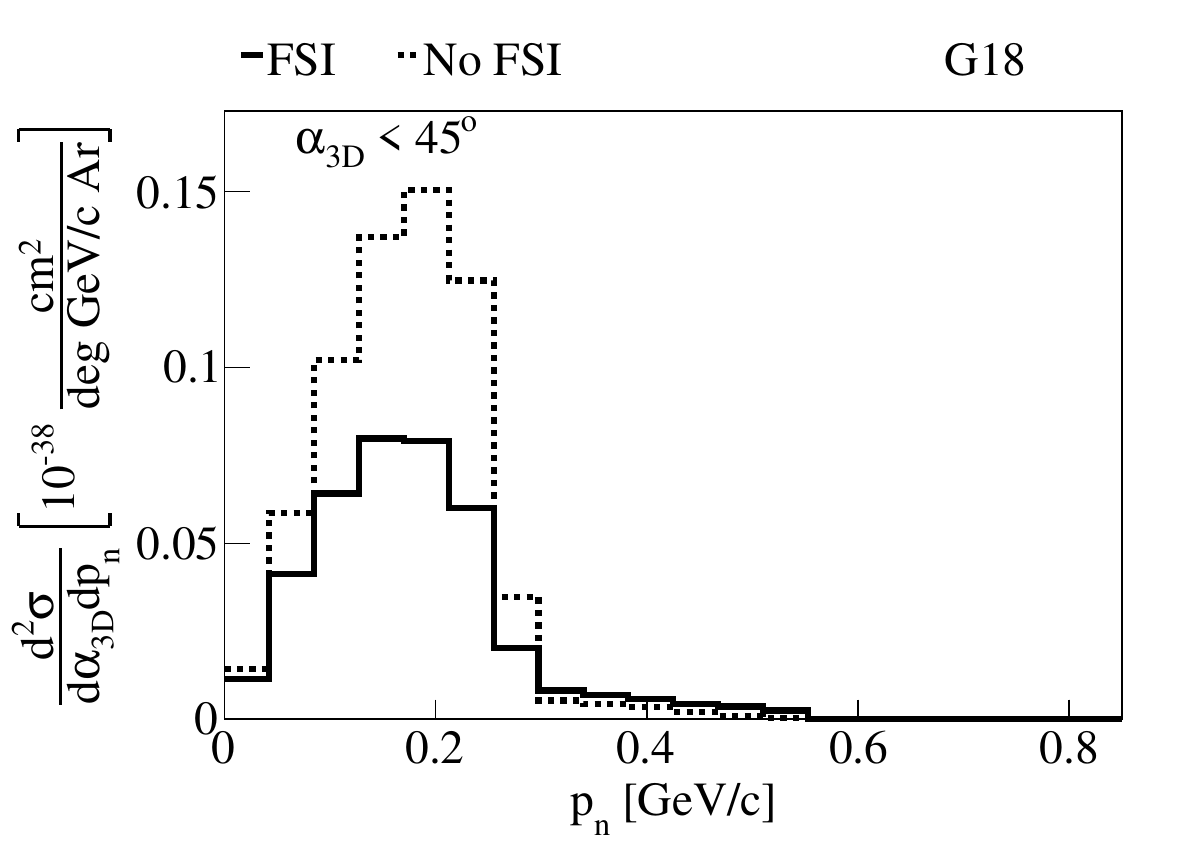}	
};
\draw (0.4, -3.2) node {(a)};	
\end{tikzpicture}
\hspace{0.05 \textwidth}
\begin{tikzpicture} \draw (0, 0) node[inner sep=0] {
\includegraphics[width=0.45\textwidth]{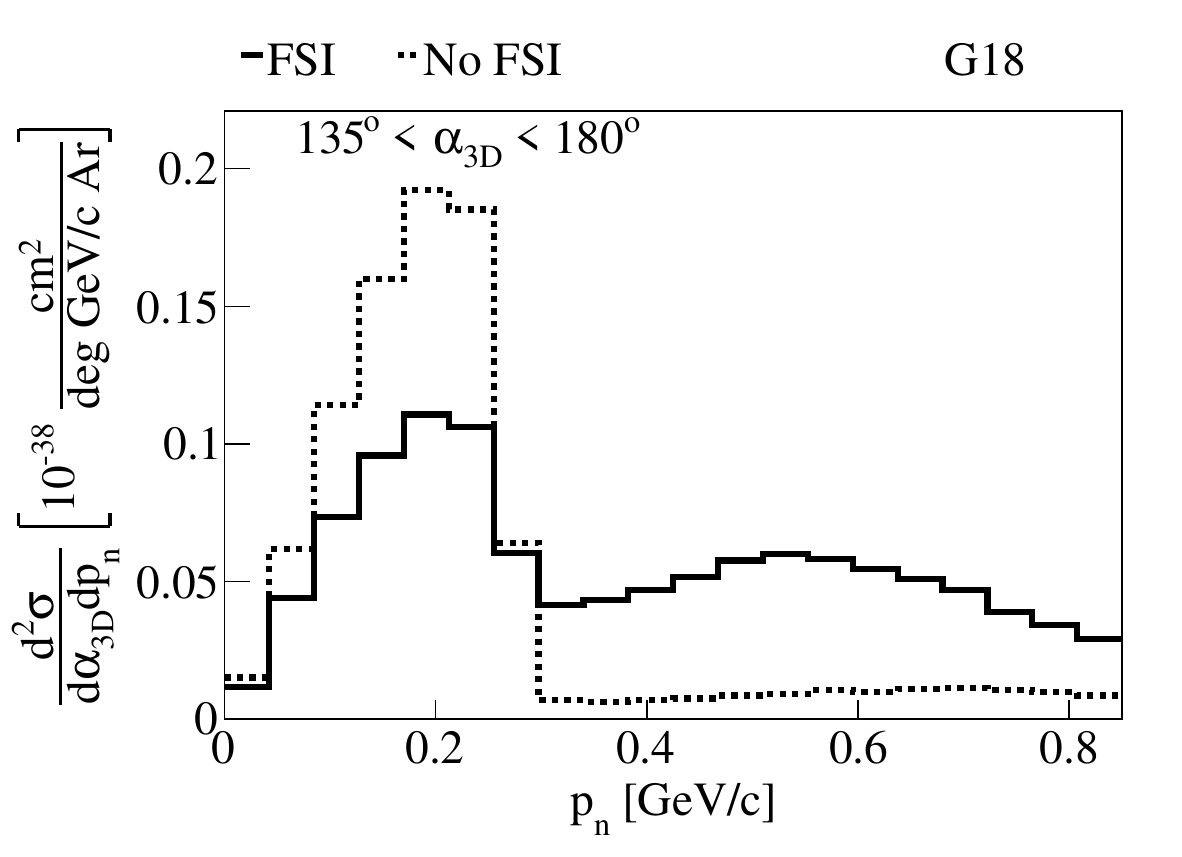}
};
\draw (0.4, -3.2) node {(b)};	
\end{tikzpicture}
\caption{
Comparison of the flux-integrated double-differential cross section as a function of $p_{n}$ for (a) $\alpha_{3D} <$ 45$^{\circ}$ and (b) 135$^{\circ}$ $< \alpha_{3D} <$ 180$^{\circ}$ with (solid) and without (dashed) FSI effects using the $\texttt{G18}$ prediction for the selected CC1p0$\pi$ events.
}
\label{fig:pn_highalpha_fsi}
\end{figure}

\begin{figure}[htb!]
\centering
\begin{tikzpicture} \draw (0, 0) node[inner sep=0] {
\includegraphics[width=0.45\textwidth]{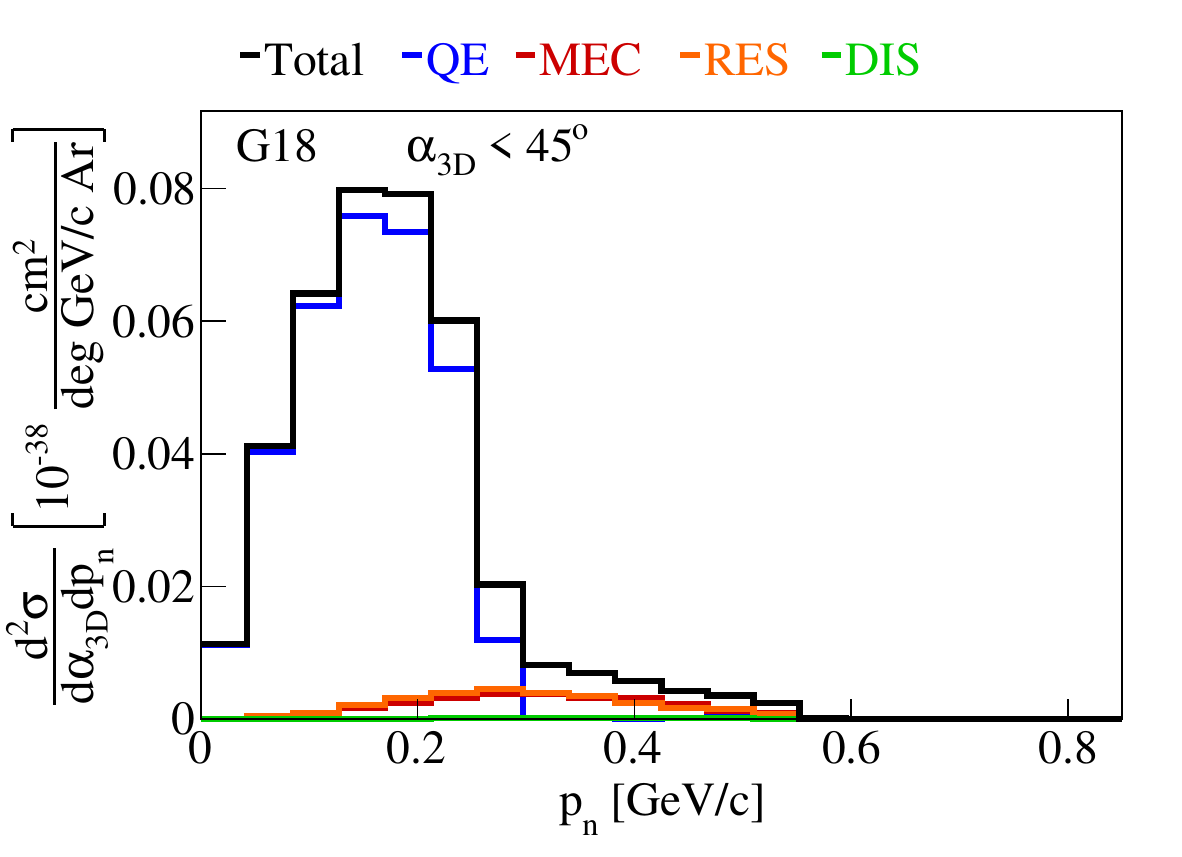}	
};
\draw (0.4, -3.2) node {(a)};	
\end{tikzpicture}
\hspace{0.05 \textwidth}
\begin{tikzpicture} \draw (0, 0) node[inner sep=0] {
\includegraphics[width=0.45\textwidth]{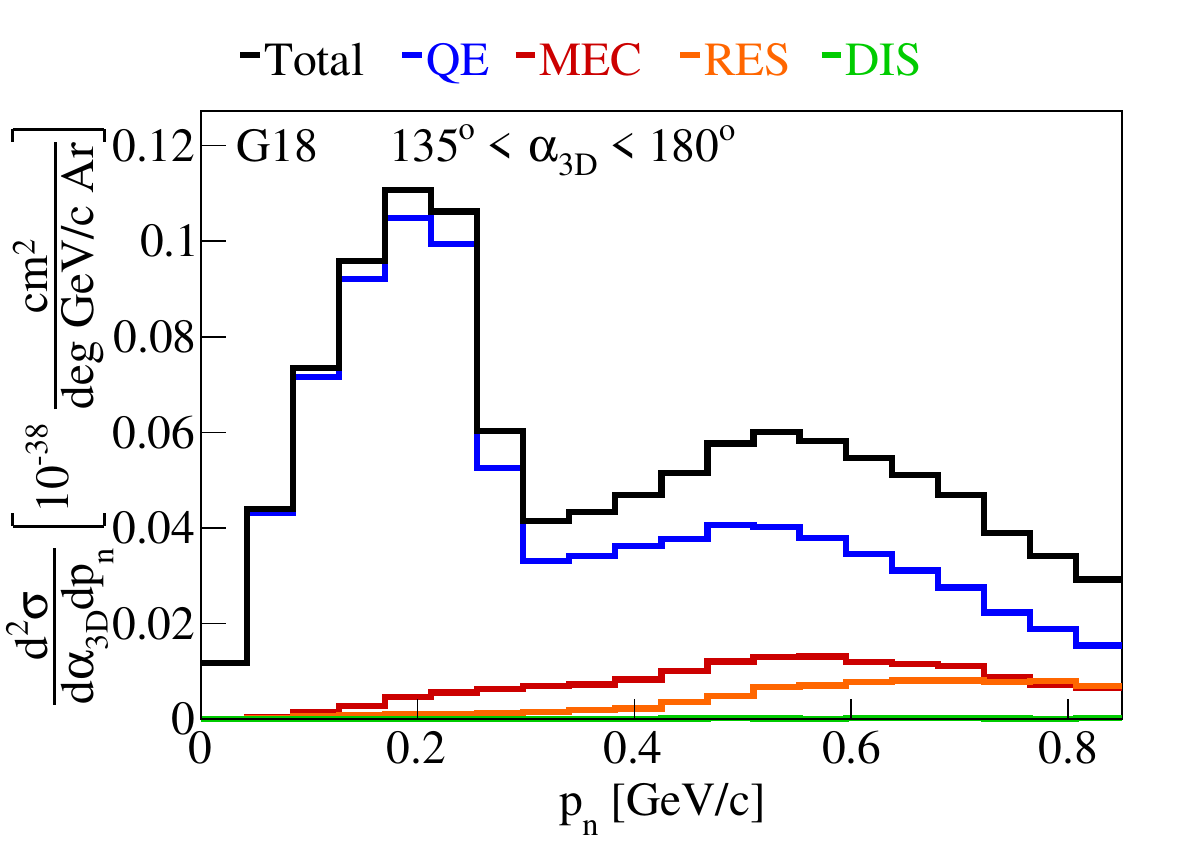}
};
\draw (0.4, -3.2) node {(b)};	
\end{tikzpicture}
\caption{
The flux-integrated double-differential cross section interaction breakdown as a function of $p_{n\,\parallel}$ for (a) $\alpha_{3D} <$ 45$^{\circ}$ and (b) 135$^{\circ}$ $< \alpha_{3D} <$ 180$^{\circ}$ for the selected CC1p0$\pi$ events.
Colored lines show the results of theoretical cross section calculations using the $\texttt{G18}$ prediction for QE (blue), MEC (red), RES (orange), and DIS (green) interactions.
}
\label{fig:pn_alphasplit}
\end{figure}

The sensitivity to nuclear effects of these variables becomes even more pronounced when performing double-differential measurements.
As mentioned earlier, the angle $\alpha_{3D}$ is sensitive to FSI, with higher values of $\alpha_{3D}$ corresponding to events that primarily undergo FSI.  
The double-differential $p_n$ cross section illustrates significant differences depending on whether FSI have been added or not (Fig.~\ref{fig:pn_highalpha_fsi}).
For events with $\alpha_{3D} <$ 45$^{\circ}$, FSI leads to a significant reduction in the normalization of the peak, while the high missing momentum tail is largely unaffected.
Conversely, events with 135$^{\circ}$ $< \alpha_{3D} <$ 180$^{\circ}$ yield a significantly enhanced high-$p_n$ tail when FSI are included.
That feature makes the high-$\alpha_{3D}$ region ideal to study the impact of FSI effects.
We can study the FSI impact on $p_{n}$ by isolating two groups of events, those with $\alpha_{3D}$ values $>$ 135$^{\circ}$ and with $\alpha_{3D}$ $<$ 45$^{\circ}$.
Figure~\ref{fig:pn_alphasplit}(a) shows the QE-dominated low-$\alpha_{3D}$ region with a tail that vanishes at $\approx 0.5$\,GeV/c. 
The primary cause of the high-$p_n$ tail in this region is interactions that produce additional undetected particles, and therefore consist mainly of non-QE interactions.
Conversely, the high-$\alpha_{3D}$ region shown in Fig.~\ref{fig:pn_alphasplit}(b) illustrates a much wider tail that extends up to $\approx 1$\,GeV/c.
This tail has a significant contribution from MEC and RES events, as well as a large population of events from QE interactions that undergo FSI and therefore yield higher $p_n$ values.

\begin{figure}[htb!]
\centering
\begin{tikzpicture} \draw (0, 0) node[inner sep=0] {
\includegraphics[width=0.45\textwidth]{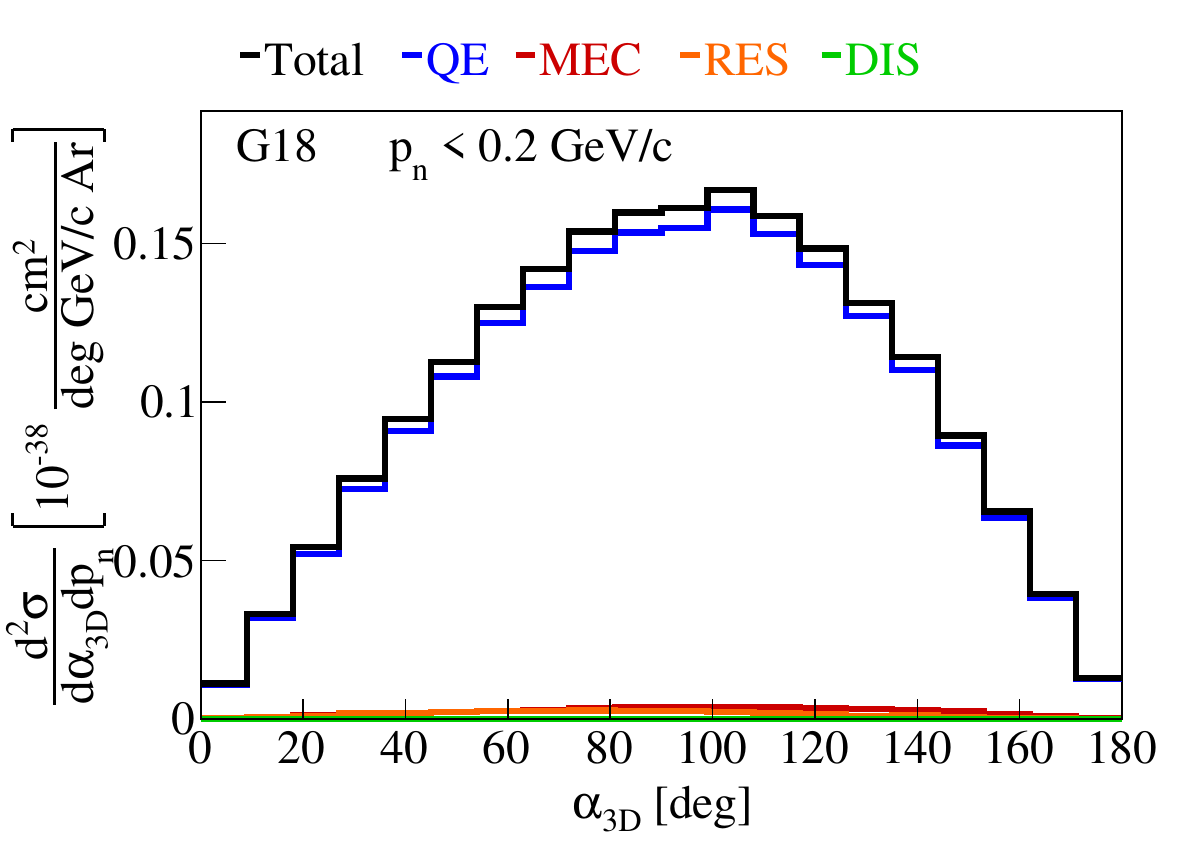}	
};
\draw (0.4, -3.2) node {(a)};	
\end{tikzpicture}
\hspace{0.05 \textwidth}
\begin{tikzpicture} \draw (0, 0) node[inner sep=0] {
\includegraphics[width=0.45\textwidth]{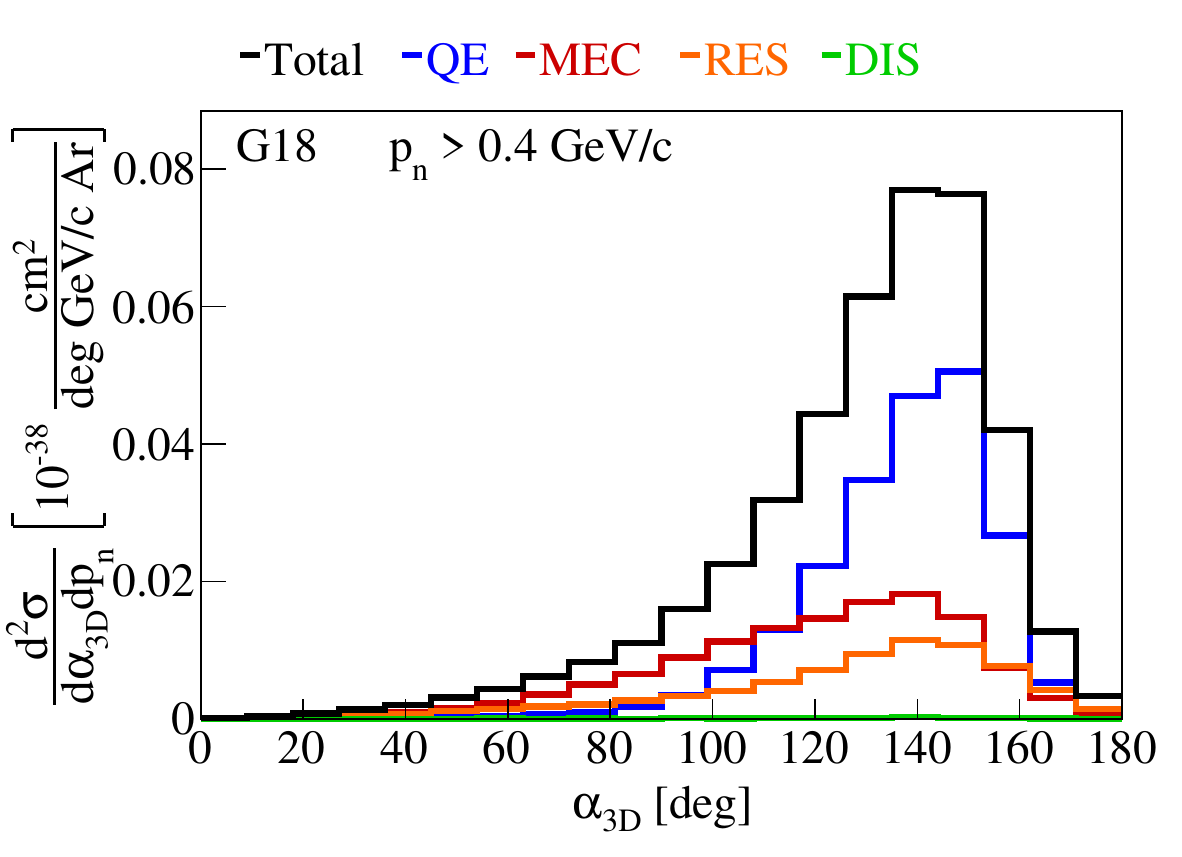}
};
\draw (0.4, -3.2) node {(b)};	
\end{tikzpicture}
\caption{
The flux-integrated double-differential cross sections as a function of $\alpha_{3D}$ for (a) $p_{n} <$ 0.2\,GeV/c and (b) $p_{n} >$ 0.4\,GeV/c for the selected CC1p0$\pi$ events.
Colored lines show the results of cross section calculations using the $\texttt{G18}$ prediction for QE (blue), MEC (red), RES (orange), and DIS (green) predictions.
}
\label{fig:alpha_pnsplit}
\end{figure}

\begin{figure}[htb!]
\centering
\begin{tikzpicture} \draw (0, 0) node[inner sep=0] {
\includegraphics[width=0.45\textwidth]{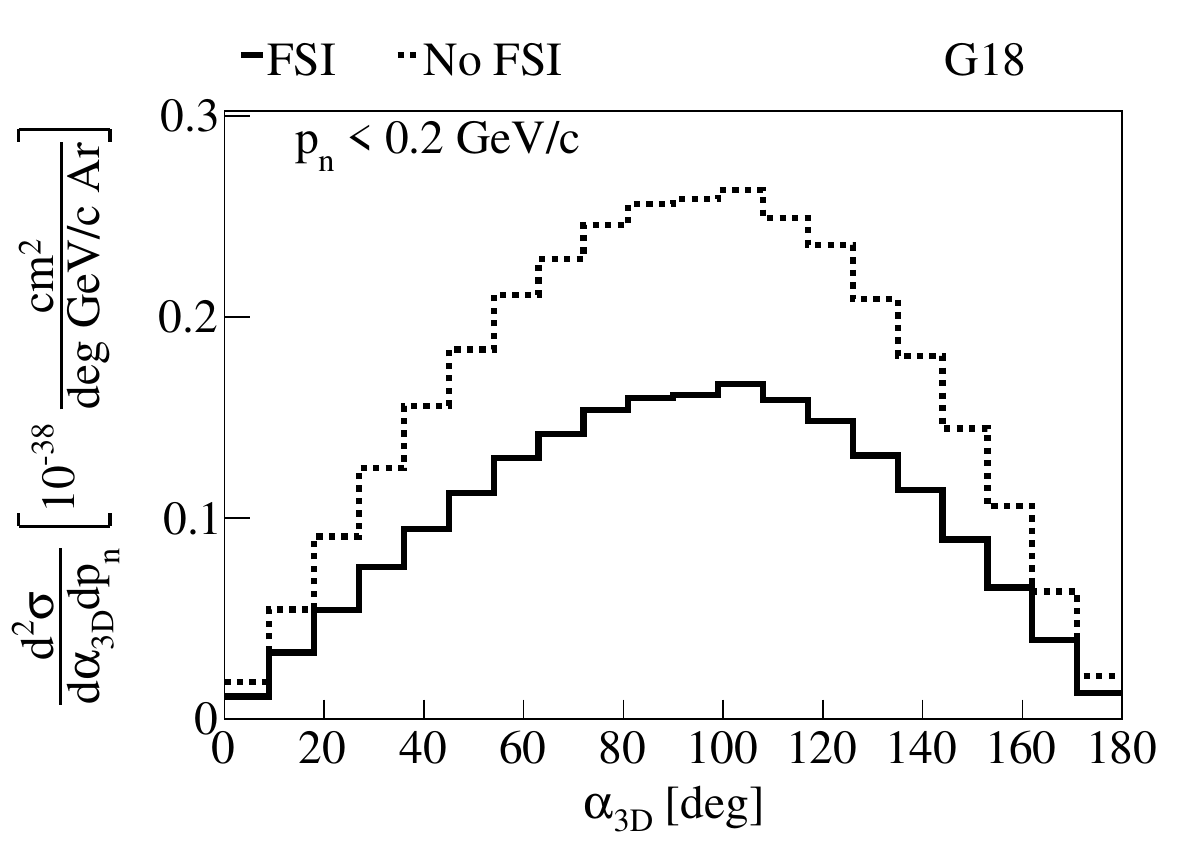}	
};
\draw (0.4, -3.2) node {(a)};	
\end{tikzpicture}
\hspace{0.05 \textwidth}
\begin{tikzpicture} \draw (0, 0) node[inner sep=0] {
\includegraphics[width=0.45\textwidth]{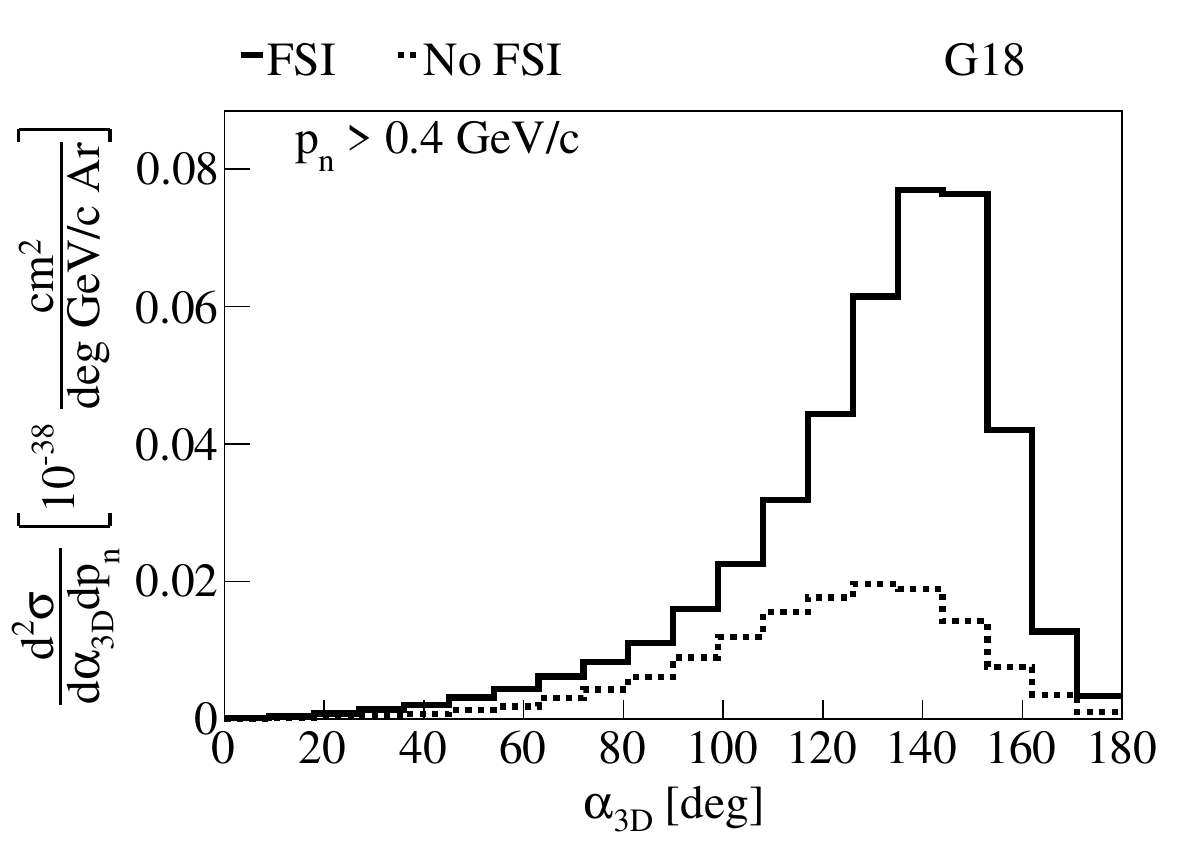}
};
\draw (0.4, -3.2) node {(b)};	
\end{tikzpicture}
\caption{
The flux-integrated double-differential cross sections as a function of $\alpha_{3D}$ for (a) $p_{n} <$ 0.2\,GeV/c and (b) $p_{n} >$ 0.4\,GeV/c with (solid) and without (dashed) FSI effects using the $\texttt{G18}$ prediction for the selected CC1p0$\pi$ events.
}
\label{fig:pn_highpn_fsi}
\end{figure}

A complementary way to group the selected events is by missing momentum ($p_n <$ 0.2\,GeV/c and $p_n >$ 0.4\,GeV/c).
The double-differential cross section in $\alpha_{3D}$ yields a distribution that approximately follows the expected sine-curve behavior when events with low missing momentum are used [Fig.~\ref{fig:alpha_pnsplit}(a)].
The same shape is observed when FSI are turned off [Fig.~\ref{fig:pn_highpn_fsi}(a)].
A different shape is observed in Fig.~\ref{fig:alpha_pnsplit}(b) for events with high missing momentum with a pronounced peak at high $\alpha_{3D}$ values, which is driven by QE events.
On the other hand, MEC and RES events result in a wider range of $\alpha_{3D}$ angles and less peaked distributions, illustrating that the missing momentum has less directional preference.
Furthermore, the region where $\alpha_{3D} <$ 90$^{\circ}$ is dominated by MEC events across all event generators due to the missing momentum introduced by undetected particles.
The prediction without FSI yields a smoother distribution [Fig.~\ref{fig:pn_highpn_fsi}(b)].

\section{MicroBooNE Cross Section Measurement} \label{sec:MicroBooNE}

To perform the first measurement of the cross section as a function of these variables, we use three years of data collected by the MicroBooNE detector.
MicroBooNE is an 85-tonne active mass liquid argon time projection chamber in the Booster Neutrino Beam (BNB).
The MicroBooNE detector is described in detail in Ref.~\cite{Acciarri:2016smi}.
We use the same event selection and measurement strategy as used in Refs.~\cite{RefPRL,RefPRD}, and the same CC1p0$\pi$ definition for the variables of interest as described in Sec.~\ref{sec:formalism}.

Data are processed by filtering noise and deconvolving the wire response to produce unipolar signals, which are then fitted with Gaussian functions to produce hits.
The $\texttt{Pandora}$ multi-algorithm pattern recognition package~\cite{pandora} is used to cluster hits, match them across wire planes, and construct tracks and showers.
Topological and optical information are used to identify and remove cosmic tracks.  
The remaining tracks and showers are grouped into neutrino candidates.
Events are selected by requiring a neutrino candidate with exactly two reconstructed tracks by applying particle identification requirements based on $dE/dx$ measurements to ensure the tracks are muon-like or proton-like, respectively.
More details on the selection can be found in Ref.~\cite{RefPRD}.


Uncertainties related to the incident neutrino flux~\cite{AguilarArevalo:2008yp}, interaction model~\cite{GENIE_tune}, particle propagation~\cite{Geant4}, and detector response~\cite{uBooNE_det_unc} are assessed separately to produce a covariance matrix describing the uncertainty of the predicted event rate.
The binning is chosen to balance resolution and statistics.  
The Wiener-Singular Value Decomposition unfolding technique~\cite{Tang_2017} is used to transform both the data measurement and covariance matrix into a regularized phase space.
The technique requires the construction of a response matrix describing the expected detector smearing and reconstruction efficiency, for which it corrects. 
The unfolding is performed for each one of the observables of interest using the $\texttt{G18T}$ model.
Each measurement is accompanied by an output additional smearing matrix $A_{C}$ which performs the conversion from the true to the regularized phase space.
The $A_{C}$ matrix is included in the Supplemental Material and needs to be applied to all theory predictions in order to compare to the data measurements, even though its effect is small. 

The robustness of the unfolding method was tested using fake data studies with alternative generator predictions, which are presented in the Supplemental Material.
Namely, we investigated three fake data samples: (a) using NuWro events, (b) by removing the weights corresponding to the MicroBooNE tune and, (c) by multiplying the weight for the MEC events by a factor of two. 
We then extracted the cross section from these fake data using our nominal MC response matrices, as well as the Wiener-SVD filter when only the covariances related to the relevant uncertainties were included.
We found that the combination of these uncertainties covered the difference between the unfolded fake data prediction and the corresponding alternative-generator theory prediction.
Additionally, the aforementioned comparisons yielded $\chi^{2}$/ndof values below unity and p-values close to 1, further supporting the robustness of our unfolding procedure.

The unfolded event rate has the predicted background subtracted before the unfolding.
It is further divided by the integrated neutrino flux and number of argon nuclei in the fiducial volume to report a differential cross section.
In the results presented below, the inner error bars on the cross sections correspond to the data statistical uncertainties.
The systematic uncertainties are decomposed into data shape- and normalization-related sources following the procedure outlined in Ref.~\cite{MatrixDecomv}.
The cross-term uncertainties are incorporated in the normalization.
The outer error bars on the reported cross sections correspond to data statistical and shape uncertainties added in quadrature.
The data normalization uncertainties are presented as a band at the bottom of each plot.
Overflow (underflow) values are included in the last (first) bin.
The degrees of freedom correspond to the number of bins.
The $\chi^{2}$/ndf data comparison for each generator shown on all the figures takes into account the total covariance matrix.
More details on the systematic uncertainties and the cross-section extraction technique can be found in Ref.~\cite{RefPRD}.



\begin{figure}[htb!]
  \centering
\begin{tikzpicture} \draw (0, 0) node[inner sep=0] {
\includegraphics[width=0.45\textwidth]{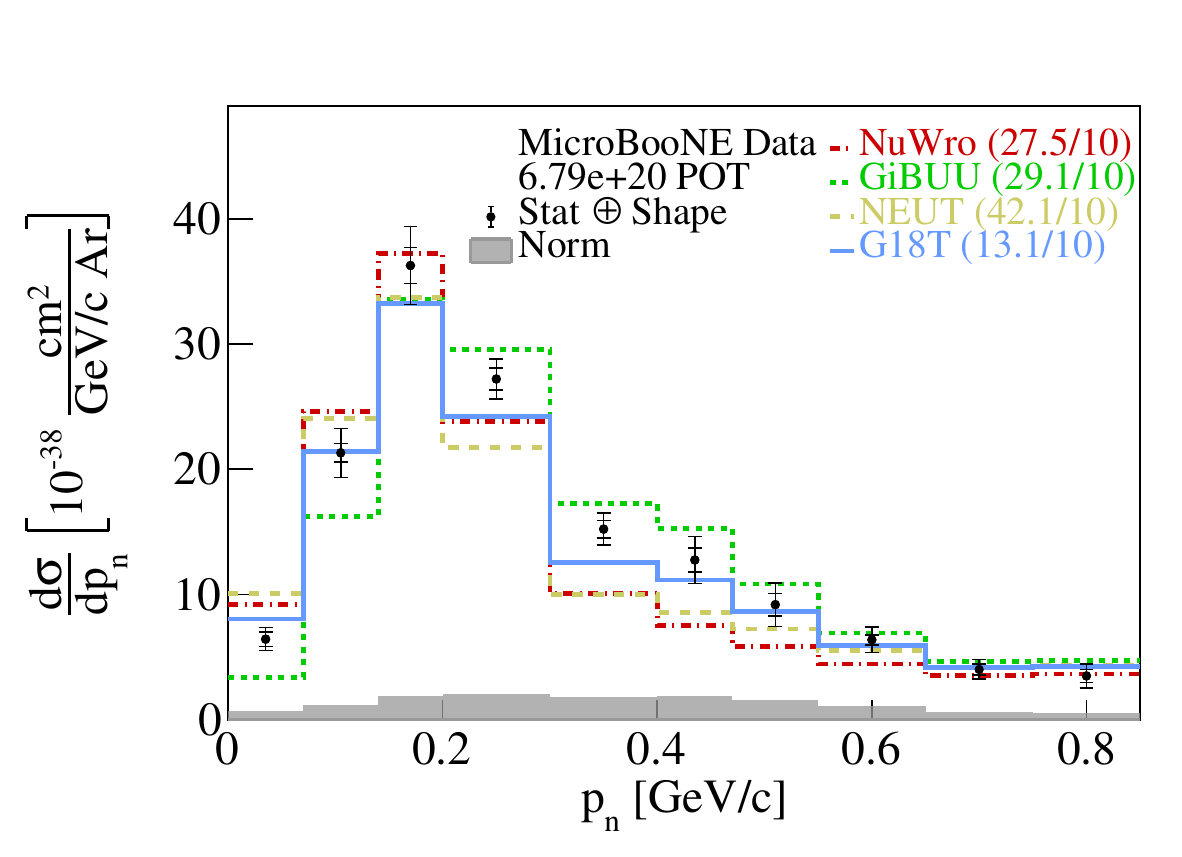}	
};
\draw (0.4, -3.2) node {(a)};	
\end{tikzpicture}
\hspace{0.05 \textwidth}
\begin{tikzpicture} \draw (0, 0) node[inner sep=0] {
\includegraphics[width=0.45\textwidth]{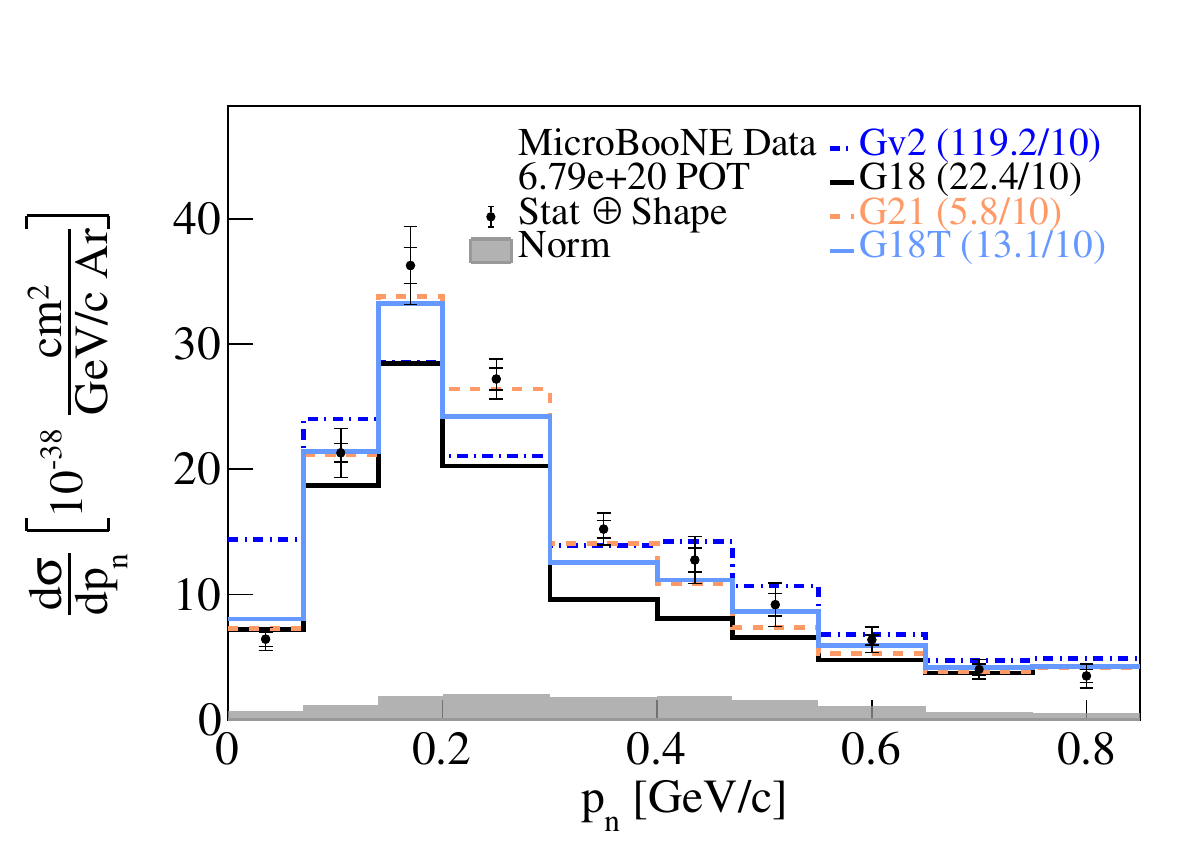}
};
\draw (0.4, -3.2) node {(b)};	
\end{tikzpicture}
  \caption{
The flux-integrated single-differential cross sections as a function of $p_{n}$ with (a) generator and (b) $\texttt{GENIE}$ configuration 
predictions compared to data. Inner and outer error bars show the statistical and total (statistical and shape systematic)
uncertainty at the 1$\sigma$, or 68\%, confidence level. The gray band shows the normalization systematic uncertainty. The numbers
in parentheses give the $\chi^{2}$/ndf calculation for each one of the predictions.
  }
  \label{fig:deltaPn_allevents}
\end{figure}

\begin{figure}[htb!]
  \centering
\begin{tikzpicture} \draw (0, 0) node[inner sep=0] {
\includegraphics[width=0.45\textwidth]{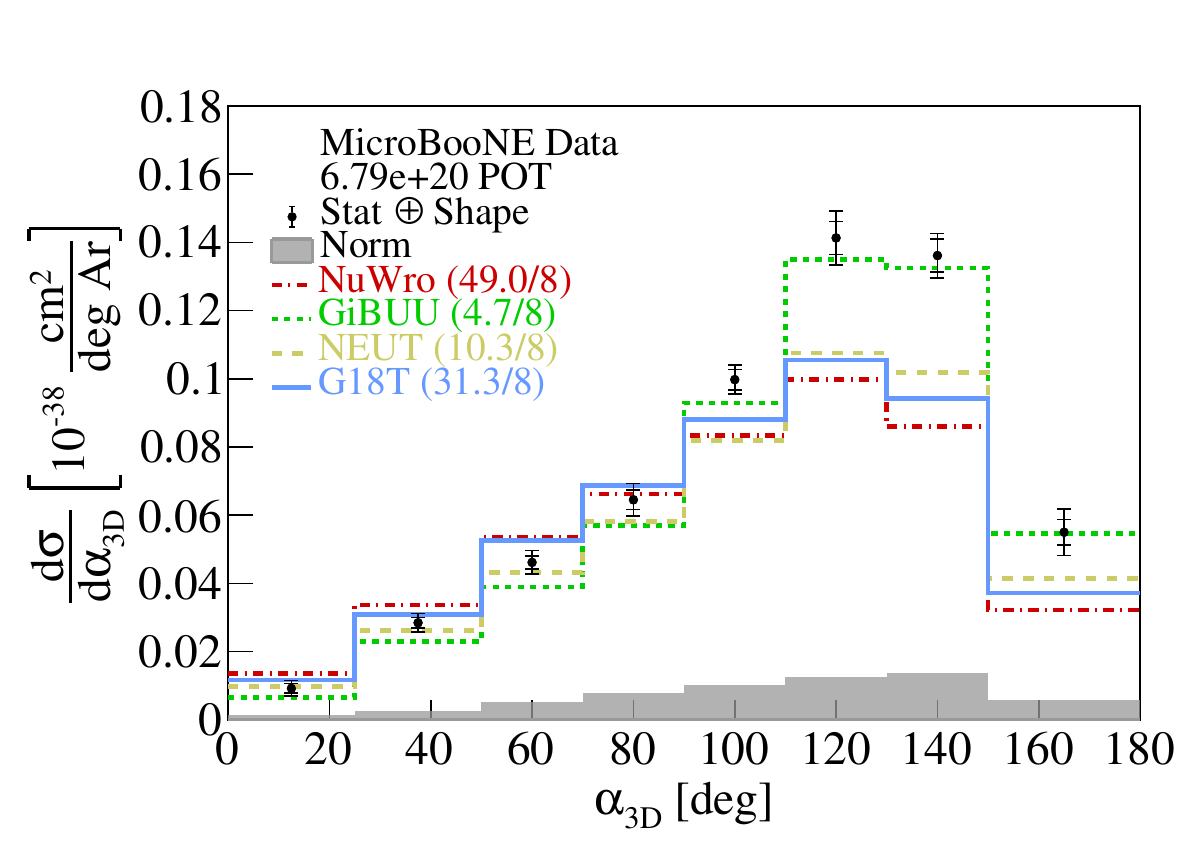}	
};
\draw (0.4, -3.2) node {(a)};	
\end{tikzpicture}
\hspace{0.05 \textwidth}
\begin{tikzpicture} \draw (0, 0) node[inner sep=0] {
\includegraphics[width=0.45\textwidth]{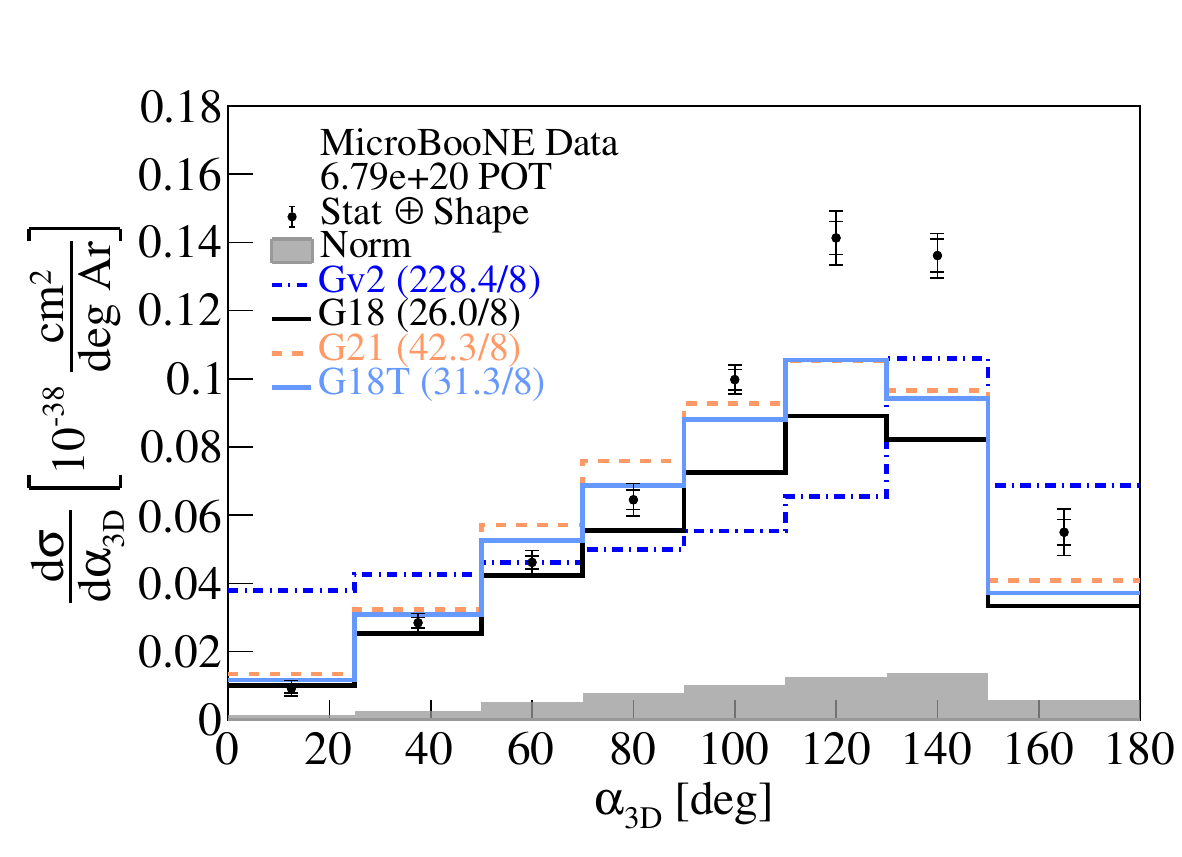}
};
\draw (0.4, -3.2) node {(b)};	
\end{tikzpicture}
  \caption{
The flux-integrated single-differential cross sections as a function of $\alpha_{3D}$ with (a) generator and (b) $\texttt{GENIE}$ configuration 
predictions compared to data. Inner and outer error bars show the statistical and total (statistical and shape systematic)
uncertainty at the 1$\sigma$, or 68\%, confidence level. The gray band shows the normalization systematic uncertainty. The numbers
in parentheses give the $\chi^{2}$/ndf calculation for each one of the predictions.
  }
  \label{fig:alpha3D_allevents}
\end{figure}

\begin{figure}[htb!]
  \centering
\begin{tikzpicture} \draw (0, 0) node[inner sep=0] {
\includegraphics[width=0.45\textwidth]{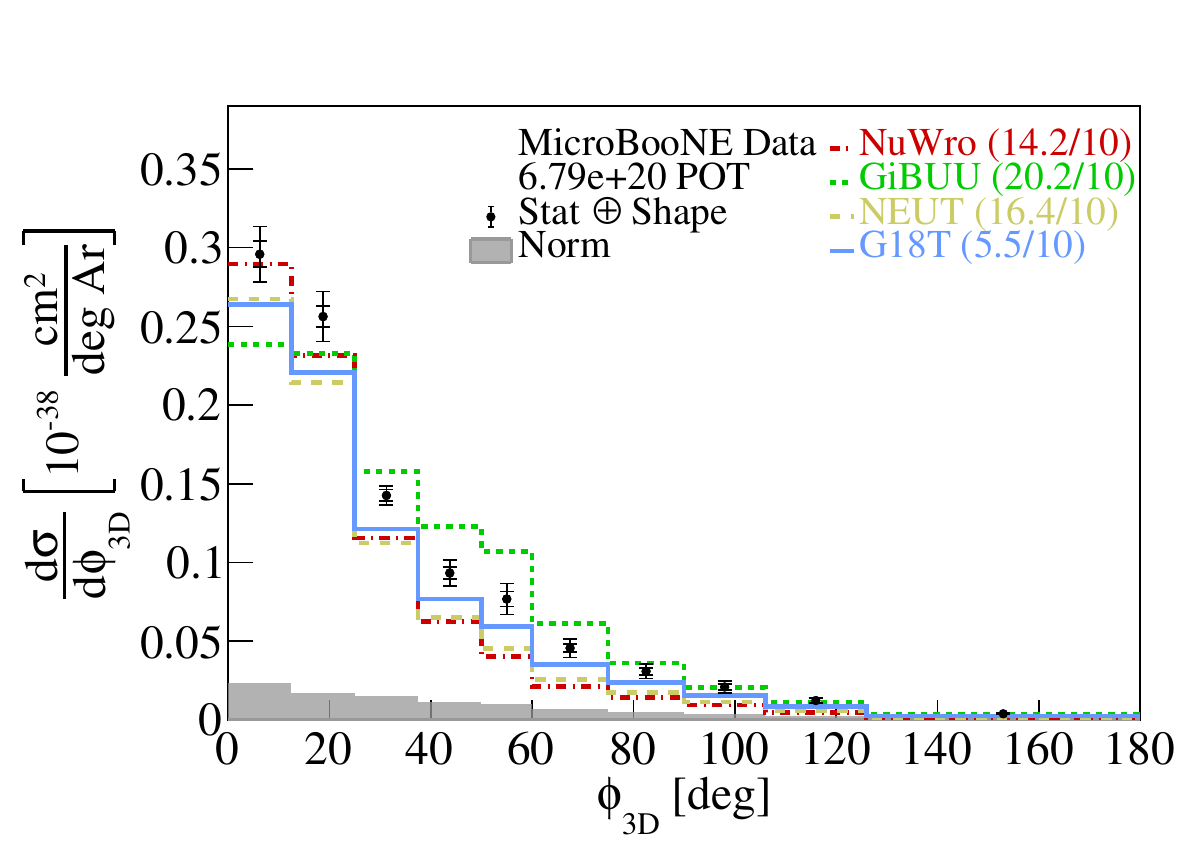}	
};
\draw (0.4, -3.2) node {(a)};	
\end{tikzpicture}
\hspace{0.05 \textwidth}
\begin{tikzpicture} \draw (0, 0) node[inner sep=0] {
\includegraphics[width=0.45\textwidth]{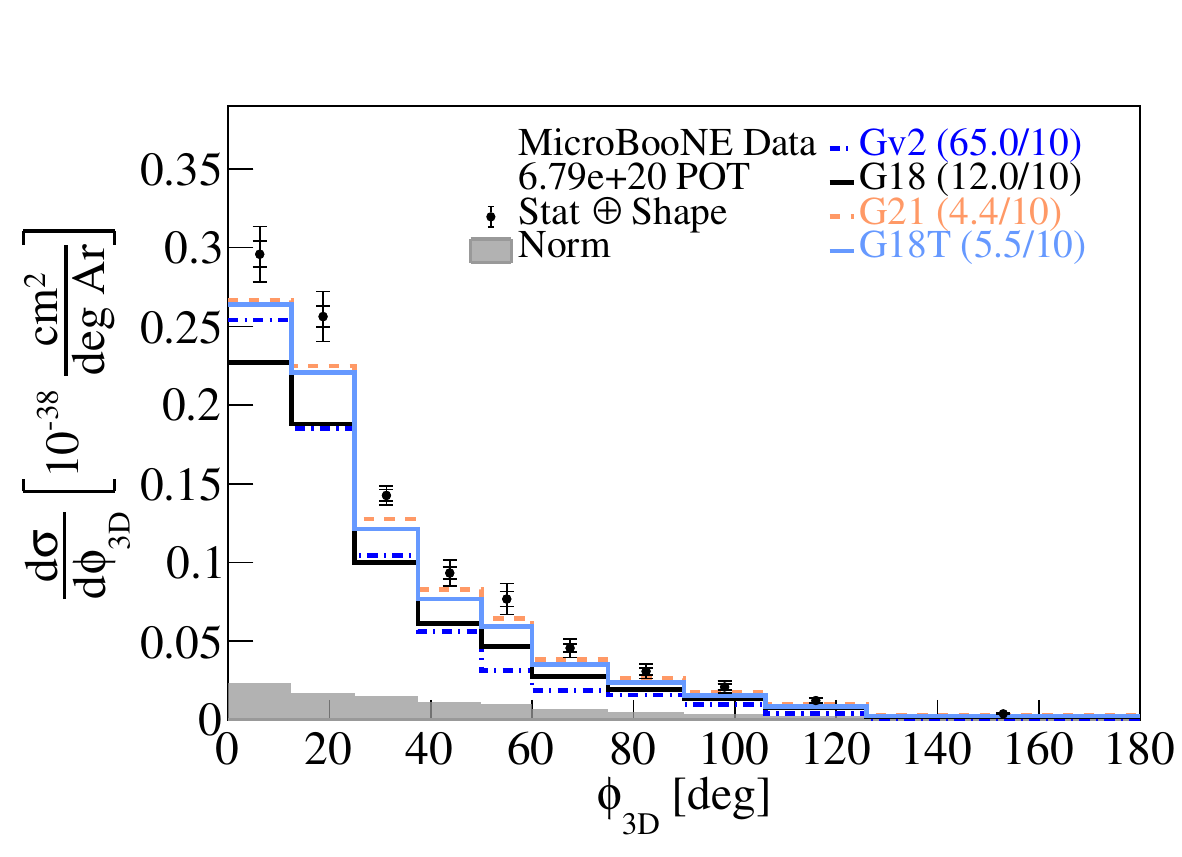}
};
\draw (0.4, -3.2) node {(b)};	
\end{tikzpicture}
  \caption{
The flux-integrated single-differential cross sections as a function of $\phi_{3D}$ with (a) generator and (b) $\texttt{GENIE}$ configuration 
predictions compared to data. Inner and outer error bars show the statistical and total (statistical and shape systematic)
uncertainty at the 1$\sigma$, or 68\%, confidence level. The gray band shows the normalization systematic uncertainty. The numbers
in parentheses give the $\chi^{2}$/ndf calculation for each one of the predictions.
  }
  \label{fig:phi3D_allevents}
\end{figure}

Figures~\ref{fig:deltaPn_allevents}-\ref{fig:phi3D_allevents} show the single-differential cross sections as a function of the $p_{n}$, $\alpha_{3D}$, and $\phi_{3D}$ compared to several predictions.
We conclude that $\texttt{Gv2}$ is a poor description of the data and results in large $\chi^{2}$ values.
Among the other generator predictions, $\texttt{GiBUU}$ provides the best description of the data in $\alpha_{3D}$.
There is a spread in $p_{n}$ and $\phi_{3D}$ with $\texttt{G21}$ describing the data best.
Unlike these variables, $\texttt{G21}$ shows a poor agreement with the data in $\alpha_{3D}$.
$\texttt{G18T}$ illustrates a similar pattern as $\texttt{G21}$ with a better agreement in $p_{n}$ and $\phi_{3D}$ and a worse performance in $\alpha_{3D}$.

\begin{figure}[htb!]
  \centering
\begin{tikzpicture} \draw (0, 0) node[inner sep=0] {
\includegraphics[width=0.45\textwidth]{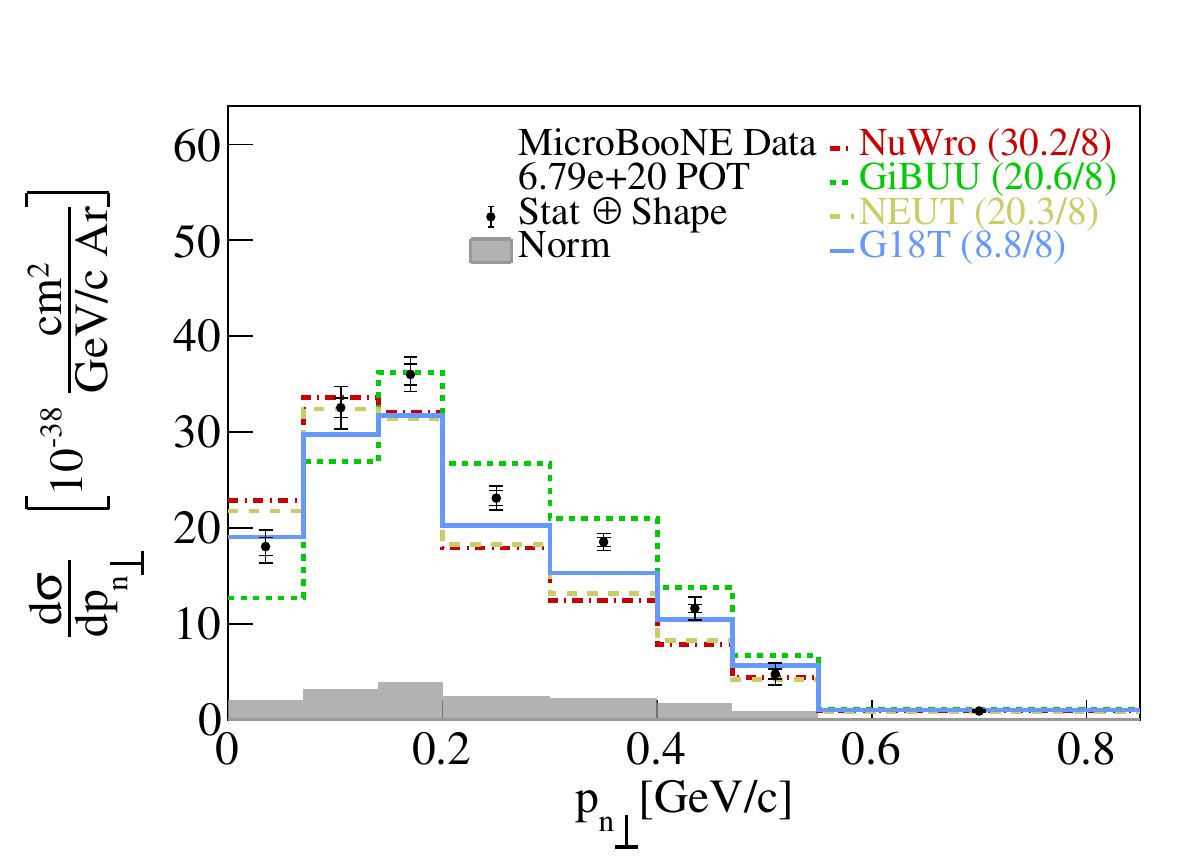}	
};
\draw (0.4, -3.2) node {(a)};	
\end{tikzpicture}
\hspace{0.05 \textwidth}
\begin{tikzpicture} \draw (0, 0) node[inner sep=0] {
\includegraphics[width=0.45\textwidth]{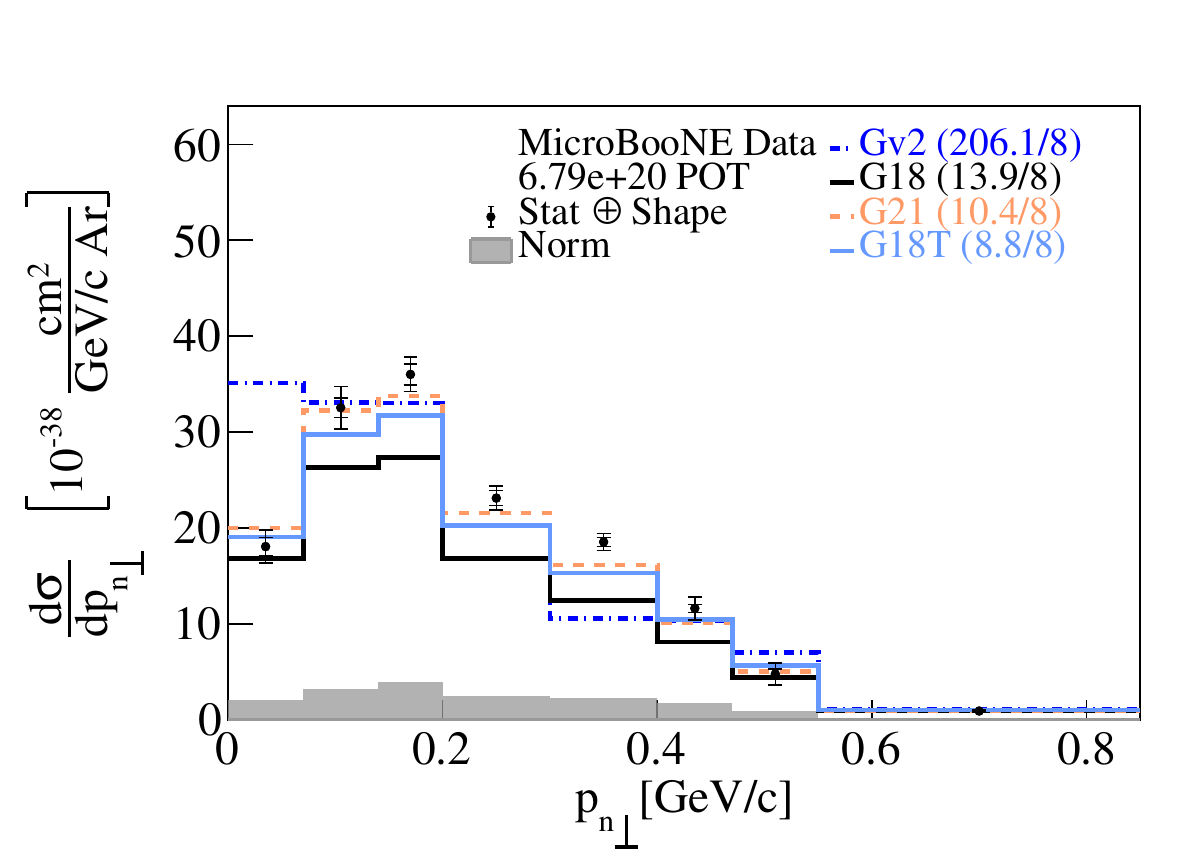}
};
\draw (0.4, -3.2) node {(b)};	
\end{tikzpicture}
  \caption{
The flux-integrated single-differential cross sections as a function of $p_{n\perp}$ with (a) generator and (b) $\texttt{GENIE}$ configuration 
predictions compared to data. Inner and outer error bars show the statistical and total (statistical and shape systematic)
uncertainty at the 1$\sigma$, or 68\%, confidence level. The gray band shows the normalization systematic uncertainty. The numbers
in parentheses give the $\chi^{2}$/ndf calculation for each one of the predictions.
  }
  \label{fig:deltaPnPerp_allevents}
\end{figure}

\begin{figure}[htb!]
    \centering 
\begin{tikzpicture} \draw (0, 0) node[inner sep=0] {
\includegraphics[width=0.45\textwidth]{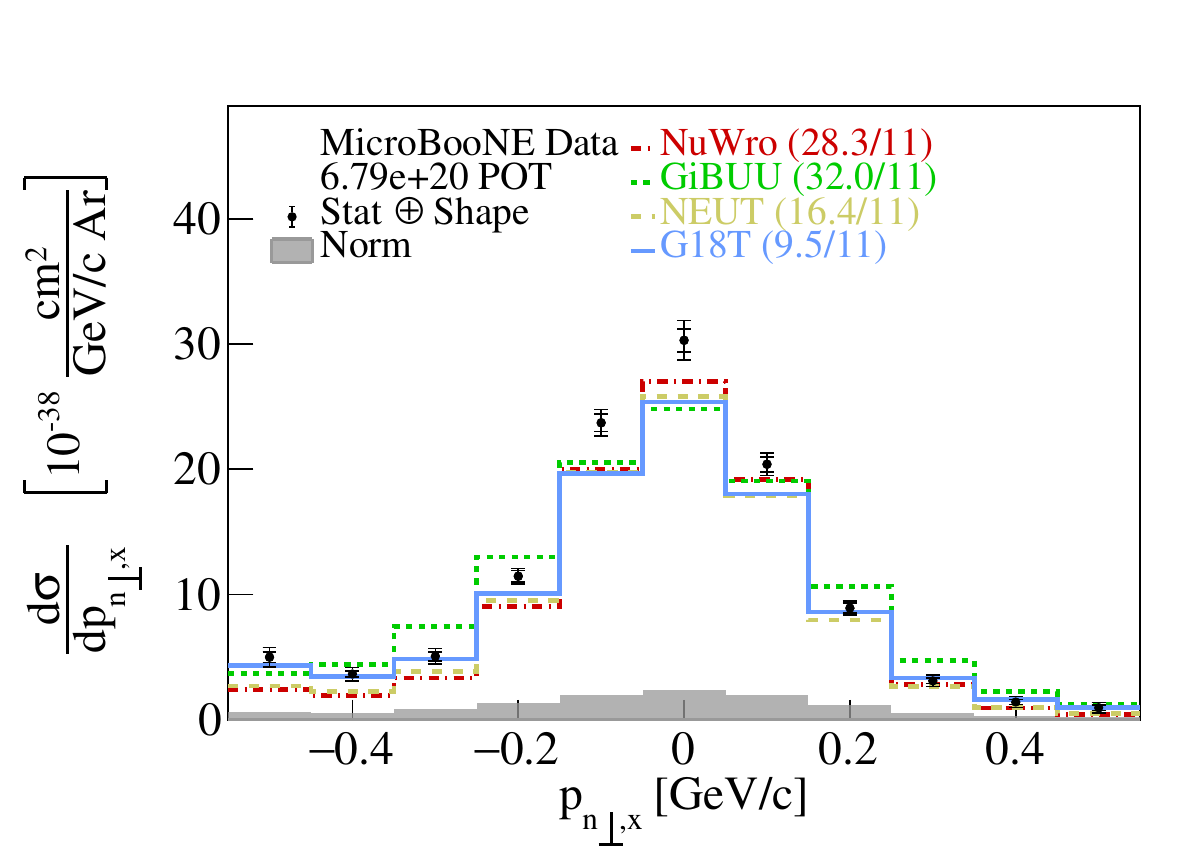}	
};
\draw (0.4, -3.2) node {(a)};	
\end{tikzpicture}
\hspace{0.05 \textwidth}
\begin{tikzpicture} \draw (0, 0) node[inner sep=0] {
\includegraphics[width=0.45\textwidth]{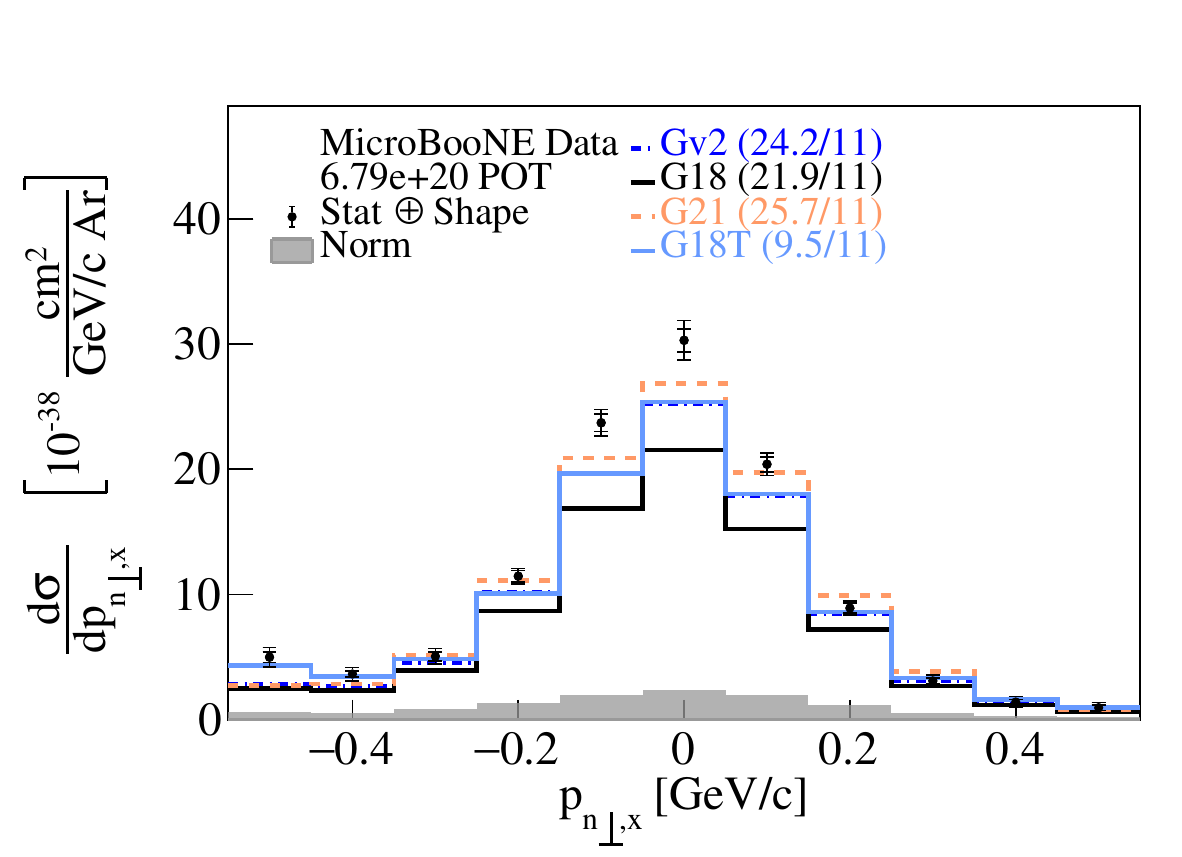}
};
\draw (0.4, -3.2) node {(b)};	
\end{tikzpicture}
  \caption{
The flux-integrated single-differential cross sections as a function of $p_{n\perp,x}$ with (a) generator and (b) $\texttt{GENIE}$ configuration 
predictions compared to data. Inner and outer error bars show the statistical and total (statistical and shape systematic)
uncertainty at the 1$\sigma$, or 68\%, confidence level. The gray band shows the normalization systematic uncertainty. The numbers
in parentheses give the $\chi^{2}$/ndf calculation for each one of the predictions.
  }    
  \label{fig:deltaPnPerpx_allevents}    
\end{figure}

\begin{figure}[htb!]
    \centering 
\begin{tikzpicture} \draw (0, 0) node[inner sep=0] {
\includegraphics[width=0.45\textwidth]{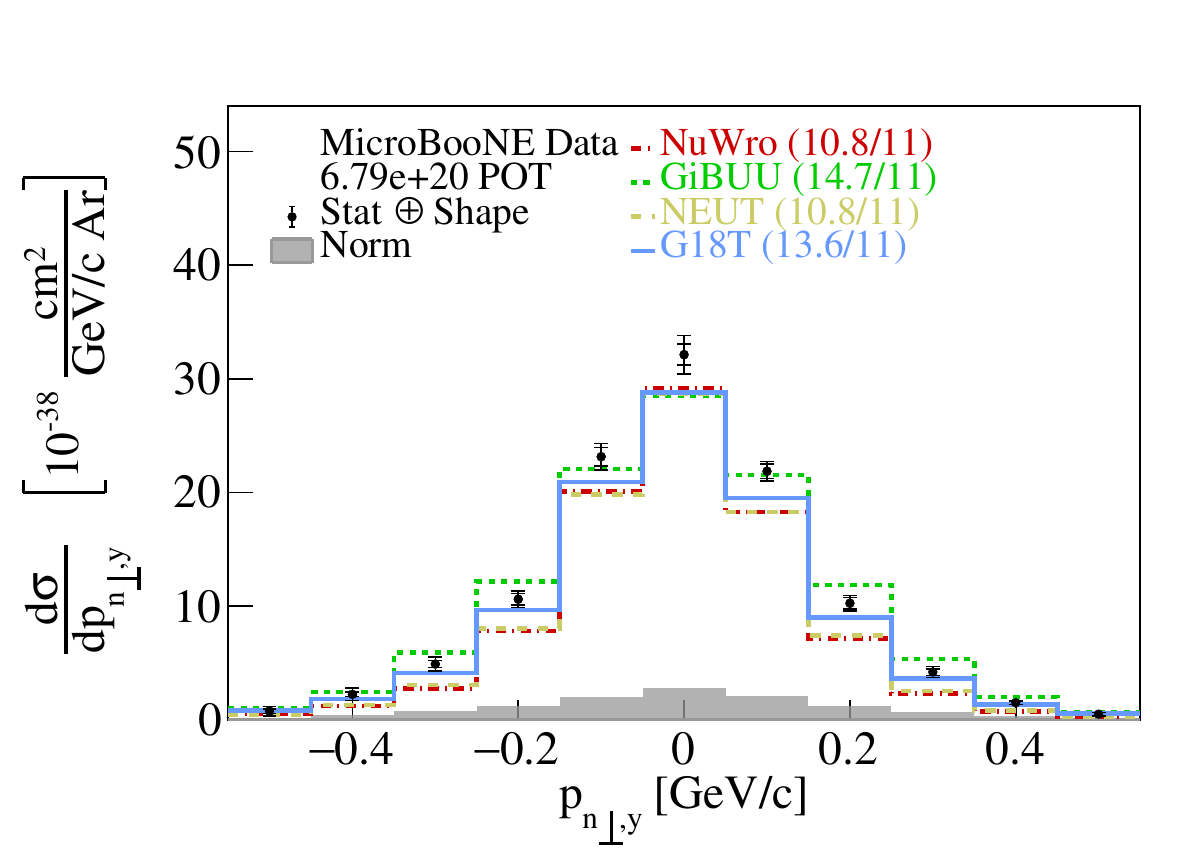}	
};
\draw (0.4, -3.2) node {(a)};	
\end{tikzpicture}
\hspace{0.05 \textwidth}
\begin{tikzpicture} \draw (0, 0) node[inner sep=0] {
\includegraphics[width=0.45\textwidth]{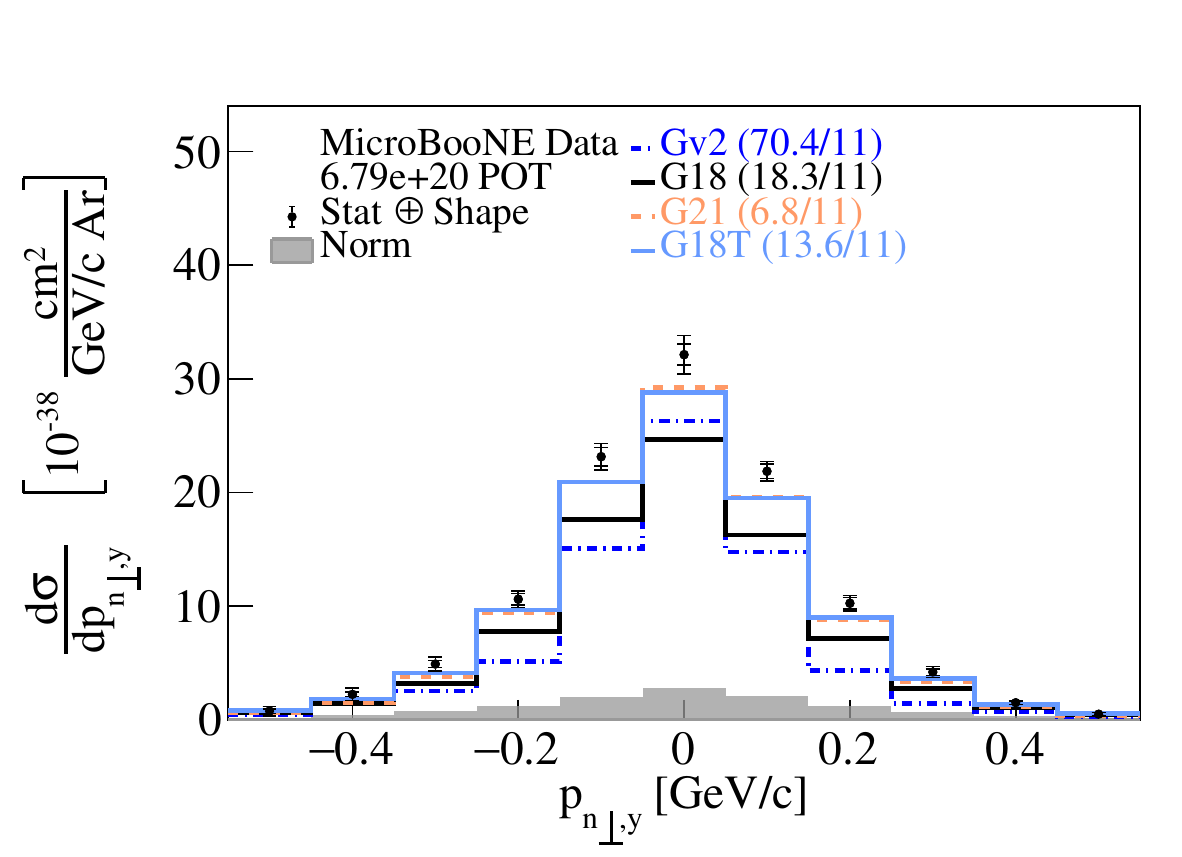}
};
\draw (0.4, -3.2) node {(b)};	
\end{tikzpicture}
  \caption{
The flux-integrated single-differential cross sections as a function of $p_{n\perp,y}$ with (a) Generator and (b) $\texttt{GENIE}$ configuration 
predictions compared to data. Inner and outer error bars show the statistical and total (statistical and shape systematic)
uncertainty at the 1$\sigma$, or 68\%, confidence level. The gray band shows the normalization systematic uncertainty. The numbers
in parentheses give the $\chi^{2}$/ndf calculation for each one of the predictions.
  }    
  \label{fig:deltaPnPerpy_allevents}    
\end{figure}

\begin{figure}[htb!]
  \centering
\begin{tikzpicture} \draw (0, 0) node[inner sep=0] {
\includegraphics[width=0.45\textwidth]{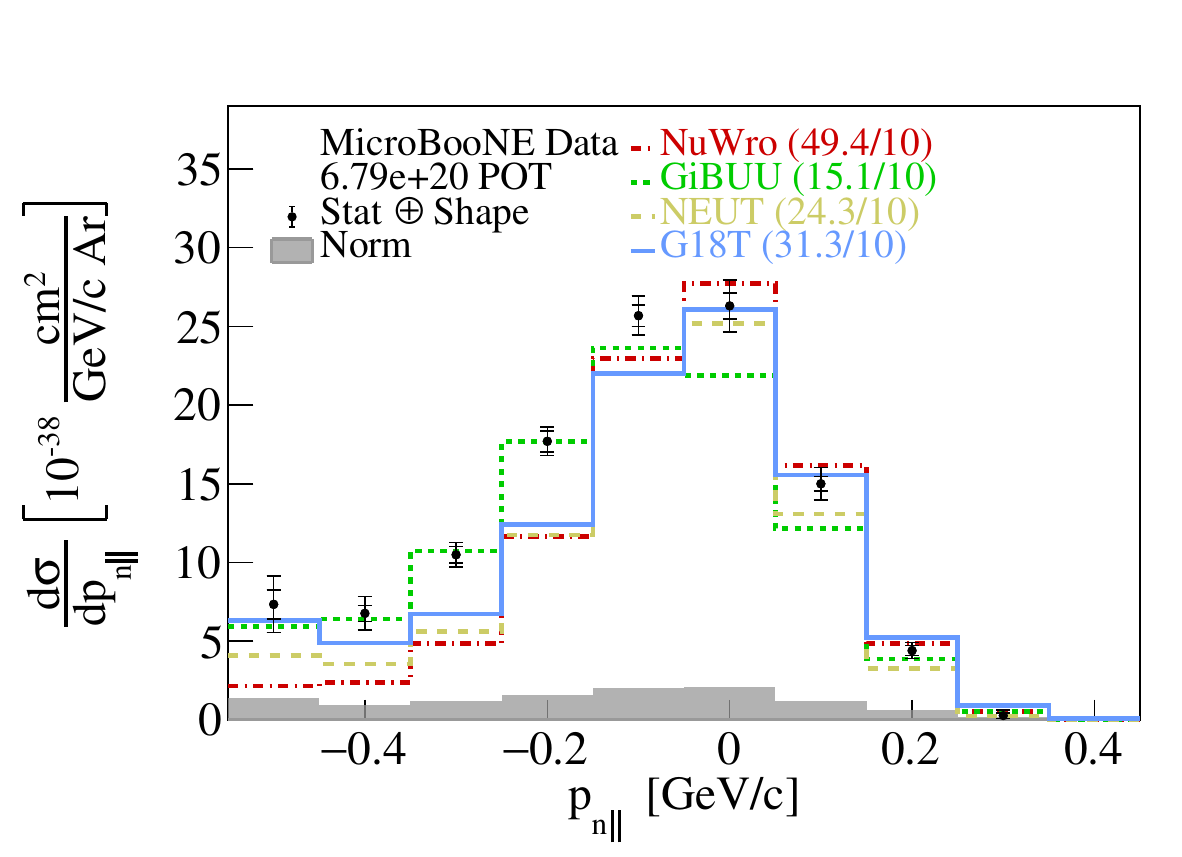}	
};
\draw (0.4, -3.2) node {(a)};	
\end{tikzpicture}
\hspace{0.05 \textwidth}
\begin{tikzpicture} \draw (0, 0) node[inner sep=0] {
\includegraphics[width=0.45\textwidth]{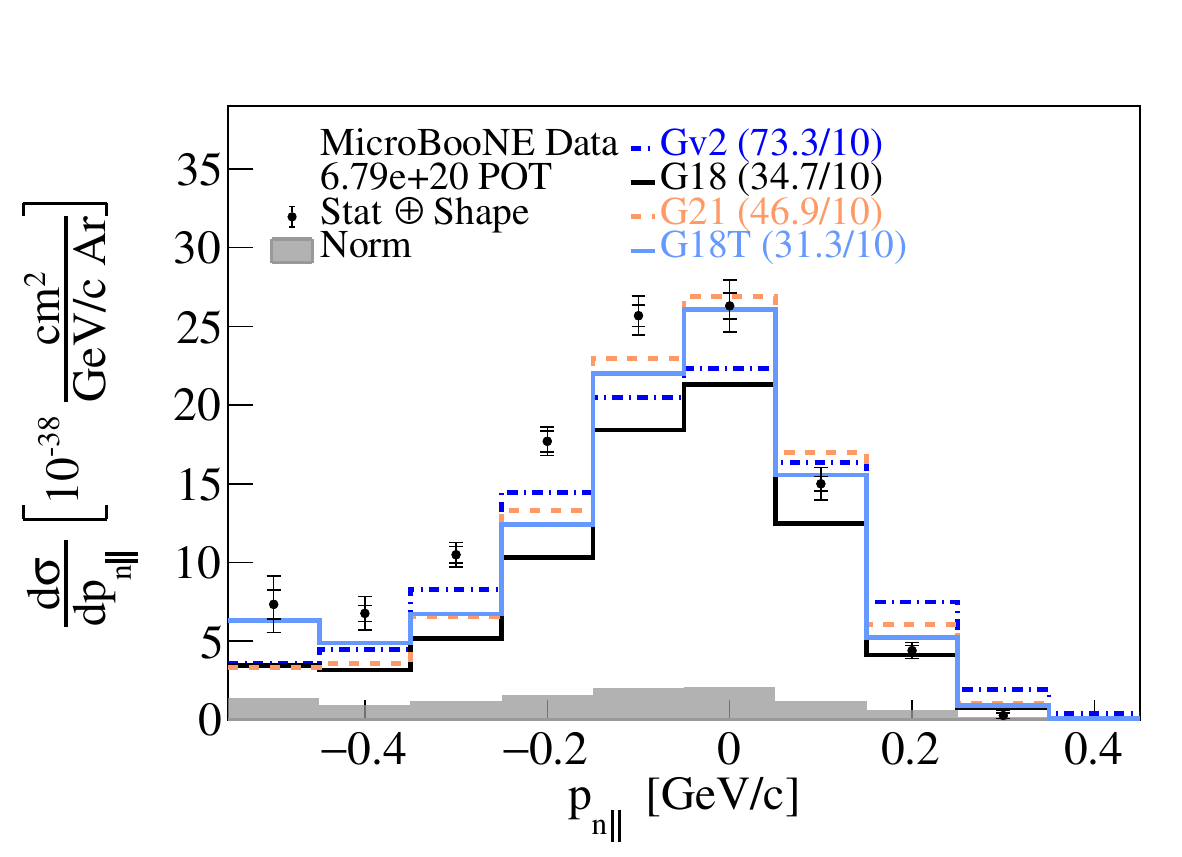}
};
\draw (0.4, -3.2) node {(b)};	
\end{tikzpicture}
  \caption{
The flux-integrated single-differential cross sections as a function of $p_{n\parallel}$ with (a) Generator and (b) $\texttt{GENIE}$ configuration 
predictions compared to data. Inner and outer error bars show the statistical and total (statistical and shape systematic)
uncertainty at the 1$\sigma$, or 68\%, confidence level. The gray band shows the normalization systematic uncertainty. The numbers
in parentheses give the $\chi^{2}$/ndf calculation for each one of the predictions.
  }
  \label{fig:deltaPnPar_allevents}
\end{figure}

The different projections of $p_{n}$ parallel and perpendicular to the momentum transfer are shown in Figs.~\ref{fig:deltaPnPerp_allevents} - \ref{fig:deltaPnPar_allevents}.
Again $\texttt{Gv2}$ provides a poor description of the data, particularly in the case of  $p_{n\perp}$.
The data show a large tail at negative values of $p_{n\parallel}$ as expected from the effects of FSI.
The predictions in this region show large variation due to the different ways of modeling FSI.
Based on the $\chi^{2}$/ndf, $\texttt{GiBUU}$ provides the best description of this data in most variables.
$\texttt{G18T}$ provides the best performance in the case of $p_\perp$ and $p_{\perp ,x}$.
Conversely, there is a spread in performance among the generators in the description of $p_{\perp ,y}$ with $\texttt{G21}$ yielding the lowest $\chi^{2}$/ndf ratio.

As discussed in Sec.~\ref{sec:generators}, the sensitivity of the generalized kinematic imbalance variables can be further enhanced when performing multi-differential measurements.  
Following the approach outlined in Refs.~\cite{RefPRL,RefPRD}, we present double-differential measurements as a function of $p_n$ and $\alpha_{3D}$.

\begin{figure}[htb!]
  \centering
\begin{tikzpicture} \draw (0, 0) node[inner sep=0] {
\includegraphics[width=0.45\textwidth]{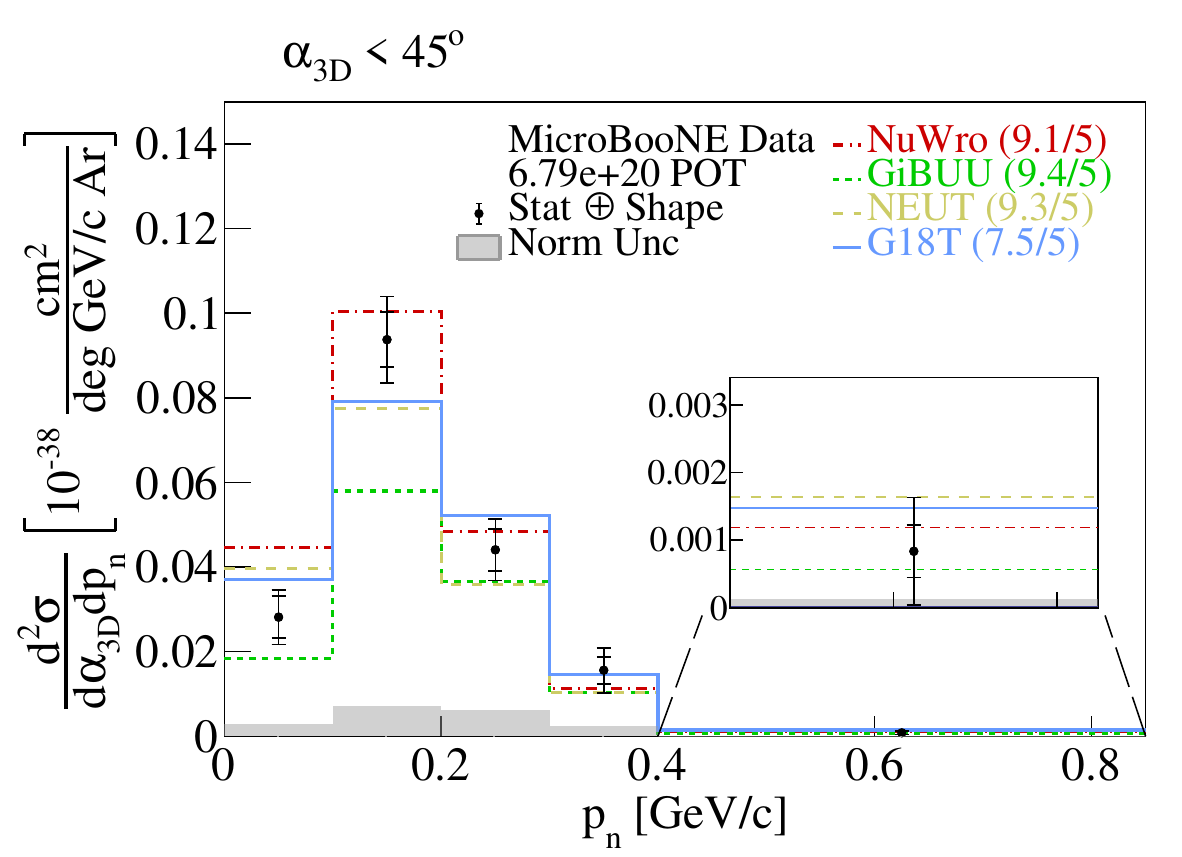}	
};
\draw (0.4, -3.2) node {(a)};	
\end{tikzpicture}
\hspace{0.05 \textwidth}
\begin{tikzpicture} \draw (0, 0) node[inner sep=0] {
\includegraphics[width=0.45\textwidth]{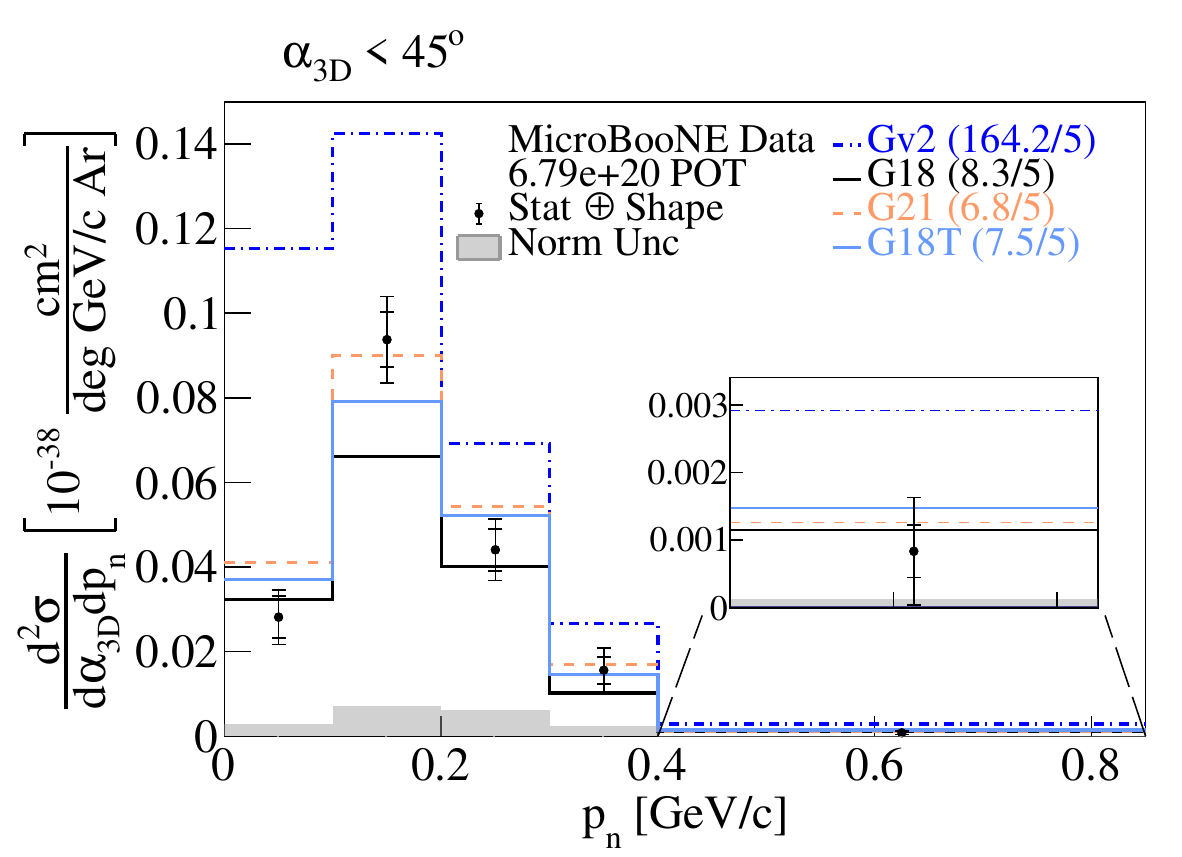}
};
\draw (0.4, -3.2) node {(b)};	
\end{tikzpicture}
  \caption{
The flux-integrated double-differential cross sections as a function of $p_{n}$ for $\alpha_{3D} <$ 45$^{\circ}$ with (a) generator and (b) $\texttt{GENIE}$ configuration 
predictions compared to data. Inner and outer error bars show the statistical and total (statistical and shape systematic)
uncertainty at the 1$\sigma$, or 68\%, confidence level. The gray band shows the normalization systematic uncertainty. The numbers
in parentheses give the $\chi^{2}$/ndf calculation for each one of the predictions.  
  }
  \label{fig:pn_lowAlpha}
\end{figure}

\begin{figure}[htb!]
  \centering
\begin{tikzpicture} \draw (0, 0) node[inner sep=0] {
\includegraphics[width=0.45\textwidth]{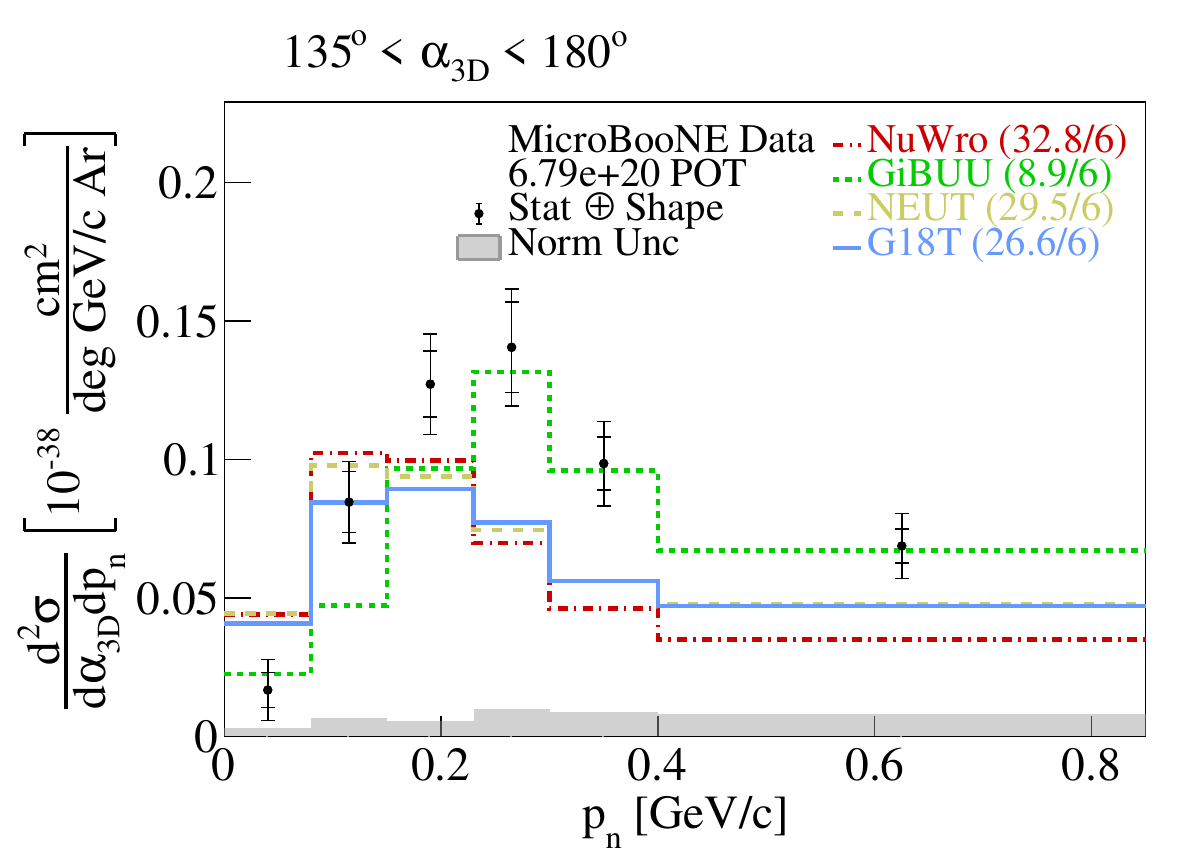}	
};
\draw (0.4, -3.2) node {(a)};	
\end{tikzpicture}
\hspace{0.05 \textwidth}
\begin{tikzpicture} \draw (0, 0) node[inner sep=0] {
\includegraphics[width=0.45\textwidth]{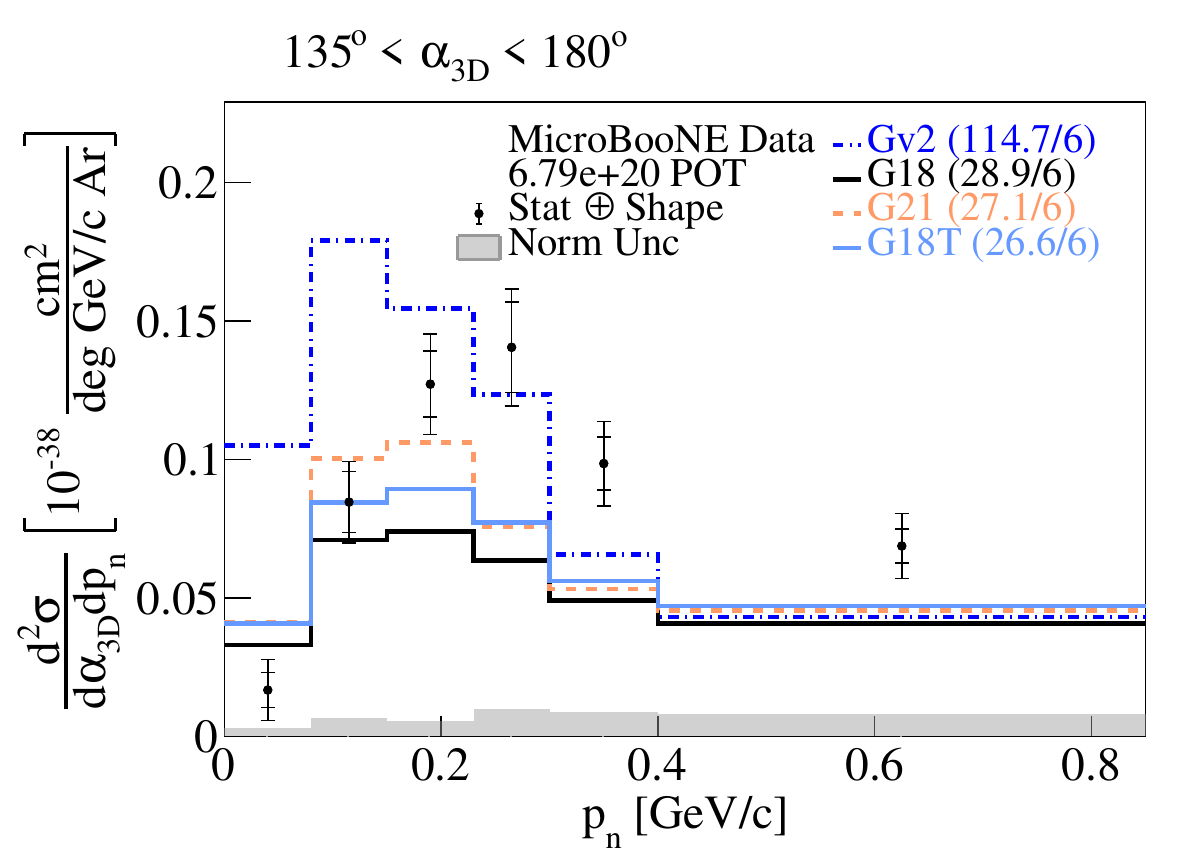}
};
\draw (0.4, -3.2) node {(b)};	
\end{tikzpicture}
  \caption{
The flux-integrated double-differential cross sections as a function of 135$^{\circ}$ $< \alpha_{3D} <$ 180$^{\circ}$ with (a) Generator and (b) $\texttt{GENIE}$ configuration 
predictions compared to data. Inner and outer error bars show the statistical and total (statistical and shape systematic)
uncertainty at the 1$\sigma$, or 68\%, confidence level. The gray band shows the normalization systematic uncertainty. The numbers
in parentheses give the $\chi^{2}$/ndf calculation for each one of the predictions. 
  }
  \label{fig:pn_highAlpha}
\end{figure}

FSI effects are minimal for the double-differential cross section of $p_{n}$ with $\alpha_{3D} < 45^{\circ}$ and, as expected, the tail of the $p_{n}$ distribution is significantly suppressed (Fig.~\ref{fig:pn_lowAlpha}).
The $\chi^{2}$/ndf is reasonably consistent across all event generators apart from $\texttt{Gv2}$.
This observation could be driven by the fact that the more modern CCQE models used by $\texttt{GENIE v3}$ and alternative event generators are very similar.
This improved picture originates from the fairly well-understood QE interaction channel that has been extensively investigated.
Conversely, the region $\alpha_{3D} > 135^{\circ}$ (Fig.~\ref{fig:pn_highAlpha}) contains a large fraction of events that undergo FSI, leading to an enhanced high-$p_{n}$ tail, which is under-predicted by most generators.
Furthermore, $\texttt{GiBUU}$ shows an offset to the right compared to other event generators and demonstrates the best agreement with the data.

\begin{figure}[htb!]
  \centering
\begin{tikzpicture} \draw (0, 0) node[inner sep=0] {
\includegraphics[width=0.45\textwidth]{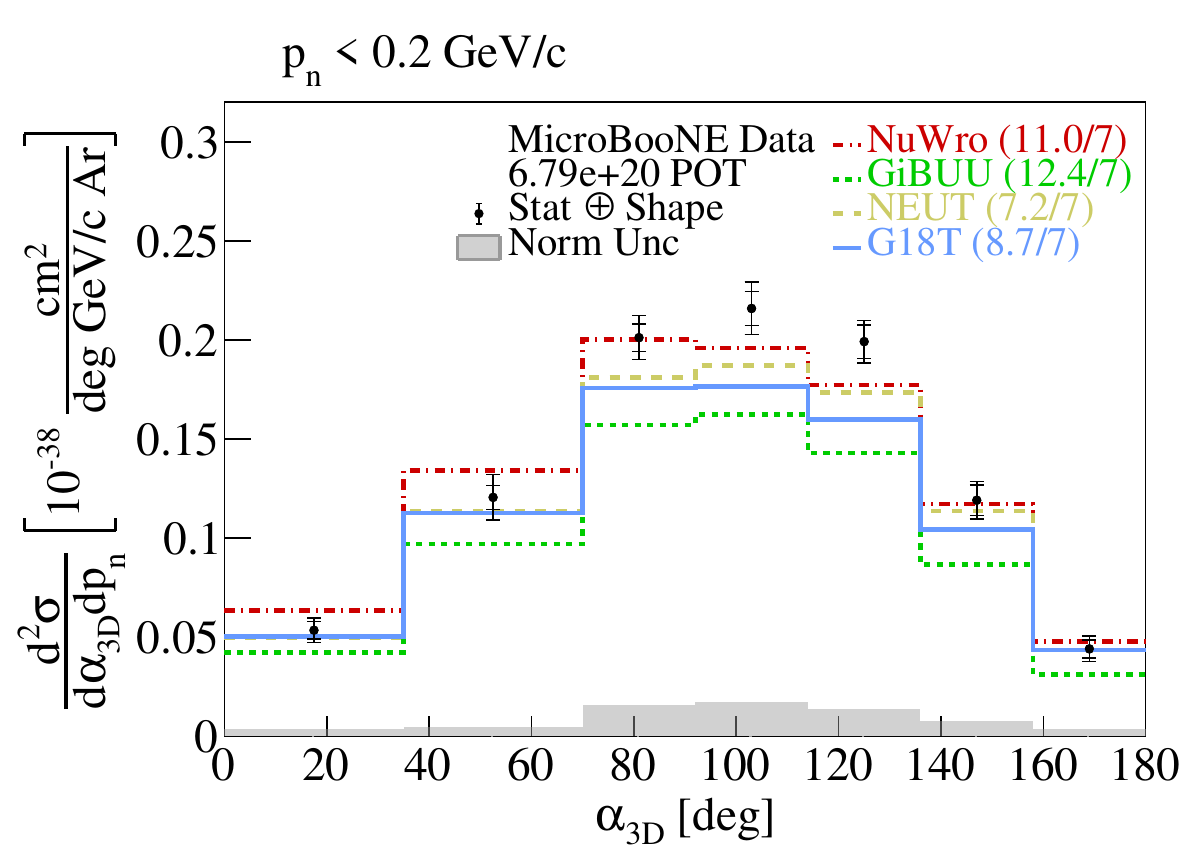}	
};
\draw (0.4, -3.2) node {(a)};	
\end{tikzpicture}
\hspace{0.05 \textwidth}
\begin{tikzpicture} \draw (0, 0) node[inner sep=0] {
\includegraphics[width=0.45\textwidth]{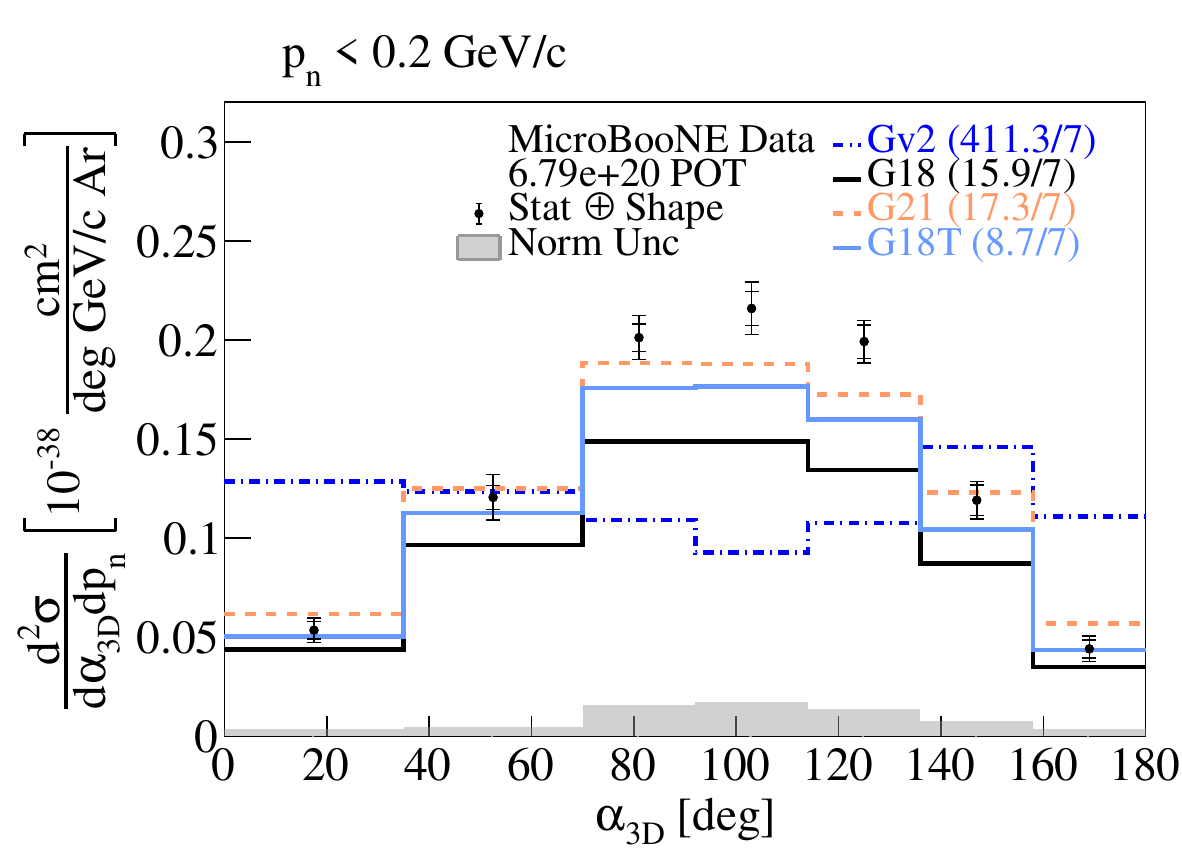}
};
\draw (0.4, -3.2) node {(b)};	
\end{tikzpicture}
  \caption{
The flux-integrated double-differential cross sections as a function of $\alpha_{3D}$ for $p_{n} <$ 0.2\,GeV/c with (a) generator and (b) $\texttt{GENIE}$ configuration 
predictions compared to data. Inner and outer error bars show the statistical and total (statistical and shape systematic)
uncertainty at the 1$\sigma$, or 68\%, confidence level. The gray band shows the normalization systematic uncertainty. The numbers
in parentheses give the $\chi^{2}$/ndf calculation for each one of the predictions. 
  }
  \label{fig:alpha3D_lowpn}
\end{figure}

\begin{figure}[H]
  \centering
\begin{tikzpicture} \draw (0, 0) node[inner sep=0] {
\includegraphics[width=0.45\textwidth]{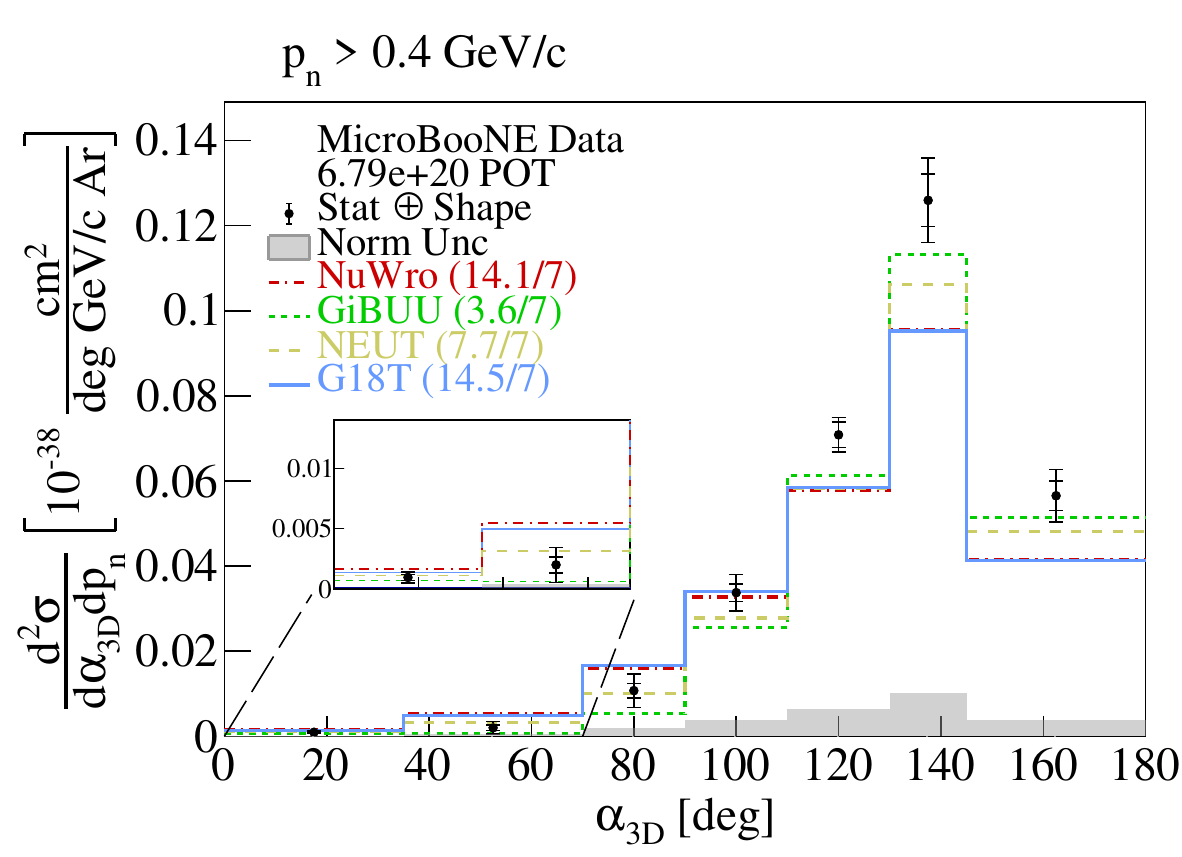}	
};
\draw (0.4, -3.2) node {(a)};	
\end{tikzpicture}
\hspace{0.05 \textwidth}
\begin{tikzpicture} \draw (0, 0) node[inner sep=0] {
\includegraphics[width=0.45\textwidth]{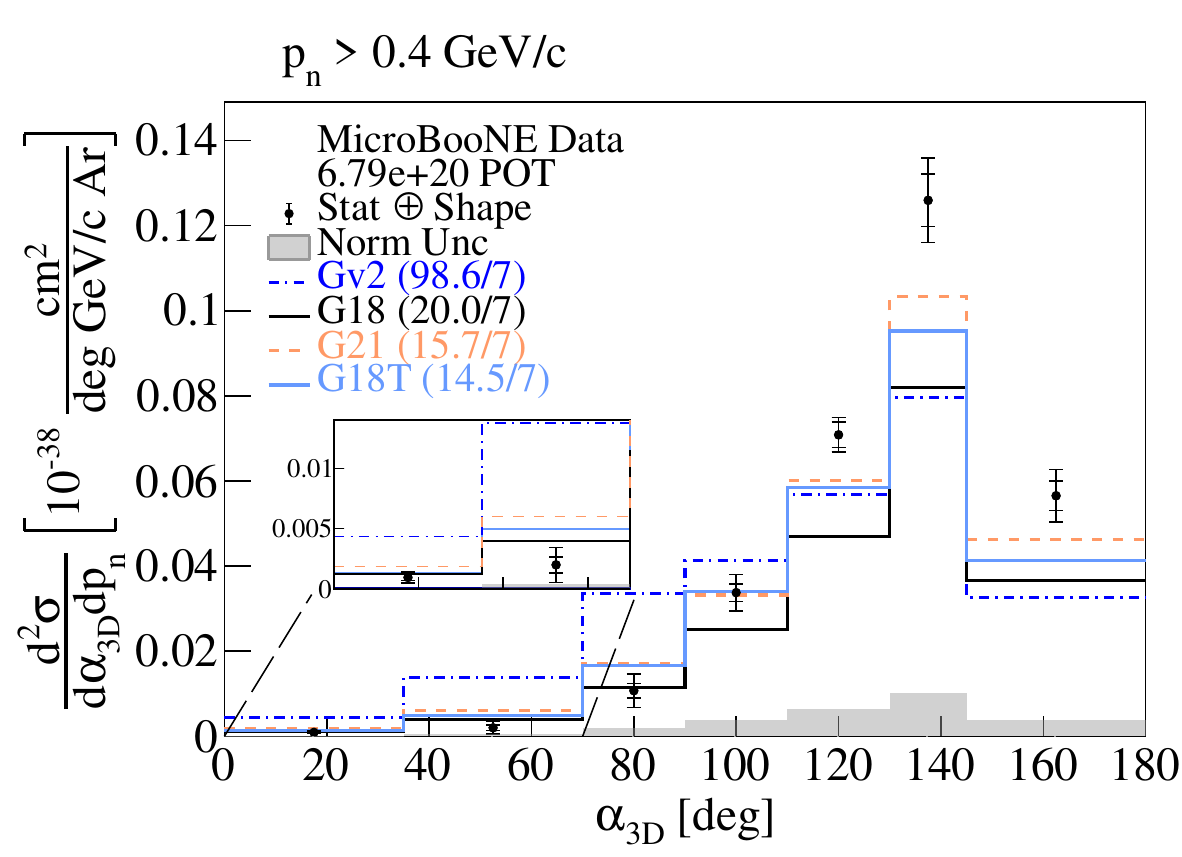}
};
\draw (0.4, -3.2) node {(b)};	
\end{tikzpicture}
  \caption{
The flux-integrated double-differential cross sections as a function of $\alpha_{3D}$ for $p_{n} >$ 0.4\,GeV/c with (a) Generator and (b) $\texttt{GENIE}$ configuration 
predictions compared to data. Inner and outer error bars show the statistical and total (statistical and shape systematic)
uncertainty at the 1$\sigma$, or 68\%, confidence level. The gray band shows the normalization systematic uncertainty. The numbers
in parentheses give the $\chi^{2}$/ndf calculation for each one of the predictions. 
  }
  \label{fig:alpha3D_highpn}
\end{figure}

Figures~\ref{fig:alpha3D_lowpn} and~\ref{fig:alpha3D_highpn} show the double-differential measurement in $\alpha_{3D}$ for events with low and high values of $p_n$, respectively.
For events with low missing momentum, the distribution is very symmetric and approximately follows the $\sin(\alpha_{3D})$ shape as expected.
FSI predominantly remove events from this region of phase space, leading to normalization differences between generators, with the exception of $\texttt{Gv2}$ which illustrates a significantly different behavior.
Conversely, events with high missing momentum (Fig.~\ref{fig:alpha3D_highpn}) show a large asymmetry with strong enhancement in the FSI-driven region at high values of $\alpha_{3D}$.
Apart from $\texttt{Gv2}$, all generator predictions show similar shapes with some normalization differences.
$\texttt{Gv2}$ predicts an enhanced tail at low values of $\alpha_{3D}$ which does not appear in the data.
This low-$\alpha_{3D}$, high-$p_n$ region contains a large contribution ($\geq$ 50\%) from MEC according to generator predictions. 
Most models provide a reasonable description of the data in this region.

A simultaneous extraction of the double-differential results across all parts of the available phase-space is presented in the Supplemental Material.


\section{Conclusions} \label{sec:discussion}

We report measurements of flux-integrated differential cross sections for event topologies with a single muon and a single proton in the final state using the Booster Neutrino Beam at  Fermi National Accelerator Laboratory.
The data are recorded with the MicroBooNE detector and studied for the first time in the form of single- and double-differential cross sections in novel generalized kinematic imbalance variables that consider not only the transverse, but also the longitudinal component of the missing momentum.
These generalized kinematic imbalance variables show sensitivity to nuclear effects, leading to some significant differences between generator predictions and data.
Some of these differences might be originating from fact that neutrinos are scattering off heavy argon nuclei.
These results, in conjunction with those on lighter targets, can provide valuable information on the evolution of nuclear effects as the target nucleus mass number increases.

The $\texttt{GENIE v2.12.10}$ ($\texttt{Gv2}$) cross section predictions are systematically a poor fit to data with significant shape and normalization differences across almost all variables of interest. 
This is in contrast to several recent measurements (e.g.,~\cite{Abe2020.11986,Abratenko2019:131801,PhysRevD.101.112007,PhysRevD.94.052005}) where $\texttt{Gv2}$ and the more modern $\texttt{GENIE v3}$ predictions show similar shapes across many variables that only depend on either the muon or the proton kinematics, whereas these kinematic imbalance variables also depend on correlations between the leptonic and hadronic system.
$\texttt{Gv2}$ is therefore no longer able to provide a good description of the data.

$\texttt{GiBUU 2021}$ ($\texttt{GiBUU}$) is able to describe most distributions well, with the exception of $\phi_{3D}$, $p_n$, and $p_{n\perp}$.  
$\texttt{GiBUU}$ agrees particularly well in areas of phase space where final state interactions have a very large effect, for example $p_{n}$ for events with $\alpha_{3D} > 135^{\circ}$, whereas all of the other generators show large disagreements in those regions of phase space.

The $\texttt{GENIE v3.0.6 G18\_10a\_02\_11a}$ cross section predictions with the MicroBooNE-specific tuning ($\texttt{G18T}$), on the other hand, fit the $\phi_{3D}$, $p_n$, and $p_{n\perp}$ data well.
This contrasts with the $\texttt{GENIE v3.0.6 G18\_10a\_02\_11a}$ configuration without additional tuning ($\texttt{G18}$) which shows a systematic deficit of $\approx 20$\%.
The MicroBooNE-specific tuning modifies the normalization of QE and MEC events, as well as the shape of MEC and RPA suppression, and none of the tuning relied on data that included hadron kinematics.
This difference indicates that the underlying physics models in $\texttt{G18}$ can describe the correlation between lepton and hadron kinematics, although the normalization of some components need to be adjusted to better agree with data.
However, in regions of phase space sensitive to FSI, this model does not perform as well as $\texttt{GiBUU}$, and the tuning does not improve the agreement.

The $\texttt{GENIE v3.2.0 G21\_11b\_00\_000}$ configuration ($\texttt{G21}$) serves as an example of a $\texttt{GENIE}$ configuration that shows good agreement with data in most variables without the need for additional tuning. 
$\texttt{NEUT}$ produces similar agreement as $\texttt{G21}$ across many distributions.
However, $\texttt{NEUT}$ provides a better description of the $\alpha_{3D}$ data, though a worse description of the $p_n$ data.
In regions of phase space sensitive to FSI, both of these models provide poor agreement with data.
In most variables and parts of the phase space, $\texttt{NuWro}$ results in $\chi^{2}$/ndf ratios fairly higher than unity.

The reported results provide precision data in new GKI variables that are more sensitive to nuclear effects for the first time.
This data could be used to benchmark and tune neutrino-nucleus interaction models, particularly the modeling of final state interactions.


\section{Acknowledgements} \label{sec:ack}

This document was prepared by the MicroBooNE collaboration using the
resources of the Fermi National Accelerator Laboratory (Fermilab), a
U.S. Department of Energy, Office of Science, HEP User Facility.
Fermilab is managed by Fermi Research Alliance, LLC (FRA), acting
under Contract No. DE-AC02-07CH11359.  
This material is based upon work supported by Laboratory Directed Research and Development (LDRD) funding from Argonne National Laboratory, provided by the Director, Office of Science, of the U.S. Department of Energy under Contract No. DE-AC02-06CH11357.
MicroBooNE is supported by the
following: the U.S. Department of Energy, Office of Science, Offices
of High Energy Physics and Nuclear Physics; the U.S. National Science
Foundation; the Swiss National Science Foundation; the Science and
Technology Facilities Council (STFC), part of the United Kingdom Research 
and Innovation; the Royal Society (United Kingdom); the UK Research 
and Innovation (UKRI) Future Leaders Fellowship; and The European 
Union’s Horizon 2020 Marie Sklodowska-Curie Actions. Additional support 
for the laser calibration system and cosmic ray tagger was provided by 
the Albert Einstein Center for Fundamental Physics, Bern, Switzerland. 
We also acknowledge the contributions of technical and scientific staff 
to the design, construction, and operation of the MicroBooNE detector 
as well as the contributions of past collaborators to the development 
of MicroBooNE analyses, without whom this work would not have been 
possible. For the purpose of open access, the authors have applied a 
Creative Commons Attribution (CC BY) license to any Author Accepted 
Manuscript version arising from this submission.


\bibliography{main}

\providecommand{\noopsort}[1]{}\providecommand{\singleletter}[1]{#1}%
\begin{thebibliography}{69}%
\makeatletter
\providecommand \@ifxundefined [1]{%
 \@ifx{#1\undefined}
}%
\providecommand \@ifnum [1]{%
 \ifnum #1\expandafter \@firstoftwo
 \else \expandafter \@secondoftwo
 \fi
}%
\providecommand \@ifx [1]{%
 \ifx #1\expandafter \@firstoftwo
 \else \expandafter \@secondoftwo
 \fi
}%
\providecommand \natexlab [1]{#1}%
\providecommand \enquote  [1]{``#1''}%
\providecommand \bibnamefont  [1]{#1}%
\providecommand \bibfnamefont [1]{#1}%
\providecommand \citenamefont [1]{#1}%
\providecommand \href@noop [0]{\@secondoftwo}%
\providecommand \href [0]{\begingroup \@sanitize@url \@href}%
\providecommand \@href[1]{\@@startlink{#1}\@@href}%
\providecommand \@@href[1]{\endgroup#1\@@endlink}%
\providecommand \@sanitize@url [0]{\catcode `\\12\catcode `\$12\catcode
  `\&12\catcode `\#12\catcode `\^12\catcode `\_12\catcode `\%12\relax}%
\providecommand \@@startlink[1]{}%
\providecommand \@@endlink[0]{}%
\providecommand \url  [0]{\begingroup\@sanitize@url \@url }%
\providecommand \@url [1]{\endgroup\@href {#1}{\urlprefix }}%
\providecommand \urlprefix  [0]{URL }%
\providecommand \Eprint [0]{\href }%
\providecommand \doibase [0]{https://doi.org/}%
\providecommand \selectlanguage [0]{\@gobble}%
\providecommand \bibinfo  [0]{\@secondoftwo}%
\providecommand \bibfield  [0]{\@secondoftwo}%
\providecommand \translation [1]{[#1]}%
\providecommand \BibitemOpen [0]{}%
\providecommand \bibitemStop [0]{}%
\providecommand \bibitemNoStop [0]{.\EOS\space}%
\providecommand \EOS [0]{\spacefactor3000\relax}%
\providecommand \BibitemShut  [1]{\csname bibitem#1\endcsname}%
\let\auto@bib@innerbib\@empty
\bibitem [{\citenamefont {Tanabashi}\ \emph {et~al.}(2018)\citenamefont
  {Tanabashi} \emph {et~al.}}]{pdg2018}%
  \BibitemOpen
  \bibfield  {author} {\bibinfo {author} {\bibfnamefont {M.}~\bibnamefont
  {Tanabashi}} \emph {et~al.} (\bibinfo {collaboration} {Particle Data
  Group}),\ }\bibfield  {title} {\bibinfo {title} {Review of particle
  physics},\ }\href {https://doi.org/10.1103/PhysRevD.98.030001} {\bibfield
  {journal} {\bibinfo  {journal} {Phys. Rev. D}\ }\textbf {\bibinfo {volume}
  {98}},\ \bibinfo {pages} {030001} (\bibinfo {year} {2018})}\BibitemShut
  {NoStop}%
\bibitem [{\citenamefont {Abe}\ \emph {et~al.}(2020{\natexlab{a}})\citenamefont
  {Abe} \emph {et~al.}}]{T2KNature20}%
  \BibitemOpen
  \bibfield  {author} {\bibinfo {author} {\bibfnamefont {K.}~\bibnamefont
  {Abe}} \emph {et~al.} (\bibinfo {collaboration} {T2K Collaboration}),\
  }\bibfield  {title} {\bibinfo {title} {{Constraint on the matter–antimatter
  symmetry-violating phase in neutrino oscillations}},\ }\href
  {https://doi.org/10.1038/s41586-020-2177-0} {\bibfield  {journal} {\bibinfo
  {journal} {Nature}\ }\textbf {\bibinfo {volume} {580}},\ \bibinfo {pages}
  {339} (\bibinfo {year} {2020}{\natexlab{a}})}\BibitemShut {NoStop}%
\bibitem [{\citenamefont {Abi}\ \emph {et~al.}(2020)\citenamefont {Abi} \emph
  {et~al.}}]{DUNE1:2016oaz}%
  \BibitemOpen
  \bibfield  {author} {\bibinfo {author} {\bibfnamefont {B.}~\bibnamefont
  {Abi}} \emph {et~al.} (\bibinfo {collaboration} {DUNE Collaboration}),\
  }\bibfield  {title} {\bibinfo {title} {{Long-baseline neutrino oscillation
  physics potential of the DUNE experiment}},\ }\href
  {https://doi.org/10.1140/epjc/s10052-020-08456-z} {\bibfield  {journal}
  {\bibinfo  {journal} {Eur. Phys. J. C}\ }\textbf {\bibinfo {volume} {80}},\
  \bibinfo {pages} {978} (\bibinfo {year} {2020})}\BibitemShut {NoStop}%
\bibitem [{\citenamefont {Abi}\ \emph {et~al.}(2021{\natexlab{a}})\citenamefont
  {Abi} \emph {et~al.}}]{DUNE2:2016oaz}%
  \BibitemOpen
  \bibfield  {author} {\bibinfo {author} {\bibfnamefont {B.}~\bibnamefont
  {Abi}} \emph {et~al.} (\bibinfo {collaboration} {DUNE Collaboration}),\
  }\bibfield  {title} {\bibinfo {title} {{Prospects for beyond the Standard
  Model physics searches at the Deep Underground Neutrino Experiment}},\ }\href
  {https://doi.org/10.1140/epjc/s10052-021-09007-w} {\bibfield  {journal}
  {\bibinfo  {journal} {Eur. Phys. J. C}\ }\textbf {\bibinfo {volume} {81}},\
  \bibinfo {pages} {322} (\bibinfo {year} {2021}{\natexlab{a}})}\BibitemShut
  {NoStop}%
\bibitem [{\citenamefont {Abi}\ \emph {et~al.}(2021{\natexlab{b}})\citenamefont
  {Abi} \emph {et~al.}}]{DUNE3:2016oaz}%
  \BibitemOpen
  \bibfield  {author} {\bibinfo {author} {\bibfnamefont {B.}~\bibnamefont
  {Abi}} \emph {et~al.} (\bibinfo {collaboration} {DUNE Collaboration}),\
  }\bibfield  {title} {\bibinfo {title} {{Supernova neutrino burst detection
  with the Deep Underground Neutrino Experiment}},\ }\href
  {https://doi.org/10.1140/epjc/s10052-021-09166-w} {\bibfield  {journal}
  {\bibinfo  {journal} {Eur. Phys. J. C}\ }\textbf {\bibinfo {volume} {81}},\
  \bibinfo {pages} {423} (\bibinfo {year} {2021}{\natexlab{b}})}\BibitemShut
  {NoStop}%
\bibitem [{\citenamefont {Abe}\ \emph {et~al.}(2018{\natexlab{a}})\citenamefont
  {Abe} \emph {et~al.}}]{HK}%
  \BibitemOpen
  \bibfield  {author} {\bibinfo {author} {\bibfnamefont {K.}~\bibnamefont
  {Abe}} \emph {et~al.} (\bibinfo {collaboration} {Hyper-Kamiokande
  Collaboration}),\ }\href@noop {} {\bibinfo {title} {{Hyper-Kamiokande Design
  Report}}} (\bibinfo {year} {2018}{\natexlab{a}}),\ \Eprint
  {https://arxiv.org/abs/1805.04163} {arXiv:1805.04163} \BibitemShut {NoStop}%
\bibitem [{\citenamefont {Nagu}\ \emph {et~al.}(2020)\citenamefont {Nagu},
  \citenamefont {Singh}, \citenamefont {Singh},\ and\ \citenamefont
  {Singh}}]{NAGU2020114888}%
  \BibitemOpen
  \bibfield  {author} {\bibinfo {author} {\bibfnamefont {S.}~\bibnamefont
  {Nagu}}, \bibinfo {author} {\bibfnamefont {J.}~\bibnamefont {Singh}},
  \bibinfo {author} {\bibfnamefont {J.}~\bibnamefont {Singh}},\ and\ \bibinfo
  {author} {\bibfnamefont {R.}~\bibnamefont {Singh}},\ }\bibfield  {title}
  {\bibinfo {title} {Impact of cross-sectional uncertainties on {DUNE}
  sensitivity due to nuclear effects},\ }\href
  {https://doi.org/https://doi.org/10.1016/j.nuclphysb.2019.114888} {\bibfield
  {journal} {\bibinfo  {journal} {Nuclear Physics B}\ }\textbf {\bibinfo
  {volume} {951}},\ \bibinfo {pages} {114888} (\bibinfo {year}
  {2020})}\BibitemShut {NoStop}%
\bibitem [{\citenamefont {Dieminger}\ \emph {et~al.}(2023)\citenamefont
  {Dieminger}, \citenamefont {Dolan}, \citenamefont {Sgalaberna}, \citenamefont
  {Nikolakopoulos}, \citenamefont {Dealtry}, \citenamefont {Bolognesi},
  \citenamefont {Pickering},\ and\ \citenamefont
  {Rubbia}}]{dieminger2023uncertainties}%
  \BibitemOpen
  \bibfield  {author} {\bibinfo {author} {\bibfnamefont {T.}~\bibnamefont
  {Dieminger}}, \bibinfo {author} {\bibfnamefont {S.}~\bibnamefont {Dolan}},
  \bibinfo {author} {\bibfnamefont {D.}~\bibnamefont {Sgalaberna}}, \bibinfo
  {author} {\bibfnamefont {A.}~\bibnamefont {Nikolakopoulos}}, \bibinfo
  {author} {\bibfnamefont {T.}~\bibnamefont {Dealtry}}, \bibinfo {author}
  {\bibfnamefont {S.}~\bibnamefont {Bolognesi}}, \bibinfo {author}
  {\bibfnamefont {L.}~\bibnamefont {Pickering}},\ and\ \bibinfo {author}
  {\bibfnamefont {A.}~\bibnamefont {Rubbia}},\ }\href@noop {} {\bibinfo {title}
  {Uncertainties on the $\nu_{\mu}$/$\nu_{e}$,
  $\bar{\nu}_{\mu}$/$\bar{\nu}_{e}$ and $\nu_{e}$/$\bar{\nu}_{e}$ cross-section
  ratio from the modelling of nuclear effects and their impact on neutrino
  oscillation experiments}} (\bibinfo {year} {2023}),\ \Eprint
  {https://arxiv.org/abs/2301.08065} {arXiv:2301.08065 [hep-ph]} \BibitemShut
  {NoStop}%
\bibitem [{\citenamefont {Avanzini}\ \emph {et~al.}(2022)\citenamefont
  {Avanzini} \emph {et~al.}}]{Tensions2019}%
  \BibitemOpen
  \bibfield  {author} {\bibinfo {author} {\bibfnamefont {M.~B.}\ \bibnamefont
  {Avanzini}} \emph {et~al.},\ }\bibfield  {title} {\bibinfo {title}
  {Comparisons and challenges of modern neutrino-scattering experiments},\
  }\href {https://doi.org/10.1103/PhysRevD.105.092004} {\bibfield  {journal}
  {\bibinfo  {journal} {Phys. Rev. D}\ }\textbf {\bibinfo {volume} {105}},\
  \bibinfo {pages} {092004} (\bibinfo {year} {2022})}\BibitemShut {NoStop}%
\bibitem [{\citenamefont {Alvarez-Ruso}\ and\ \citenamefont
  {othersr}(2018)}]{ALVAREZRUSO20181}%
  \BibitemOpen
  \bibfield  {author} {\bibinfo {author} {\bibfnamefont {L.}~\bibnamefont
  {Alvarez-Ruso}}\ and\ \bibinfo {author} {\bibnamefont {othersr}},\ }\bibfield
   {title} {\bibinfo {title} {Nustec white paper: Status and challenges of
  neutrino–nucleus scattering},\ }\href
  {https://doi.org/https://doi.org/10.1016/j.ppnp.2018.01.006} {\bibfield
  {journal} {\bibinfo  {journal} {Progress in Particle and Nuclear Physics}\
  }\textbf {\bibinfo {volume} {100}},\ \bibinfo {pages} {1} (\bibinfo {year}
  {2018})}\BibitemShut {NoStop}%
\bibitem [{\citenamefont {Betancourt}\ \emph {et~al.}(2018)\citenamefont
  {Betancourt} \emph {et~al.}}]{BETANCOURT20181}%
  \BibitemOpen
  \bibfield  {author} {\bibinfo {author} {\bibfnamefont {M.}~\bibnamefont
  {Betancourt}} \emph {et~al.},\ }\bibfield  {title} {\bibinfo {title}
  {Comparisons and challenges of modern neutrino scattering experiments
  (tensions2016 report)},\ }\href
  {https://doi.org/https://doi.org/10.1016/j.physrep.2018.08.003} {\bibfield
  {journal} {\bibinfo  {journal} {Physics Reports}\ }\textbf {\bibinfo {volume}
  {773-774}},\ \bibinfo {pages} {1} (\bibinfo {year} {2018})},\ \bibinfo {note}
  {comparisons and challenges of modern neutrino scattering experiments
  (TENSIONS2016 report)}\BibitemShut {NoStop}%
\bibitem [{\citenamefont {Abe}\ \emph {et~al.}(2021)\citenamefont {Abe} \emph
  {et~al.}}]{PhysRevD.103.112009}%
  \BibitemOpen
  \bibfield  {author} {\bibinfo {author} {\bibfnamefont {K.}~\bibnamefont
  {Abe}} \emph {et~al.} (\bibinfo {collaboration} {T2K Collaboration}),\
  }\bibfield  {title} {\bibinfo {title} {First {T2K} measurement of transverse
  kinematic imbalance in the muon-neutrino charged-current
  single-${\ensuremath{\pi}}^{+}$ production channel containing at least one
  proton},\ }\href {https://doi.org/10.1103/PhysRevD.103.112009} {\bibfield
  {journal} {\bibinfo  {journal} {Phys. Rev. D}\ }\textbf {\bibinfo {volume}
  {103}},\ \bibinfo {pages} {112009} (\bibinfo {year} {2021})}\BibitemShut
  {NoStop}%
\bibitem [{\citenamefont {Abratenko}\ \emph
  {et~al.}(2023{\natexlab{a}})\citenamefont {Abratenko} \emph
  {et~al.}}]{RefPRL}%
  \BibitemOpen
  \bibfield  {author} {\bibinfo {author} {\bibfnamefont {P.}~\bibnamefont
  {Abratenko}} \emph {et~al.} (\bibinfo {collaboration} {MicroBooNE
  Collaboration}),\ }\href@noop {} {\bibinfo {title} {{First
  double-differential measurement of kinematic imbalance in neutrino
  interactions with the MicroBooNE detector}}} (\bibinfo {year}
  {2023}{\natexlab{a}}),\ \Eprint {https://arxiv.org/abs/2301.03706}
  {arXiv:2301.03706} \BibitemShut {NoStop}%
\bibitem [{\citenamefont {Abratenko}\ \emph
  {et~al.}(2023{\natexlab{b}})\citenamefont {Abratenko} \emph
  {et~al.}}]{RefPRD}%
  \BibitemOpen
  \bibfield  {author} {\bibinfo {author} {\bibfnamefont {P.}~\bibnamefont
  {Abratenko}} \emph {et~al.} (\bibinfo {collaboration} {MicroBooNE
  Collaboration}),\ }\href@noop {} {\bibinfo {title} {{Multi-Differential Cross
  Section Measurements of Muon-Neutrino-Argon Quasielastic-like Reactions with
  the MicroBooNE Detector}}} (\bibinfo {year} {2023}{\natexlab{b}}),\ \Eprint
  {https://arxiv.org/abs/2301.03700} {arXiv:2301.03700} \BibitemShut {NoStop}%
\bibitem [{\citenamefont {Acciarri}\ \emph {et~al.}(2017)\citenamefont
  {Acciarri} \emph {et~al.}}]{Acciarri:2016smi}%
  \BibitemOpen
  \bibfield  {author} {\bibinfo {author} {\bibfnamefont {R.}~\bibnamefont
  {Acciarri}} \emph {et~al.} (\bibinfo {collaboration} {MicroBooNE
  Collaboration}),\ }\bibfield  {title} {\bibinfo {title} {{Design and
  Construction of the MicroBooNE Detector}},\ }\href
  {https://doi.org/10.1088/1748-0221/12/02/P02017} {\bibfield  {journal}
  {\bibinfo  {journal} {J. Instrum.}\ }\textbf {\bibinfo {volume} {12}}\bibinfo
   {number} { (02)},\ \bibinfo {pages} {P02017}}\BibitemShut {NoStop}%
\bibitem [{\citenamefont {Aguilar-Arevalo}\ \emph {et~al.}(2009)\citenamefont
  {Aguilar-Arevalo} \emph {et~al.}}]{AguilarArevalo:2008yp}%
  \BibitemOpen
\bibfield  {number} {  }\bibfield  {author} {\bibinfo {author} {\bibfnamefont
  {A.}~\bibnamefont {Aguilar-Arevalo}} \emph {et~al.} (\bibinfo {collaboration}
  {MiniBooNE Collaboration}),\ }\bibfield  {title} {\bibinfo {title} {{The
  Neutrino Flux prediction at MiniBooNE}},\ }\href
  {https://doi.org/10.1103/PhysRevD.79.072002} {\bibfield  {journal} {\bibinfo
  {journal} {Phys. Rev. D}\ }\textbf {\bibinfo {volume} {79}},\ \bibinfo
  {pages} {072002} (\bibinfo {year} {2009})}\BibitemShut {NoStop}%
\bibitem [{\citenamefont {Lu}\ \emph {et~al.}(2016)\citenamefont {Lu} \emph
  {et~al.}}]{PhysRevC.94.015503}%
  \BibitemOpen
  \bibfield  {author} {\bibinfo {author} {\bibfnamefont {X.-G.}\ \bibnamefont
  {Lu}} \emph {et~al.},\ }\bibfield  {title} {\bibinfo {title} {Measurement of
  nuclear effects in neutrino interactions with minimal dependence on neutrino
  energy},\ }\href {https://doi.org/10.1103/PhysRevC.94.015503} {\bibfield
  {journal} {\bibinfo  {journal} {Phys. Rev. C}\ }\textbf {\bibinfo {volume}
  {94}},\ \bibinfo {pages} {015503} (\bibinfo {year} {2016})}\BibitemShut
  {NoStop}%
\bibitem [{\citenamefont {Lu}\ \emph {et~al.}(2018)\citenamefont {Lu} \emph
  {et~al.}}]{PhysRevLett.121.022504}%
  \BibitemOpen
  \bibfield  {author} {\bibinfo {author} {\bibfnamefont {X.-G.}\ \bibnamefont
  {Lu}} \emph {et~al.} (\bibinfo {collaboration} {MINERvA Collaboration}),\
  }\bibfield  {title} {\bibinfo {title} {Measurement of final-state
  correlations in neutrino muon-proton mesonless production on hydrocarbon at
  $\langle {E}_{\ensuremath{\nu}}\rangle=3\,\mathrm{GeV}$},\ }\href
  {https://doi.org/10.1103/PhysRevLett.121.022504} {\bibfield  {journal}
  {\bibinfo  {journal} {Phys. Rev. Lett.}\ }\textbf {\bibinfo {volume} {121}},\
  \bibinfo {pages} {022504} (\bibinfo {year} {2018})}\BibitemShut {NoStop}%
\bibitem [{\citenamefont {Cai}\ \emph {et~al.}(2020)\citenamefont {Cai} \emph
  {et~al.}}]{PhysRevD.101.092001}%
  \BibitemOpen
  \bibfield  {author} {\bibinfo {author} {\bibfnamefont {T.}~\bibnamefont
  {Cai}} \emph {et~al.} (\bibinfo {collaboration} {MINERvA Collaboration}),\
  }\bibfield  {title} {\bibinfo {title} {Nucleon binding energy and transverse
  momentum imbalance in neutrino-nucleus reactions},\ }\href
  {https://doi.org/10.1103/PhysRevD.101.092001} {\bibfield  {journal} {\bibinfo
   {journal} {Phys. Rev. D}\ }\textbf {\bibinfo {volume} {101}},\ \bibinfo
  {pages} {092001} (\bibinfo {year} {2020})}\BibitemShut {NoStop}%
\bibitem [{\citenamefont {Bathe-Peters}\ \emph {et~al.}(2022)\citenamefont
  {Bathe-Peters}, \citenamefont {Gardiner},\ and\ \citenamefont
  {Guenette}}]{Bathe-Peters:2022kkj}%
  \BibitemOpen
  \bibfield  {author} {\bibinfo {author} {\bibfnamefont {L.}~\bibnamefont
  {Bathe-Peters}}, \bibinfo {author} {\bibfnamefont {S.}~\bibnamefont
  {Gardiner}},\ and\ \bibinfo {author} {\bibfnamefont {R.}~\bibnamefont
  {Guenette}},\ }\href@noop {} {\bibinfo {title} {{Comparing generator
  predictions of transverse kinematic imbalance in neutrino-argon scattering}}}
  (\bibinfo {year} {2022}),\ \Eprint {https://arxiv.org/abs/2201.04664}
  {arXiv:2201.04664} \BibitemShut {NoStop}%
\bibitem [{\citenamefont {Abe}\ \emph {et~al.}(2018{\natexlab{b}})\citenamefont
  {Abe} \emph {et~al.}}]{Abe:2018pwo}%
  \BibitemOpen
  \bibfield  {author} {\bibinfo {author} {\bibfnamefont {K.}~\bibnamefont
  {Abe}} \emph {et~al.} (\bibinfo {collaboration} {T2K Collaboration}),\
  }\bibfield  {title} {\bibinfo {title} {{Characterization of nuclear effects
  in muon-neutrino scattering on hydrocarbon with a measurement of final-state
  kinematics and correlations in charged-current pionless interactions at
  {T2K}}},\ }\href {https://doi.org/10.1103/PhysRevD.98.032003} {\bibfield
  {journal} {\bibinfo  {journal} {Phys.\ Rev.\ D}\ }\textbf {\bibinfo {volume}
  {98}},\ \bibinfo {pages} {032003} (\bibinfo {year}
  {2018}{\natexlab{b}})}\BibitemShut {NoStop}%
\bibitem [{\citenamefont {Coplowe}\ \emph {et~al.}(2020)\citenamefont {Coplowe}
  \emph {et~al.}}]{PhysRevD.102.072007}%
  \BibitemOpen
  \bibfield  {author} {\bibinfo {author} {\bibfnamefont {D.}~\bibnamefont
  {Coplowe}} \emph {et~al.} (\bibinfo {collaboration}
  {$\mathrm{MINER}\ensuremath{\nu}\mathrm{A}$ Collaboration}),\ }\bibfield
  {title} {\bibinfo {title} {Probing nuclear effects with neutrino-induced
  charged-current neutral pion production},\ }\href
  {https://doi.org/10.1103/PhysRevD.102.072007} {\bibfield  {journal} {\bibinfo
   {journal} {Phys. Rev. D}\ }\textbf {\bibinfo {volume} {102}},\ \bibinfo
  {pages} {072007} (\bibinfo {year} {2020})}\BibitemShut {NoStop}%
\bibitem [{\citenamefont {Furmanski}\ and\ \citenamefont
  {Sobczyk}(2017)}]{Furmanski2016}%
  \BibitemOpen
  \bibfield  {author} {\bibinfo {author} {\bibfnamefont {A.~P.}\ \bibnamefont
  {Furmanski}}\ and\ \bibinfo {author} {\bibfnamefont {J.~T.}\ \bibnamefont
  {Sobczyk}},\ }\bibfield  {title} {\bibinfo {title} {Neutrino energy
  reconstruction from one-muon and one-proton events},\ }\href
  {https://doi.org/10.1103/PhysRevC.95.065501} {\bibfield  {journal} {\bibinfo
  {journal} {Phys. Rev. C}\ }\textbf {\bibinfo {volume} {95}},\ \bibinfo
  {pages} {065501} (\bibinfo {year} {2017})}\BibitemShut {NoStop}%
\bibitem [{\citenamefont {Lu}\ and\ \citenamefont
  {Sobczyk}(2019)}]{PhysRevC.99.055504}%
  \BibitemOpen
  \bibfield  {author} {\bibinfo {author} {\bibfnamefont {X.}~\bibnamefont
  {Lu}}\ and\ \bibinfo {author} {\bibfnamefont {J.~T.}\ \bibnamefont
  {Sobczyk}},\ }\bibfield  {title} {\bibinfo {title} {Identification of nuclear
  effects in neutrino and antineutrino interactions on nuclei using generalized
  final-state correlations},\ }\href
  {https://doi.org/10.1103/PhysRevC.99.055504} {\bibfield  {journal} {\bibinfo
  {journal} {Phys. Rev. C}\ }\textbf {\bibinfo {volume} {99}},\ \bibinfo
  {pages} {055504} (\bibinfo {year} {2019})}\BibitemShut {NoStop}%
\bibitem [{\citenamefont {Lu}(2018)}]{WCLu}%
  \BibitemOpen
  \bibfield  {author} {\bibinfo {author} {\bibfnamefont {X.}~\bibnamefont
  {Lu}},\ }\href@noop {} {\bibinfo {title} {{Neutrino Shadow Play-Kinematic
  determination of nuclear effects at MINERvA}}},\ \bibinfo {howpublished}
  {https://minerva-docdb.fnal.gov/cgi-bin/sso/ShowDocument?docid=17864}
  (\bibinfo {year} {2018})\BibitemShut {NoStop}%
\bibitem [{\citenamefont {Gran}\ \emph {et~al.}(2006)\citenamefont {Gran} \emph
  {et~al.}}]{PhysRevD.74.052002}%
  \BibitemOpen
  \bibfield  {author} {\bibinfo {author} {\bibfnamefont {R.}~\bibnamefont
  {Gran}} \emph {et~al.} (\bibinfo {collaboration} {K2K Collaboration}),\
  }\bibfield  {title} {\bibinfo {title} {Measurement of the quasielastic axial
  vector mass in neutrino interactions on oxygen},\ }\href
  {https://doi.org/10.1103/PhysRevD.74.052002} {\bibfield  {journal} {\bibinfo
  {journal} {Phys. Rev. D}\ }\textbf {\bibinfo {volume} {74}},\ \bibinfo
  {pages} {052002} (\bibinfo {year} {2006})}\BibitemShut {NoStop}%
\bibitem [{\citenamefont {Bodek}\ and\ \citenamefont
  {Cai}(2019)}]{BodekCai2019}%
  \BibitemOpen
  \bibfield  {author} {\bibinfo {author} {\bibfnamefont {A.}~\bibnamefont
  {Bodek}}\ and\ \bibinfo {author} {\bibfnamefont {T.}~\bibnamefont {Cai}},\
  }\bibfield  {title} {\bibinfo {title} {Removal energies and final state
  interaction in lepton nucleus scattering},\ }\href
  {https://doi.org/https://doi.org/10.1140/epjc/s10052-019-6750-3} {\bibfield
  {journal} {\bibinfo  {journal} {Eur. Phys. J. C}\ }\textbf {\bibinfo {volume}
  {79}},\ \bibinfo {pages} {293} (\bibinfo {year} {2019})}\BibitemShut
  {NoStop}%
\bibitem [{\citenamefont {Bourguille}\ \emph {et~al.}(2021)\citenamefont
  {Bourguille}, \citenamefont {Nieves},\ and\ \citenamefont {Sánchez}}]{PWIA}%
  \BibitemOpen
  \bibfield  {author} {\bibinfo {author} {\bibfnamefont {B.}~\bibnamefont
  {Bourguille}}, \bibinfo {author} {\bibfnamefont {J.}~\bibnamefont {Nieves}},\
  and\ \bibinfo {author} {\bibfnamefont {F.}~\bibnamefont {Sánchez}},\
  }\bibfield  {title} {\bibinfo {title} {Inclusive and exclusive
  neutrino-nucleus cross sections and the reconstruction of the interaction
  kinematics},\ }\href {https://doi.org/10.1007/JHEP04(2021)004} {\bibfield
  {journal} {\bibinfo  {journal} {J. High Energ. Phys.}\ }\textbf {\bibinfo
  {volume} {2021}},\ \bibinfo {pages} {153}}\BibitemShut {NoStop}%
\bibitem [{\citenamefont {{P. Abratenko et
  al.}}(2022{\natexlab{a}})}]{GENIE_tune}%
  \BibitemOpen
  \bibfield  {author} {\bibinfo {author} {\bibnamefont {{P. Abratenko et al.}}}
  (\bibinfo {collaboration} {MicroBooNE Collaboration}),\ }\bibfield  {title}
  {\bibinfo {title} {New $\mathrm{CC}0\ensuremath{\pi}$ genie model tune for
  microboone},\ }\href {https://doi.org/10.1103/PhysRevD.105.072001} {\bibfield
   {journal} {\bibinfo  {journal} {Phys. Rev. D}\ }\textbf {\bibinfo {volume}
  {105}},\ \bibinfo {pages} {072001} (\bibinfo {year}
  {2022}{\natexlab{a}})}\BibitemShut {NoStop}%
\bibitem [{\citenamefont {Stowell}\ \emph {et~al.}(2017)\citenamefont
  {Stowell}, \citenamefont {Wret}, \citenamefont {Wilkinson}, \citenamefont
  {Pickering}, \citenamefont {Cartwright}, \citenamefont {Hayato},
  \citenamefont {Mahn}, \citenamefont {McFarland}, \citenamefont {Sobczyk},
  \citenamefont {Terri}, \citenamefont {Thompson}, \citenamefont {Wascko},\
  and\ \citenamefont {Uchida}}]{Stowell_2017}%
  \BibitemOpen
  \bibfield  {author} {\bibinfo {author} {\bibfnamefont {P.}~\bibnamefont
  {Stowell}}, \bibinfo {author} {\bibfnamefont {C.}~\bibnamefont {Wret}},
  \bibinfo {author} {\bibfnamefont {C.}~\bibnamefont {Wilkinson}}, \bibinfo
  {author} {\bibfnamefont {L.}~\bibnamefont {Pickering}}, \bibinfo {author}
  {\bibfnamefont {S.}~\bibnamefont {Cartwright}}, \bibinfo {author}
  {\bibfnamefont {Y.}~\bibnamefont {Hayato}}, \bibinfo {author} {\bibfnamefont
  {K.}~\bibnamefont {Mahn}}, \bibinfo {author} {\bibfnamefont {K.}~\bibnamefont
  {McFarland}}, \bibinfo {author} {\bibfnamefont {J.}~\bibnamefont {Sobczyk}},
  \bibinfo {author} {\bibfnamefont {R.}~\bibnamefont {Terri}}, \bibinfo
  {author} {\bibfnamefont {L.}~\bibnamefont {Thompson}}, \bibinfo {author}
  {\bibfnamefont {M.}~\bibnamefont {Wascko}},\ and\ \bibinfo {author}
  {\bibfnamefont {Y.}~\bibnamefont {Uchida}},\ }\bibfield  {title} {\bibinfo
  {title} {{NUISANCE}: a neutrino cross-section generator tuning and comparison
  framework},\ }\href {https://doi.org/10.1088/1748-0221/12/01/p01016}
  {\bibfield  {journal} {\bibinfo  {journal} {J. Instrum.}\ }\textbf {\bibinfo
  {volume} {12}}\bibinfo  {number} { (01)},\ \bibinfo {pages}
  {P01016}}\BibitemShut {NoStop}%
\bibitem [{\citenamefont {Andreopoulos}\ \emph {et~al.}(2010)\citenamefont
  {Andreopoulos} \emph {et~al.}}]{Andreopoulos:2009rq}%
  \BibitemOpen
\bibfield  {number} {  }\bibfield  {author} {\bibinfo {author} {\bibfnamefont
  {C.}~\bibnamefont {Andreopoulos}} \emph {et~al.},\ }\bibfield  {title}
  {\bibinfo {title} {{The GENIE Neutrino Monte Carlo Generator}},\ }\href
  {https://doi.org/10.1016/j.nima.2009.12.009} {\bibfield  {journal} {\bibinfo
  {journal} {Nucl.\ Instrum.\ Meth.\ A}\ }\textbf {\bibinfo {volume} {614}},\
  \bibinfo {pages} {87} (\bibinfo {year} {2010})}\BibitemShut {NoStop}%
\bibitem [{\citenamefont {Andreopoulos}\ \emph {et~al.}(2015)\citenamefont
  {Andreopoulos} \emph {et~al.}}]{Andreopoulos:2015wxa}%
  \BibitemOpen
  \bibfield  {author} {\bibinfo {author} {\bibfnamefont {C.}~\bibnamefont
  {Andreopoulos}} \emph {et~al.},\ }\href@noop {} {\bibinfo {title} {{The GENIE
  Neutrino Monte Carlo Generator: Physics and User Manual}}} (\bibinfo {year}
  {2015}),\ \Eprint {https://arxiv.org/abs/1510.05494} {arXiv:1510.05494}
  \BibitemShut {NoStop}%
\bibitem [{\citenamefont {Alvarez-Ruso}\ \emph {et~al.}(2021)\citenamefont
  {Alvarez-Ruso} \emph {et~al.}}]{geniev3highlights}%
  \BibitemOpen
  \bibfield  {author} {\bibinfo {author} {\bibfnamefont {L.}~\bibnamefont
  {Alvarez-Ruso}} \emph {et~al.} (\bibinfo {collaboration} {GENIE
  Collaboration}),\ }\bibfield  {title} {\bibinfo {title} {{Recent highlights
  from GENIE v3}},\ }\href {https://doi.org/10.1140/epjs/s11734-021-00295-7}
  {\bibfield  {journal} {\bibinfo  {journal} {Eur. Phys. J. ST}\ }\textbf
  {\bibinfo {volume} {230}},\ \bibinfo {pages} {4449} (\bibinfo {year}
  {2021})}\BibitemShut {NoStop}%
\bibitem [{\citenamefont {Mosel}(2019)}]{Mosel:2019vhx}%
  \BibitemOpen
  \bibfield  {author} {\bibinfo {author} {\bibfnamefont {U.}~\bibnamefont
  {Mosel}},\ }\bibfield  {title} {\bibinfo {title} {Neutrino event generators:
  foundation, status and future},\ }\href
  {https://doi.org/10.1088/1361-6471/ab3830} {\bibfield  {journal} {\bibinfo
  {journal} {Phys. Rev. G}\ }\textbf {\bibinfo {volume} {46}},\ \bibinfo
  {pages} {113001} (\bibinfo {year} {2019})}\BibitemShut {NoStop}%
\bibitem [{\citenamefont {Golan}\ \emph {et~al.}(2012)\citenamefont {Golan}
  \emph {et~al.}}]{GolanNuWro:2008yp}%
  \BibitemOpen
  \bibfield  {author} {\bibinfo {author} {\bibfnamefont {T.}~\bibnamefont
  {Golan}} \emph {et~al.},\ }\bibfield  {title} {\bibinfo {title} {{NuWro: the
  Wroclaw Monte Carlo Generator of Neutrino Interactions}},\ }\href@noop {}
  {\bibfield  {journal} {\bibinfo  {journal} {Nucl.Phys.Proc.Suppl.}\ }\textbf
  {\bibinfo {volume} {499}},\ \bibinfo {pages} {229} (\bibinfo {year}
  {2012})}\BibitemShut {NoStop}%
\bibitem [{\citenamefont {Hayato}(2009)}]{Hayato:2008yp}%
  \BibitemOpen
  \bibfield  {author} {\bibinfo {author} {\bibfnamefont {Y.}~\bibnamefont
  {Hayato}},\ }\bibfield  {title} {\bibinfo {title} {{A neutrino interaction
  simulation program library NEUT}},\ }\href@noop {} {\bibfield  {journal}
  {\bibinfo  {journal} {Acta Phys. Polon.}\ }\textbf {\bibinfo {volume}
  {B40}},\ \bibinfo {pages} {2477} (\bibinfo {year} {2009})}\BibitemShut
  {NoStop}%
\bibitem [{\citenamefont {Abratenko}\ \emph
  {et~al.}(2020{\natexlab{a}})\citenamefont {Abratenko} \emph
  {et~al.}}]{PhysRevLett.125.201803}%
  \BibitemOpen
  \bibfield  {author} {\bibinfo {author} {\bibfnamefont {P.}~\bibnamefont
  {Abratenko}} \emph {et~al.} (\bibinfo {collaboration} {MicroBooNE
  Collaboration}),\ }\bibfield  {title} {\bibinfo {title} {First measurement of
  differential charged current quasielasticlike
  ${\ensuremath{\nu}}_{\ensuremath{\mu}}$-argon scattering cross sections with
  the microboone detector},\ }\href
  {https://doi.org/10.1103/PhysRevLett.125.201803} {\bibfield  {journal}
  {\bibinfo  {journal} {Phys. Rev. Lett.}\ }\textbf {\bibinfo {volume} {125}},\
  \bibinfo {pages} {201803} (\bibinfo {year} {2020}{\natexlab{a}})}\BibitemShut
  {NoStop}%
\bibitem [{\citenamefont {Abratenko}\ \emph
  {et~al.}(2020{\natexlab{b}})\citenamefont {Abratenko} \emph
  {et~al.}}]{PhysRevD.102.112013}%
  \BibitemOpen
  \bibfield  {author} {\bibinfo {author} {\bibfnamefont {P.}~\bibnamefont
  {Abratenko}} \emph {et~al.} (\bibinfo {collaboration} {MicroBooNE
  Collaboration}),\ }\bibfield  {title} {\bibinfo {title} {Measurement of
  differential cross sections for ${\ensuremath{\nu}}_{\ensuremath{\mu}}$-ar
  charged-current interactions with protons and no pions in the final state
  with the microboone detector},\ }\href
  {https://doi.org/10.1103/PhysRevD.102.112013} {\bibfield  {journal} {\bibinfo
   {journal} {Phys. Rev. D}\ }\textbf {\bibinfo {volume} {102}},\ \bibinfo
  {pages} {112013} (\bibinfo {year} {2020}{\natexlab{b}})}\BibitemShut
  {NoStop}%
\bibitem [{\citenamefont {Llewellyn~Smith}(1972)}]{LlewellynSmith:1971uhs}%
  \BibitemOpen
  \bibfield  {author} {\bibinfo {author} {\bibfnamefont {C.}~\bibnamefont
  {Llewellyn~Smith}},\ }\bibfield  {title} {\bibinfo {title} {{Neutrino
  Reactions at Accelerator Energies}},\ }\href
  {https://doi.org/10.1016/0370-1573(72)90010-5} {\bibfield  {journal}
  {\bibinfo  {journal} {Phys.\ Rept.}\ }\textbf {\bibinfo {volume} {3}},\
  \bibinfo {pages} {261} (\bibinfo {year} {1972})}\BibitemShut {NoStop}%
\bibitem [{\citenamefont {Katori}(2015)}]{Katori:2013eoa}%
  \BibitemOpen
  \bibfield  {author} {\bibinfo {author} {\bibfnamefont {T.}~\bibnamefont
  {Katori}},\ }\bibfield  {title} {\bibinfo {title} {{Meson Exchange Current
  (MEC) Models in Neutrino Interaction Generators}},\ }\href
  {https://doi.org/10.1063/1.4919465} {\bibfield  {journal} {\bibinfo
  {journal} {AIP Conf.\ Proc.}\ }\textbf {\bibinfo {volume} {1663}},\ \bibinfo
  {pages} {030001} (\bibinfo {year} {2015})}\BibitemShut {NoStop}%
\bibitem [{\citenamefont {Rein}\ and\ \citenamefont
  {Sehgal}(1981)}]{Rein:1980wg}%
  \BibitemOpen
  \bibfield  {author} {\bibinfo {author} {\bibfnamefont {D.}~\bibnamefont
  {Rein}}\ and\ \bibinfo {author} {\bibfnamefont {L.}~\bibnamefont {Sehgal}},\
  }\bibfield  {title} {\bibinfo {title} {{Neutrino Excitation of Baryon
  Resonances and Single Pion Production}},\ }\href
  {https://doi.org/10.1016/0003-4916(81)90242-6} {\bibfield  {journal}
  {\bibinfo  {journal} {Annals Phys.}\ }\textbf {\bibinfo {volume} {133}},\
  \bibinfo {pages} {79} (\bibinfo {year} {1981})}\BibitemShut {NoStop}%
\bibitem [{\citenamefont {Yang}\ and\ \citenamefont
  {Bodek}(1999)}]{PhysRevLett.82.2467}%
  \BibitemOpen
  \bibfield  {author} {\bibinfo {author} {\bibfnamefont {U.~K.}\ \bibnamefont
  {Yang}}\ and\ \bibinfo {author} {\bibfnamefont {A.}~\bibnamefont {Bodek}},\
  }\bibfield  {title} {\bibinfo {title} {Parton distributions,
  $\mathit{d}/\mathit{u}$, and higher twist effects at high $\mathit{x}$},\
  }\href {https://doi.org/10.1103/PhysRevLett.82.2467} {\bibfield  {journal}
  {\bibinfo  {journal} {Phys. Rev. Lett.}\ }\textbf {\bibinfo {volume} {82}},\
  \bibinfo {pages} {2467} (\bibinfo {year} {1999})}\BibitemShut {NoStop}%
\bibitem [{\citenamefont {Sjostrand}\ \emph {et~al.}(2006)\citenamefont
  {Sjostrand}, \citenamefont {Mrenna},\ and\ \citenamefont
  {Skands}}]{Sjostrand:2006za}%
  \BibitemOpen
  \bibfield  {author} {\bibinfo {author} {\bibfnamefont {T.}~\bibnamefont
  {Sjostrand}}, \bibinfo {author} {\bibfnamefont {S.}~\bibnamefont {Mrenna}},\
  and\ \bibinfo {author} {\bibfnamefont {P.~Z.}\ \bibnamefont {Skands}},\
  }\bibfield  {title} {\bibinfo {title} {{PYTHIA 6.4 Physics and Manual}},\
  }\href {https://doi.org/10.1088/1126-6708/2006/05/026} {\bibfield  {journal}
  {\bibinfo  {journal} {J. High Energ. Phys.}\ }\textbf {\bibinfo {volume}
  {2006}}\bibinfo  {number} { (05)},\ \bibinfo {pages} {026}}\BibitemShut
  {NoStop}%
\bibitem [{\citenamefont {Mashnik}\ \emph {et~al.}(2006)\citenamefont
  {Mashnik}, \citenamefont {Sierk}, \citenamefont {Gudima},\ and\ \citenamefont
  {Baznat}}]{Mashnik:2005ay}%
  \BibitemOpen
\bibfield  {number} {  }\bibfield  {author} {\bibinfo {author} {\bibfnamefont
  {S.}~\bibnamefont {Mashnik}}, \bibinfo {author} {\bibfnamefont
  {A.}~\bibnamefont {Sierk}}, \bibinfo {author} {\bibfnamefont
  {K.}~\bibnamefont {Gudima}},\ and\ \bibinfo {author} {\bibfnamefont
  {M.}~\bibnamefont {Baznat}},\ }\bibfield  {title} {\bibinfo {title} {{CEM03
  and LAQGSM03: New modeling tools for nuclear applications}},\ }\href
  {https://doi.org/10.1088/1742-6596/41/1/037} {\bibfield  {journal} {\bibinfo
  {journal} {J.\ Phys.\ Conf.\ Ser.}\ }\textbf {\bibinfo {volume} {41}},\
  \bibinfo {pages} {340} (\bibinfo {year} {2006})}\BibitemShut {NoStop}%
\bibitem [{\citenamefont {Carrasco}\ and\ \citenamefont
  {Oset}(1992)}]{Carrasco:1989vq}%
  \BibitemOpen
  \bibfield  {author} {\bibinfo {author} {\bibfnamefont {R.}~\bibnamefont
  {Carrasco}}\ and\ \bibinfo {author} {\bibfnamefont {E.}~\bibnamefont
  {Oset}},\ }\bibfield  {title} {\bibinfo {title} {{Interaction of Real Photons
  With Nuclei From 100-{MeV} to 500-{MeV}}},\ }\href
  {https://doi.org/10.1016/0375-9474(92)90109-W} {\bibfield  {journal}
  {\bibinfo  {journal} {Nucl.\ Phys.\ A}\ }\textbf {\bibinfo {volume} {536}},\
  \bibinfo {pages} {445} (\bibinfo {year} {1992})}\BibitemShut {NoStop}%
\bibitem [{\citenamefont {Nieves}\ \emph {et~al.}(2012)\citenamefont {Nieves},
  \citenamefont {Sanchez}, \citenamefont {Ruiz~Simo},\ and\ \citenamefont
  {Vicente~Vacas}}]{Nieves:2012yz}%
  \BibitemOpen
  \bibfield  {author} {\bibinfo {author} {\bibfnamefont {J.}~\bibnamefont
  {Nieves}}, \bibinfo {author} {\bibfnamefont {F.}~\bibnamefont {Sanchez}},
  \bibinfo {author} {\bibfnamefont {I.}~\bibnamefont {Ruiz~Simo}},\ and\
  \bibinfo {author} {\bibfnamefont {M.}~\bibnamefont {Vicente~Vacas}},\
  }\bibfield  {title} {\bibinfo {title} {{Neutrino Energy Reconstruction and
  the Shape of the CCQE-like Total Cross Section}},\ }\href
  {https://doi.org/10.1103/PhysRevD.85.113008} {\bibfield  {journal} {\bibinfo
  {journal} {Phys.\ Rev.\ D}\ }\textbf {\bibinfo {volume} {85}},\ \bibinfo
  {pages} {113008} (\bibinfo {year} {2012})}\BibitemShut {NoStop}%
\bibitem [{\citenamefont {Engel}(1998)}]{Engel:1997fy}%
  \BibitemOpen
  \bibfield  {author} {\bibinfo {author} {\bibfnamefont {J.}~\bibnamefont
  {Engel}},\ }\bibfield  {title} {\bibinfo {title} {{Approximate treatment of
  lepton distortion in charged current neutrino scattering from nuclei}},\
  }\href {https://doi.org/10.1103/PhysRevC.57.2004} {\bibfield  {journal}
  {\bibinfo  {journal} {Phys. Rev. C}\ }\textbf {\bibinfo {volume} {57}},\
  \bibinfo {pages} {2004} (\bibinfo {year} {1998})}\BibitemShut {NoStop}%
\bibitem [{\citenamefont {Nieves}\ \emph {et~al.}(2004)\citenamefont {Nieves},
  \citenamefont {Amaro},\ and\ \citenamefont {Valverde}}]{RPA}%
  \BibitemOpen
  \bibfield  {author} {\bibinfo {author} {\bibfnamefont {J.}~\bibnamefont
  {Nieves}}, \bibinfo {author} {\bibfnamefont {J.~E.}\ \bibnamefont {Amaro}},\
  and\ \bibinfo {author} {\bibfnamefont {M.}~\bibnamefont {Valverde}},\
  }\bibfield  {title} {\bibinfo {title} {Inclusive quasielastic charged-current
  neutrino-nucleus reactions},\ }\href
  {https://doi.org/10.1103/PhysRevC.70.055503} {\bibfield  {journal} {\bibinfo
  {journal} {Phys. Rev. C}\ }\textbf {\bibinfo {volume} {70}},\ \bibinfo
  {pages} {055503} (\bibinfo {year} {2004})}\BibitemShut {NoStop}%
\bibitem [{\citenamefont {Schwehr}\ \emph {et~al.}(2016)\citenamefont
  {Schwehr}, \citenamefont {Cherdack},\ and\ \citenamefont
  {Gran}}]{Schwehr:2016pvn}%
  \BibitemOpen
  \bibfield  {author} {\bibinfo {author} {\bibfnamefont {J.}~\bibnamefont
  {Schwehr}}, \bibinfo {author} {\bibfnamefont {D.}~\bibnamefont {Cherdack}},\
  and\ \bibinfo {author} {\bibfnamefont {R.}~\bibnamefont {Gran}},\ }\href@noop
  {} {\bibinfo {title} {{GENIE implementation of IFIC Valencia model for
  QE-like 2p2h neutrino-nucleus cross section}}} (\bibinfo {year} {2016}),\
  \Eprint {https://arxiv.org/abs/1601.02038} {arXiv:1601.02038} \BibitemShut
  {NoStop}%
\bibitem [{\citenamefont {Berger}\ and\ \citenamefont
  {Sehgal}(2007)}]{Berger:2007rq}%
  \BibitemOpen
  \bibfield  {author} {\bibinfo {author} {\bibfnamefont {C.}~\bibnamefont
  {Berger}}\ and\ \bibinfo {author} {\bibfnamefont {L.}~\bibnamefont
  {Sehgal}},\ }\bibfield  {title} {\bibinfo {title} {{Lepton mass effects in
  single pion production by neutrinos}},\ }\href
  {https://doi.org/10.1103/PhysRevD.76.113004} {\bibfield  {journal} {\bibinfo
  {journal} {Phys.\ Rev.\ D}\ }\textbf {\bibinfo {volume} {76}},\ \bibinfo
  {pages} {113004} (\bibinfo {year} {2007})}\BibitemShut {NoStop}%
\bibitem [{\citenamefont {Tena-Vidal}\ \emph {et~al.}(2021)\citenamefont
  {Tena-Vidal} \emph {et~al.}}]{PhysRevD.104.072009}%
  \BibitemOpen
  \bibfield  {author} {\bibinfo {author} {\bibfnamefont {J.}~\bibnamefont
  {Tena-Vidal}} \emph {et~al.} (\bibinfo {collaboration} {GENIE
  Collaboration}),\ }\bibfield  {title} {\bibinfo {title} {Neutrino-nucleon
  cross-section model tuning in genie v3},\ }\href
  {https://doi.org/10.1103/PhysRevD.104.072009} {\bibfield  {journal} {\bibinfo
   {journal} {Phys. Rev. D}\ }\textbf {\bibinfo {volume} {104}},\ \bibinfo
  {pages} {072009} (\bibinfo {year} {2021})}\BibitemShut {NoStop}%
\bibitem [{\citenamefont {Berger}\ and\ \citenamefont
  {Sehgal}(2009)}]{Berger:2008xs}%
  \BibitemOpen
  \bibfield  {author} {\bibinfo {author} {\bibfnamefont {C.}~\bibnamefont
  {Berger}}\ and\ \bibinfo {author} {\bibfnamefont {L.}~\bibnamefont
  {Sehgal}},\ }\bibfield  {title} {\bibinfo {title} {{PCAC and coherent pion
  production by low energy neutrinos}},\ }\href
  {https://doi.org/10.1103/PhysRevD.79.053003} {\bibfield  {journal} {\bibinfo
  {journal} {Phys. Rev. D}\ }\textbf {\bibinfo {volume} {79}},\ \bibinfo
  {pages} {053003} (\bibinfo {year} {2009})}\BibitemShut {NoStop}%
\bibitem [{\citenamefont {Ashery}\ \emph {et~al.}(1981)\citenamefont {Ashery},
  \citenamefont {Navon}, \citenamefont {Azuelos}, \citenamefont {Walter},
  \citenamefont {Pfeiffer},\ and\ \citenamefont {Schleputz}}]{Ashery:1981tq}%
  \BibitemOpen
  \bibfield  {author} {\bibinfo {author} {\bibfnamefont {D.}~\bibnamefont
  {Ashery}}, \bibinfo {author} {\bibfnamefont {I.}~\bibnamefont {Navon}},
  \bibinfo {author} {\bibfnamefont {G.}~\bibnamefont {Azuelos}}, \bibinfo
  {author} {\bibfnamefont {H.}~\bibnamefont {Walter}}, \bibinfo {author}
  {\bibfnamefont {H.}~\bibnamefont {Pfeiffer}},\ and\ \bibinfo {author}
  {\bibfnamefont {F.}~\bibnamefont {Schleputz}},\ }\bibfield  {title} {\bibinfo
  {title} {{True Absorption and Scattering of Pions on Nuclei}},\ }\href
  {https://doi.org/10.1103/PhysRevC.23.2173} {\bibfield  {journal} {\bibinfo
  {journal} {Phys.\ Rev.\ C}\ }\textbf {\bibinfo {volume} {23}},\ \bibinfo
  {pages} {2173} (\bibinfo {year} {1981})}\BibitemShut {NoStop}%
\bibitem [{\citenamefont {Dolan}\ \emph {et~al.}(2020)\citenamefont {Dolan},
  \citenamefont {Megias},\ and\ \citenamefont
  {Bolognesi}}]{PhysRevD.101.033003}%
  \BibitemOpen
  \bibfield  {author} {\bibinfo {author} {\bibfnamefont {S.}~\bibnamefont
  {Dolan}}, \bibinfo {author} {\bibfnamefont {G.~D.}\ \bibnamefont {Megias}},\
  and\ \bibinfo {author} {\bibfnamefont {S.}~\bibnamefont {Bolognesi}},\
  }\bibfield  {title} {\bibinfo {title} {Implementation of the susav2-meson
  exchange current 1p1h and 2p2h models in genie and analysis of nuclear
  effects in {T2K} measurements},\ }\href
  {https://doi.org/10.1103/PhysRevD.101.033003} {\bibfield  {journal} {\bibinfo
   {journal} {Phys. Rev. D}\ }\textbf {\bibinfo {volume} {101}},\ \bibinfo
  {pages} {033003} (\bibinfo {year} {2020})}\BibitemShut {NoStop}%
\bibitem [{\citenamefont {Dytman}\ \emph {et~al.}(2021)\citenamefont {Dytman},
  \citenamefont {Hayato}, \citenamefont {Raboanary}, \citenamefont {Sobczyk},
  \citenamefont {Tena-Vidal},\ and\ \citenamefont {Vololoniaina}}]{hN2018}%
  \BibitemOpen
  \bibfield  {author} {\bibinfo {author} {\bibfnamefont {S.}~\bibnamefont
  {Dytman}}, \bibinfo {author} {\bibfnamefont {Y.}~\bibnamefont {Hayato}},
  \bibinfo {author} {\bibfnamefont {R.}~\bibnamefont {Raboanary}}, \bibinfo
  {author} {\bibfnamefont {J.~T.}\ \bibnamefont {Sobczyk}}, \bibinfo {author}
  {\bibfnamefont {J.}~\bibnamefont {Tena-Vidal}},\ and\ \bibinfo {author}
  {\bibfnamefont {N.}~\bibnamefont {Vololoniaina}},\ }\bibfield  {title}
  {\bibinfo {title} {Comparison of validation methods of simulations for final
  state interactions in hadron production experiments},\ }\href
  {https://doi.org/10.1103/PhysRevD.104.053006} {\bibfield  {journal} {\bibinfo
   {journal} {Phys. Rev. D}\ }\textbf {\bibinfo {volume} {104}},\ \bibinfo
  {pages} {053006} (\bibinfo {year} {2021})}\BibitemShut {NoStop}%
\bibitem [{\citenamefont {Leitner}\ \emph {et~al.}(2006)\citenamefont
  {Leitner}, \citenamefont {Alvarez-Ruso},\ and\ \citenamefont
  {Mosel}}]{Leitner:2006ww}%
  \BibitemOpen
  \bibfield  {author} {\bibinfo {author} {\bibfnamefont {T.}~\bibnamefont
  {Leitner}}, \bibinfo {author} {\bibfnamefont {L.}~\bibnamefont
  {Alvarez-Ruso}},\ and\ \bibinfo {author} {\bibfnamefont {U.}~\bibnamefont
  {Mosel}},\ }\bibfield  {title} {\bibinfo {title} {{Charged current neutrino
  nucleus interactions at intermediate energies}},\ }\href
  {https://doi.org/10.1103/PhysRevC.73.065502} {\bibfield  {journal} {\bibinfo
  {journal} {Phys. Rev. C}\ }\textbf {\bibinfo {volume} {73}},\ \bibinfo
  {pages} {065502} (\bibinfo {year} {2006})}\BibitemShut {NoStop}%
\bibitem [{\citenamefont {Nieves}\ \emph {et~al.}(2011)\citenamefont {Nieves},
  \citenamefont {Simo},\ and\ \citenamefont {Vacas}}]{ValenciaModel}%
  \BibitemOpen
  \bibfield  {author} {\bibinfo {author} {\bibfnamefont {J.}~\bibnamefont
  {Nieves}}, \bibinfo {author} {\bibfnamefont {I.~R.}\ \bibnamefont {Simo}},\
  and\ \bibinfo {author} {\bibfnamefont {M.~J.~V.}\ \bibnamefont {Vacas}},\
  }\bibfield  {title} {\bibinfo {title} {Inclusive charged-current
  neutrino-nucleus reactions},\ }\href
  {https://doi.org/10.1103/PhysRevC.83.045501} {\bibfield  {journal} {\bibinfo
  {journal} {Phys. Rev. C}\ }\textbf {\bibinfo {volume} {83}},\ \bibinfo
  {pages} {045501} (\bibinfo {year} {2011})}\BibitemShut {NoStop}%
\bibitem [{\citenamefont {Graczyk}\ and\ \citenamefont
  {Sobczyk}(2008)}]{Graczyk:2007bc}%
  \BibitemOpen
  \bibfield  {author} {\bibinfo {author} {\bibfnamefont {K.~M.}\ \bibnamefont
  {Graczyk}}\ and\ \bibinfo {author} {\bibfnamefont {J.~T.}\ \bibnamefont
  {Sobczyk}},\ }\bibfield  {title} {\bibinfo {title} {{Form Factors in the
  Quark Resonance Model}},\ }\href {https://doi.org/10.1103/PhysRevD.79.079903}
  {\bibfield  {journal} {\bibinfo  {journal} {Phys.\ Rev.\ D}\ }\textbf
  {\bibinfo {volume} {77}},\ \bibinfo {pages} {053001} (\bibinfo {year}
  {2008})},\ \bibinfo {note} {[Erratum: Phys. Rev. D 79, 079903
  (2009)]}\BibitemShut {NoStop}%
\bibitem [{\citenamefont {Hayato}\ and\ \citenamefont
  {Pickering}(2021)}]{neut}%
  \BibitemOpen
  \bibfield  {author} {\bibinfo {author} {\bibfnamefont {Y.}~\bibnamefont
  {Hayato}}\ and\ \bibinfo {author} {\bibfnamefont {L.}~\bibnamefont
  {Pickering}},\ }\bibfield  {title} {\bibinfo {title} {{The NEUT neutrino
  interaction simulation program library}},\ }\href
  {https://doi.org/10.1140/epjs/s11734-021-00287-7} {\bibfield  {journal}
  {\bibinfo  {journal} {Eur. Phys. J. ST}\ }\textbf {\bibinfo {volume} {230}},\
  \bibinfo {pages} {4469} (\bibinfo {year} {2021})}\BibitemShut {NoStop}%
\bibitem [{\citenamefont {Auchincloss}\ \emph {et~al.}(1990)\citenamefont
  {Auchincloss} \emph {et~al.}}]{Auchincloss}%
  \BibitemOpen
  \bibfield  {author} {\bibinfo {author} {\bibfnamefont {P.~S.}\ \bibnamefont
  {Auchincloss}} \emph {et~al.},\ }\bibfield  {title} {\bibinfo {title}
  {{Measurement of the inclusive charged-current cross section for neutrino and
  antineutrino scattering on isoscalar nucleons}},\ }\href
  {https://doi.org/10.1007/BF01572022} {\bibfield  {journal} {\bibinfo
  {journal} {Zeitschrift für Physik C Particles and Fields}\ }\textbf
  {\bibinfo {volume} {1007}},\ \bibinfo {pages} {1431} (\bibinfo {year}
  {1990})}\BibitemShut {NoStop}%
\bibitem [{\citenamefont {{R. Acciari et al.}}(2018)}]{pandora}%
  \BibitemOpen
  \bibfield  {author} {\bibinfo {author} {\bibnamefont {{R. Acciari et al.}}}
  (\bibinfo {collaboration} {MicroBooNE Collaboration}),\ }\bibfield  {title}
  {\bibinfo {title} {The pandora multi-algorithm approach to automated pattern
  recognition of cosmic-ray muon and neutrino events in the microboone
  detector},\ }\href
  {https://doi.org/https://doi.org/10.1140/epjc/s10052-017-5481-6} {\bibfield
  {journal} {\bibinfo  {journal} {Eur. Phys. J. C}\ ,\ \bibinfo {pages} {82}}
  (\bibinfo {year} {2018})}\BibitemShut {NoStop}%
\bibitem [{\citenamefont {Agostinelli}\ \emph {et~al.}(2003)\citenamefont
  {Agostinelli} \emph {et~al.}}]{Geant4}%
  \BibitemOpen
  \bibfield  {author} {\bibinfo {author} {\bibfnamefont {S.}~\bibnamefont
  {Agostinelli}} \emph {et~al.} (\bibinfo {collaboration} {GEANT4}),\
  }\bibfield  {title} {\bibinfo {title} {{GEANT4--a simulation toolkit}},\
  }\href {https://doi.org/10.1016/S0168-9002(03)01368-8} {\bibfield  {journal}
  {\bibinfo  {journal} {Nucl. Instrum. Meth. A}\ }\textbf {\bibinfo {volume}
  {506}},\ \bibinfo {pages} {250} (\bibinfo {year} {2003})}\BibitemShut
  {NoStop}%
\bibitem [{\citenamefont {{P. Abratenko et
  al.}}(2022{\natexlab{b}})}]{uBooNE_det_unc}%
  \BibitemOpen
  \bibfield  {author} {\bibinfo {author} {\bibnamefont {{P. Abratenko et al.}}}
  (\bibinfo {collaboration} {MicroBooNE Collaboration}),\ }\bibfield  {title}
  {\bibinfo {title} {Novel approach for evaluating detector-related
  uncertainties in a lartpc using microboone data},\ }\href
  {https://doi.org/https://doi.org/10.1140/epjc/s10052-022-10270-8} {\bibfield
  {journal} {\bibinfo  {journal} {Eur. Phys. J. C}\ ,\ \bibinfo {pages} {454}}
  (\bibinfo {year} {2022}{\natexlab{b}})}\BibitemShut {NoStop}%
\bibitem [{\citenamefont {Tang}\ \emph {et~al.}(2017)\citenamefont {Tang},
  \citenamefont {Li}, \citenamefont {Qian}, \citenamefont {Wei},\ and\
  \citenamefont {Zhang}}]{Tang_2017}%
  \BibitemOpen
  \bibfield  {author} {\bibinfo {author} {\bibfnamefont {W.}~\bibnamefont
  {Tang}}, \bibinfo {author} {\bibfnamefont {X.}~\bibnamefont {Li}}, \bibinfo
  {author} {\bibfnamefont {X.}~\bibnamefont {Qian}}, \bibinfo {author}
  {\bibfnamefont {H.}~\bibnamefont {Wei}},\ and\ \bibinfo {author}
  {\bibfnamefont {C.}~\bibnamefont {Zhang}},\ }\bibfield  {title} {\bibinfo
  {title} {Data unfolding with wiener-svd method},\ }\href
  {https://doi.org/10.1088/1748-0221/12/10/p10002} {\bibfield  {journal}
  {\bibinfo  {journal} {J. Instrum.}\ }\textbf {\bibinfo {volume} {12}}\bibinfo
   {number} { (10)},\ \bibinfo {pages} {P10002–P10002}}\BibitemShut {NoStop}%
\bibitem [{\citenamefont {Mahn}(2009)}]{MatrixDecomv}%
  \BibitemOpen
\bibfield  {number} {  }\bibfield  {author} {\bibinfo {author} {\bibfnamefont
  {K.}~\bibnamefont {Mahn}},\ }\emph {\bibinfo {title} {{A search for muon
  neutrino and antineutrino disappearance in the Booster Neutrino Beam}}},\
  \href {https://doi.org/10.2172/959057} {Ph.D. thesis},\ \bibinfo  {school}
  {Columbia University} (\bibinfo {year} {2009})\BibitemShut {NoStop}%
\bibitem [{\citenamefont {Abe}\ \emph {et~al.}(2020{\natexlab{b}})\citenamefont
  {Abe} \emph {et~al.}}]{Abe2020.11986}%
  \BibitemOpen
  \bibfield  {author} {\bibinfo {author} {\bibfnamefont {K.}~\bibnamefont
  {Abe}} \emph {et~al.},\ }\bibfield  {title} {\bibinfo {title} {Measurement of
  the charged-current electron (anti-)neutrino inclusive cross-sections at the
  t2k off-axis near detector},\ }\href
  {https://doi.org/https://doi.org/10.1007/JHEP10(2020)114} {\bibfield
  {journal} {\bibinfo  {journal} {J. High Energ. Phys}\ }\textbf {\bibinfo
  {volume} {2020}},\ \bibinfo {pages} {114} (\bibinfo {year}
  {2020}{\natexlab{b}})}\BibitemShut {NoStop}%
\bibitem [{\citenamefont {Abratenko}\ \emph {et~al.}(2019)\citenamefont
  {Abratenko} \emph {et~al.}}]{Abratenko2019:131801}%
  \BibitemOpen
  \bibfield  {author} {\bibinfo {author} {\bibfnamefont {P.}~\bibnamefont
  {Abratenko}} \emph {et~al.} (\bibinfo {collaboration} {MicroBooNE
  Collaboration}),\ }\bibfield  {title} {\bibinfo {title} {First measurement of
  inclusive muon neutrino charged current differential cross sections on argon
  at ${E}_{\ensuremath{\nu}}\ensuremath{\sim}0.8\text{ }\text{ }\mathrm{GeV}$
  with the microboone detector},\ }\href
  {https://doi.org/10.1103/PhysRevLett.123.131801} {\bibfield  {journal}
  {\bibinfo  {journal} {Phys. Rev. Lett.}\ }\textbf {\bibinfo {volume} {123}},\
  \bibinfo {pages} {131801} (\bibinfo {year} {2019})}\BibitemShut {NoStop}%
\bibitem [{\citenamefont {Filkins}\ \emph {et~al.}(2020)\citenamefont {Filkins}
  \emph {et~al.}}]{PhysRevD.101.112007}%
  \BibitemOpen
  \bibfield  {author} {\bibinfo {author} {\bibfnamefont {A.}~\bibnamefont
  {Filkins}} \emph {et~al.} (\bibinfo {collaboration} {MINER$\ensuremath{\nu}$A
  Collaboration}),\ }\bibfield  {title} {\bibinfo {title} {Double-differential
  inclusive charged-current ${\ensuremath{\nu}}_{\ensuremath{\mu}}$ cross
  sections on hydrocarbon in minerva at
  $\langle{E}_{\ensuremath{\nu}}\rangle\ensuremath{\sim}3.5\text{ }\text{
  }\mathrm{GeV}$},\ }\href {https://doi.org/10.1103/PhysRevD.101.112007}
  {\bibfield  {journal} {\bibinfo  {journal} {Phys. Rev. D}\ }\textbf {\bibinfo
  {volume} {101}},\ \bibinfo {pages} {112007} (\bibinfo {year}
  {2020})}\BibitemShut {NoStop}%
\bibitem [{\citenamefont {McGivern}\ \emph {et~al.}(2016)\citenamefont
  {McGivern} \emph {et~al.}}]{PhysRevD.94.052005}%
  \BibitemOpen
  \bibfield  {author} {\bibinfo {author} {\bibfnamefont {C.~L.}\ \bibnamefont
  {McGivern}} \emph {et~al.} (\bibinfo {collaboration} {MINERvA
  Collaboration}),\ }\bibfield  {title} {\bibinfo {title} {Cross sections for
  ${\ensuremath{\nu}}_{\ensuremath{\mu}}$ and
  ${\overline{\ensuremath{\nu}}}_{\ensuremath{\mu}}$ induced pion production on
  hydrocarbon in the few-{GeV} region using {MINERvA}},\ }\href
  {https://doi.org/10.1103/PhysRevD.94.052005} {\bibfield  {journal} {\bibinfo
  {journal} {Phys. Rev. D}\ }\textbf {\bibinfo {volume} {94}},\ \bibinfo
  {pages} {052005} (\bibinfo {year} {2016})}\BibitemShut {NoStop}%
\end{thebibliography}%


\end{document}


\centering
\large{Measurement of Nuclear Effects in neutrino-argon interactions using Generalised Kinematic Imbalance Variables with the MicroBooNE Detector}

(Dated: \today)



\justify
\section{Data Release}\label{datarel}

Overflow (underflow) values are included in the last (first) bin.
The additional smearing matrix $A_{C}$ should first be applied to the event distribution of an independent theoretical prediction when a comparison is performed to the data reported, and then divided by the bin width.
The $A_{C}$ matrices are dimensionless.
The double-differential cross sections include correlations between the phase-space slices.
The data release with the data results, the covariance matrices, and the additional smearing matrices is included in the DataRelease.root file.
These are also included in the Supplemental Material in Sec.~\ref{datarel}, Sec.~\ref{cov}, and Sec.~\ref{smear}, respectively.
Instructions on how to use the data release and the description of the binning scheme are included in the README file.

\raggedbottom

\begin{table}[H]
\raggedright
\begin{adjustbox}{width=\textwidth}
\small
\begin{tabular}{ |c|c|c|c|c| }
\hline
\multicolumn{5}{|c|}{Cross Section $p_{n}$, $All\,events$} \\
\hline
\hline
Bin \# & Low edge [GeV/\textit{c}] & High edge [GeV/\textit{c}] & Cross Section [$10^{-38}\frac{cm^{2}}{(GeV/\textit{c})\,^{40}Ar}$] & Uncertainty [$10^{-38}\frac{cm^{2}}{(GeV/\textit{c})\,^{40}Ar}$] \\
\hline
\hline
1 & 0 & 0.07 & 6.4406 & 1.1679\\
2 & 0.07 & 0.14 & 21.314 & 2.2968\\
3 & 0.14 & 0.2 & 36.266 & 3.6505\\
4 & 0.2 & 0.3 & 27.206 & 2.6118\\
5 & 0.3 & 0.4 & 15.223 & 2.2399\\
6 & 0.4 & 0.47 & 12.758 & 2.6894\\
7 & 0.47 & 0.55 & 9.1936 & 2.3617\\
8 & 0.55 & 0.65 & 6.3845 & 1.4767\\
9 & 0.65 & 0.75 & 4.0277 & 0.99182\\
10 & 0.75 & 0.85 & 3.5047 & 1.0892\\
\hline
\end{tabular}
\end{adjustbox}
\caption{Single differential cross section measurement as a function of $p_{n}$.}
\end{table}

\begin{table}[H]
\raggedright
\begin{adjustbox}{width=\textwidth}
\small
\begin{tabular}{ |c|c|c|c|c| }
\hline
\multicolumn{5}{|c|}{Cross Section $\alpha_{3D}$, $All\,events$} \\
\hline
\hline
Bin \# & Low edge [deg] & High edge [deg] & Cross Section [$10^{-38}\frac{cm^{2}}{(deg)\,^{40}Ar}$] & Uncertainty [$10^{-38}\frac{cm^{2}}{(deg)\,^{40}Ar}$] \\
\hline
\hline
1 & 0 & 25 & 0.0092428 & 0.0026782\\
2 & 25 & 50 & 0.028455 & 0.0037331\\
3 & 50 & 70 & 0.046157 & 0.0061733\\
4 & 70 & 90 & 0.064527 & 0.0091935\\
5 & 90 & 110 & 0.099771 & 0.011189\\
6 & 110 & 130 & 0.1413 & 0.014805\\
7 & 130 & 150 & 0.13613 & 0.015257\\
8 & 150 & 180 & 0.05499 & 0.0088809\\
\hline
\end{tabular}
\end{adjustbox}
\caption{Single differential cross section measurement as a function of $\alpha_{3D}$.}
\end{table}

\begin{table}[H]
\raggedright
\begin{adjustbox}{width=\textwidth}
\small
\begin{tabular}{ |c|c|c|c|c| }
\hline
\multicolumn{5}{|c|}{Cross Section $\phi_{3D}$, $All\,events$} \\
\hline
\hline
Bin \# & Low edge [deg] & High edge [deg] & Cross Section [$10^{-38}\frac{cm^{2}}{(deg)\,^{40}Ar}$] & Uncertainty [$10^{-38}\frac{cm^{2}}{(deg)\,^{40}Ar}$] \\
\hline
\hline
1 & 0 & 12.5 & 0.29579 & 0.029271\\
2 & 12.5 & 25 & 0.2562 & 0.023545\\
3 & 25 & 37.5 & 0.14256 & 0.016022\\
4 & 37.5 & 50 & 0.09325 & 0.014113\\
5 & 50 & 60 & 0.076748 & 0.013939\\
6 & 60 & 75 & 0.045579 & 0.0089236\\
7 & 75 & 90 & 0.030827 & 0.0066006\\
8 & 90 & 106 & 0.02087 & 0.0050552\\
9 & 106 & 126 & 0.012219 & 0.0034782\\
10 & 126 & 180 & 0.0037828 & 0.0011726\\
\hline
\end{tabular}
\end{adjustbox}
\caption{Single differential cross section measurement as a function of $\phi_{3D}$.}
\end{table}

\begin{table}[H]
\raggedright
\begin{adjustbox}{width=\textwidth}
\small
\begin{tabular}{ |c|c|c|c|c| }
\hline
\multicolumn{5}{|c|}{Cross Section $p_{n\perp}$, $All\,events$} \\
\hline
\hline
Bin \# & Low edge [GeV/\textit{c}] & High edge [GeV/\textit{c}] & Cross Section [$10^{-38}\frac{cm^{2}}{(GeV/\textit{c})\,^{40}Ar}$] & Uncertainty [$10^{-38}\frac{cm^{2}}{(GeV/\textit{c})\,^{40}Ar}$] \\
\hline
\hline
1 & 0 & 0.07 & 18.062 & 2.6968\\
2 & 0.07 & 0.14 & 32.544 & 3.8819\\
3 & 0.14 & 0.2 & 36.013 & 4.2935\\
4 & 0.2 & 0.3 & 23.116 & 2.813\\
5 & 0.3 & 0.4 & 18.544 & 2.4476\\
6 & 0.4 & 0.47 & 11.629 & 2.0983\\
7 & 0.47 & 0.55 & 4.7832 & 1.4637\\
8 & 0.55 & 0.85 & 0.91941 & 0.34749\\
\hline
\end{tabular}
\end{adjustbox}
\caption{Single differential cross section measurement as a function of $p_{n\perp}$.}
\end{table}

\begin{table}[H]
\raggedright
\begin{adjustbox}{width=\textwidth}
\small
\begin{tabular}{ |c|c|c|c|c| }
\hline
\multicolumn{5}{|c|}{Cross Section $p_{n\perp,x}$, $All\,events$} \\
\hline
\hline
Bin \# & Low edge [GeV/\textit{c}] & High edge [GeV/\textit{c}] & Cross Section [$10^{-38}\frac{cm^{2}}{(GeV/\textit{c})\,^{40}Ar}$] & Uncertainty [$10^{-38}\frac{cm^{2}}{(GeV/\textit{c})\,^{40}Ar}$] \\
\hline
\hline
1 & -0.55 & -0.45 & 5.0018 & 1.0085\\
2 & -0.45 & -0.35 & 3.6489 & 0.7749\\
3 & -0.35 & -0.25 & 5.0634 & 1.047\\
4 & -0.25 & -0.15 & 11.465 & 1.4888\\
5 & -0.15 & -0.05 & 23.711 & 2.2797\\
6 & -0.05 & 0.05 & 30.292 & 2.8632\\
7 & 0.05 & 0.15 & 20.399 & 2.1419\\
8 & 0.15 & 0.25 & 8.914 & 1.2885\\
9 & 0.25 & 0.35 & 3.112 & 0.7141\\
10 & 0.35 & 0.45 & 1.422 & 0.52351\\
11 & 0.45 & 0.55 & 0.96175 & 0.47493\\
\hline
\end{tabular}
\end{adjustbox}
\caption{Single differential cross section measurement as a function of $p_{n\perp,x}$.}
\end{table}

\begin{table}[H]
\raggedright
\begin{adjustbox}{width=\textwidth}
\small
\begin{tabular}{ |c|c|c|c|c| }
\hline
\multicolumn{5}{|c|}{Cross Section $p_{n\perp,y}$, $All\,events$} \\
\hline
\hline
Bin \# & Low edge [GeV/\textit{c}] & High edge [GeV/\textit{c}] & Cross Section [$10^{-38}\frac{cm^{2}}{(GeV/\textit{c})\,^{40}Ar}$] & Uncertainty [$10^{-38}\frac{cm^{2}}{(GeV/\textit{c})\,^{40}Ar}$] \\
\hline
\hline
1 & -0.55 & -0.45 & 0.76402 & 0.47042\\
2 & -0.45 & -0.35 & 2.2345 & 0.69819\\
3 & -0.35 & -0.25 & 4.9074 & 0.9922\\
4 & -0.25 & -0.15 & 10.61 & 1.419\\
5 & -0.15 & -0.05 & 23.149 & 2.342\\
6 & -0.05 & 0.05 & 32.125 & 3.2785\\
7 & 0.05 & 0.15 & 21.864 & 2.2828\\
8 & 0.15 & 0.25 & 10.255 & 1.4101\\
9 & 0.25 & 0.35 & 4.1982 & 0.84425\\
10 & 0.35 & 0.45 & 1.5029 & 0.52401\\
11 & 0.45 & 0.55 & 0.49323 & 0.36266\\
\hline
\end{tabular}
\end{adjustbox}
\caption{Single differential cross section measurement as a function of $p_{n\perp,y}$.}
\end{table}

\begin{table}[H]
\raggedright
\begin{adjustbox}{width=\textwidth}
\small
\begin{tabular}{ |c|c|c|c|c| }
\hline
\multicolumn{5}{|c|}{Cross Section $p_{n\parallel}$, $All\,events$} \\
\hline
\hline
Bin \# & Low edge [GeV/\textit{c}] & High edge [GeV/\textit{c}] & Cross Section [$10^{-38}\frac{cm^{2}}{(GeV/\textit{c})\,^{40}Ar}$] & Uncertainty [$10^{-38}\frac{cm^{2}}{(GeV/\textit{c})\,^{40}Ar}$] \\
\hline
\hline
1 & -0.55 & -0.45 & 7.3354 & 2.263\\
2 & -0.45 & -0.35 & 6.7636 & 1.4412\\
3 & -0.35 & -0.25 & 10.489 & 1.4262\\
4 & -0.25 & -0.15 & 17.696 & 1.7962\\
5 & -0.15 & -0.05 & 25.676 & 2.3889\\
6 & -0.05 & 0.05 & 26.294 & 2.6477\\
7 & 0.05 & 0.15 & 14.997 & 1.5763\\
8 & 0.15 & 0.25 & 4.3977 & 0.81678\\
9 & 0.25 & 0.35 & 0.26795 & 0.41179\\
10 & 0.35 & 0.45 & -0.22201 & 0.19504\\
\hline
\end{tabular}
\end{adjustbox}
\caption{Single differential cross section measurement as a function of $p_{n\parallel}$.}
\end{table}

\begin{table}[H]
\raggedright
\begin{adjustbox}{width=\textwidth}
\small
\begin{tabular}{ |c|c|c|c|c| }
\hline
\multicolumn{5}{|c|}{Cross Section $p_{n}$, $\alpha_{3D}\,<\,45^{o}$} \\
\hline
\hline
Bin \# & Low edge [GeV/\textit{c}] & High edge [GeV/\textit{c}] & Cross Section [$10^{-38}\frac{cm^{2}}{(GeV/\textit{c})\,^{40}Ar}$] & Uncertainty [$10^{-38}\frac{cm^{2}}{(GeV/\textit{c})\,^{40}Ar}$] \\
\hline
\hline
1 & 0 & 0.1 & 0.028159 & 0.0070451\\
2 & 0.1 & 0.2 & 0.09376 & 0.012414\\
3 & 0.2 & 0.3 & 0.04408 & 0.009544\\
4 & 0.3 & 0.4 & 0.015599 & 0.0058132\\
5 & 0.4 & 0.85 & 0.00083573 & 0.0008078\\
\hline
\end{tabular}
\end{adjustbox}
\caption{Double differential cross section measurement as a function of $p_{n}$ for $\alpha_{3D}\,<\,45^{o}$.}
\end{table}

\begin{table}[H]
\raggedright
\begin{adjustbox}{width=\textwidth}
\small
\begin{tabular}{ |c|c|c|c|c| }
\hline
\multicolumn{5}{|c|}{Cross Section $p_{n}$, $135^{o}\,<\,\alpha_{3D}\,<\,180^{o}$} \\
\hline
\hline
Bin \# & Low edge [GeV/\textit{c}] & High edge [GeV/\textit{c}] & Cross Section [$10^{-38}\frac{cm^{2}}{(GeV/\textit{c})\,^{40}Ar}$] & Uncertainty [$10^{-38}\frac{cm^{2}}{(GeV/\textit{c})\,^{40}Ar}$] \\
\hline
\hline
1 & 0 & 0.08 & 0.0167 & 0.011364\\
2 & 0.08 & 0.15 & 0.08459 & 0.016098\\
3 & 0.15 & 0.23 & 0.12713 & 0.019017\\
4 & 0.23 & 0.3 & 0.14042 & 0.02328\\
5 & 0.3 & 0.4 & 0.098465 & 0.017631\\
6 & 0.4 & 0.85 & 0.068701 & 0.014215\\
\hline
\end{tabular}
\end{adjustbox}
\caption{Double differential cross section measurement as a function of $p_{n}$ for $135^{o}\,<\,\alpha_{3D}\,<\,180^{o}$.}
\end{table}

\begin{table}[H]
\raggedright
\begin{adjustbox}{width=\textwidth}
\small
\begin{tabular}{ |c|c|c|c|c| }
\hline
\multicolumn{5}{|c|}{Cross Section $\alpha_{3D}$, $p_{n}\,<\,0.2\,GeV/c$} \\
\hline
\hline
Bin \# & Low edge [deg] & High edge [deg] & Cross Section [$10^{-38}\frac{cm^{2}}{(deg)\,^{40}Ar}$] & Uncertainty [$10^{-38}\frac{cm^{2}}{(deg)\,^{40}Ar}$] \\
\hline
\hline
1 & 0 & 35 & 0.053491 & 0.0071818\\
2 & 35 & 70 & 0.12055 & 0.012314\\
3 & 70 & 92 & 0.20117 & 0.019441\\
4 & 92 & 114 & 0.21589 & 0.021971\\
5 & 114 & 136 & 0.19916 & 0.017257\\
6 & 136 & 158 & 0.11911 & 0.012186\\
7 & 158 & 180 & 0.044109 & 0.0072957\\
\hline
\end{tabular}
\end{adjustbox}
\caption{Double differential cross section measurement as a function of $\alpha_{3D}$ for $p_{n}\,<\,0.2\,GeV/c$.}
\end{table}

\begin{table}[H]
\raggedright
\begin{adjustbox}{width=\textwidth}
\small
\begin{tabular}{ |c|c|c|c|c| }
\hline
\multicolumn{5}{|c|}{Cross Section $\alpha_{3D}$, $p_{n}\,>\,0.4\,GeV/c$} \\
\hline
\hline
Bin \# & Low edge [deg] & High edge [deg] & Cross Section [$10^{-38}\frac{cm^{2}}{(deg)\,^{40}Ar}$] & Uncertainty [$10^{-38}\frac{cm^{2}}{(deg)\,^{40}Ar}$] \\
\hline
\hline
1 & 0 & 35 & 0.00095038 & 0.00045408\\
2 & 35 & 70 & 0.0019962 & 0.0015217\\
3 & 70 & 90 & 0.01073 & 0.0044479\\
4 & 90 & 110 & 0.03376 & 0.0056656\\
5 & 110 & 130 & 0.070846 & 0.0076066\\
6 & 130 & 145 & 0.12594 & 0.014246\\
7 & 145 & 180 & 0.056519 & 0.0072774\\
\hline
\end{tabular}
\end{adjustbox}
\caption{Double differential cross section measurement as a function of $\alpha_{3D}$ for $p_{n}\,>\,0.4\,GeV/c$.}
\end{table}


\section{Acceptance-Rejection curves}\label{accrej}

\begin{figure}[htb!]
\centering
\begin{tikzpicture} \draw (0, 0) node[inner sep=0] {
\includegraphics[width=0.45\textwidth]{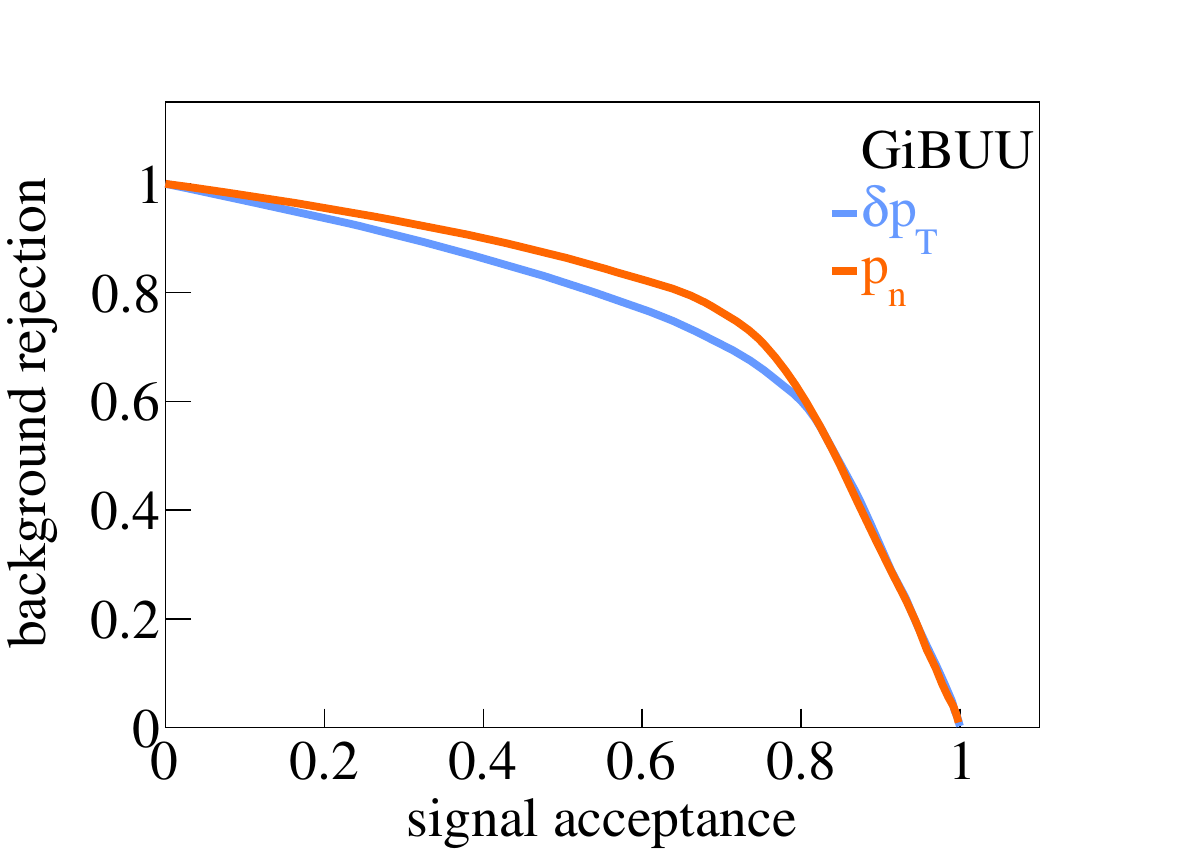}	
};
\draw (0.4, -3.2) node {(a)};	
\end{tikzpicture}
\hspace{0.05 \textwidth}
\begin{tikzpicture} \draw (0, 0) node[inner sep=0] {
\includegraphics[width=0.45\textwidth]{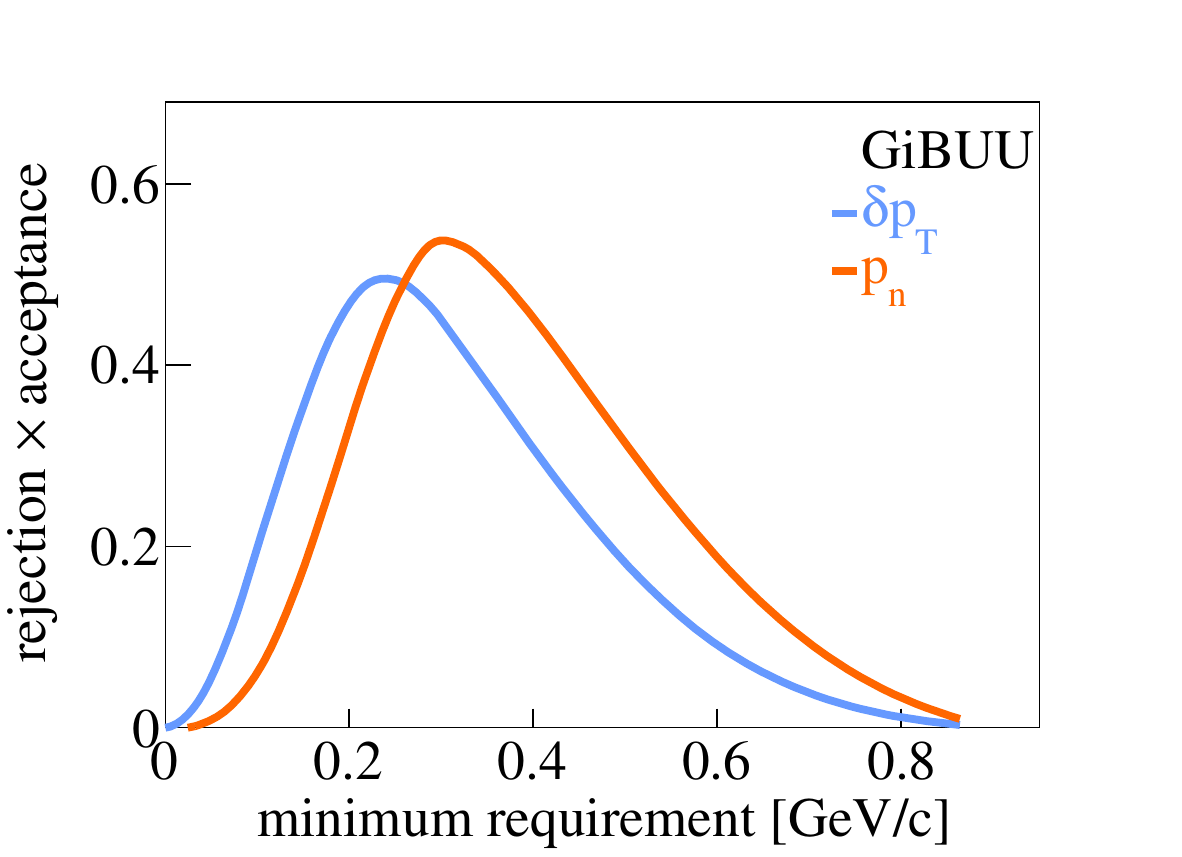}
};
\draw (0.4, -3.2) node {(b)};	
\end{tikzpicture}
\caption{
(a) Signal acceptance fraction vs background rejection fraction as a function of the cut on the reconstructed missing momentum $p_{n}$ (orange) and the transverse missing momentum $\delta p_{T}$ (blue) for CC1p0$\pi$ events using the GiBUU prediction.
(b) Evolution of the product between the signal acceptance and the background rejection denoted as ``rejection $\times$ acceptance'' as a function of the cut value.
}
\label{fig:rejectionGiBUU}
\end{figure}

\begin{figure}[htb!]
\centering
\begin{tikzpicture} \draw (0, 0) node[inner sep=0] {
\includegraphics[width=0.45\textwidth]{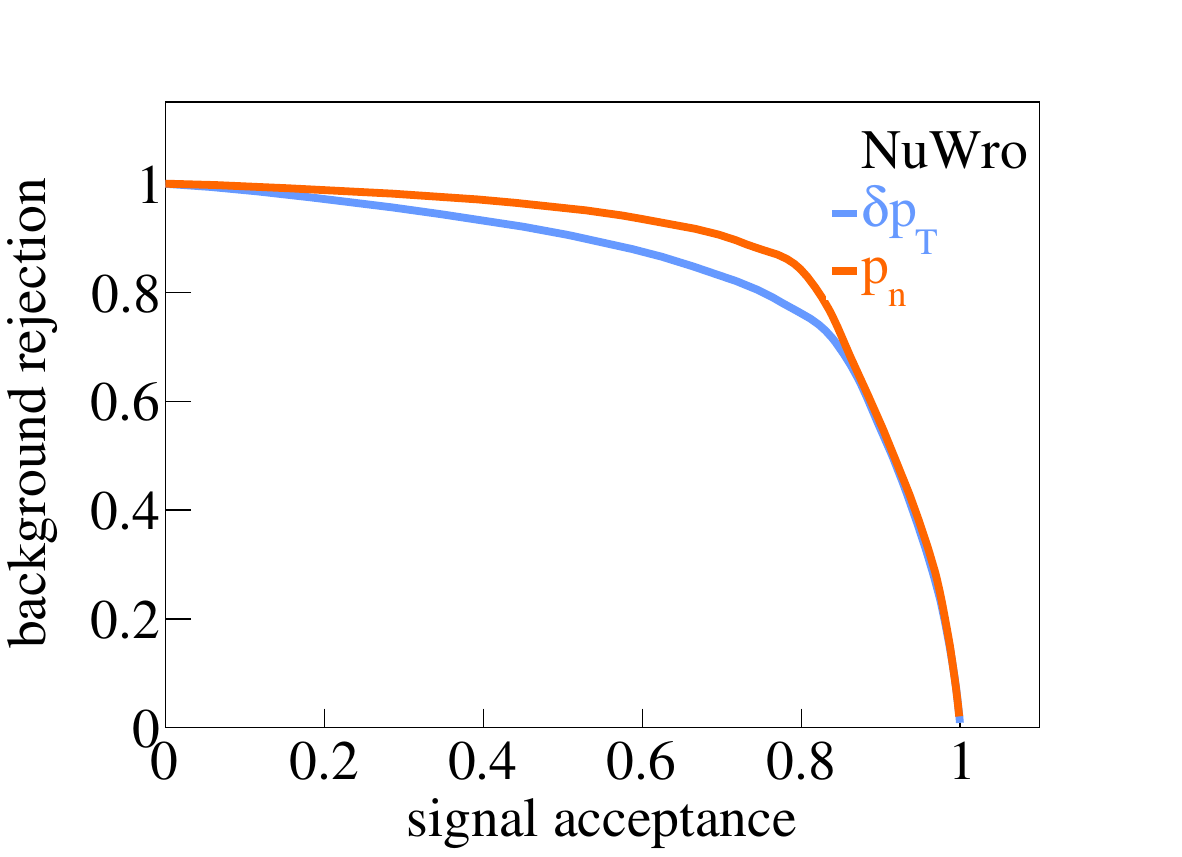}	
};
\draw (0.4, -3.2) node {(a)};	
\end{tikzpicture}
\hspace{0.05 \textwidth}
\begin{tikzpicture} \draw (0, 0) node[inner sep=0] {
\includegraphics[width=0.45\textwidth]{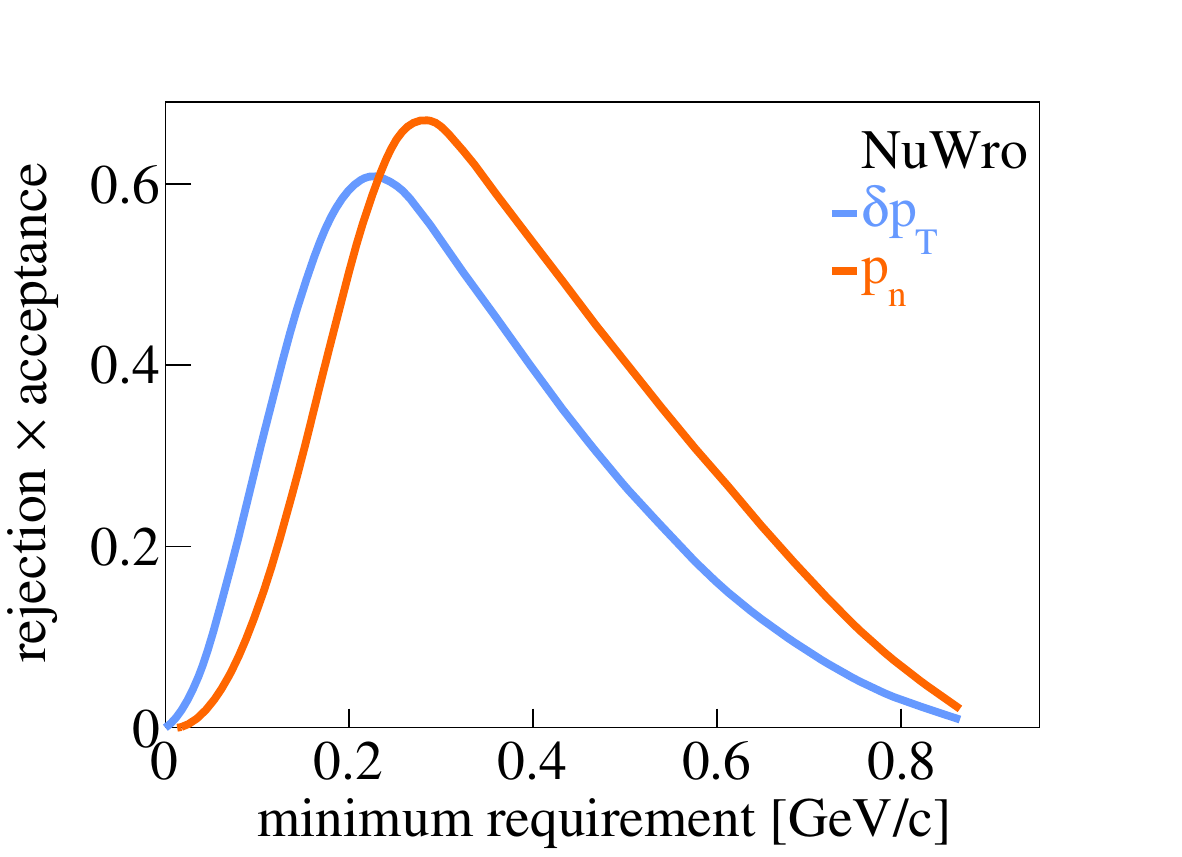}
};
\draw (0.4, -3.2) node {(b)};	
\end{tikzpicture}
\caption{
(a) Signal acceptance fraction vs background rejection fraction as a function of the cut on the reconstructed missing momentum $p_{n}$ (orange) and the transverse missing momentum $\delta p_{T}$ (blue) for CC1p0$\pi$ events using the NuWro prediction.
(b) Evolution of the product between the signal acceptance and the background rejection denoted as ``rejection $\times$ acceptance'' as a function of the cut value.
}
\label{fig:rejectionNuWro}
\end{figure}

\begin{figure}[H]
\centering
\begin{tikzpicture} \draw (0, 0) node[inner sep=0] {
\includegraphics[width=0.45\textwidth]{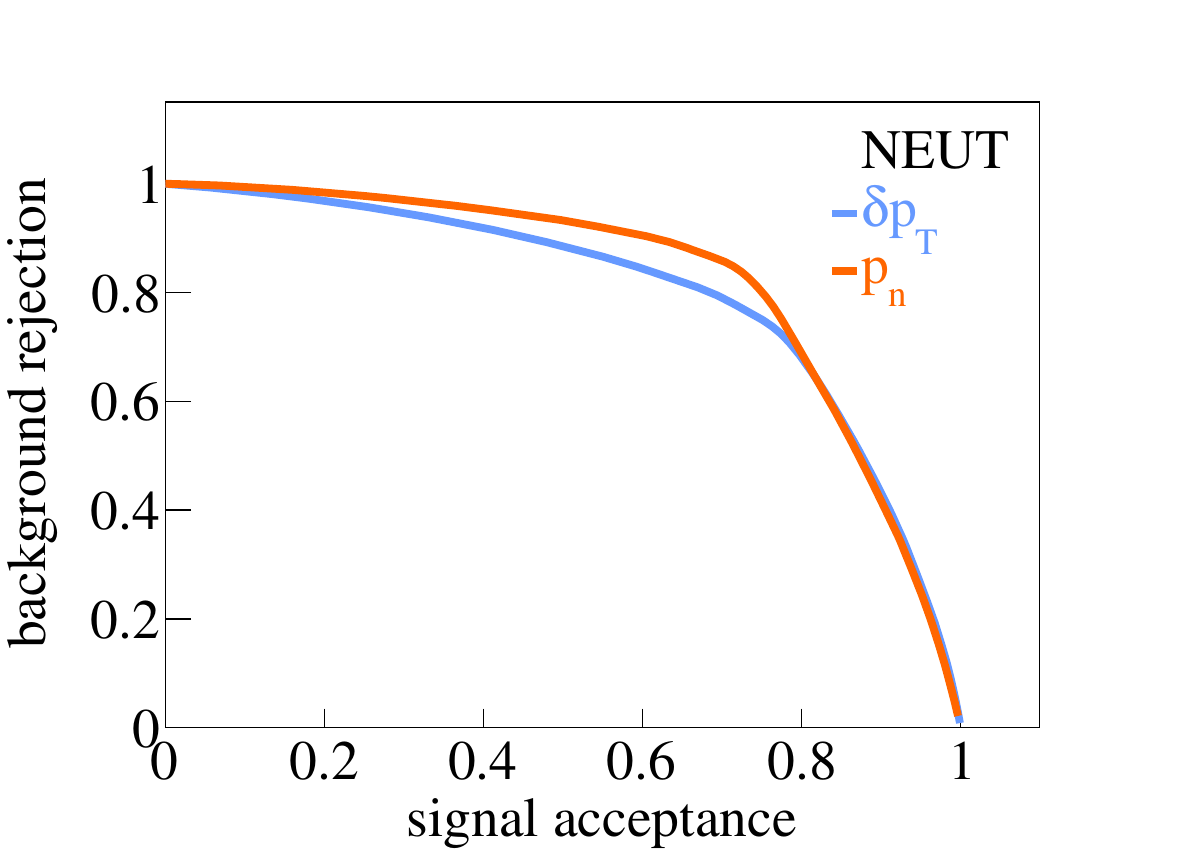}	
};
\draw (0.4, -3.2) node {(a)};	
\end{tikzpicture}
\hspace{0.05 \textwidth}
\begin{tikzpicture} \draw (0, 0) node[inner sep=0] {
\includegraphics[width=0.45\textwidth]{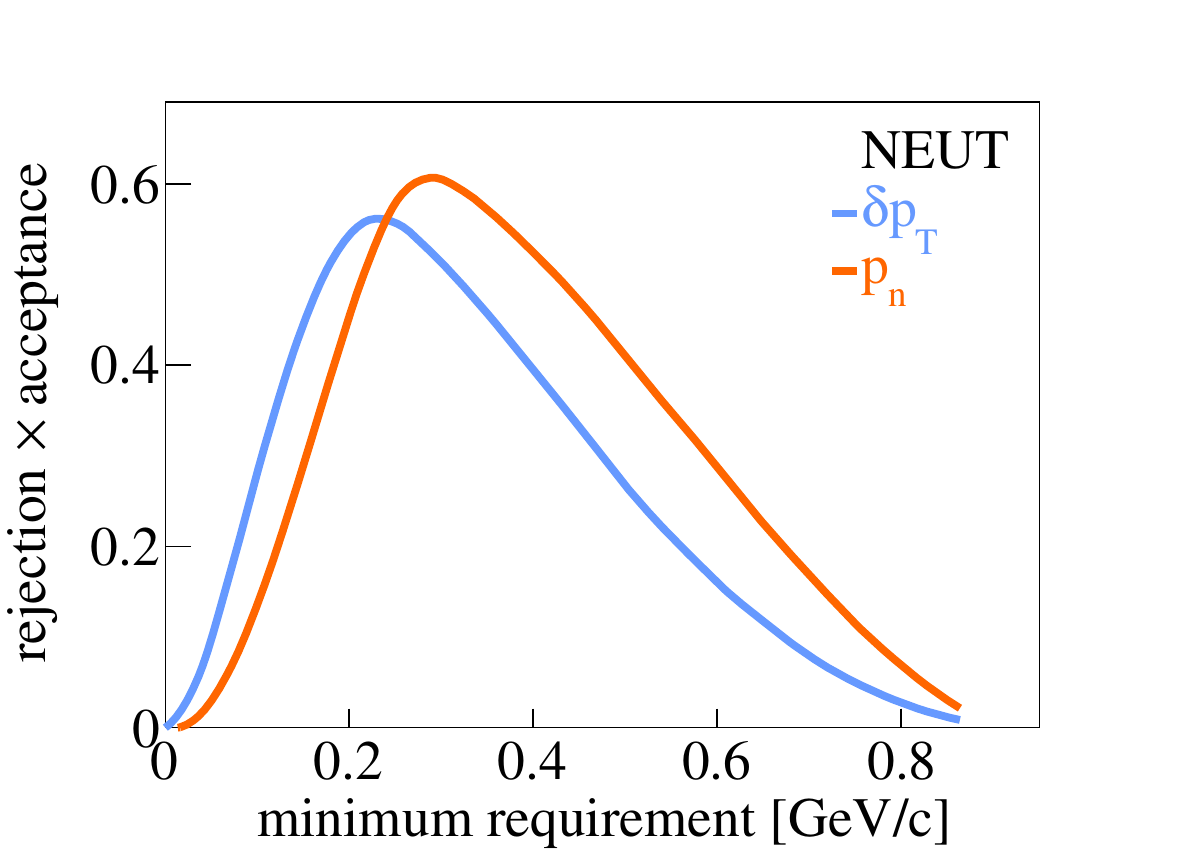}
};
\draw (0.4, -3.2) node {(b)};	
\end{tikzpicture}
\caption{
(a) Signal acceptance fraction vs background rejection fraction as a function of the cut on the reconstructed missing momentum $p_{n}$ (orange) and the transverse missing momentum $\delta p_{T}$ (blue) for CC1p0$\pi$ events using the NEUT prediction.
(b) Evolution of the product between the signal acceptance and the background rejection denoted as ``rejection $\times$ acceptance'' as a function of the cut value.
}
\label{fig:rejectionNEUT}
\end{figure}


\section{Generalized and Transverse Kinematic Imbalance Cross Sections without FSI}\label{GKINoFSI}

Both the GKI and the TKI variables are presented below in the absence of FSI effects.

\begin{figure}[htb!]
\centering
\begin{tikzpicture} \draw (0, 0) node[inner sep=0] {
\includegraphics[width=0.45\textwidth]{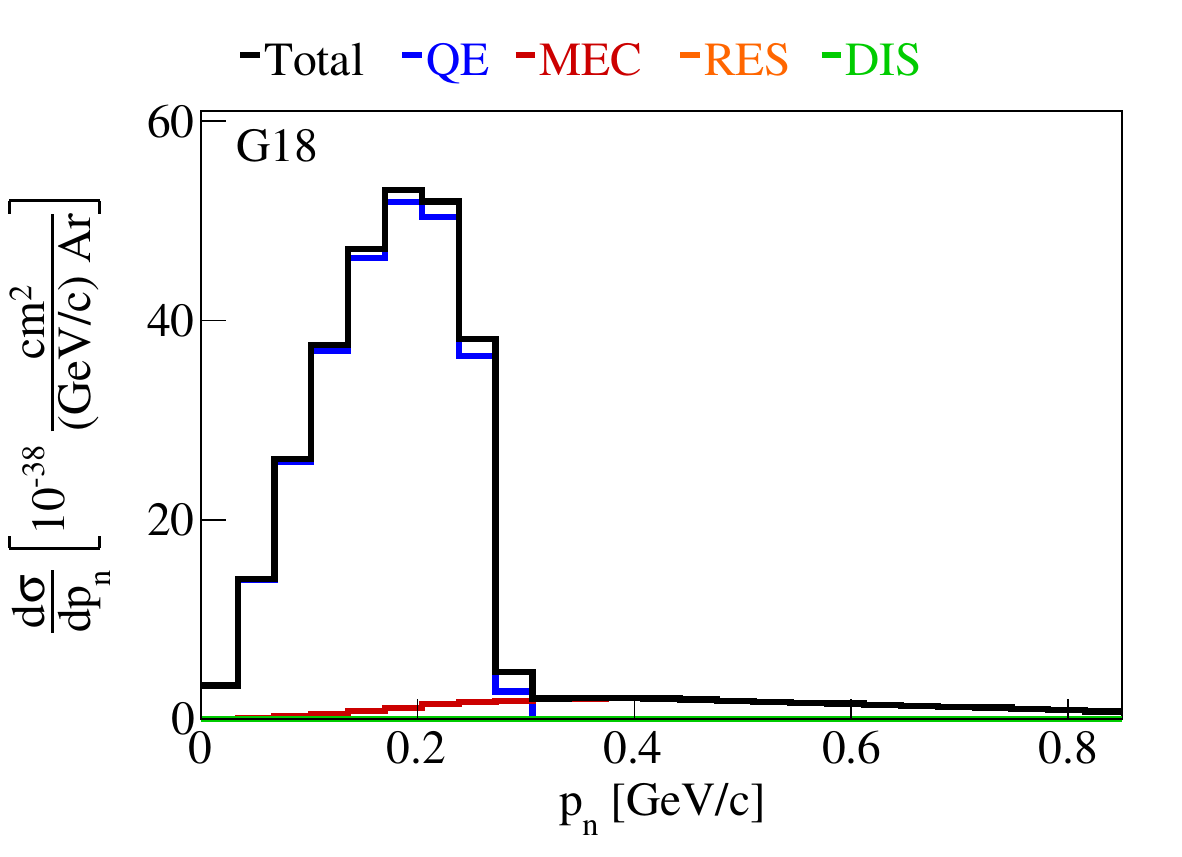}	
};
\draw (0.4, -3.2) node {(a)};	
\end{tikzpicture}
\hspace{0.05 \textwidth}
\begin{tikzpicture} \draw (0, 0) node[inner sep=0] {
\includegraphics[width=0.45\textwidth]{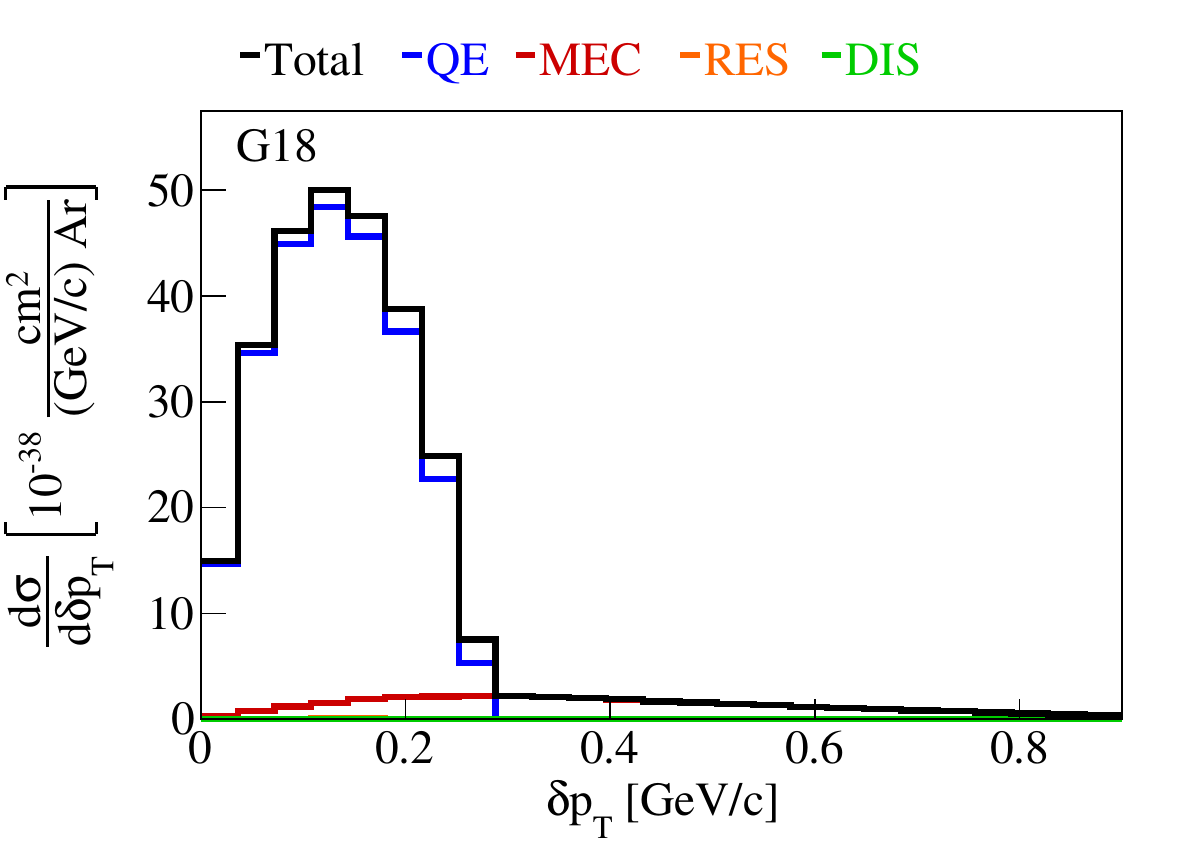}
};
\draw (0.4, -3.2) node {(b)};	
\end{tikzpicture}
\caption{The flux-integrated single-differential cross section interaction breakdown as a function of (left) $p_{n}$ and (right) $\delta p_{T}$.
Colored lines show the results of theoretical cross section calculations using the $\texttt{G18}$ prediction without FSI for QE (blue), MEC (red), RES (orange), and DIS (green) interactions.}
\label{fig:nofsideltapt}
\end{figure}

\begin{figure}[htb!]
\centering
\begin{tikzpicture} \draw (0, 0) node[inner sep=0] {
\includegraphics[width=0.45\textwidth]{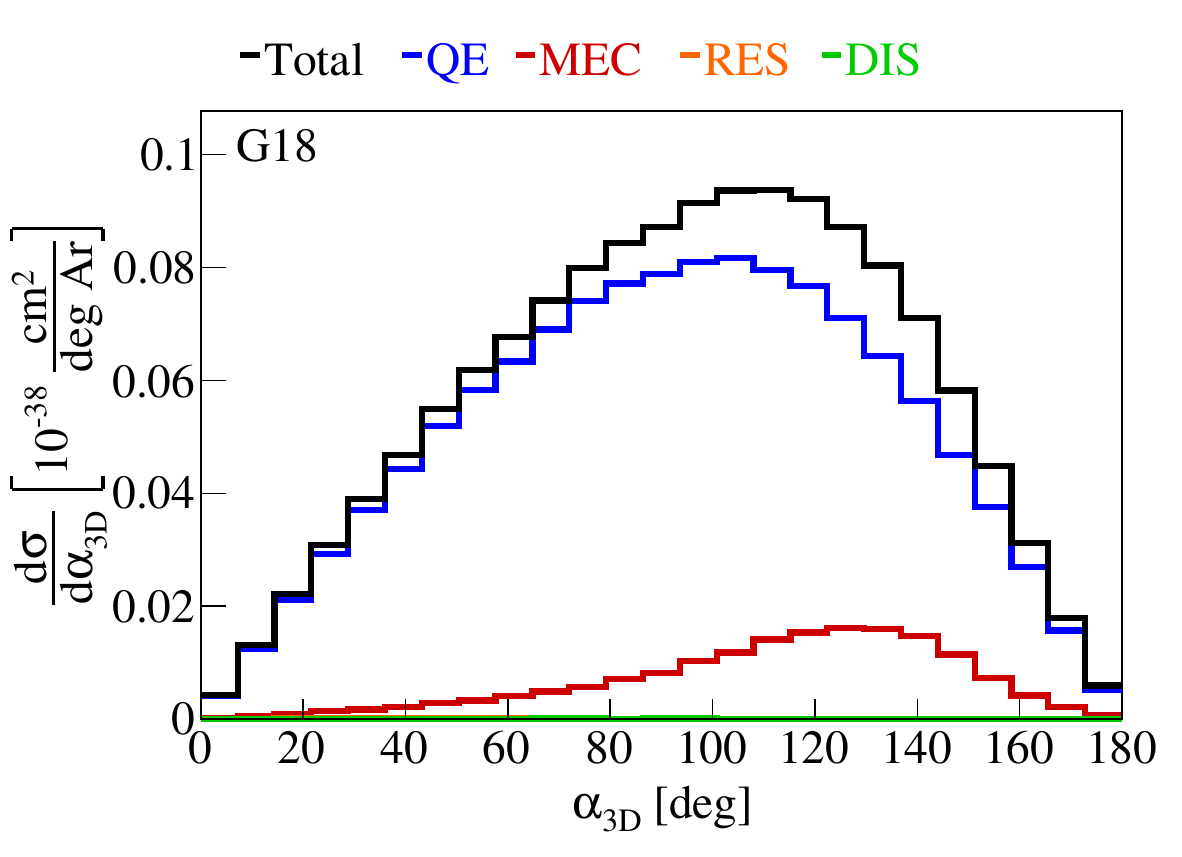}	
};
\draw (0.4, -3.2) node {(a)};	
\end{tikzpicture}
\hspace{0.05 \textwidth}
\begin{tikzpicture} \draw (0, 0) node[inner sep=0] {
\includegraphics[width=0.45\textwidth]{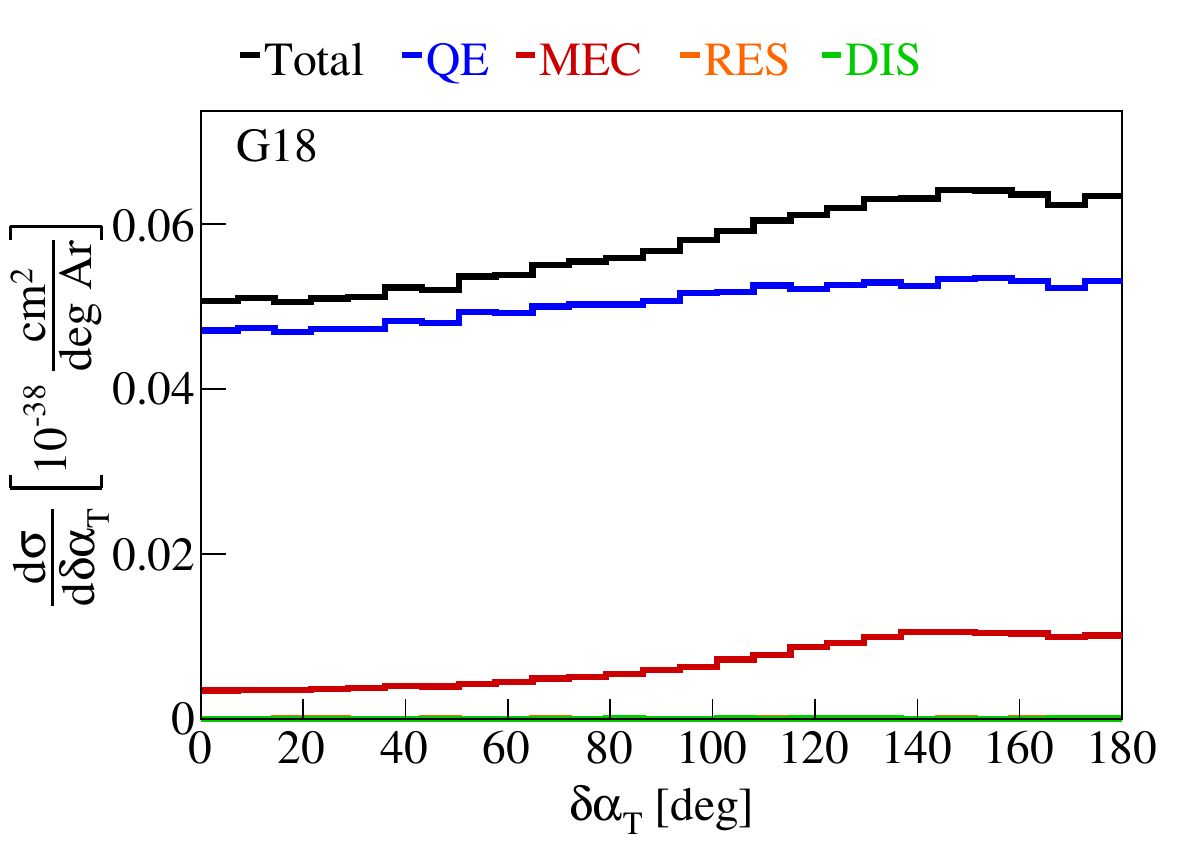}
};
\draw (0.4, -3.2) node {(b)};	
\end{tikzpicture}
\caption{The flux-integrated single-differential cross section interaction breakdown as a function of (left) $\alpha_{3D}$ and (right) $\delta\alpha_{T}$.
Colored lines show the results of theoretical cross section calculations using the $\texttt{G18}$ prediction without FSI for QE (blue), MEC (red), RES (orange), and DIS (green) interactions.}
\label{fig:nofsideltaalphat}
\end{figure}

\begin{figure}[htb!]
\centering
\begin{tikzpicture} \draw (0, 0) node[inner sep=0] {
\includegraphics[width=0.45\textwidth]{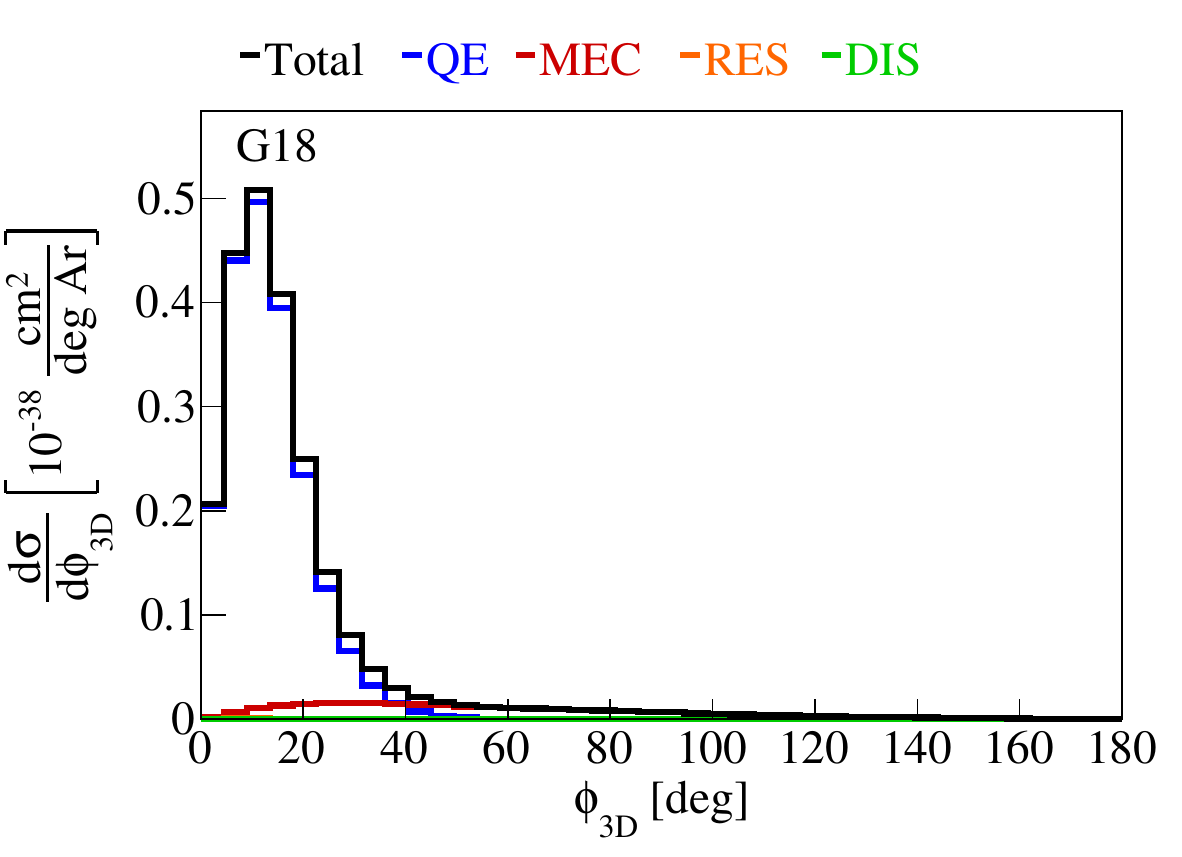}	
};
\draw (0.4, -3.2) node {(a)};	
\end{tikzpicture}
\hspace{0.05 \textwidth}
\begin{tikzpicture} \draw (0, 0) node[inner sep=0] {
\includegraphics[width=0.45\textwidth]{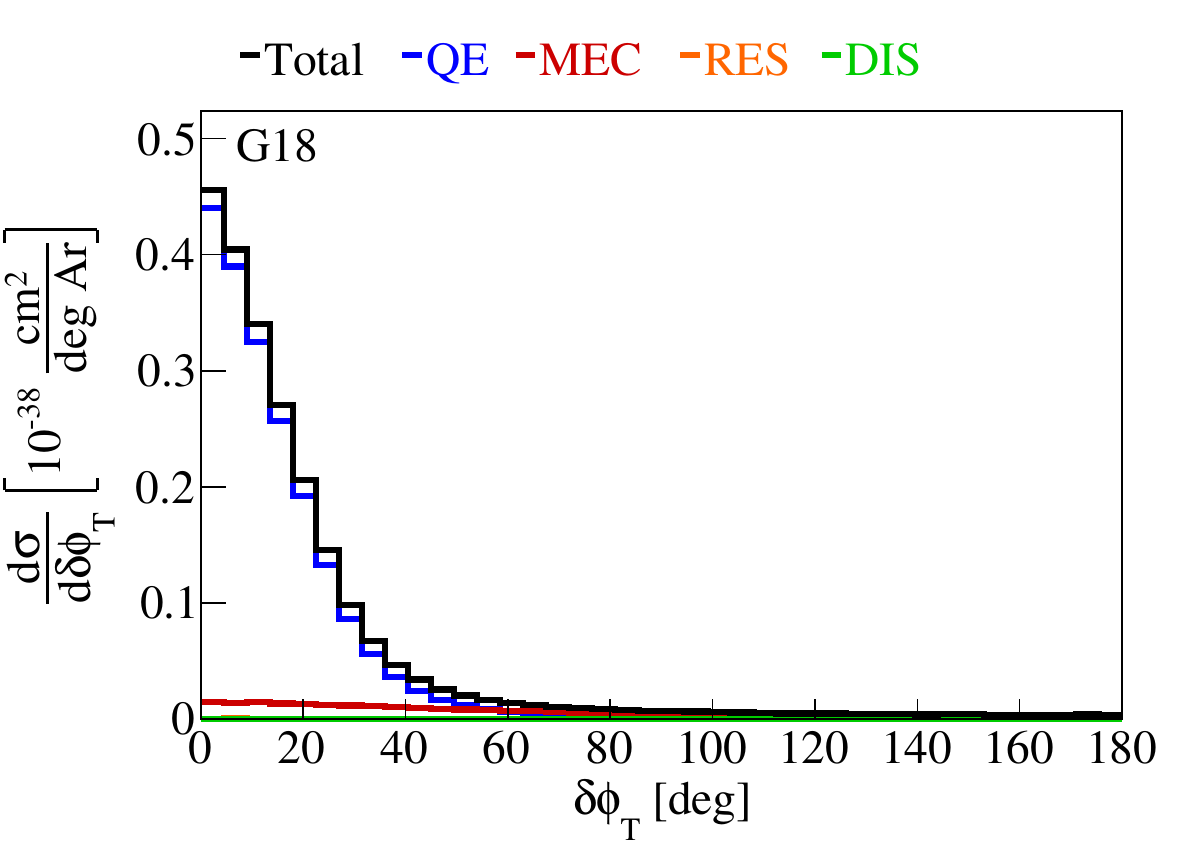}
};
\draw (0.4, -3.2) node {(b)};	
\end{tikzpicture}
\caption{The flux-integrated single-differential cross section interaction breakdown as a function of (left) $\phi_{3D}$ and (right) $\delta\phi_{T}$.
Colored lines show the results of theoretical cross section calculations using the $\texttt{G18}$ prediction without FSI for QE (blue), MEC (red), RES (orange), and DIS (green) interactions.}
\label{fig:nofsideltaphit}
\end{figure}

\begin{figure}[htb!]
\centering
\begin{tikzpicture} \draw (0, 0) node[inner sep=0] {
\includegraphics[width=0.45\textwidth]{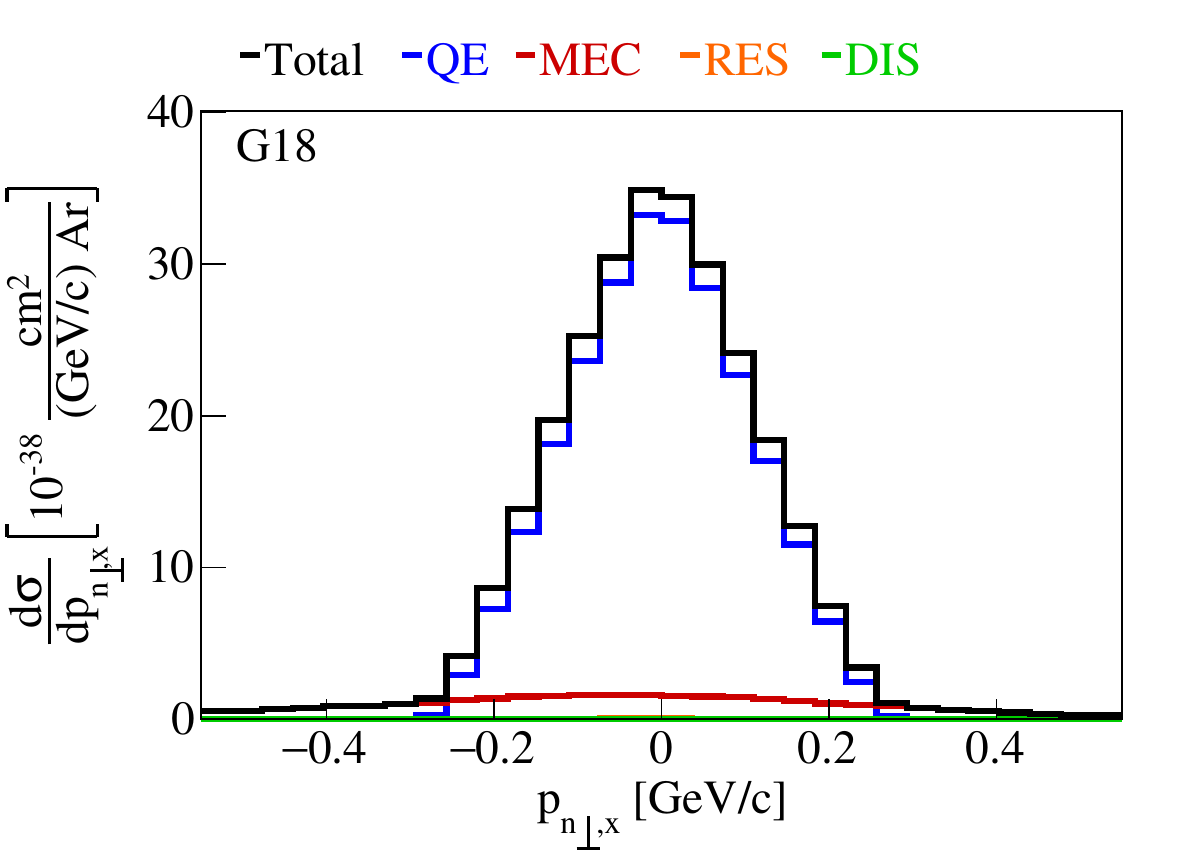}	
};
\draw (0.4, -3.2) node {(a)};	
\end{tikzpicture}
\hspace{0.05 \textwidth}
\begin{tikzpicture} \draw (0, 0) node[inner sep=0] {
\includegraphics[width=0.45\textwidth]{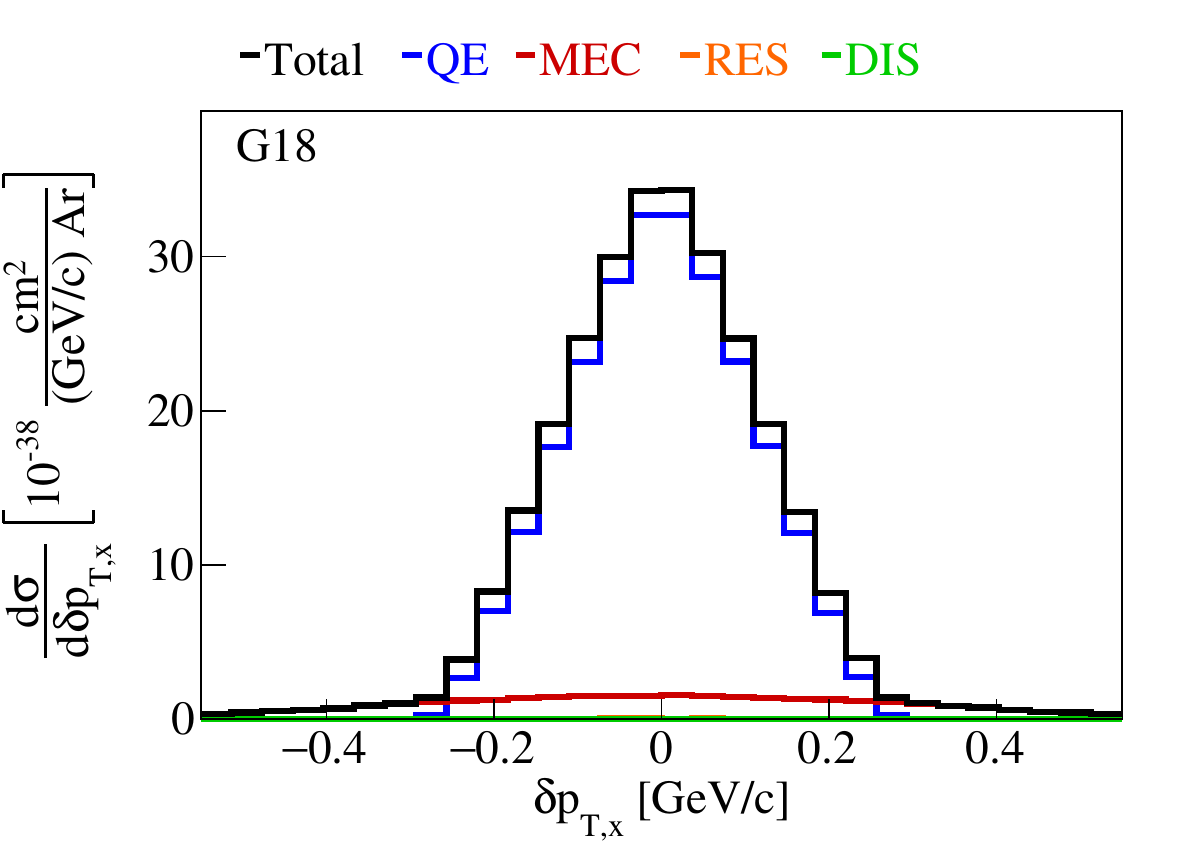}
};
\draw (0.4, -3.2) node {(b)};	
\end{tikzpicture}
\caption{The flux-integrated single-differential cross section interaction breakdown as a function of (left) $p_{n,\perp x}$ and (right) $p_{T,x}$.
Colored lines show the results of theoretical cross section calculations using the $\texttt{G18}$ prediction without FSI for QE (blue), MEC (red), RES (orange), and DIS (green) interactions.}
\label{fig:nofsideltaptx}
\end{figure}

\begin{figure}[htb!]
\centering
\begin{tikzpicture} \draw (0, 0) node[inner sep=0] {
\includegraphics[width=0.45\textwidth]{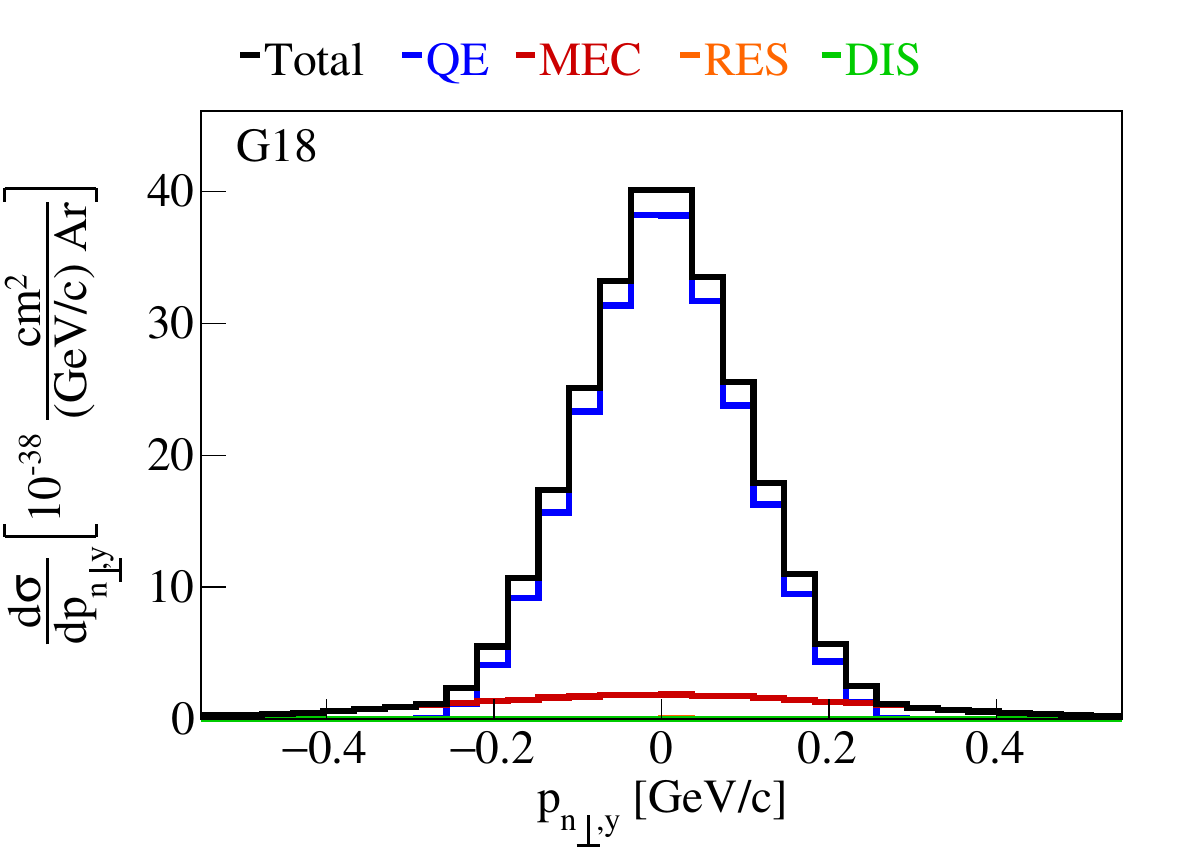}	
};
\draw (0.4, -3.2) node {(a)};	
\end{tikzpicture}
\hspace{0.05 \textwidth}
\begin{tikzpicture} \draw (0, 0) node[inner sep=0] {
\includegraphics[width=0.45\textwidth]{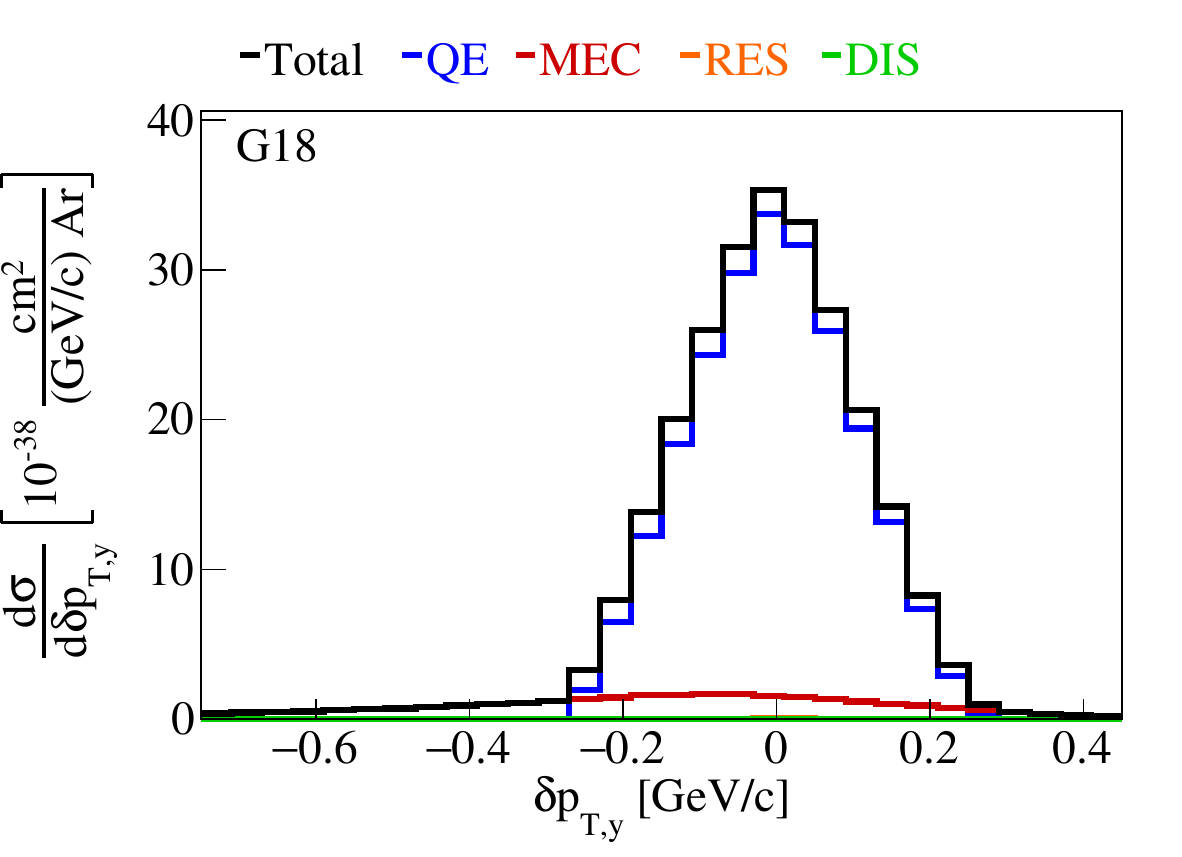}
};
\draw (0.4, -3.2) node {(b)};	
\end{tikzpicture}
\caption{The flux-integrated single-differential cross section interaction breakdown as a function of (left) $p_{n,\perp y}$ and (right) $p_{T,y}$.
Colored lines show the results of theoretical cross section calculations using the $\texttt{G18}$ prediction without FSI for QE (blue), MEC (red), RES (orange), and DIS (green) interactions.}
\label{fig:nofsideltapty}
\end{figure}


\clearpage
\section{Two-dimensional Simultaneous Cross Section Extraction}\label{TwoD}

\begin{figure}[htb!]
  \centering
\begin{tikzpicture} \draw (0, 0) node[inner sep=0] {
\includegraphics[width=0.45\textwidth]{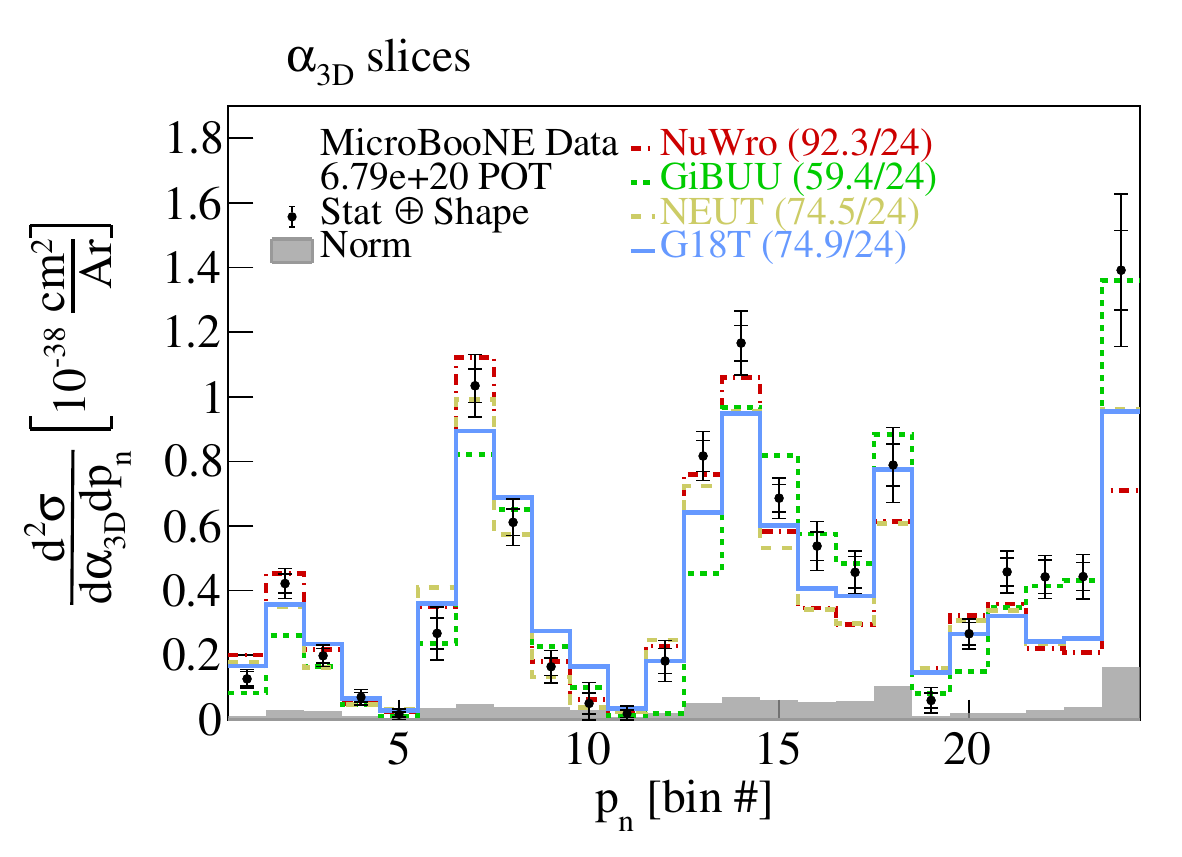}	
};
\draw (0.4, -3.2) node {(a)};	
\end{tikzpicture}
\hspace{0.05 \textwidth}
\begin{tikzpicture} \draw (0, 0) node[inner sep=0] {
\includegraphics[width=0.45\textwidth]{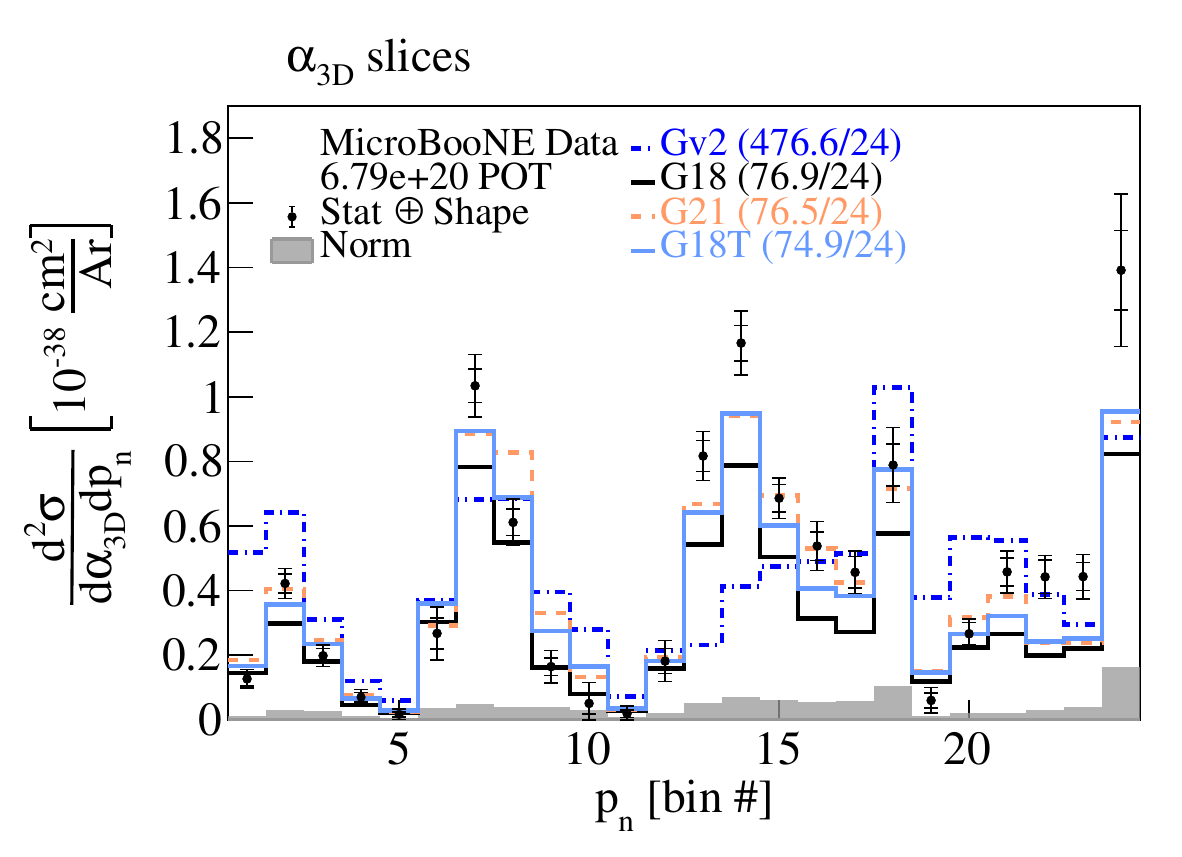}
};
\draw (0.4, -3.2) node {(b)};	
\end{tikzpicture}
  \caption{
The flux-integrated double-differential cross sections as a function of the $p_{n}$ bin number. (Left) Generator and (right) $\texttt{GENIE}$ configuration 
predictions are compared to data. Inner and outer error bars show the statistical and total (statistical and shape systematic)
uncertainty at the 1$\sigma$, or 68\%, confidence level. The gray band shows the normalization systematic uncertainty. The numbers
in parentheses show the $\chi^{2}$/ndf calculation for each one of the predictions.  
  }
  \label{fig:serialpn_deltaalpha3D}
\end{figure}

\begin{figure}[htb!]
  \centering
\begin{tikzpicture} \draw (0, 0) node[inner sep=0] {
\includegraphics[width=0.45\textwidth]{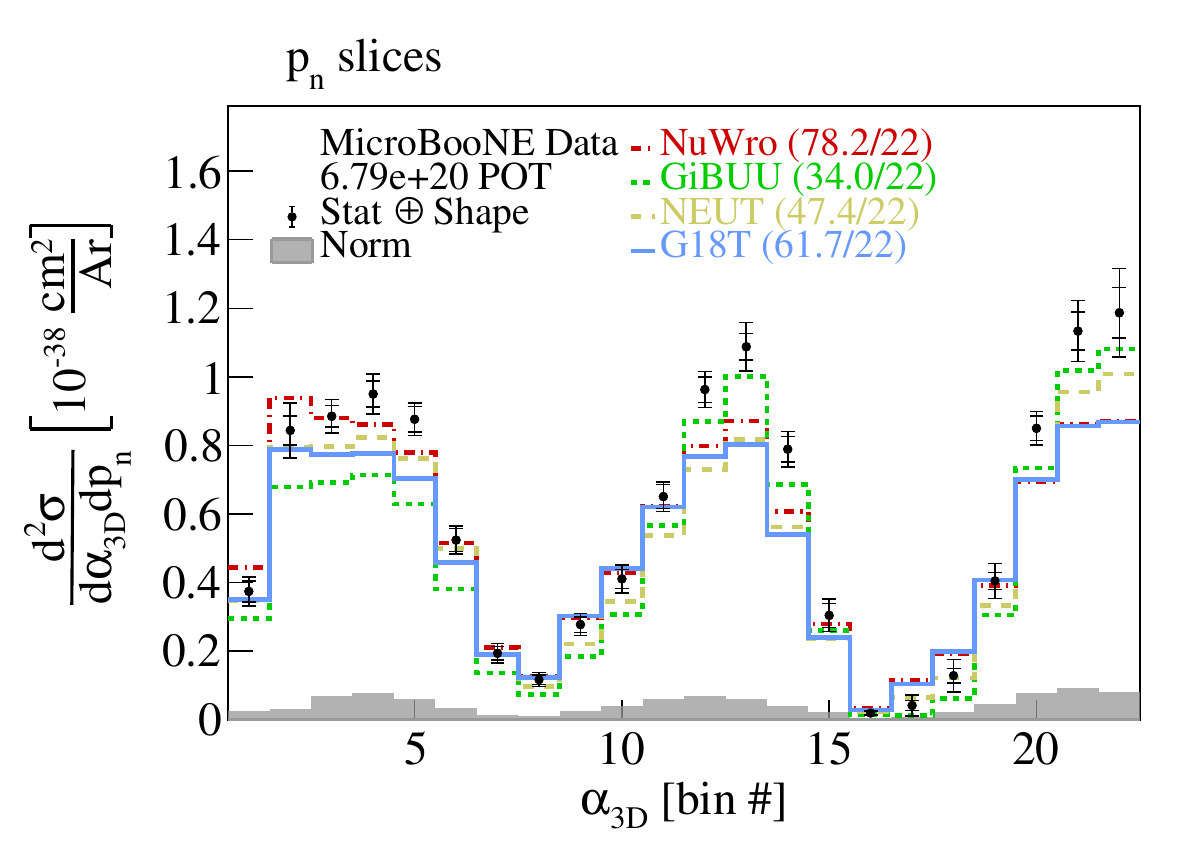}	
};
\draw (0.4, -3.2) node {(a)};	
\end{tikzpicture}
\hspace{0.05 \textwidth}
\begin{tikzpicture} \draw (0, 0) node[inner sep=0] {
\includegraphics[width=0.45\textwidth]{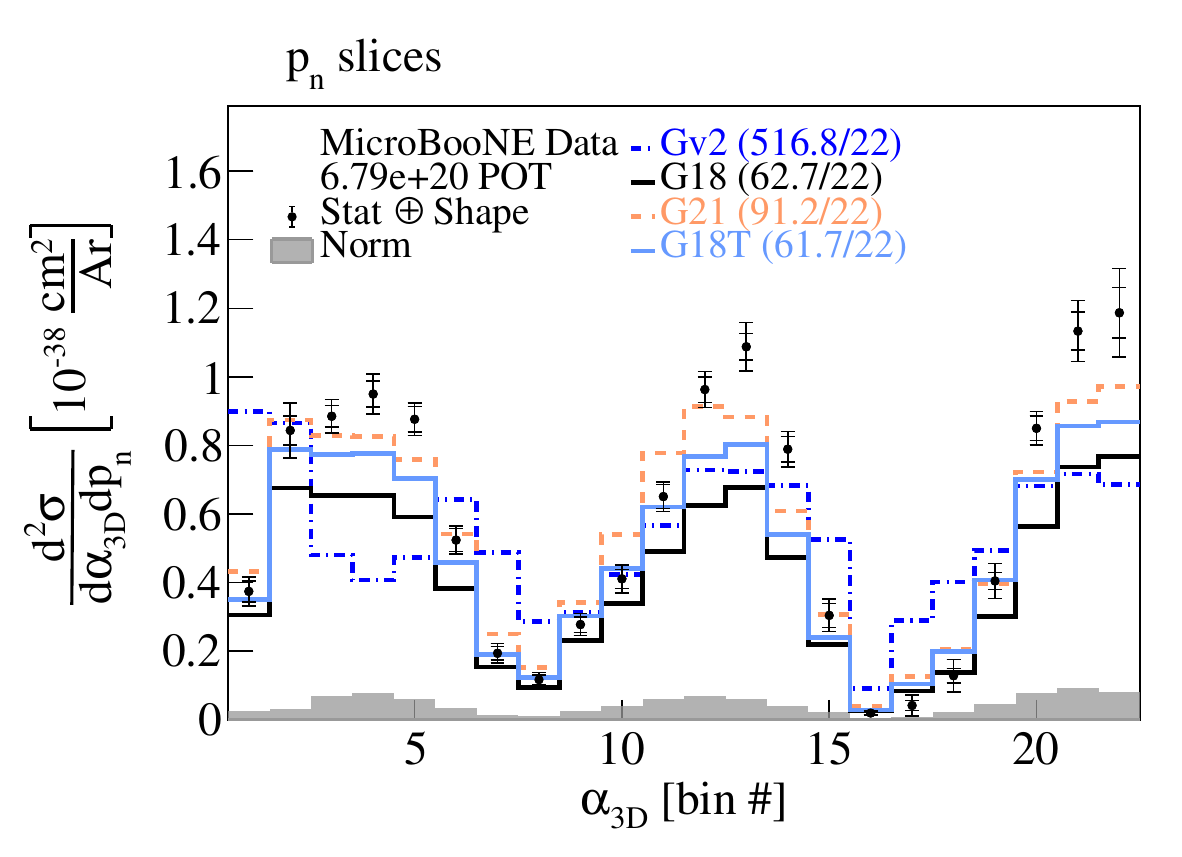}
};
\draw (0.4, -3.2) node {(b)};	
\end{tikzpicture}
  \caption{
The flux-integrated double-differential cross sections as a function of the $\delta \alpha_{3D}$ bin number. (Left) Generator and (right) $\texttt{GENIE}$ configuration 
predictions are compared to data. Inner and outer error bars show the statistical and total (statistical and shape systematic)
uncertainty at the 1$\sigma$, or 68\%, confidence level. The gray band shows the normalization systematic uncertainty. The numbers
in parentheses show the $\chi^{2}$/ndf calculation for each one of the predictions.  
  }
  \label{fig:serialdeltaalpha3d_deltapn}
\end{figure}


\clearpage
\section{Fake Data Studies}\label{FakeData}

In order to ensure that the unfolded data results are not biased due to the choice of the original MC, we performed fake data studies and treated the signal reconstructed events from an alternative model as if they were real data. 
We investigated three samples: (a) using NuWro events, (b) by removing the weights corresponding to the MicroBooNE tune (NoTune) and, (c) by multiplying the weight for the MEC events by 2 (TwiceMEC).
We then extracted the cross section from these fake data using our nominal MC response and covariance matrices and the Wiener-SVD filter both for the single- and double-differential cross sections. 
Given that we have the true cross section information from the fake data, we compared the fake data extracted cross section to the corresponding truth-level cross section predictions.  
The outcome agreed within an  expected  uncertainty.   
In  the  case  of  all  these  variations,  uncertainties  due  to  beam exposure, number-of-targets, detector, fluxes, and re-interactions are irrelevant and were not included in the fake-data uncertainty for comparison.
Only the uncorrelated uncertainties (cross section, statistical, MC statistical, alternative generator) were used for the minimization procedure and are shown on the plots to account for the unfolding process and the change in the event generator.
The combination of these uncertainties covers the differences between the unfolded fake data points (black) and the alternative-generator theory results (orange).
The relevant $A_{C}$ matrices have been applied to the generator predictions.

\begin{figure*}[htb!]
\centering 
\includegraphics[width=0.32\linewidth]{\figures 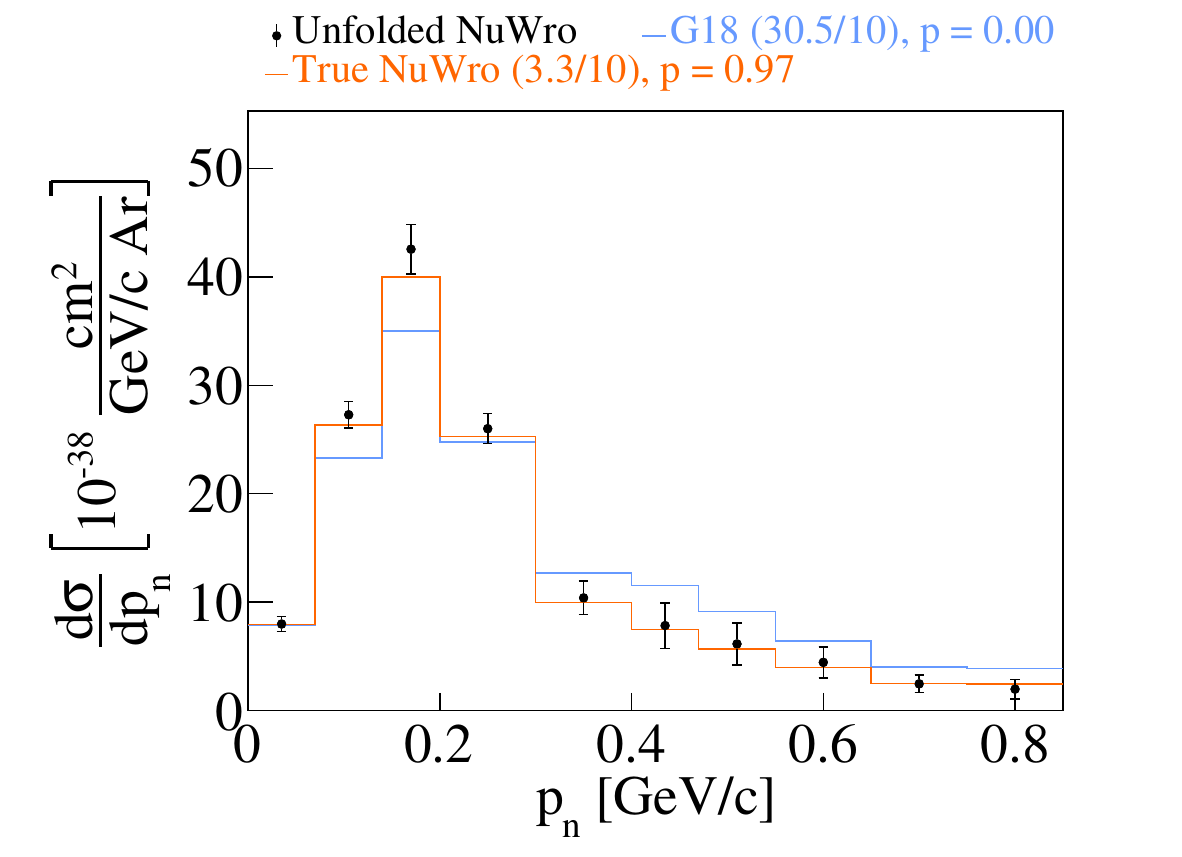}
\includegraphics[width=0.32\linewidth]{\figures 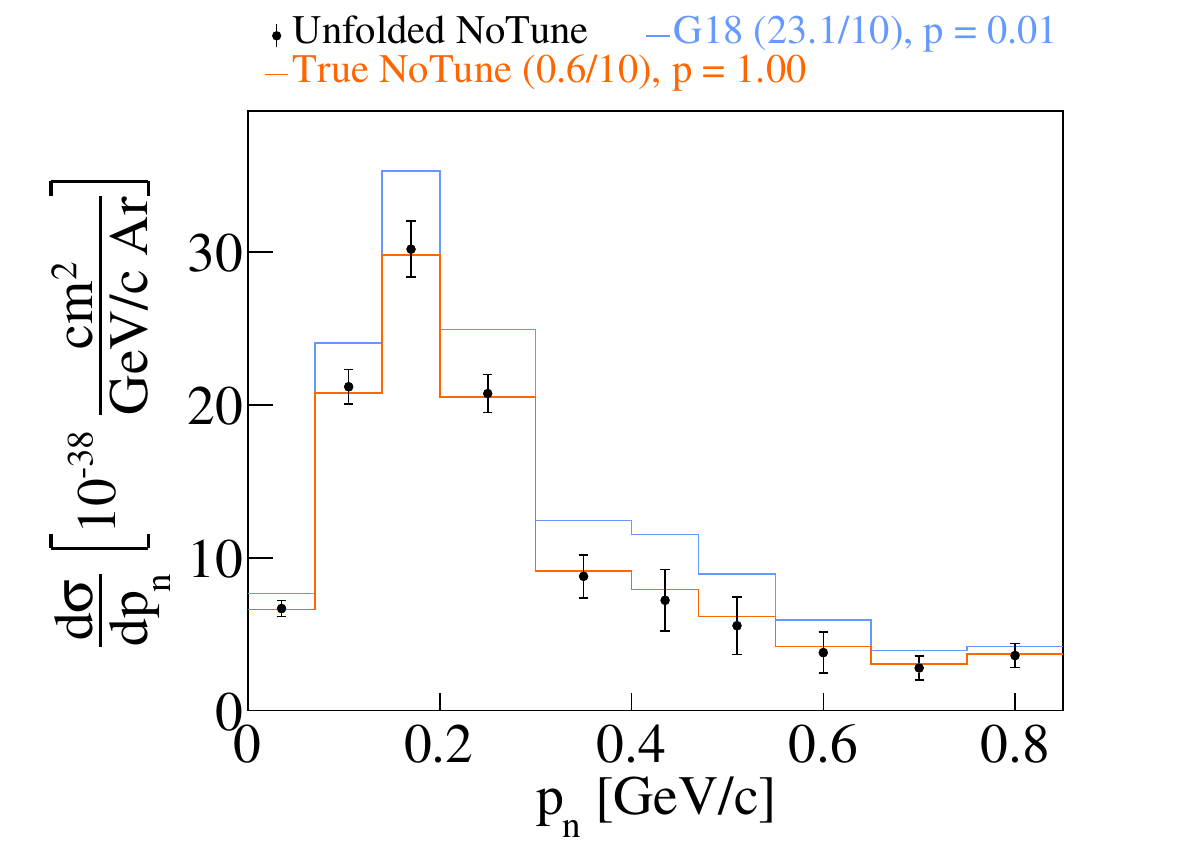}
\includegraphics[width=0.32\linewidth]{\figures 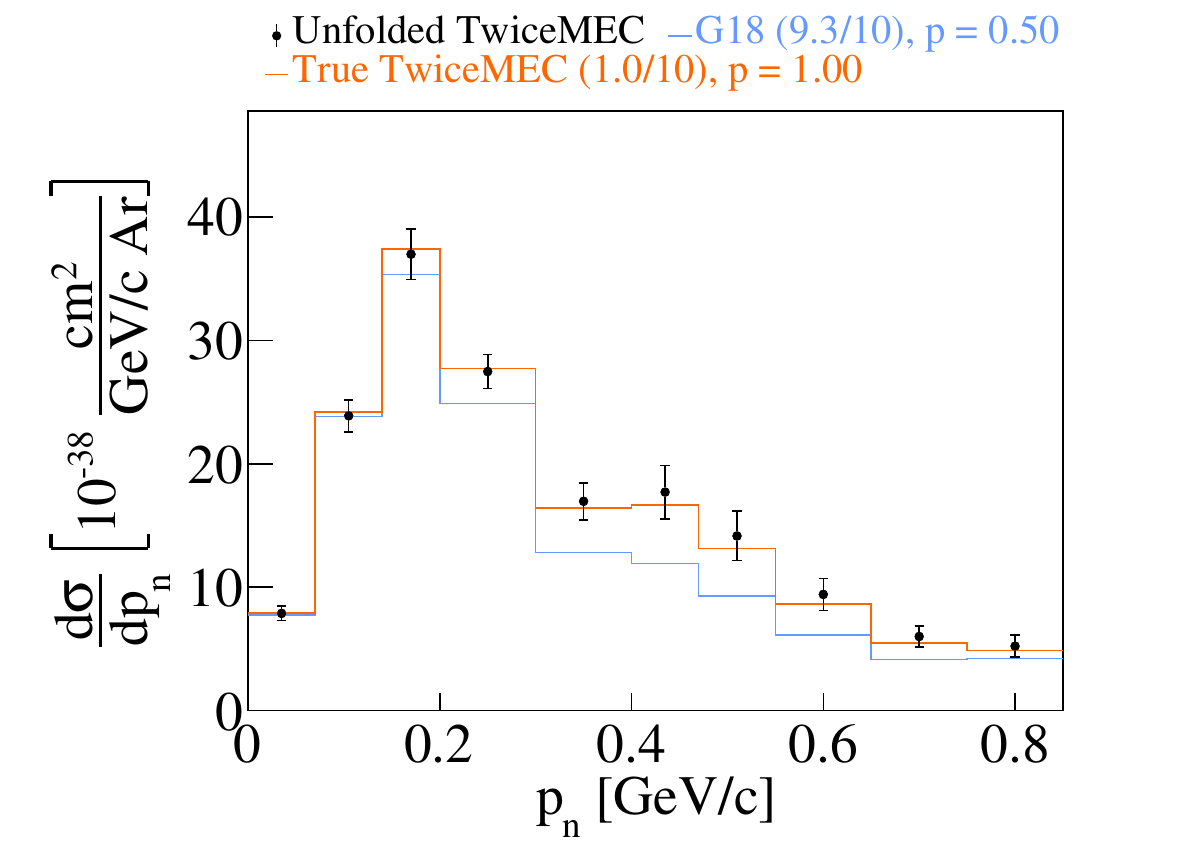}
\caption{
Fake data studies for $p_{n}$ using (left) NuWro, (center) GENIE without the MicroBooNE tune (NoTune), and (right) twice the weights for MEC events (TwiceMEC) as fake data samples.
}
\label{DeltaPnFakeData}
\end{figure*}

\begin{figure*}[htb!]
\centering 
\includegraphics[width=0.32\linewidth]{\figures 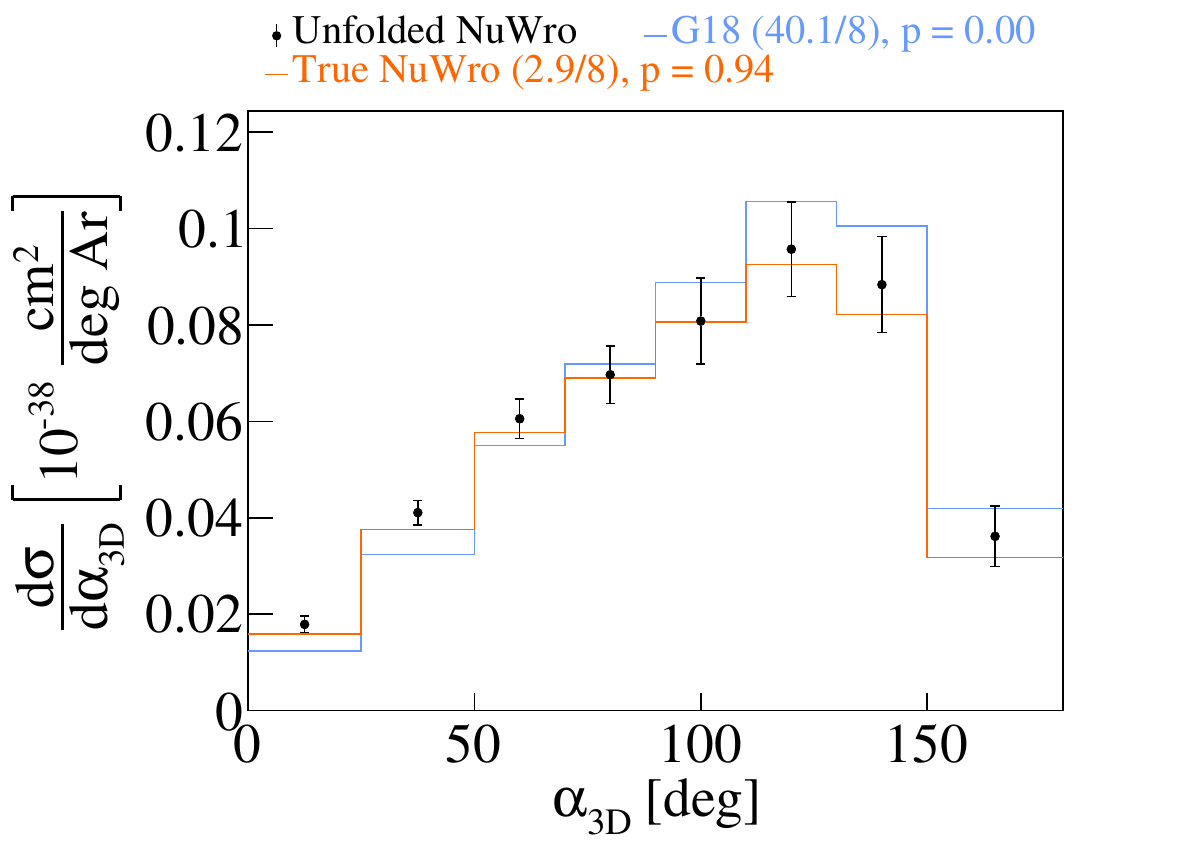}
\includegraphics[width=0.32\linewidth]{\figures 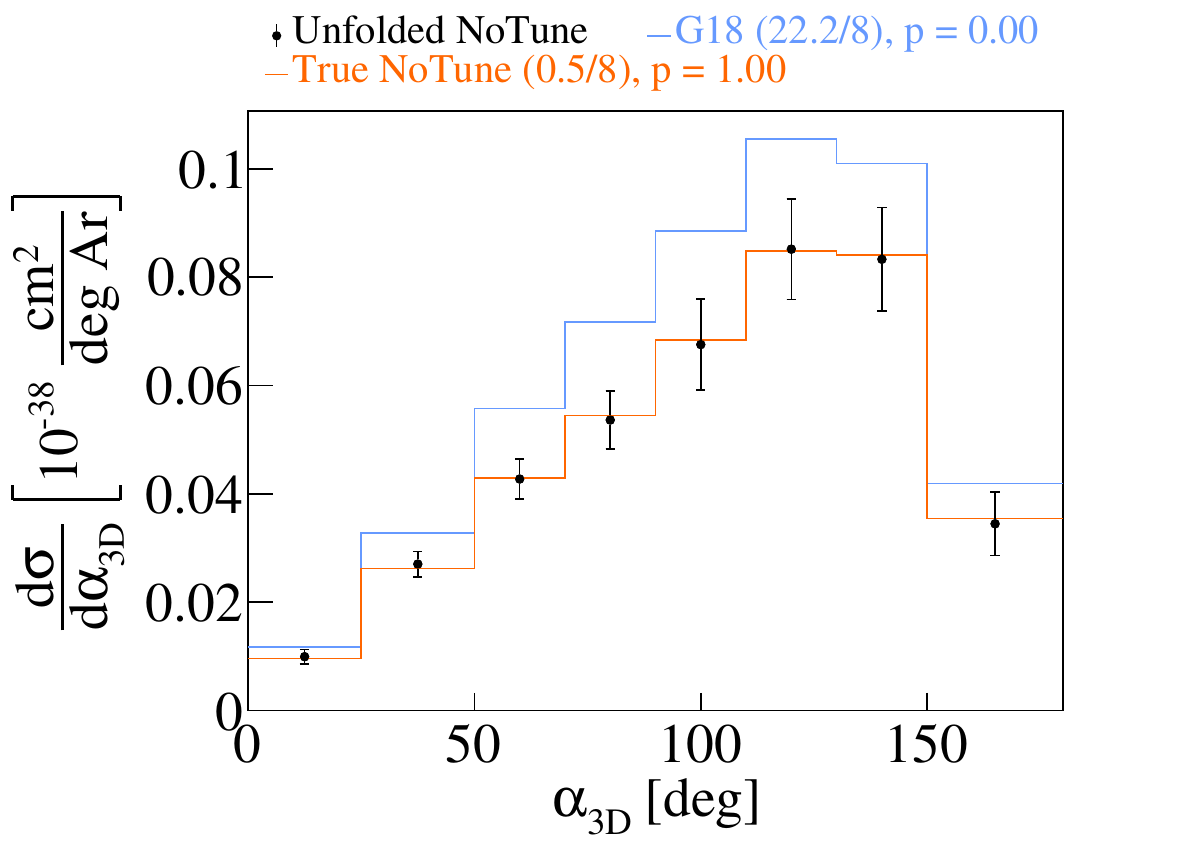}
\includegraphics[width=0.32\linewidth]{\figures 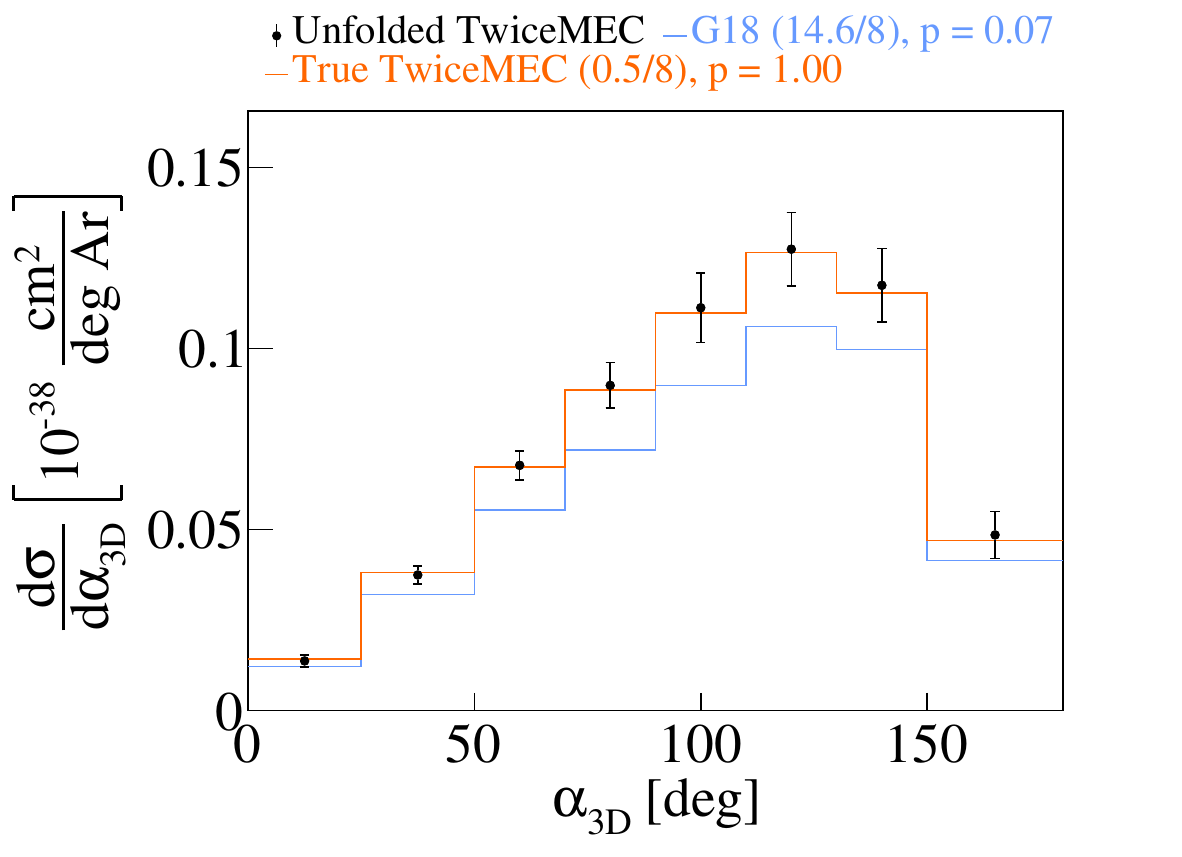}
\caption{
Fake data studies for $\alpha_{3D}$ using (left) NuWro, (center) GENIE without the MicroBooNE tune (NoTune), and (right) twice the weights for MEC events (TwiceMEC) as fake data samples.
}
\label{DeltaAlpha3DqFakeData}
\end{figure*}

\begin{figure*}[htb!]
\centering 
\includegraphics[width=0.32\linewidth]{\figures 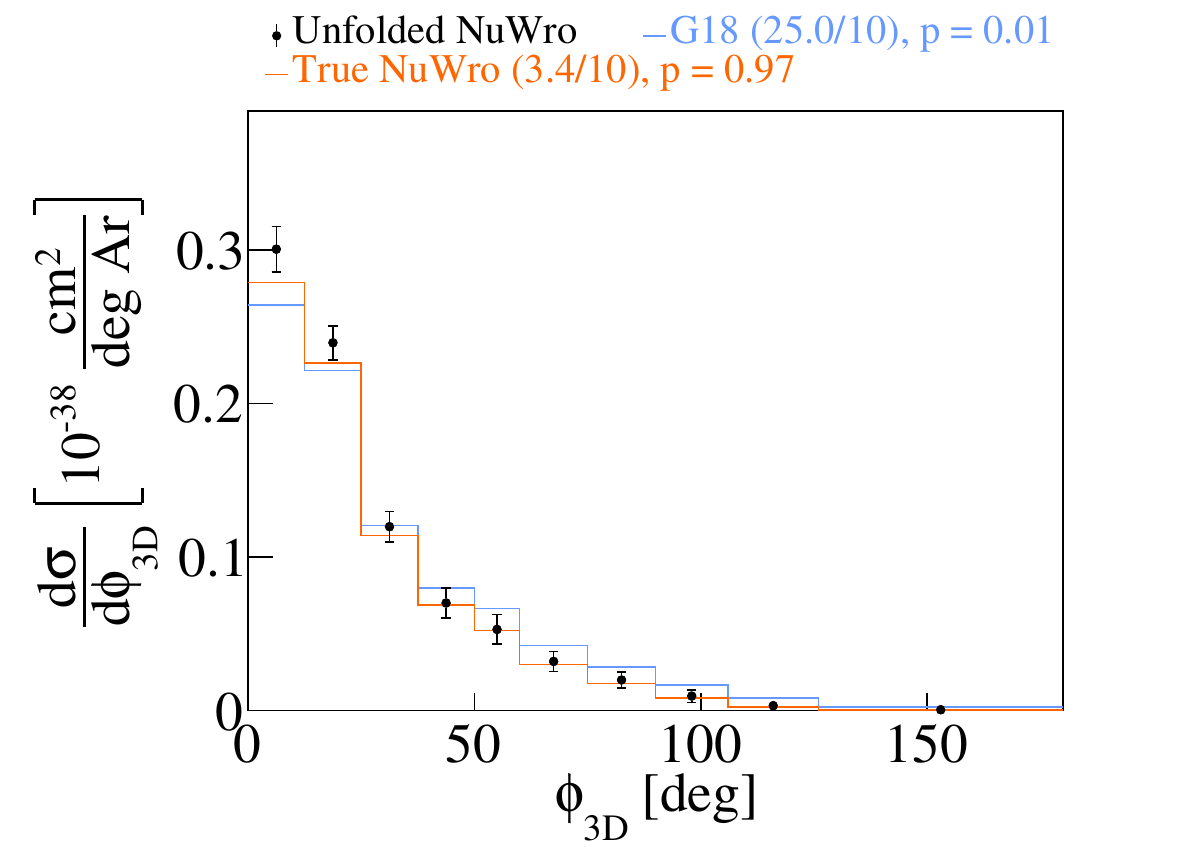}
\includegraphics[width=0.32\linewidth]{\figures 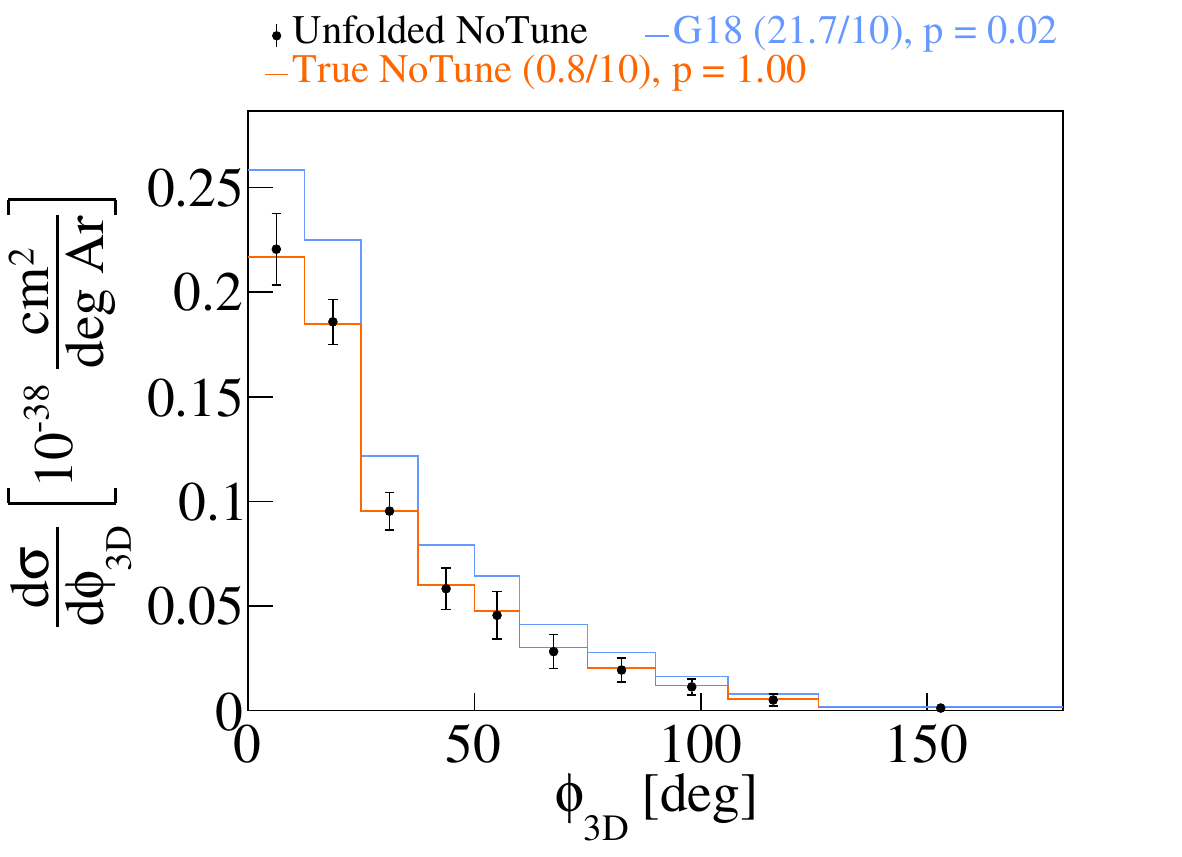}
\includegraphics[width=0.32\linewidth]{\figures 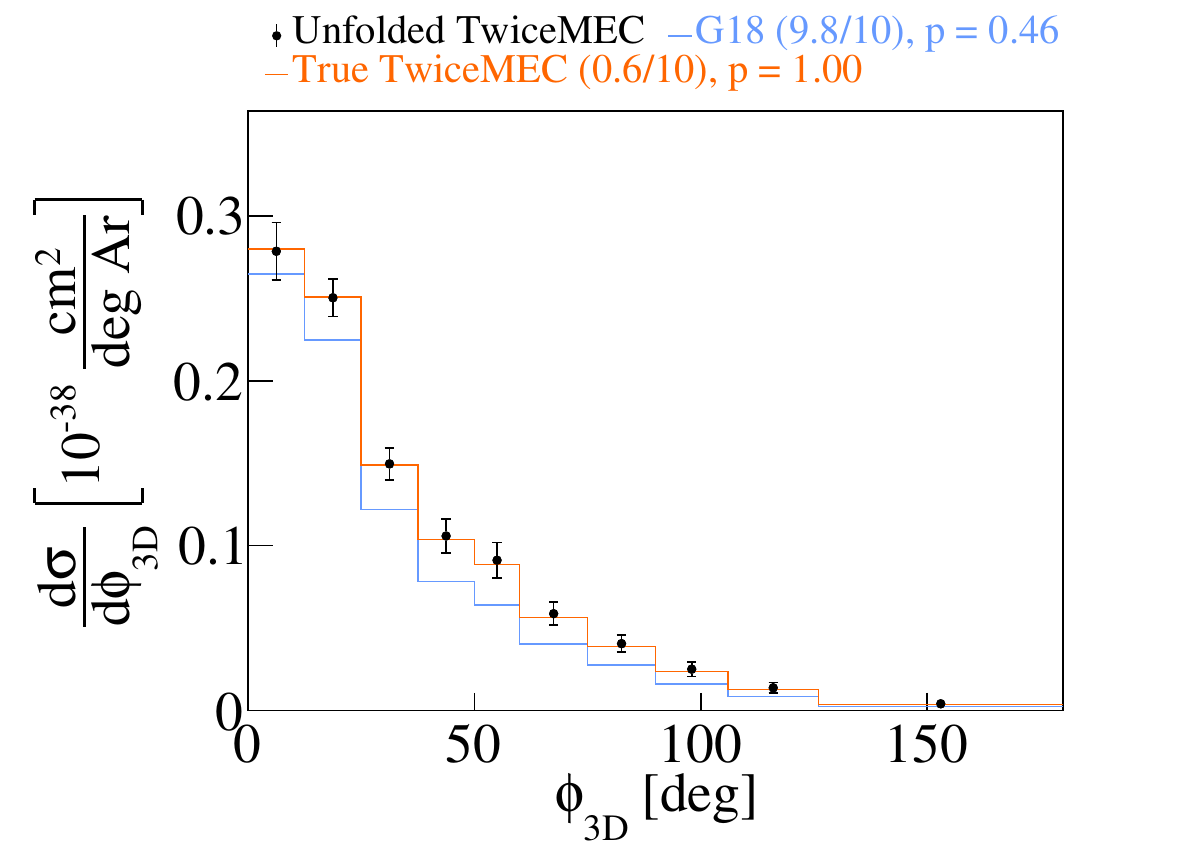}
\caption{
Fake data studies for $\phi_{3D}$ using (left) NuWro, (center) GENIE without the MicroBooNE tune (NoTune), and (right) twice the weights for MEC events (TwiceMEC) as fake data samples.
}
\label{DeltaPhi3DFakeData}
\end{figure*}

\begin{figure*}[htb!]
\centering 
\includegraphics[width=0.32\linewidth]{\figures 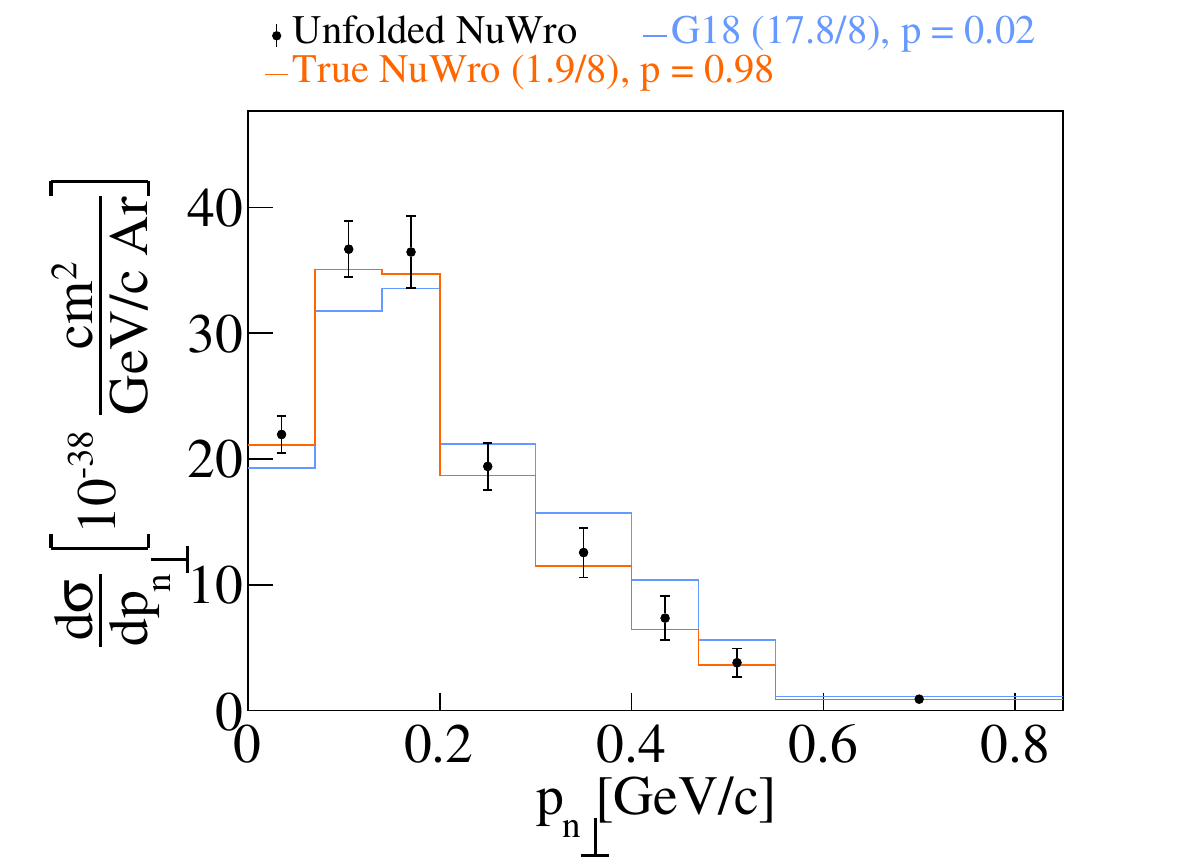}
\includegraphics[width=0.32\linewidth]{\figures 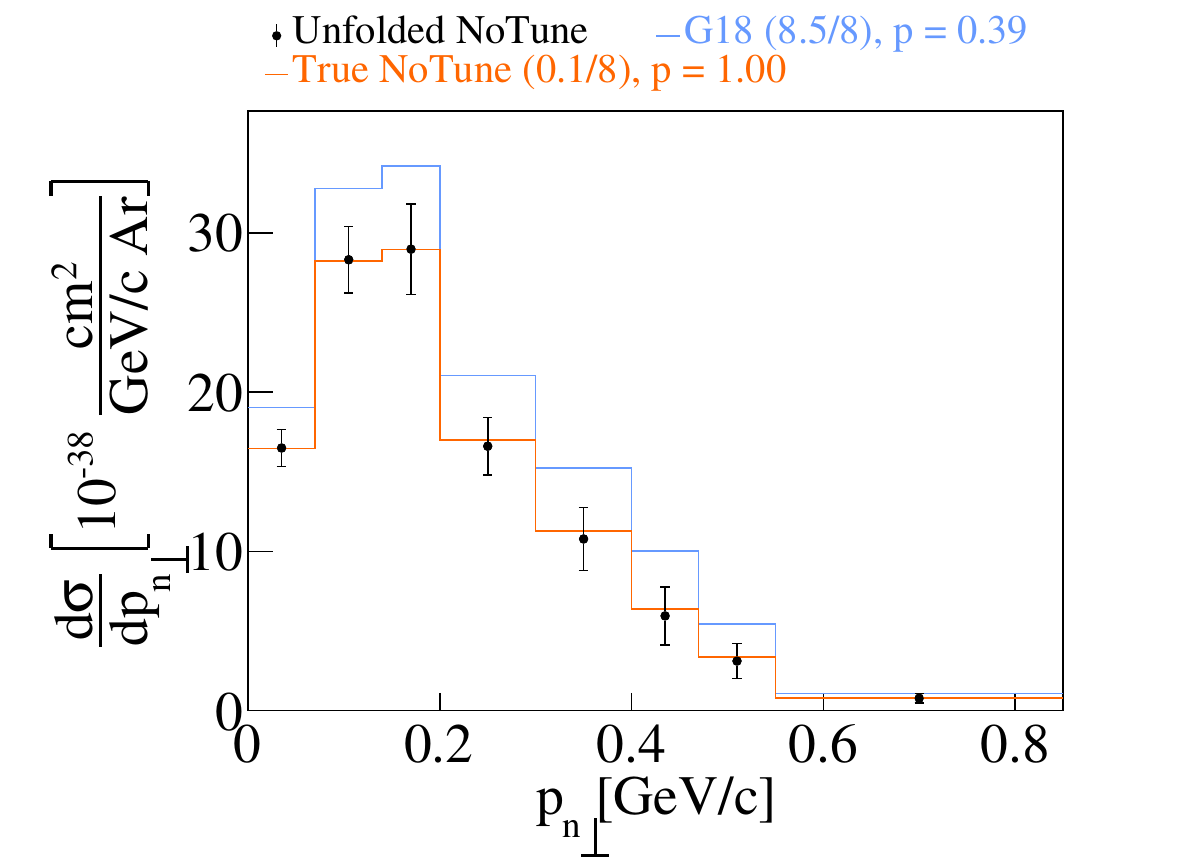}
\includegraphics[width=0.32\linewidth]{\figures 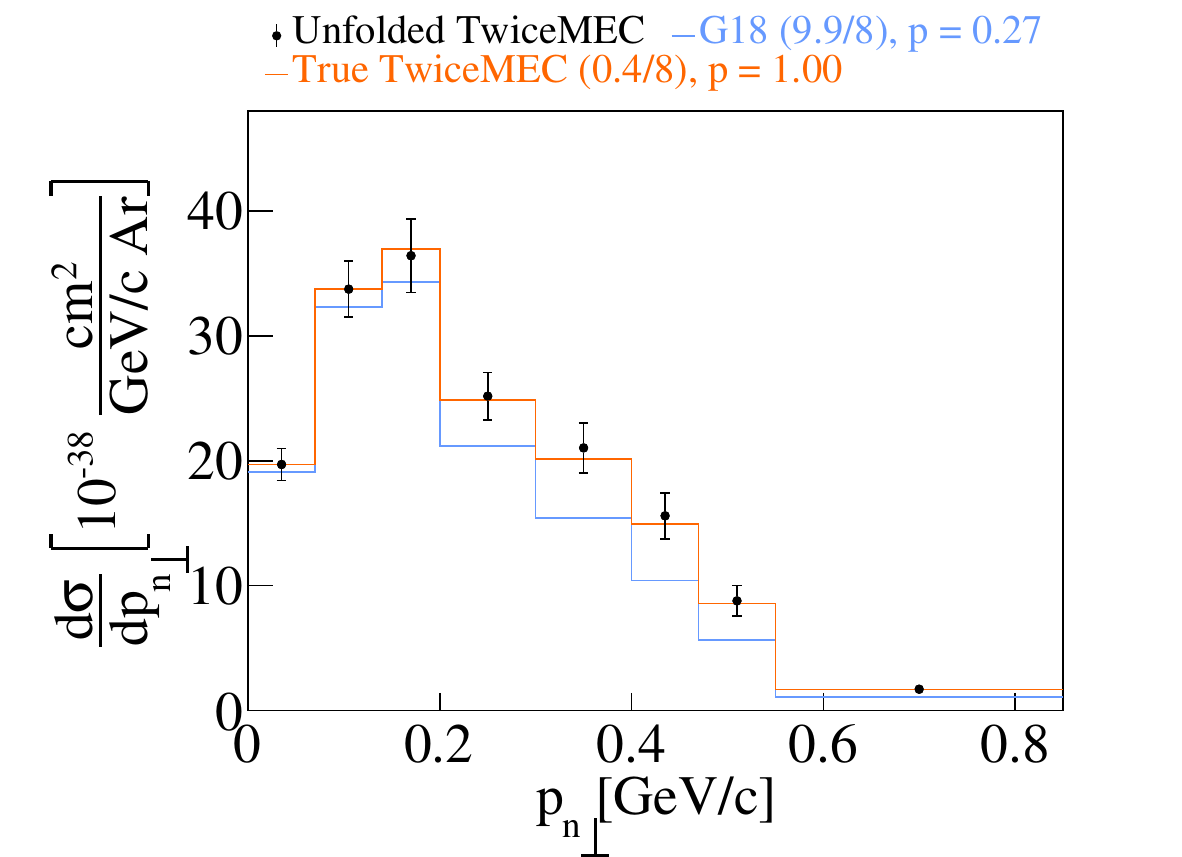}
\caption{
Fake data studies for $p_{n\perp}$ using (left) NuWro, (center) GENIE without the MicroBooNE tune (NoTune), and (right) twice the weights for MEC events (TwiceMEC) as fake data samples.
}
\label{DeltaPnPerpFakeData}
\end{figure*}

\begin{figure*}[htb!]
\centering 
\includegraphics[width=0.32\linewidth]{\figures 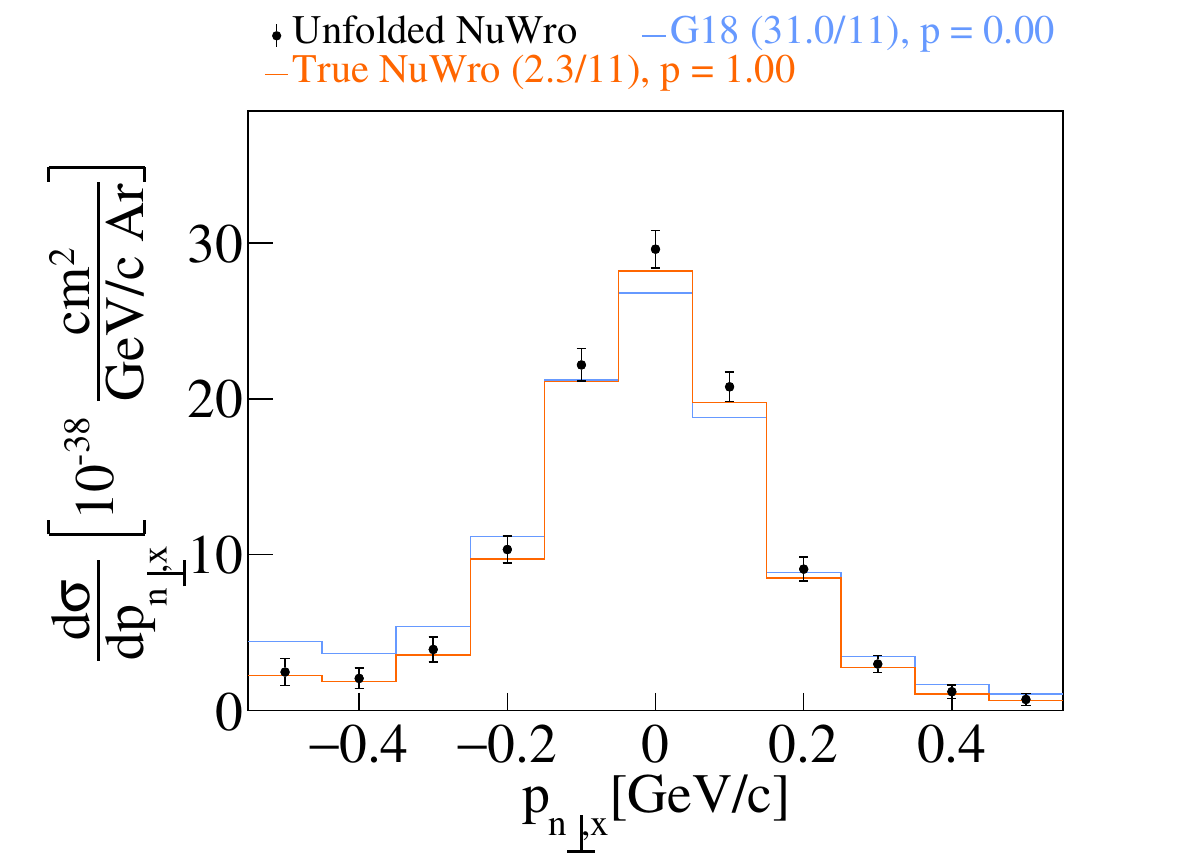}
\includegraphics[width=0.32\linewidth]{\figures 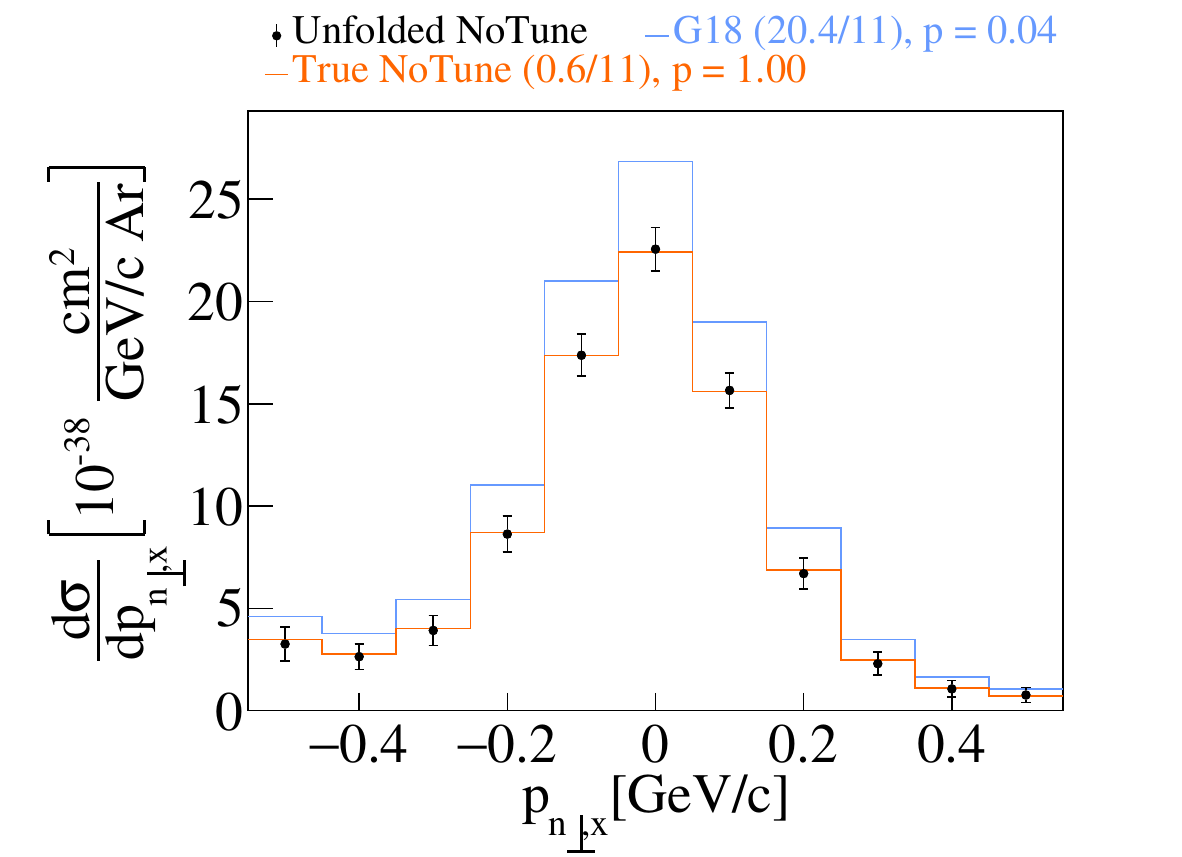}
\includegraphics[width=0.32\linewidth]{\figures 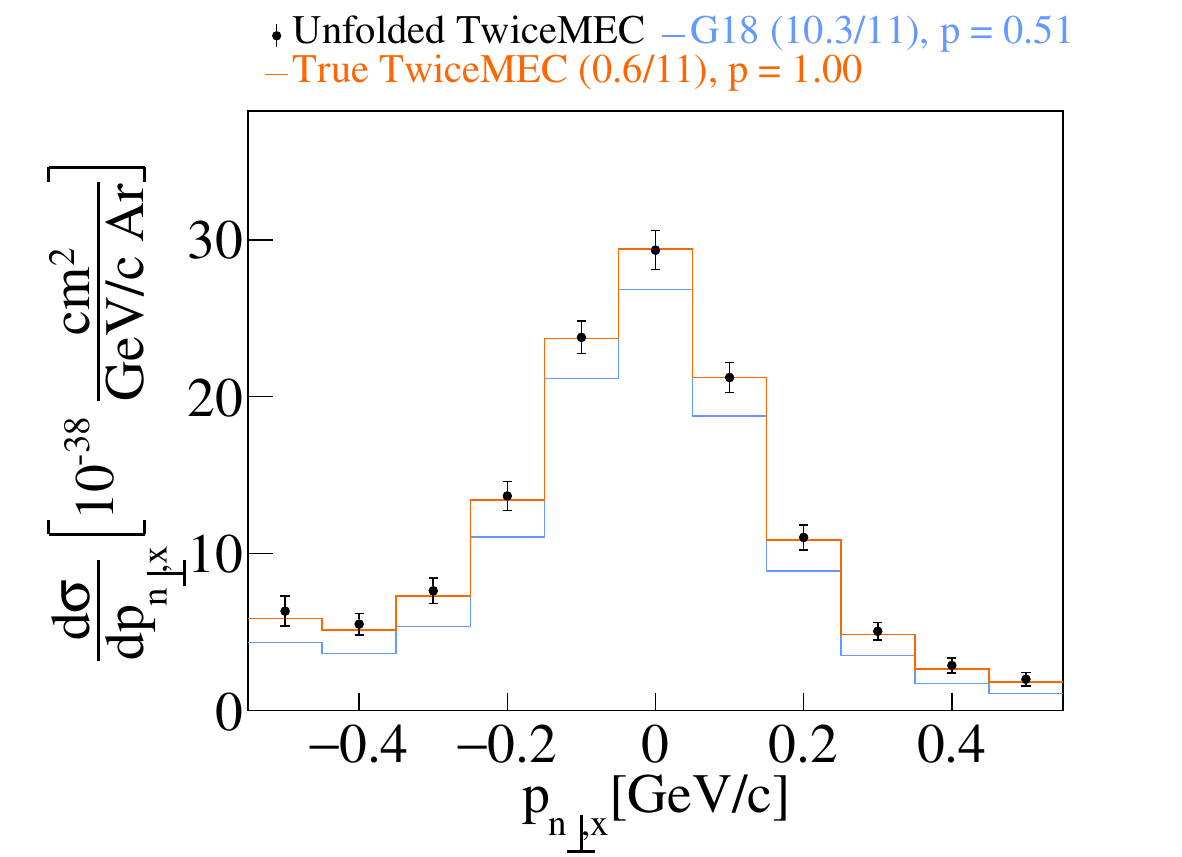}
\caption{
Fake data studies for $p_{n\perp,x}$ using (left) NuWro, (center) GENIE without the MicroBooNE tune (NoTune), and (right) twice the weights for MEC events (TwiceMEC) as fake data samples.
}
\label{DeltaPnPerpxFakeData}
\end{figure*}

\begin{figure*}[htb!]
\centering 
\includegraphics[width=0.32\linewidth]{\figures 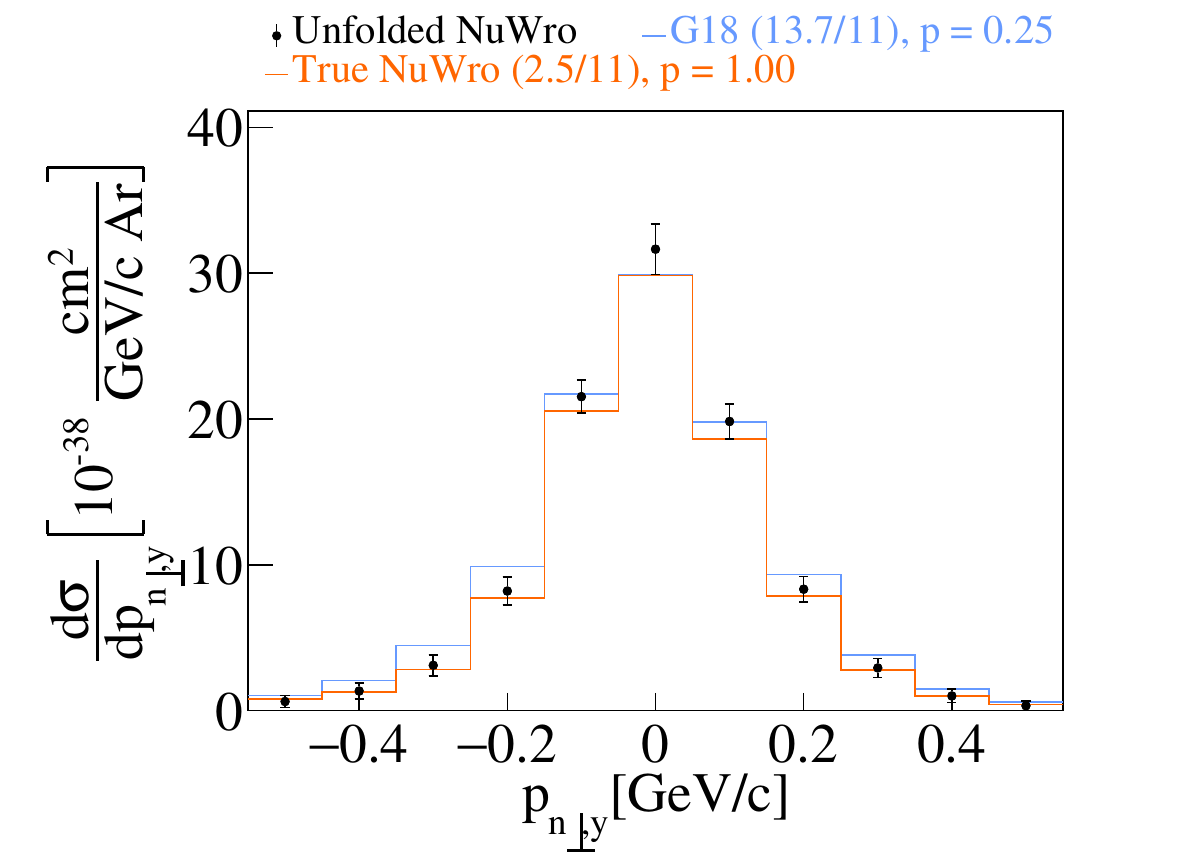}
\includegraphics[width=0.32\linewidth]{\figures 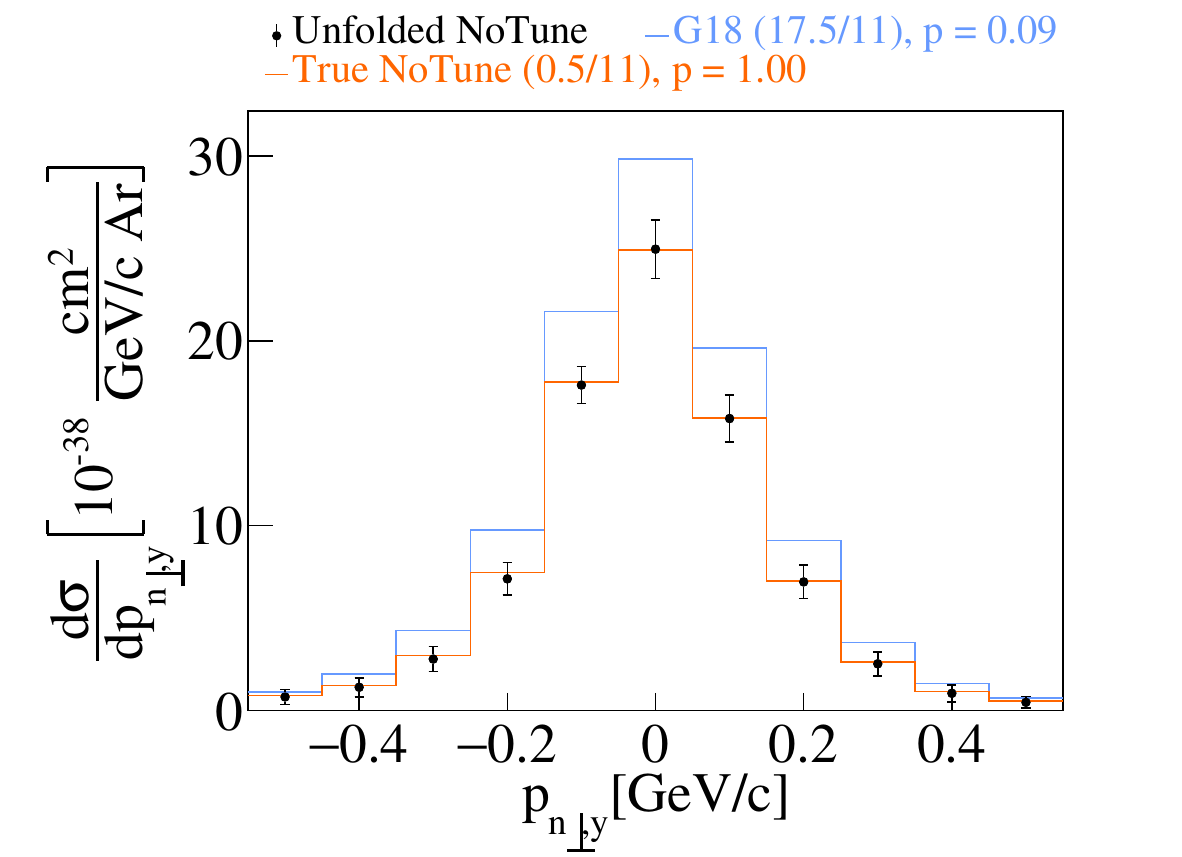}
\includegraphics[width=0.32\linewidth]{\figures 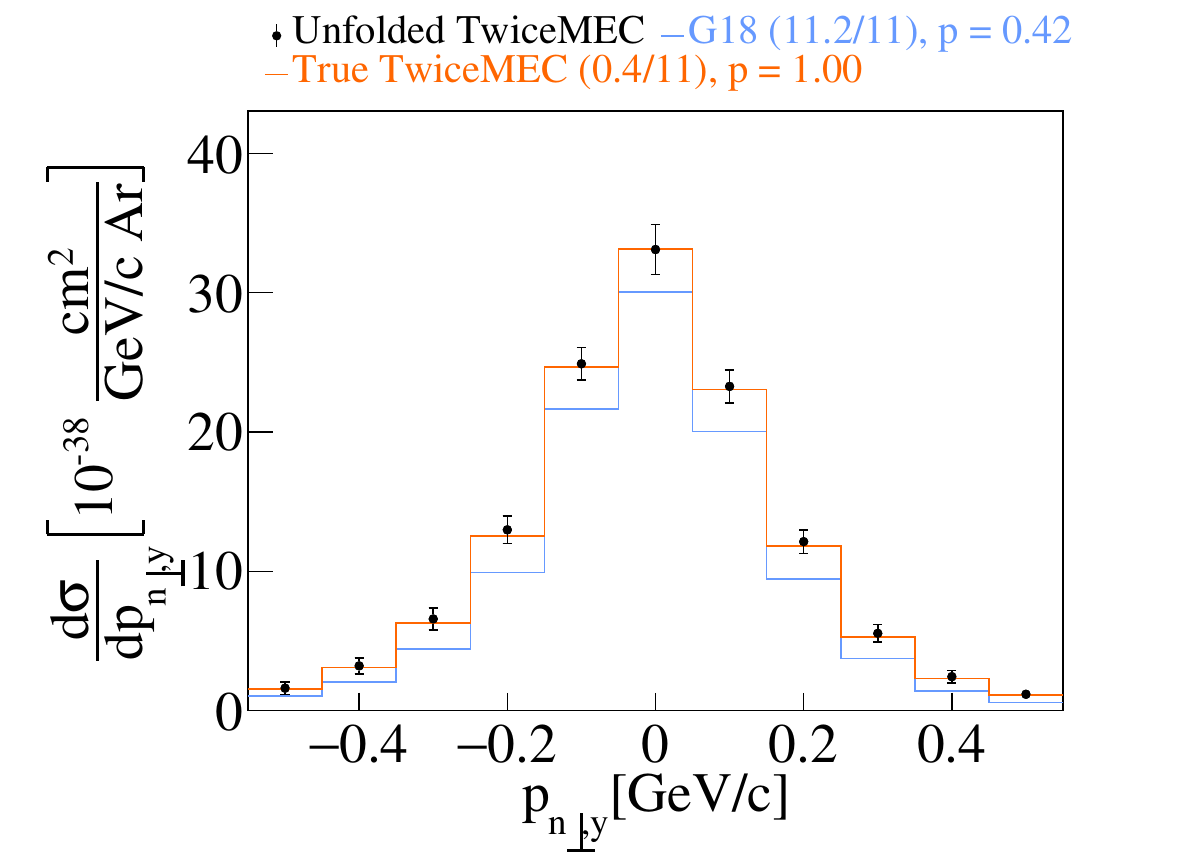}
\caption{
Fake data studies for $p_{n\perp,y}$ using (left) NuWro, (center) GENIE without the MicroBooNE tune (NoTune), and (right) twice the weights for MEC events (TwiceMEC) as fake data samples.
}
\label{DeltaPnPerpyFakeData}
\end{figure*}

\begin{figure*}[htb!]
\centering 
\includegraphics[width=0.32\linewidth]{\figures 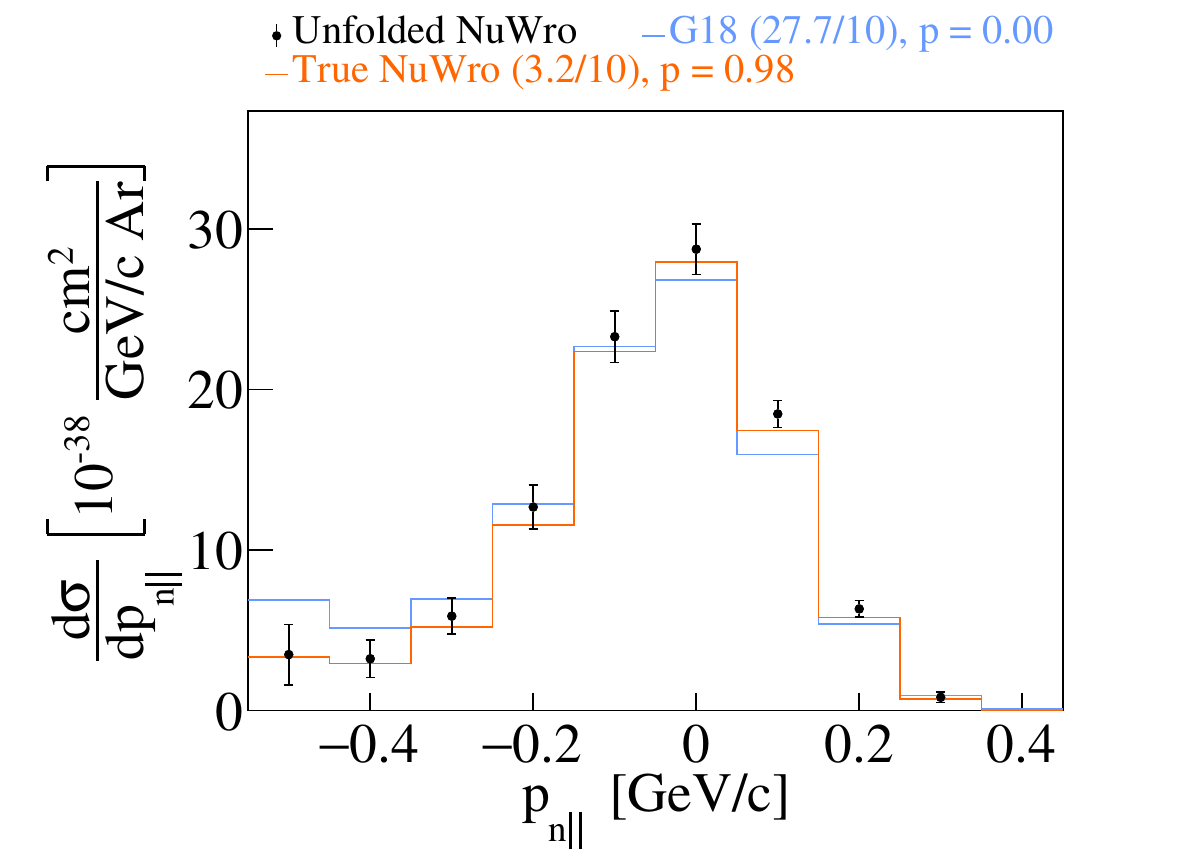}
\includegraphics[width=0.32\linewidth]{\figures 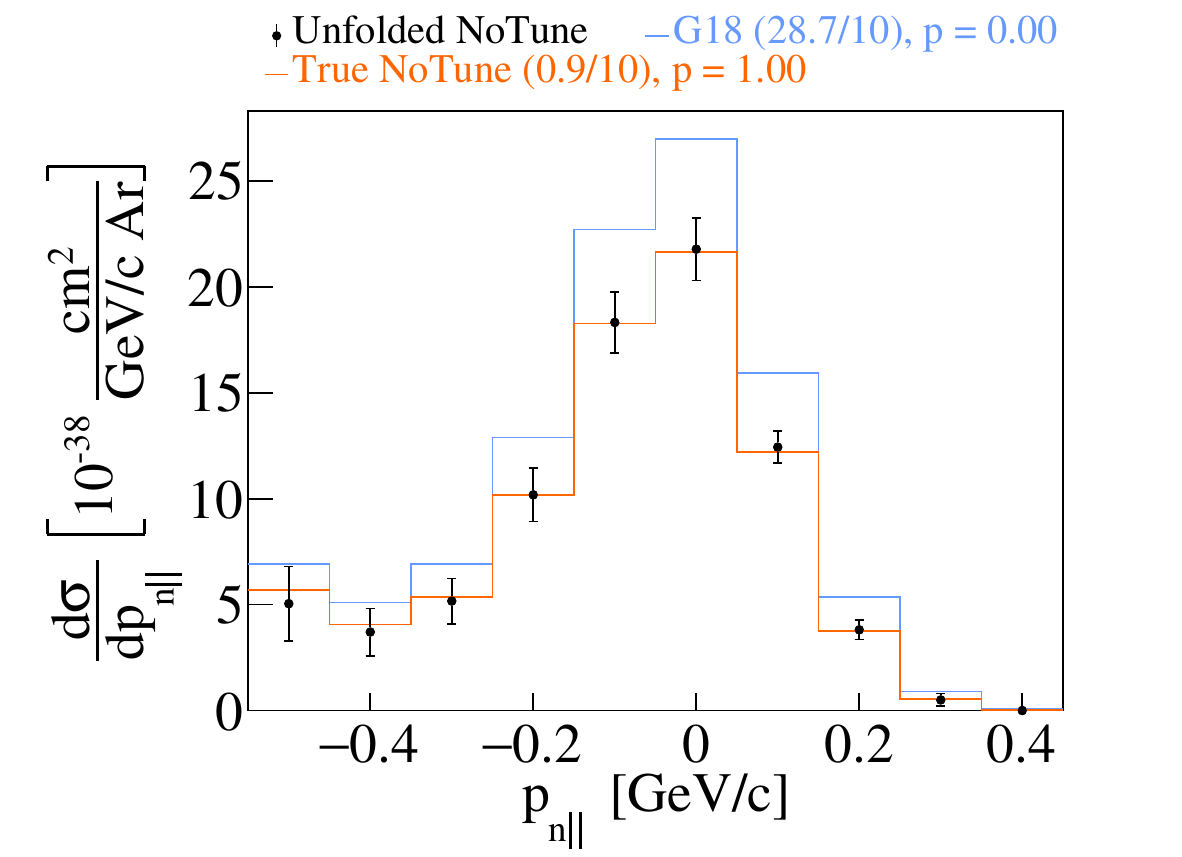}
\includegraphics[width=0.32\linewidth]{\figures 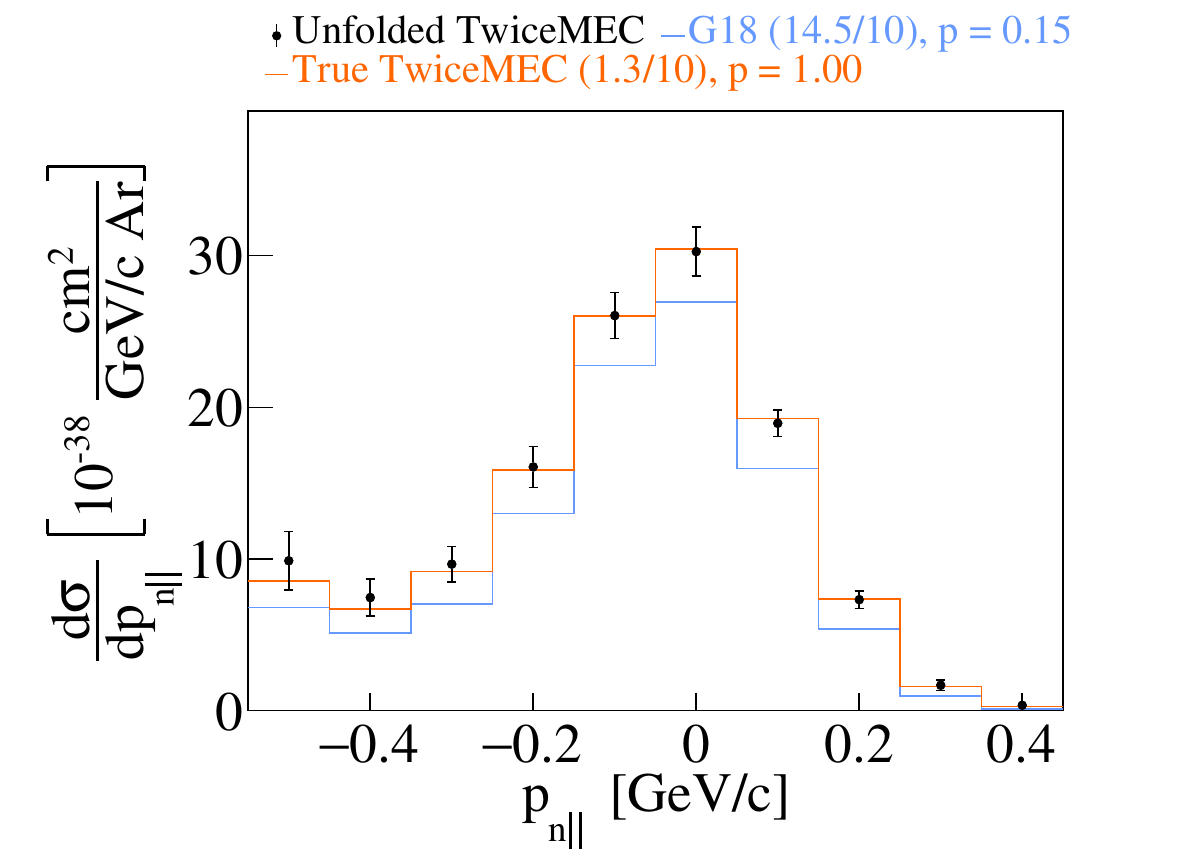}
\caption{
Fake data studies for $p_{n\parallel}$ using (left) NuWro, (center) GENIE without the MicroBooNE tune (NoTune), and (right) twice the weights for MEC events (TwiceMEC) as fake data samples.
}
\label{DeltaPnParFakeData}
\end{figure*}

\begin{figure*}[htb!]
\centering 
\includegraphics[width=0.32\linewidth]{\figures 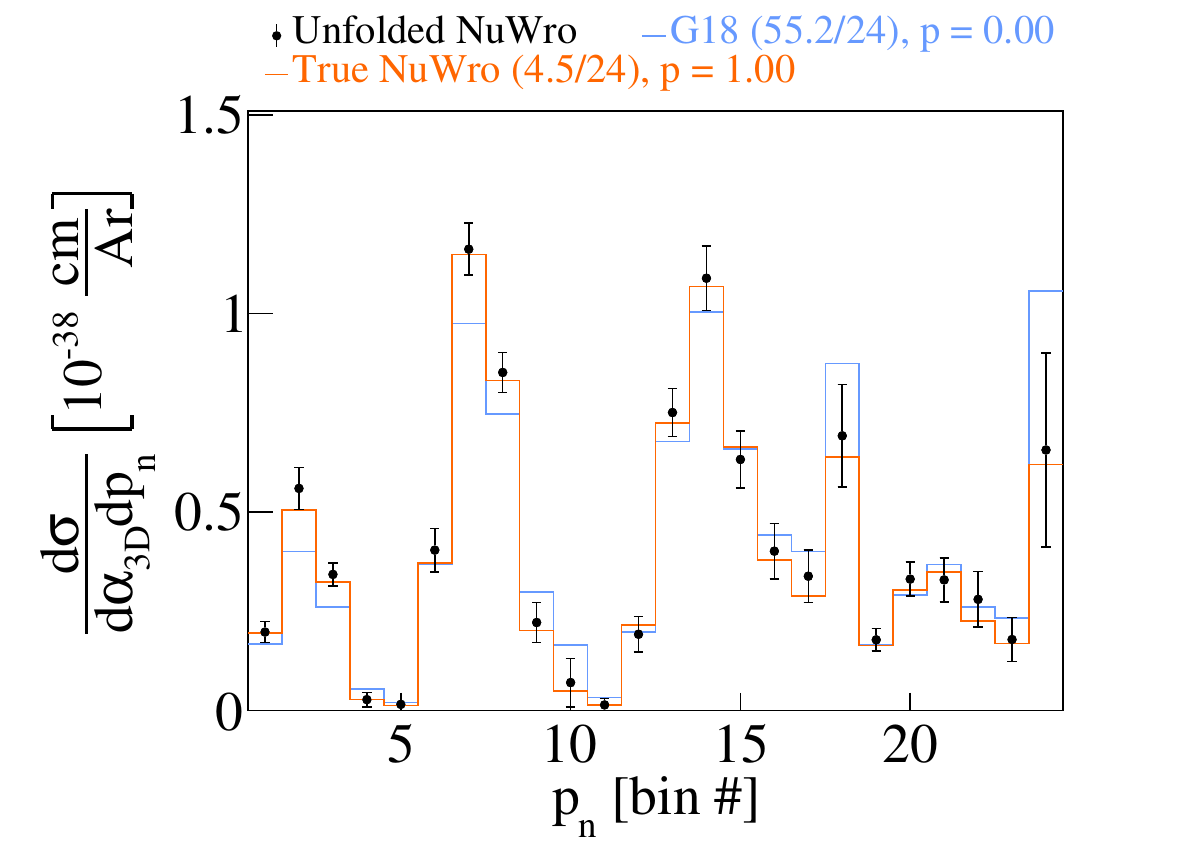}
\includegraphics[width=0.32\linewidth]{\figures 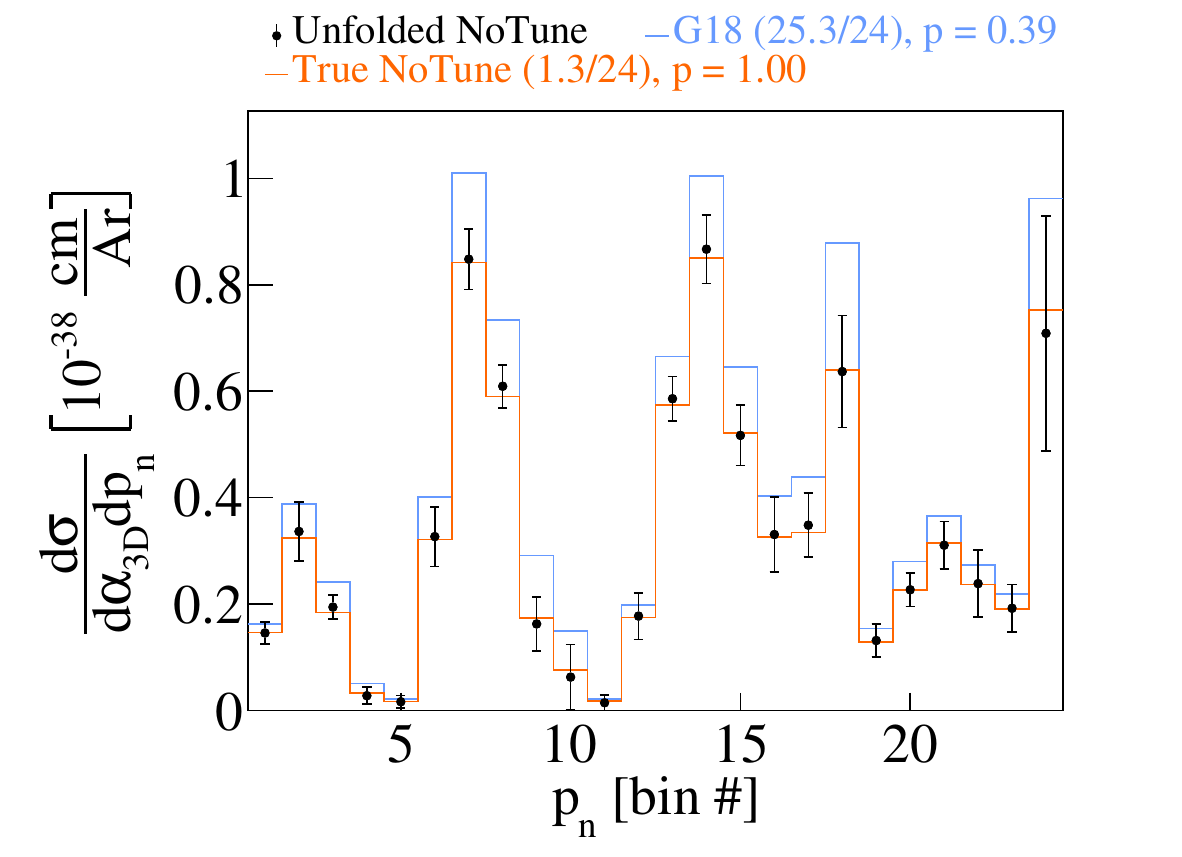}
\includegraphics[width=0.32\linewidth]{\figures 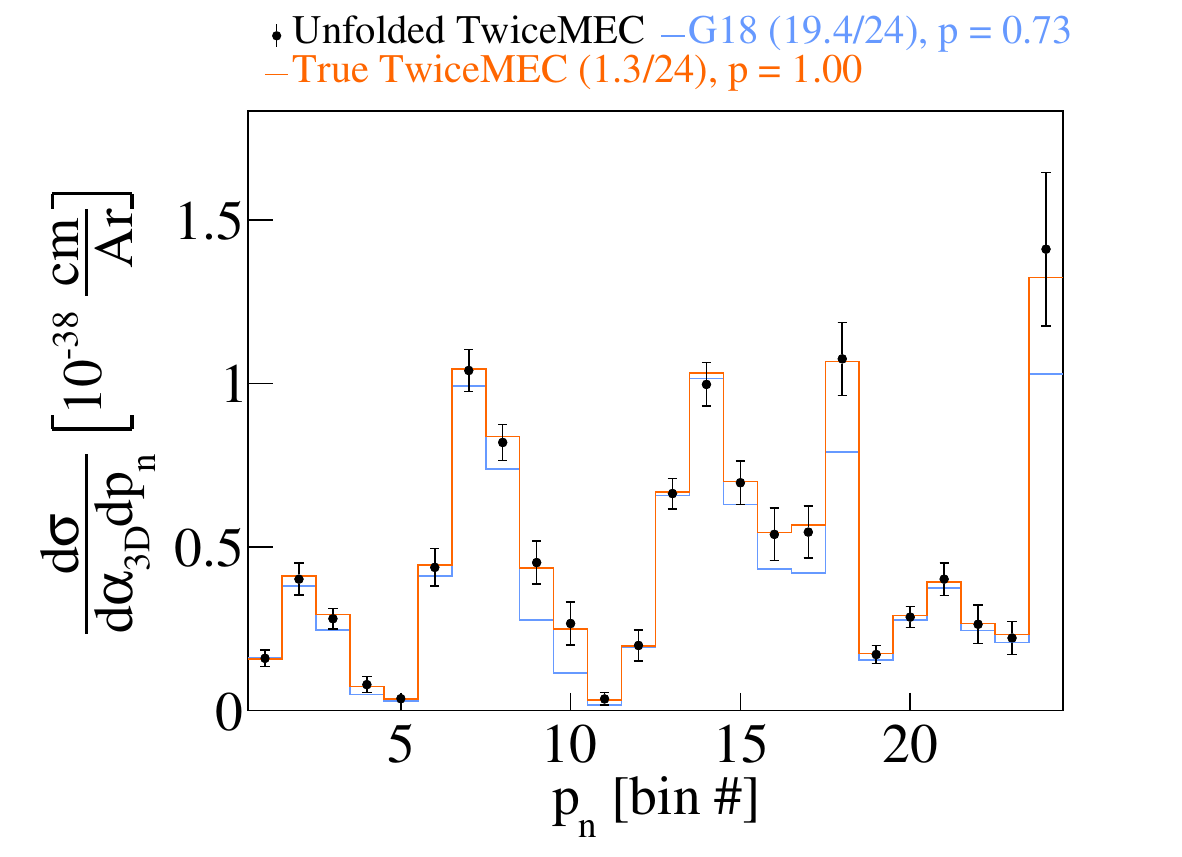}
\caption{
Fake data studies for $p_{n\parallel}$ in $\alpha_{3D}$ slices using (left) NuWro, (center) GENIE without the MicroBooNE tune (NoTune), and (right) twice the weights for MEC events (TwiceMEC) as fake data samples.
}
\label{DeltaPnDeltaAlpha3DFakeData}
\end{figure*}

\begin{figure*}[htb!]
\centering 
\includegraphics[width=0.32\linewidth]{\figures 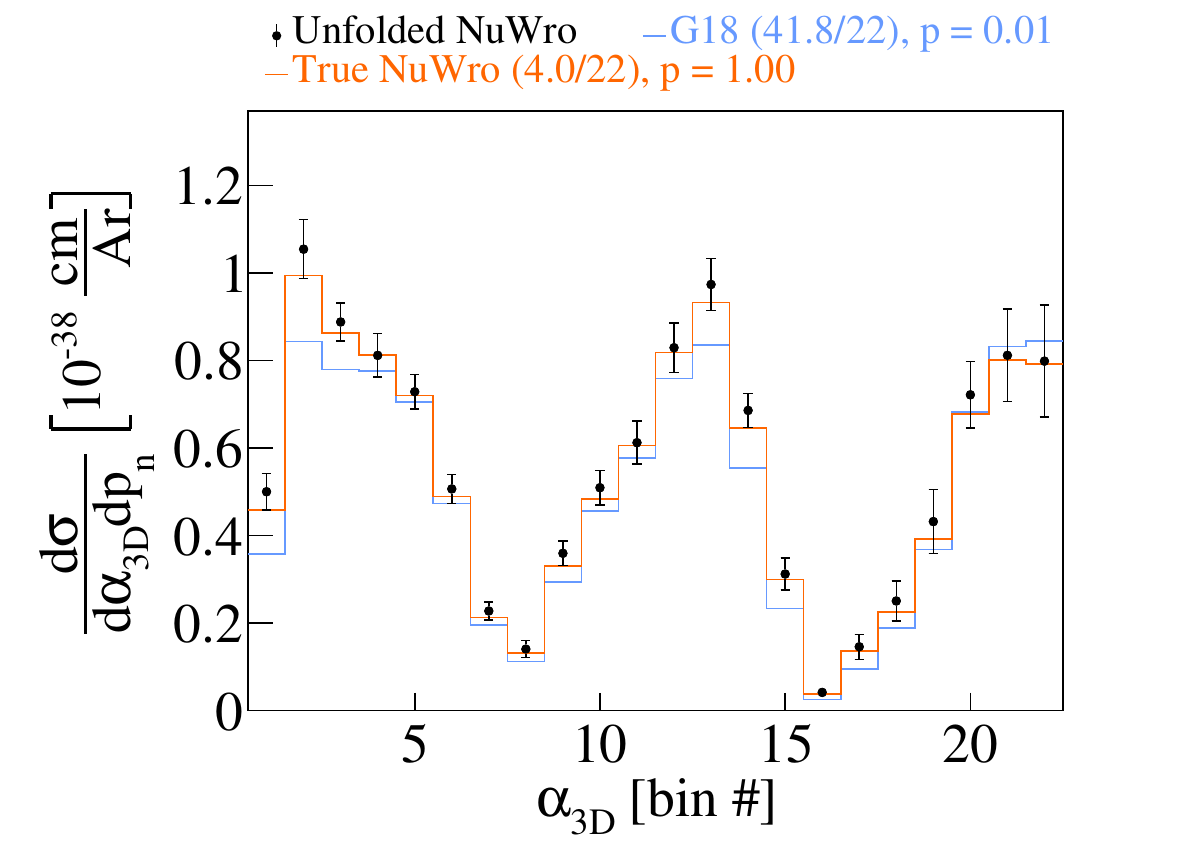}
\includegraphics[width=0.32\linewidth]{\figures 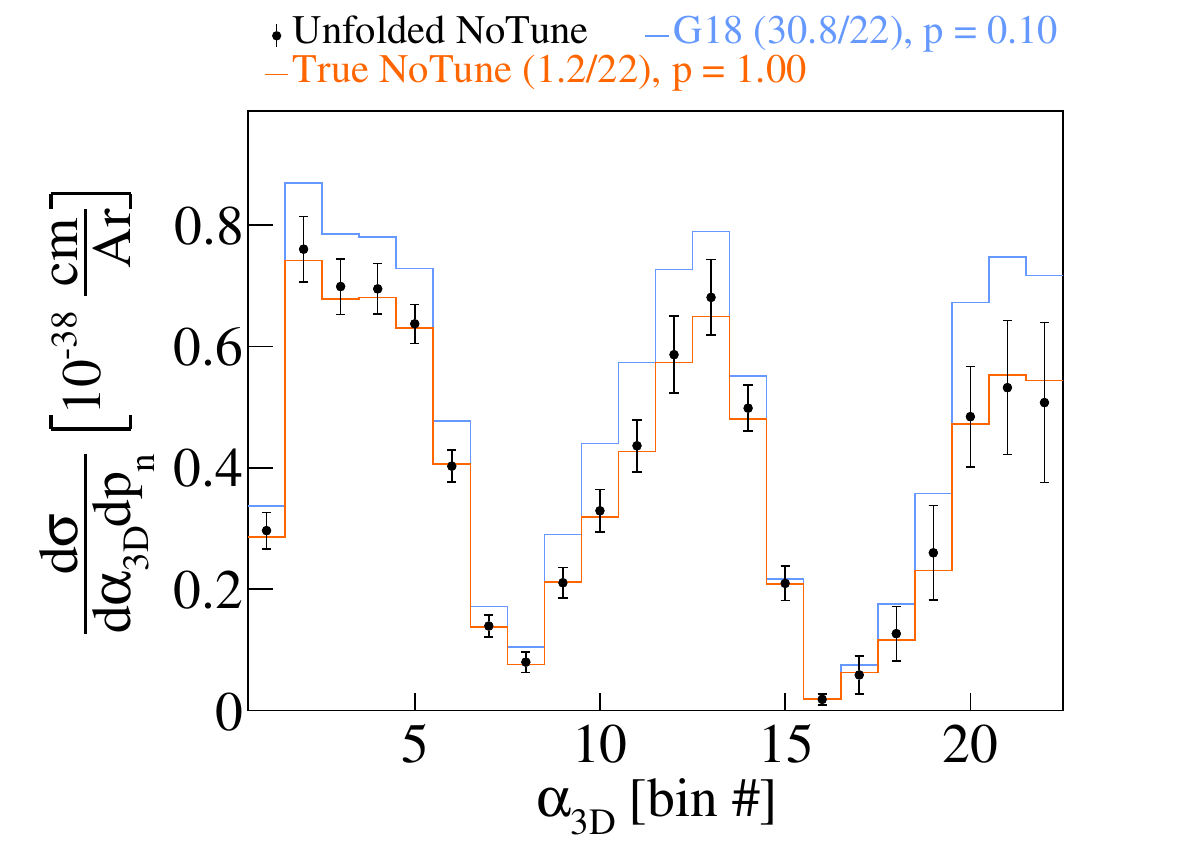}
\includegraphics[width=0.32\linewidth]{\figures 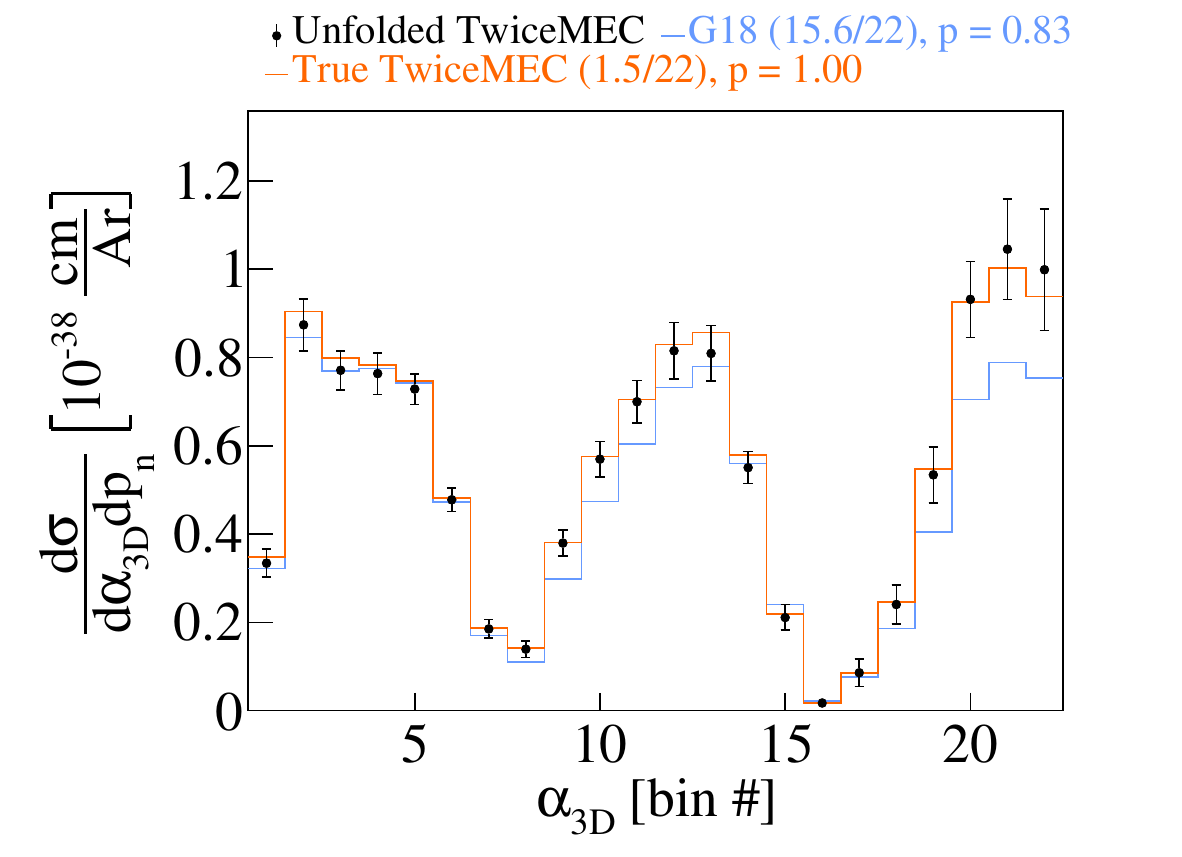}
\caption{
Fake data studies for $\alpha_{3D}$ in $p_{n\parallel}$ slices using (left) NuWro, (center) GENIE without the MicroBooNE tune (NoTune), and (right) twice the weights for MEC events (TwiceMEC) as fake data samples.
}
\label{DeltaAlpha3DDeltaPnFakeData}
\end{figure*}


\clearpage
\section{Muon \& Proton Efficiencies}\label{eff}

\begin{figure*}[htb!]
\centering 
\includegraphics[width=0.48\linewidth]{\figures 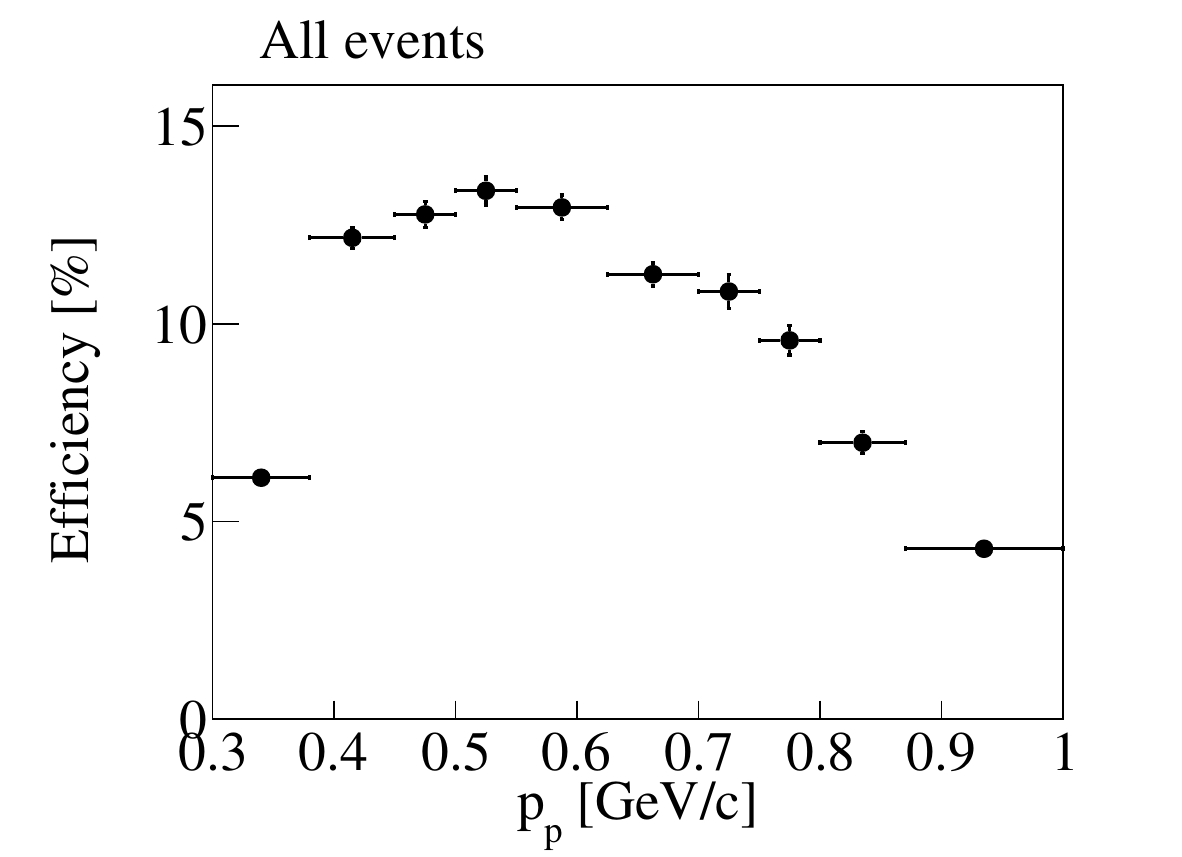}
\includegraphics[width=0.48\linewidth]{\figures 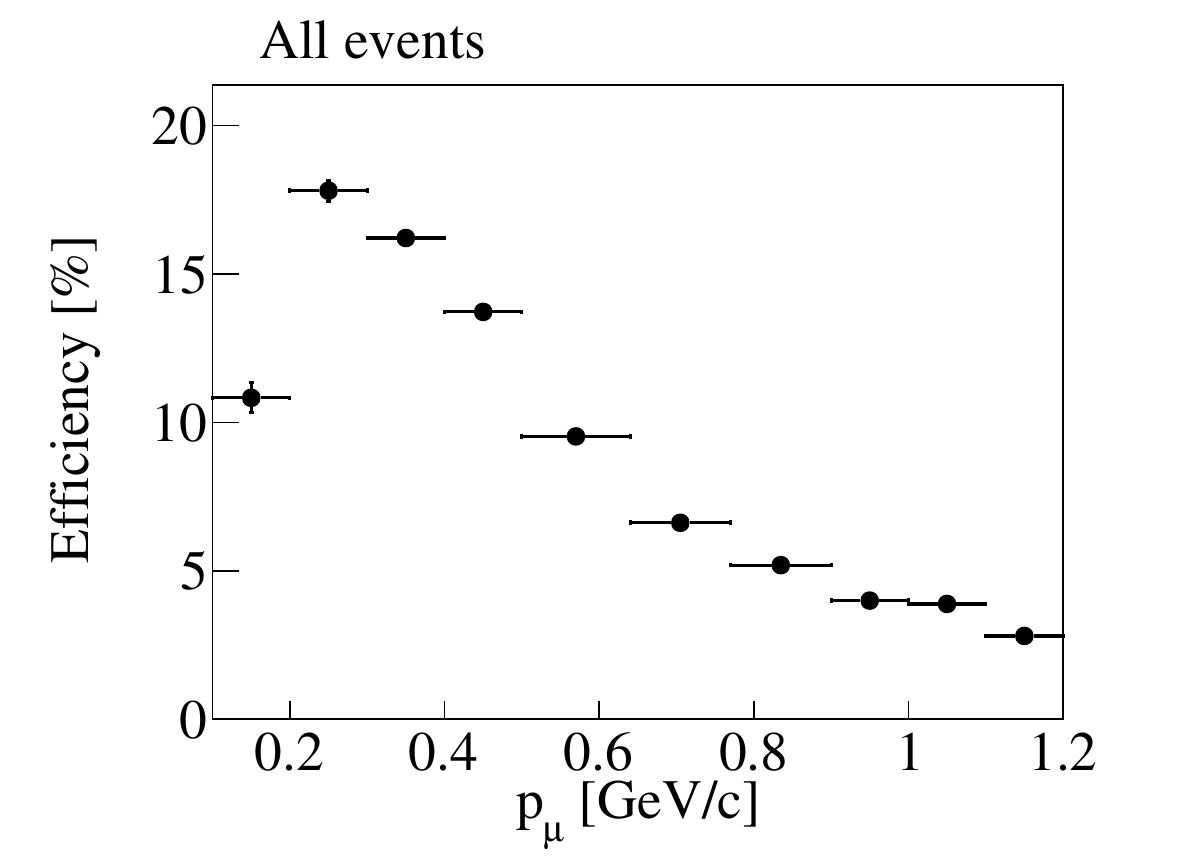}
\caption{
Efficiency as a function of the (left) proton and (right) muon momenta.
}
\label{EffMom}
\end{figure*}

\begin{figure*}[htb!]
\centering 
\includegraphics[width=0.48\linewidth]{\figures 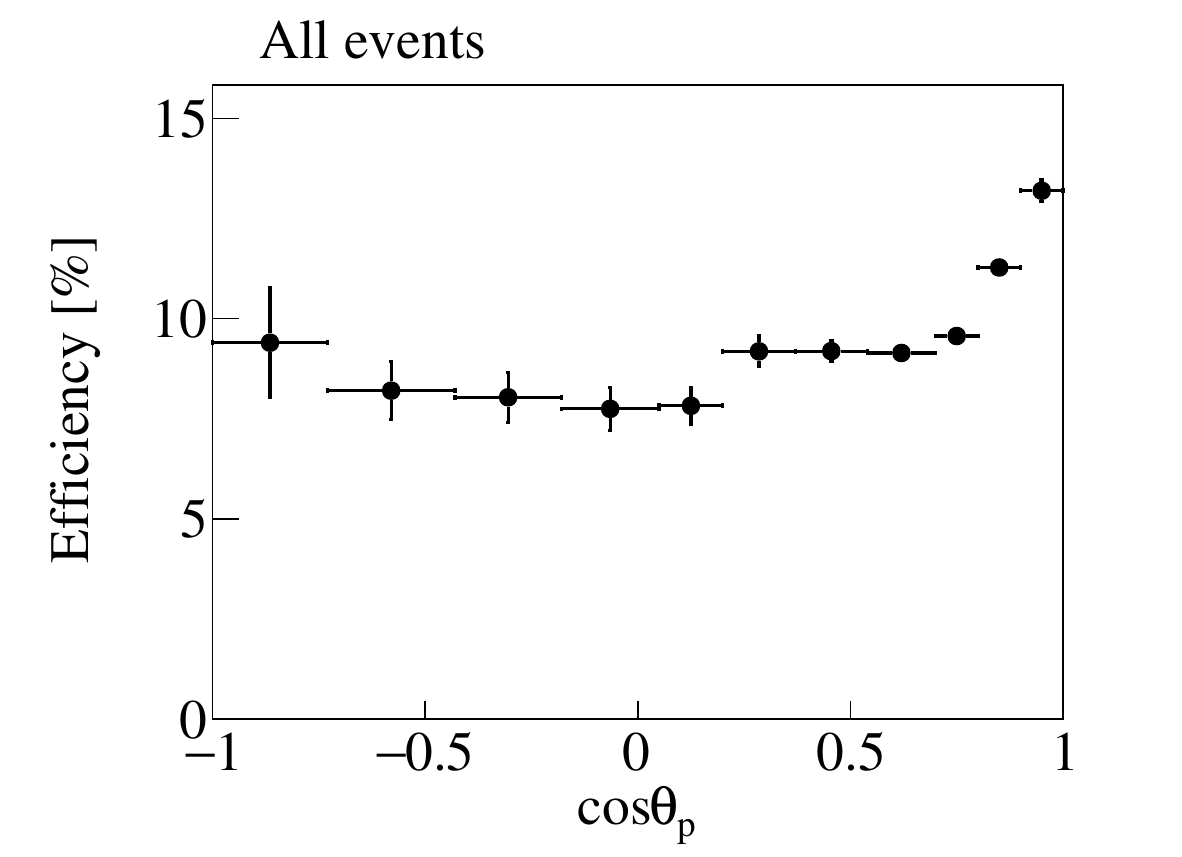}
\includegraphics[width=0.48\linewidth]{\figures 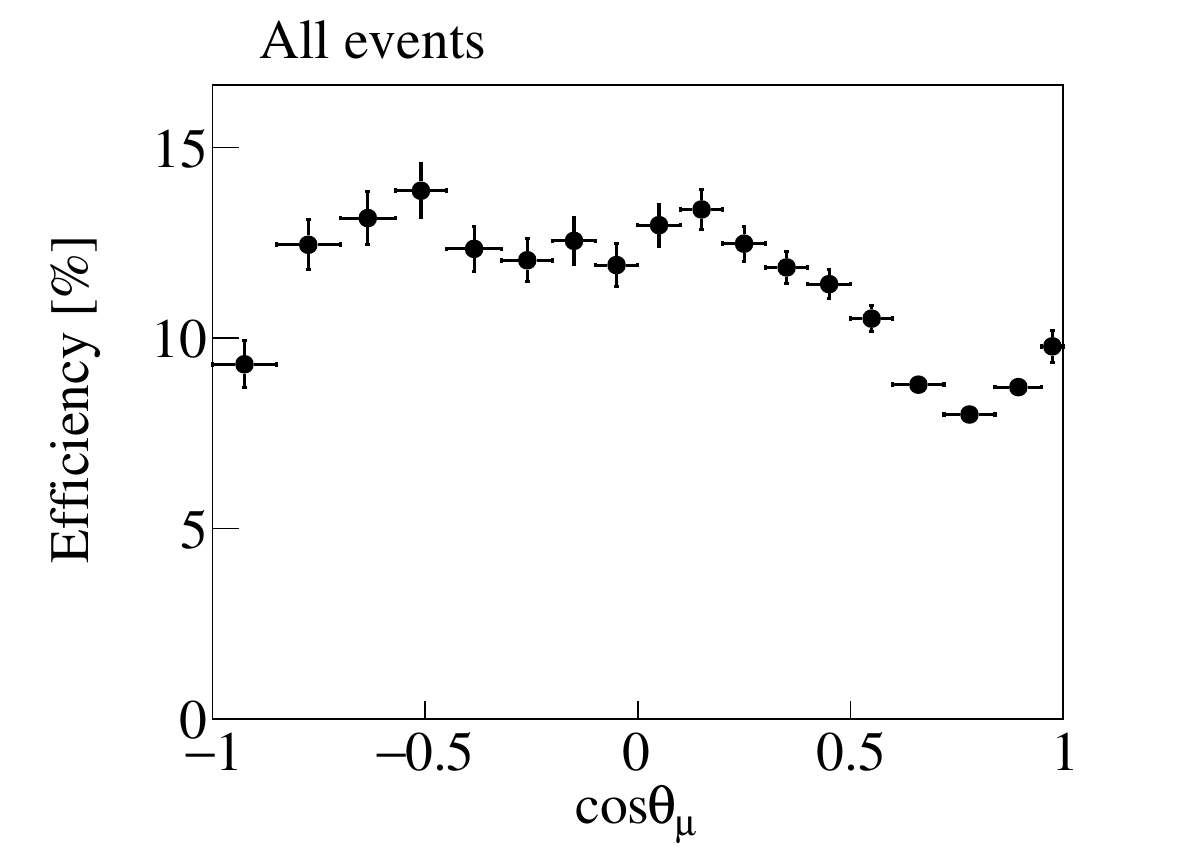}
\caption{
Efficiency as a function of the (left) proton and (right) muon cos$\theta$.
}
\label{EffAngle}
\end{figure*}

Figure~\ref{EffMom} shows the proton and muon efficiencies in the momenta range of interest given the CC1p0$\pi$ signal definition, namely 0.1 $< p_{\mu}<$ 1.2\,GeV/$c$ and 0.3 $< p_{p} <$ 1\,GeV/$c$.
Figure~\ref{EffAngle} shows the proton and muon efficiencies as a function of cos$\theta$, demonstrating the full angular coverage in the MicroBooNE LArTPC detector.

\clearpage
\section{Covariance Matrices}\label{cov}

The figures below present the covariance matrices for the cross section results presented in this work.
They are also included in the DataRelease.root file.
More details on how to manipulate the covariance matrices in order to calculate a $\chi^{2}$ GoF metric can be found in the README file.

\begin{figure*}[htb!]
\centering 
\includegraphics[width=0.48\linewidth]{\figures 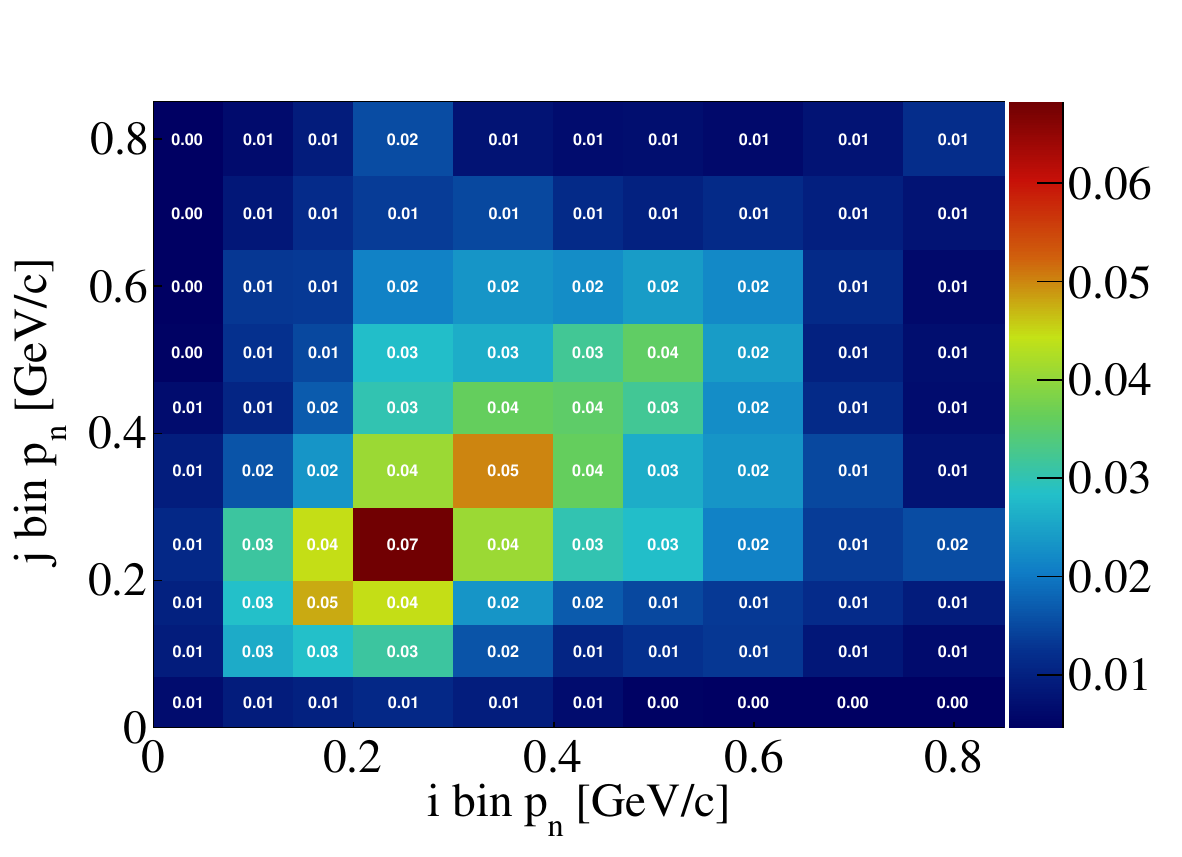}
\caption{
Covariance matrix for $p_{n}$.
}
\label{CovDeltaPn}
\end{figure*}

\begin{figure*}[htb!]
\centering 
\includegraphics[width=0.48\linewidth]{\figures 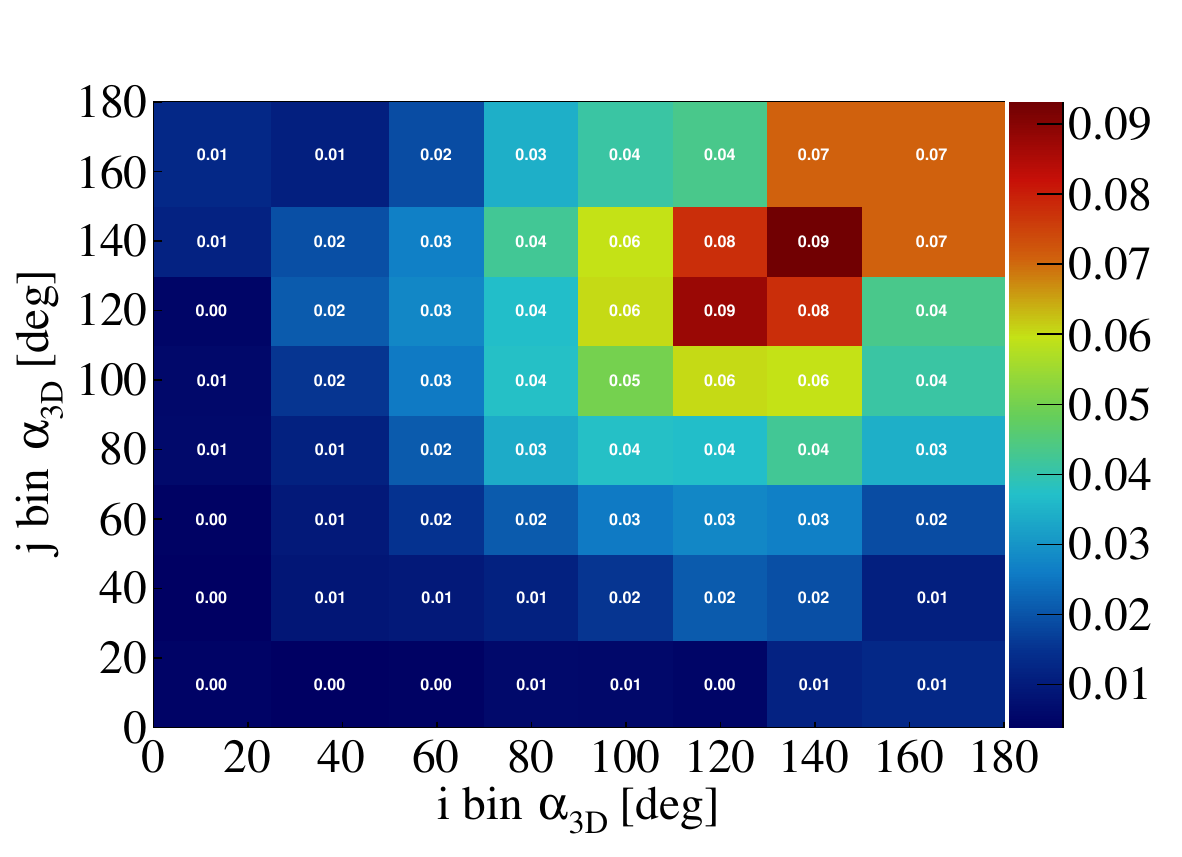}
\includegraphics[width=0.48\linewidth]{\figures 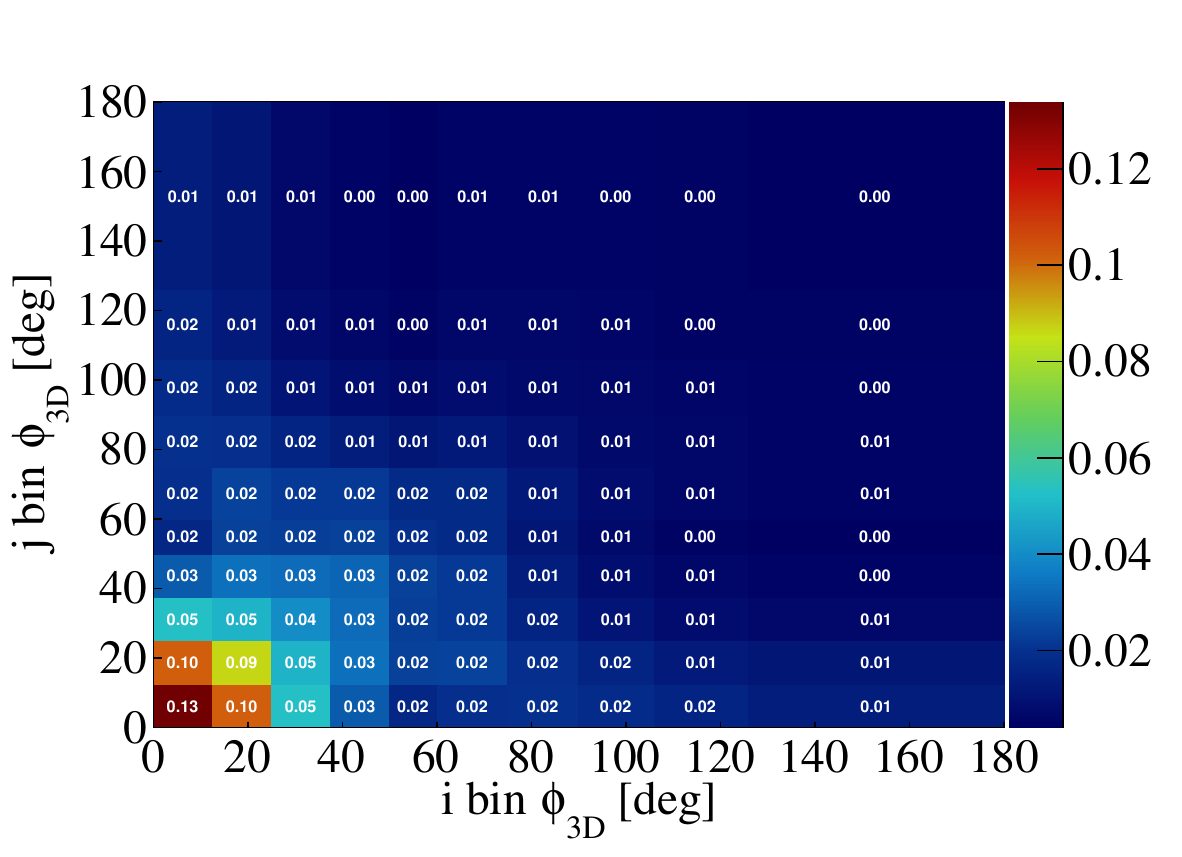}
\caption{
Covariance matrix for (left) $\alpha_{3D}$ and (right) $\phi_{3D}$.
}
\label{CovDeltaAlpha3DPhi3D}
\end{figure*}

\begin{figure*}[htb!]
\centering 
\includegraphics[width=0.48\linewidth]{\figures 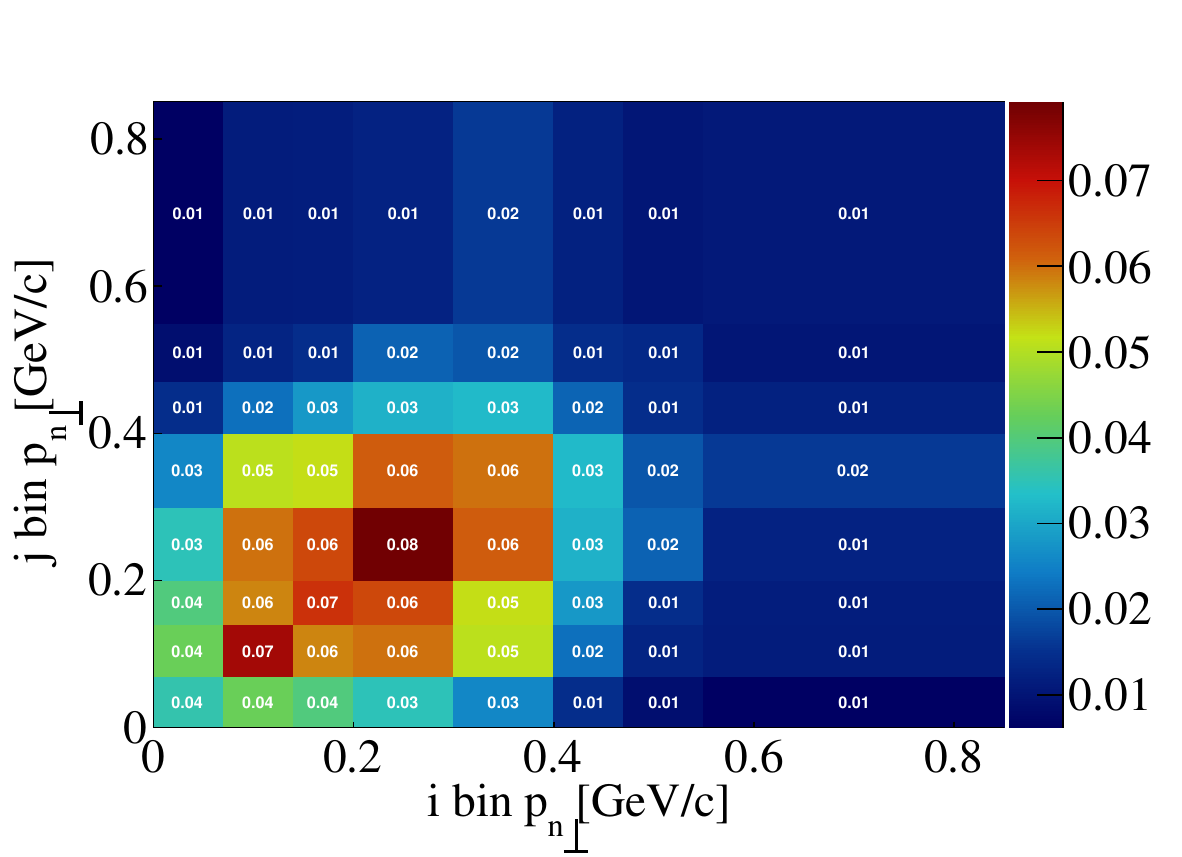}
\includegraphics[width=0.48\linewidth]{\figures 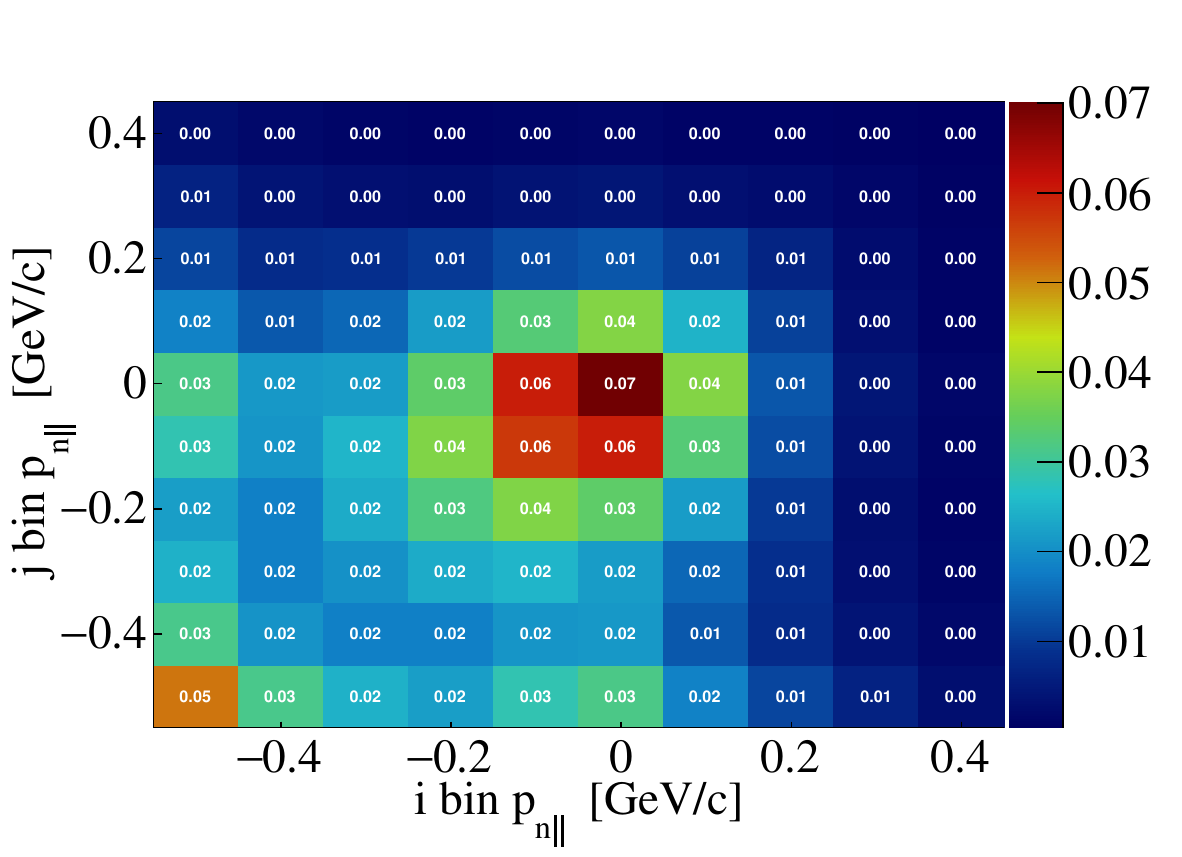}
\caption{
Covariance matrix for (left) $p_{n\perp}$ and (right) $p_{n\parallel}$.
}
\label{CovDeltaPnPerpPar}
\end{figure*}

\begin{figure*}[htb!]
\centering 
\includegraphics[width=0.48\linewidth]{\figures 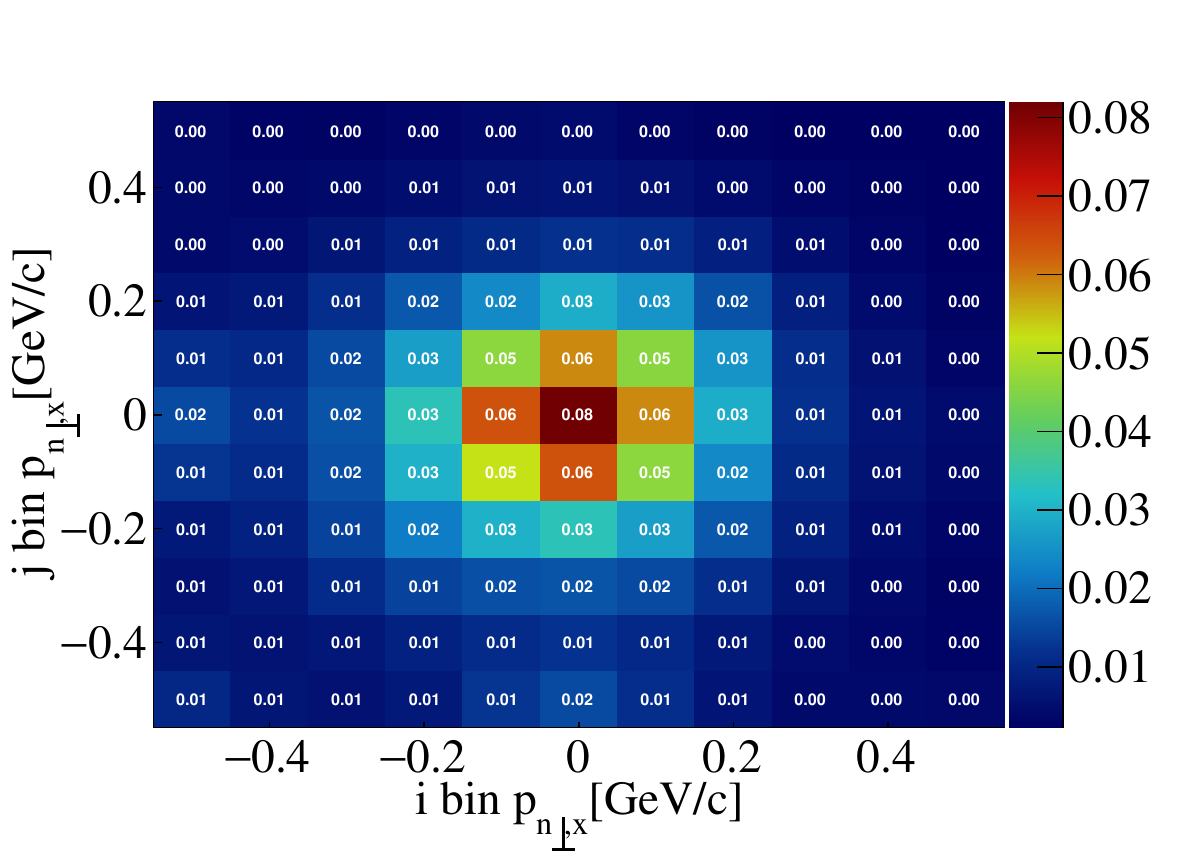}
\includegraphics[width=0.48\linewidth]{\figures 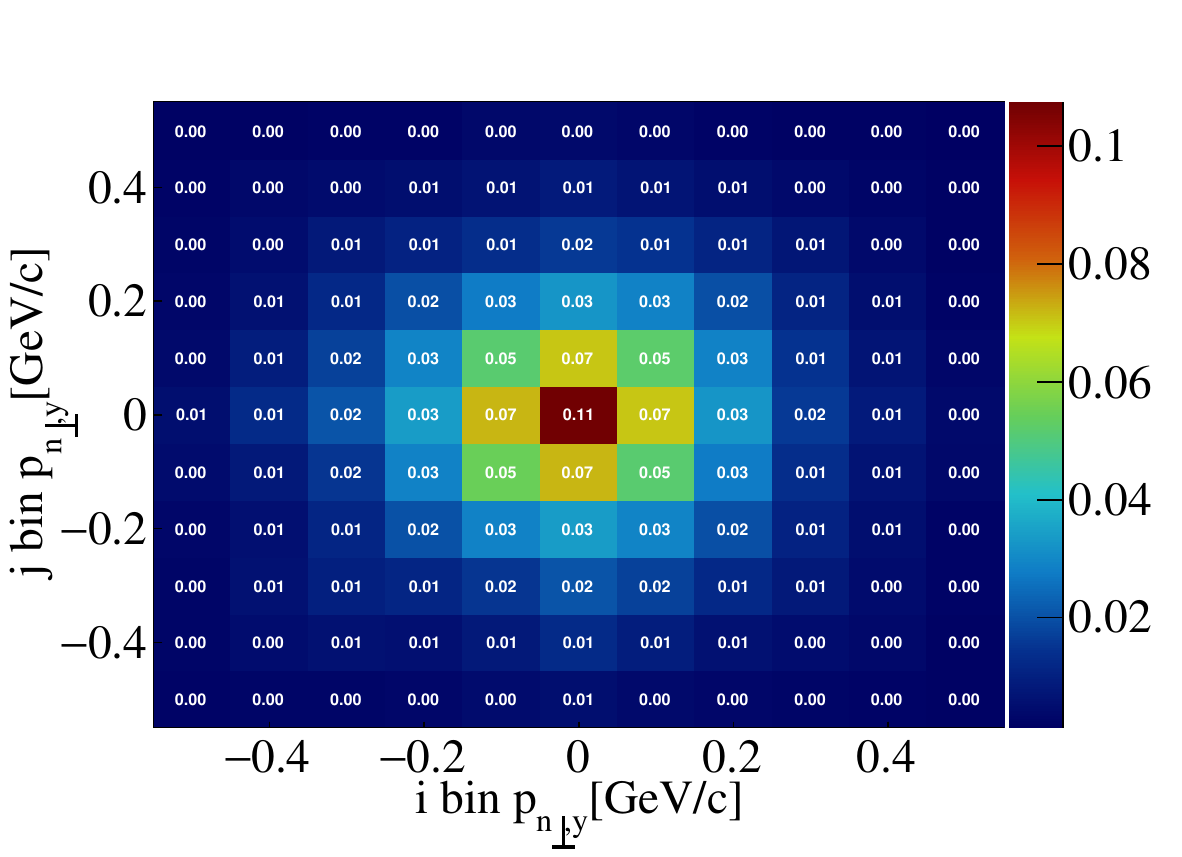}
\caption{
Covariance matrix for (left) $p_{n\perp,x}$ and (right) $p_{n\perp,y}$.
}
\label{CovDeltaPnPerpxy}
\end{figure*}

\begin{figure*}[htb!]
\centering 
\includegraphics[width=0.48\linewidth]{\figures 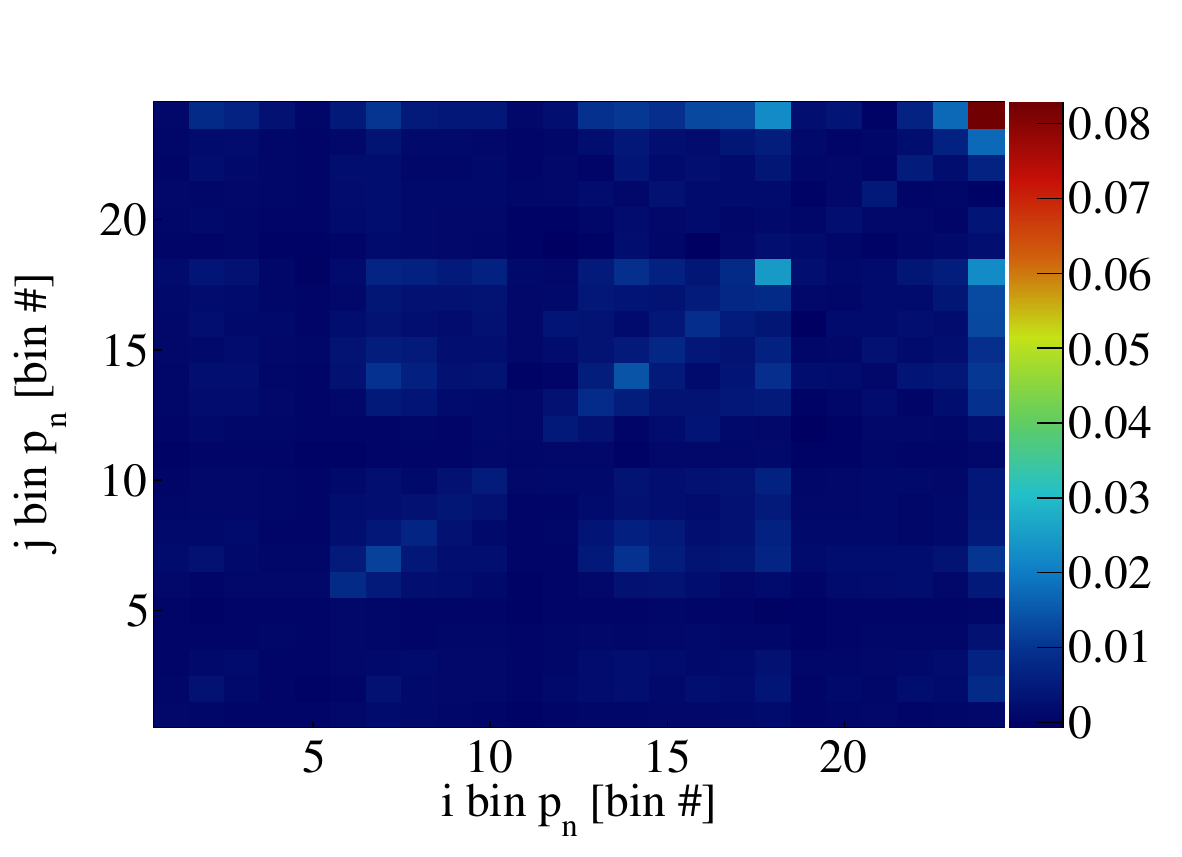}
\includegraphics[width=0.48\linewidth]{\figures 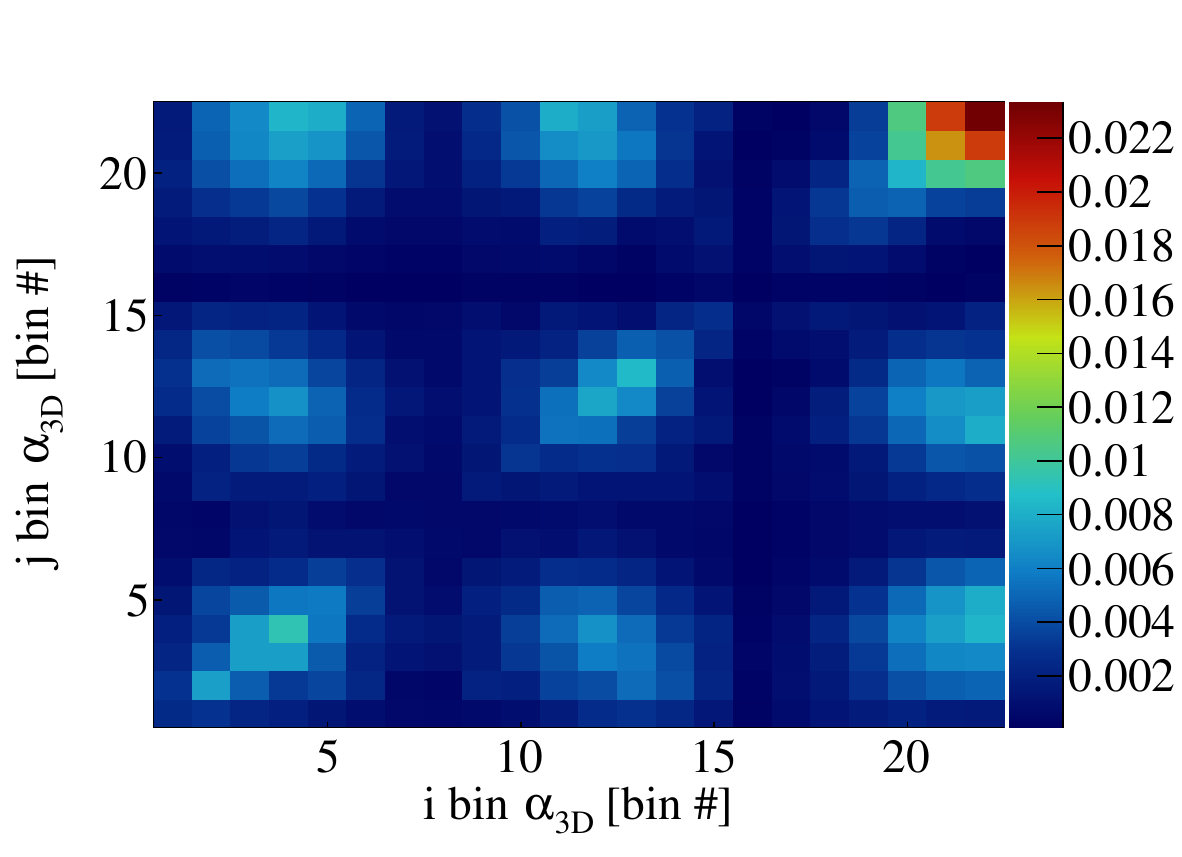}
\caption{
Covariance matrix for (left) $p_{n}$ in $\alpha_{3D}$ bins and (right) $\alpha_{3D}$ in $p_{n}$ bins.
}
\label{Cov2D}
\end{figure*}


\clearpage
\section{Additional Smearing Matrices}\label{smear}

The figures below present the additional smearing matrices for the cross section results presented in this work.
They are also included in the DataRelease.root file.

\begin{figure*}[htb!]
\centering 
\includegraphics[width=0.48\linewidth]{\figures 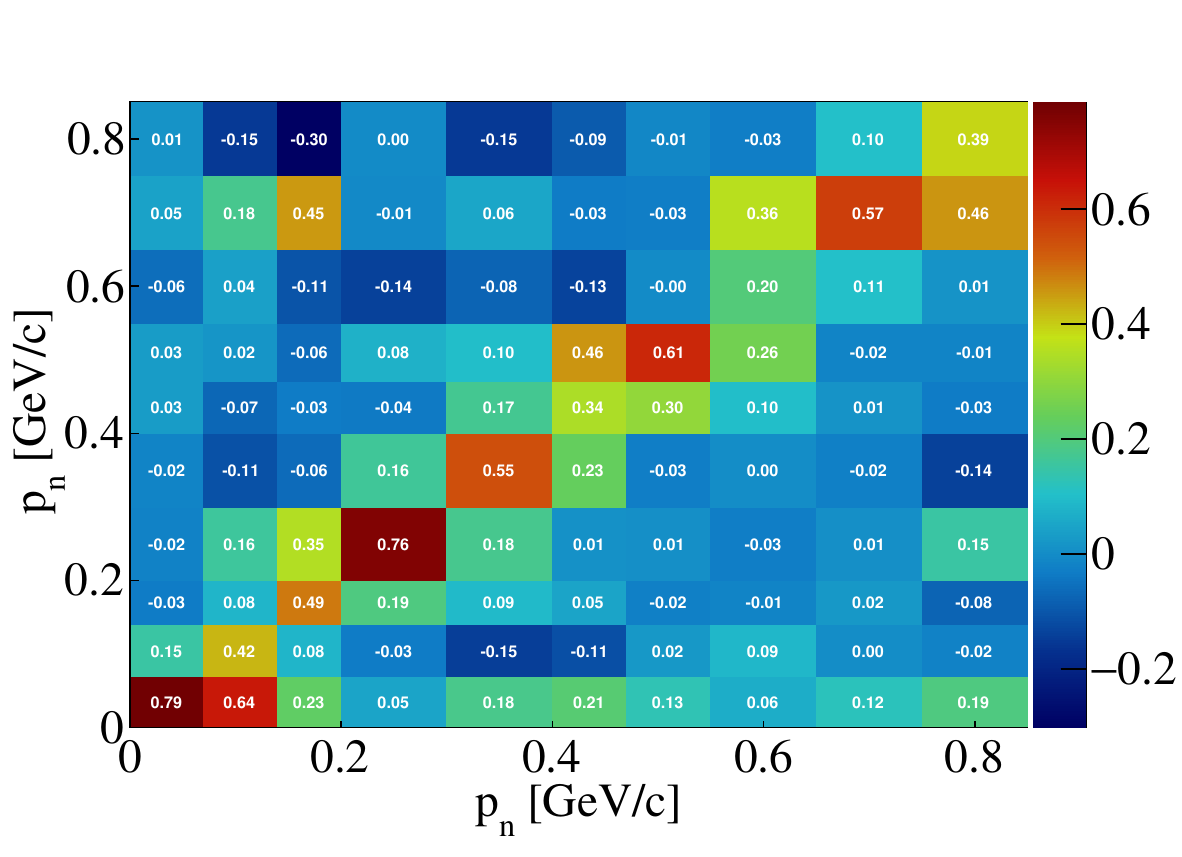}
\caption{
Additional smearing matrix for $p_{n}$.
}
\label{AcDeltaPn}
\end{figure*}

\begin{figure*}[htb!]
\centering 
\includegraphics[width=0.48\linewidth]{\figures 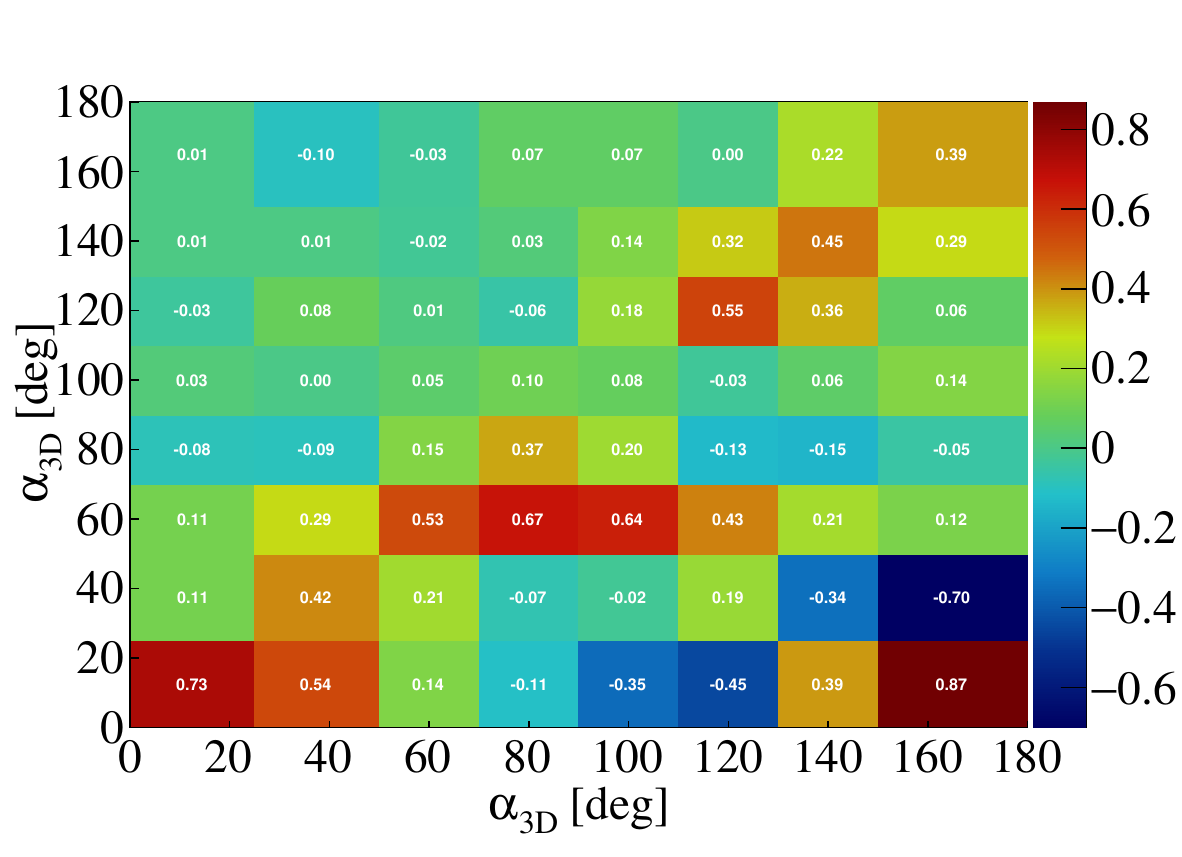}
\includegraphics[width=0.48\linewidth]{\figures 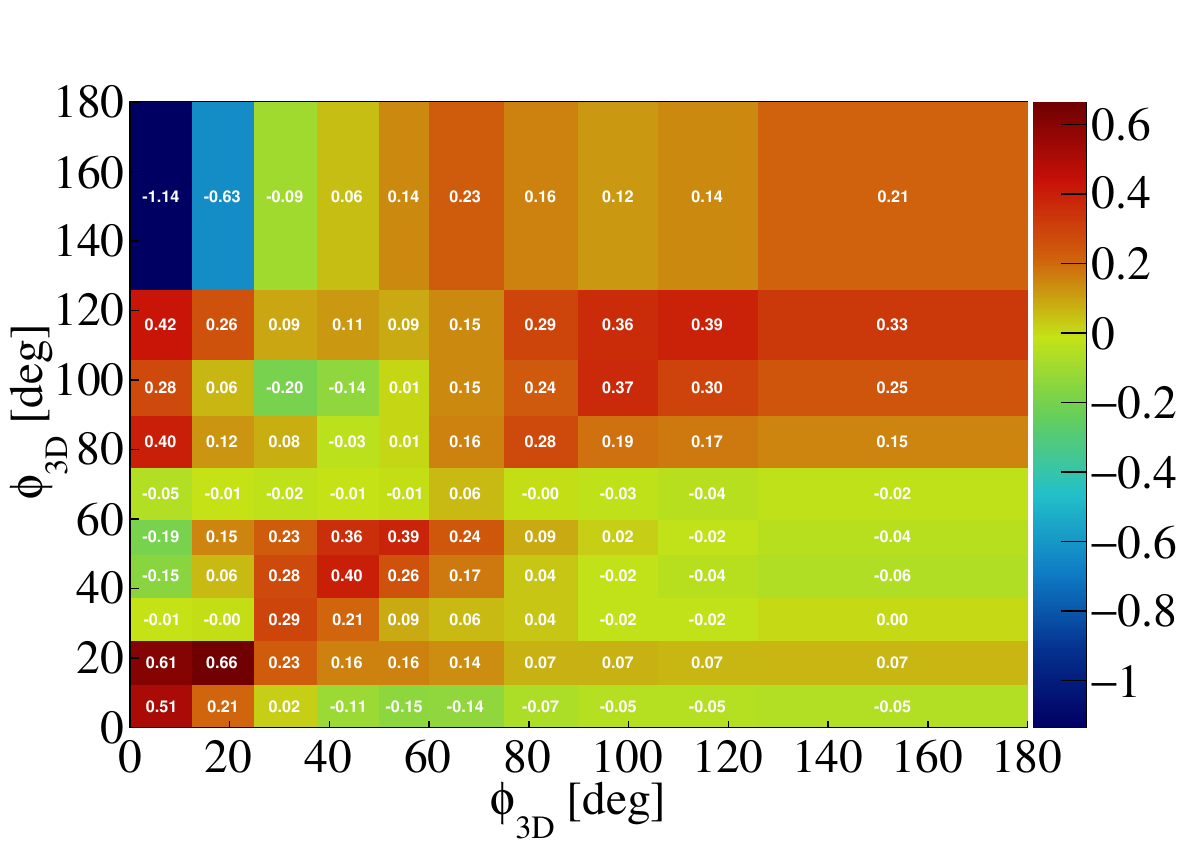}
\caption{
Additional smearing matrix for (left) $\alpha_{3D}$ and (right) $\phi_{3D}$.
}
\label{AcDeltaAlpha3DPhi3D}
\end{figure*}

\begin{figure*}[htb!]
\centering 
\includegraphics[width=0.48\linewidth]{\figures 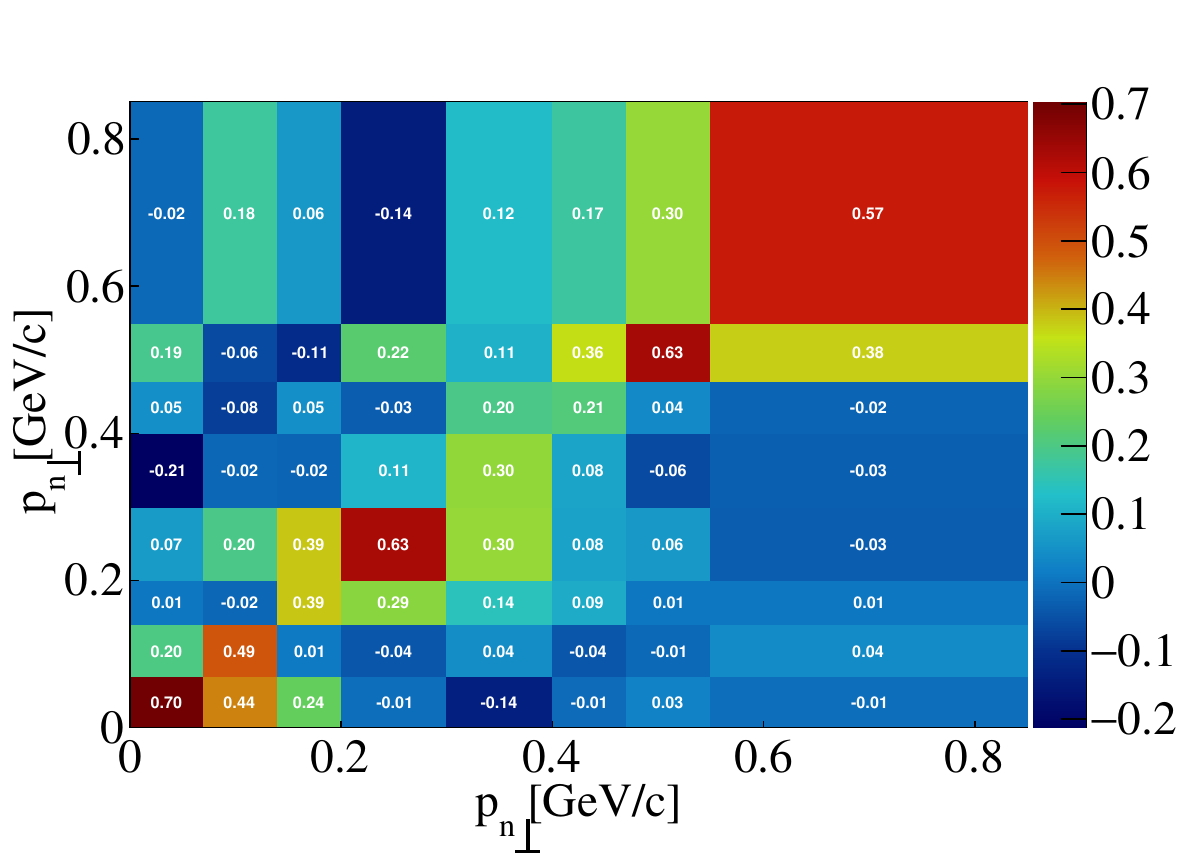}
\includegraphics[width=0.48\linewidth]{\figures 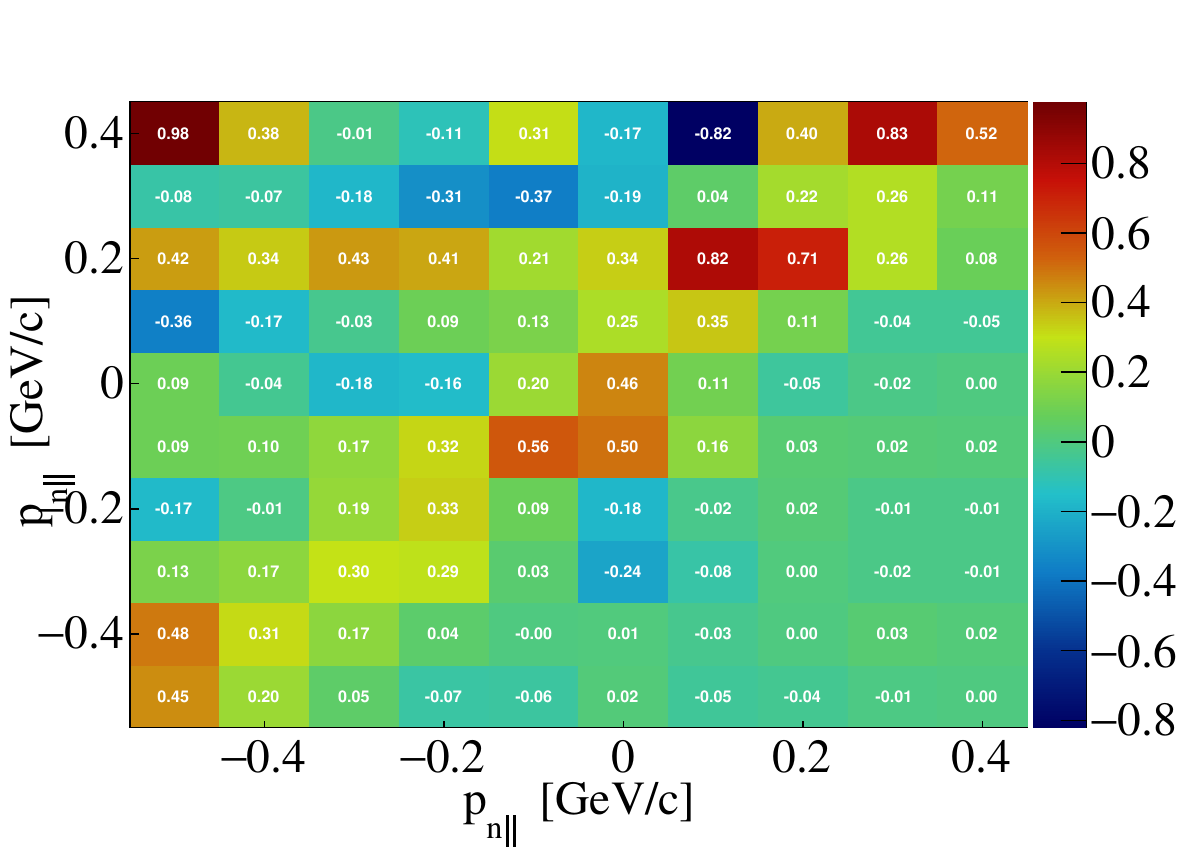}
\caption{
Additional smearing matrix for (left) $p_{n\perp}$ and (right) $p_{n\parallel}$.
}
\label{AcDeltaPnPerpPar}
\end{figure*}

\begin{figure*}[htb!]
\centering 
\includegraphics[width=0.48\linewidth]{\figures 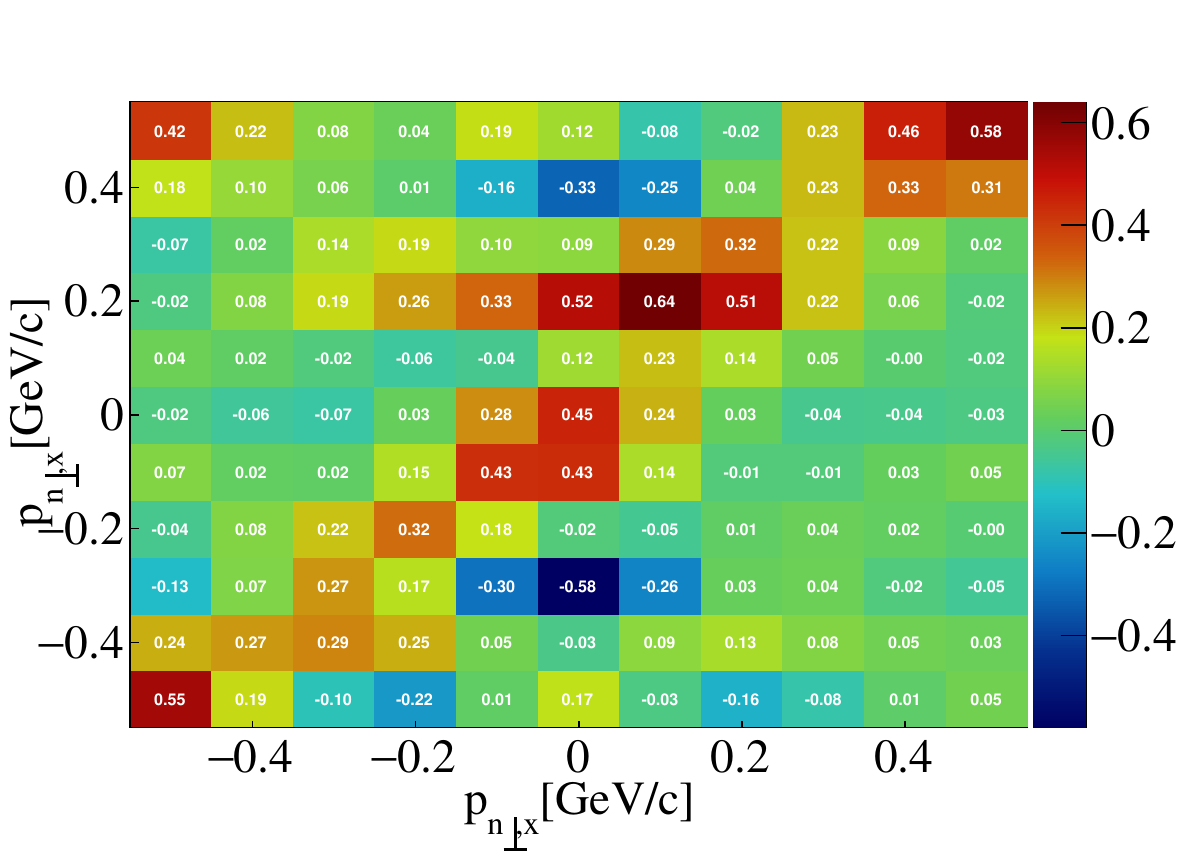}
\includegraphics[width=0.48\linewidth]{\figures 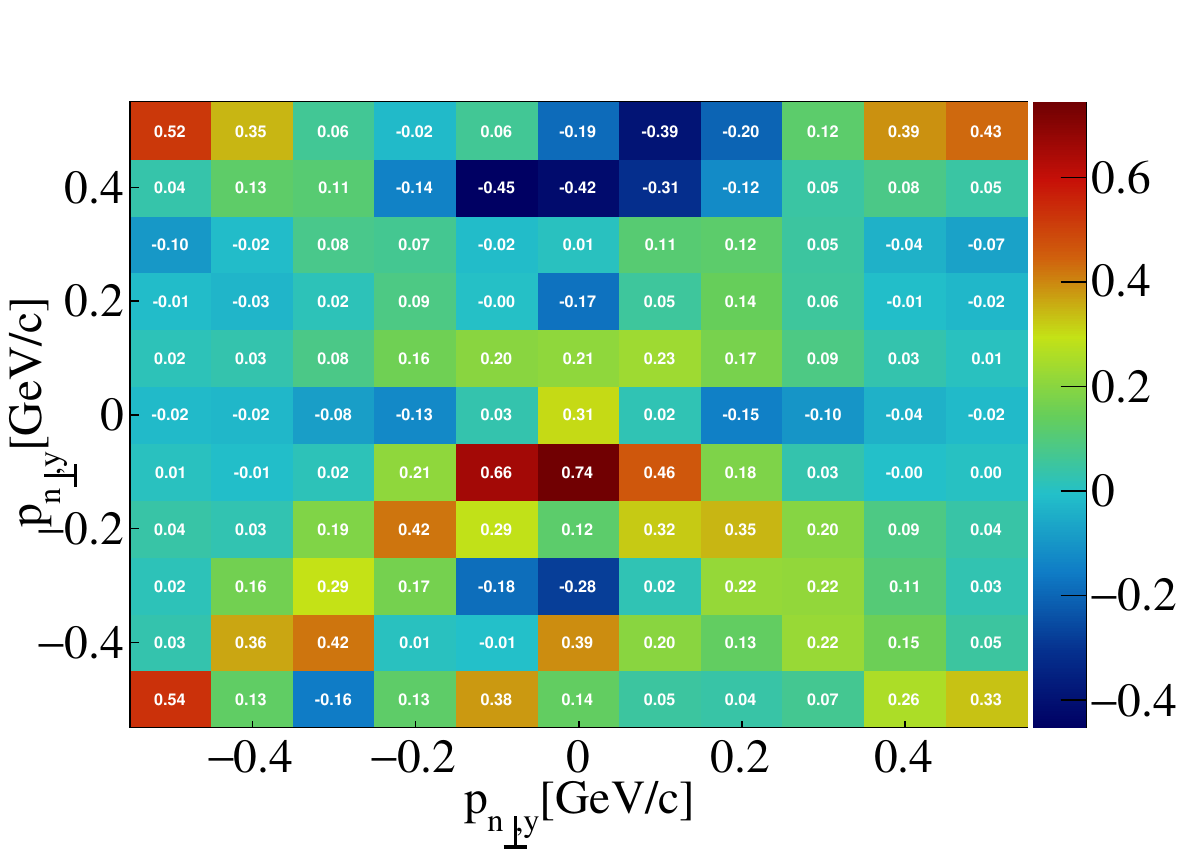}
\caption{
Additional smearing matrix for (left) $p_{n\perp,x}$ and (right) $p_{n\perp,y}$.
}
\label{AcDeltaPnPerpxy}
\end{figure*}

\begin{figure*}[htb!]
\centering 
\includegraphics[width=0.48\linewidth]{\figures 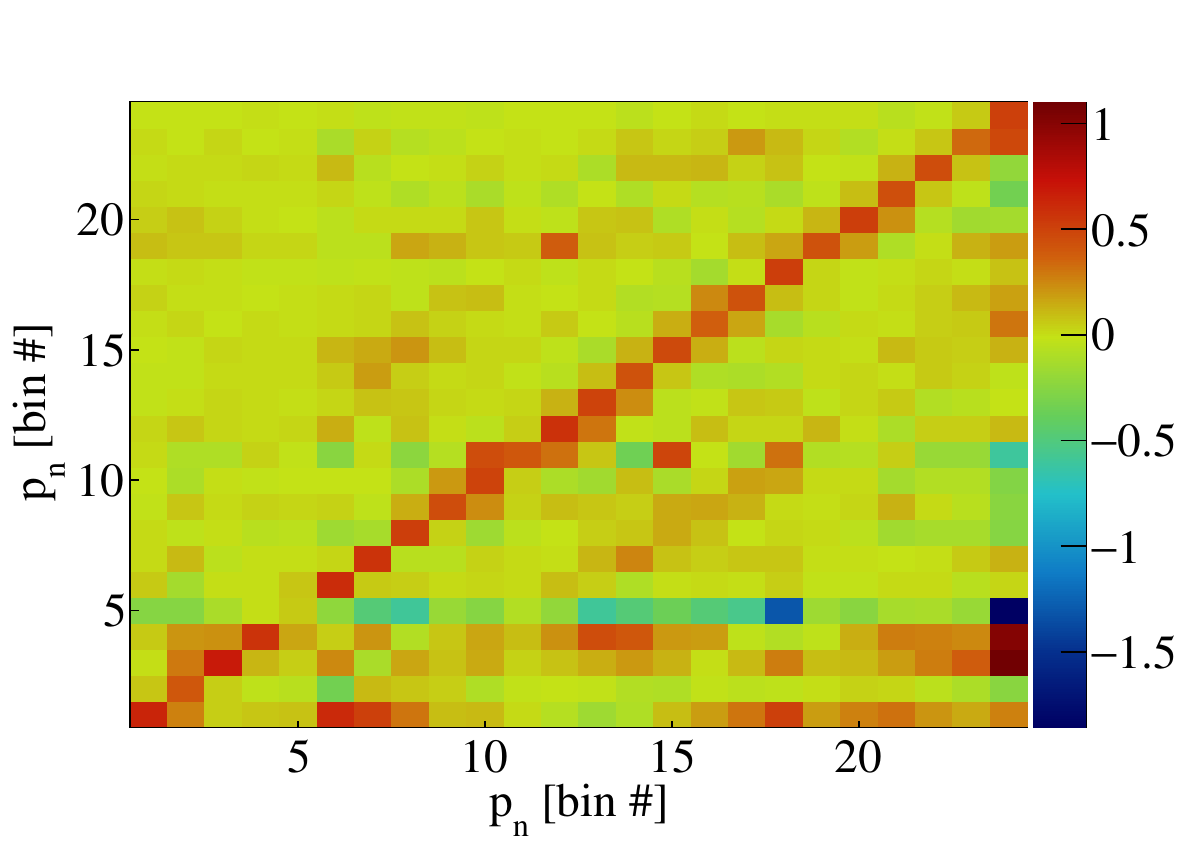}
\includegraphics[width=0.48\linewidth]{\figures 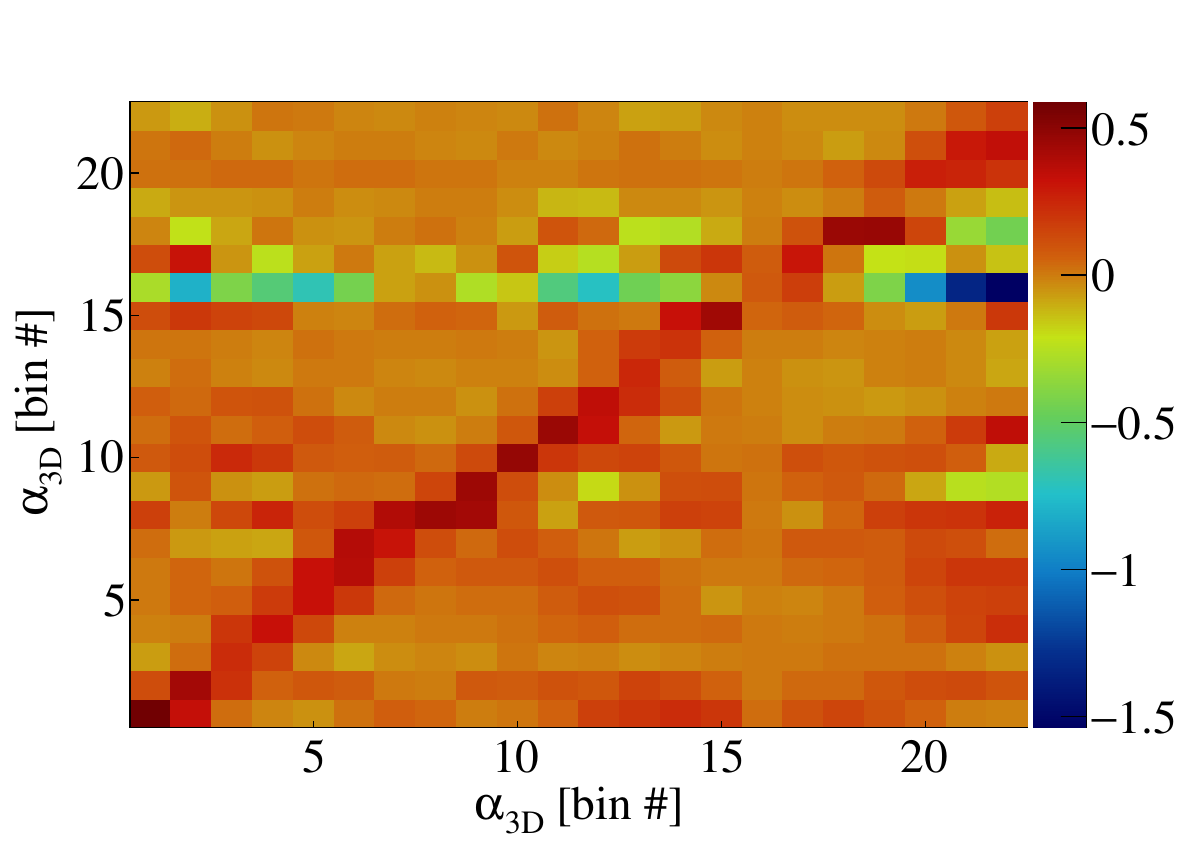}
\caption{
Additional smearing matrix for (left) $p_{n}$ in $\alpha_{3D}$ bins and (right) $\alpha_{3D}$ in $p_{n}$ bins.
}
\label{Ac2D}
\end{figure*}


